\documentclass[12pt]{report}
\usepackage[english]{babel}
\usepackage[cp1250]{inputenc}
\usepackage{amsmath}
\usepackage[toc,page]{appendix}
\usepackage{bbm}
\usepackage{dsfont}
\usepackage{nameref}
\usepackage[OT4]{fontenc}
\usepackage{enumerate}
\usepackage[a4paper,left=2.0cm,right=2.0cm,top=2.0cm,bottom=2.0cm,includefoot=false,includehead=false]{geometry}
\usepackage{graphicx}
\usepackage{wrapfig}
\usepackage{amsfonts}
\usepackage{mathrsfs}
\usepackage[bottom]{footmisc}
\usepackage{color}
\usepackage{multicol}
\usepackage{appendix}
\usepackage{footnote}
\usepackage{hyphenat}
\usepackage{braket}
\usepackage{indentfirst}
\usepackage[colorlinks=true,linkcolor=blue,urlcolor=red,citecolor=red]{hyperref} 
\makeatletter
\newcommand{\vast}{\bBigg@{3}}
\newcommand{\Vast}{\bBigg@{4}}
\makeatother

\begin{document}

\pagestyle{empty}

\vspace{0.2cm}

\begin{center}
\begin{Huge}Jagiellonian University in Krakow
\end{Huge}
\end{center}
\vspace{0.015cm}

\begin{center}
\begin{Large}Faculty of Physics, Astronomy and Applied Computer Science
\end{Large}
\end{center}
\begin{center}\includegraphics[width=6cm,angle=0]{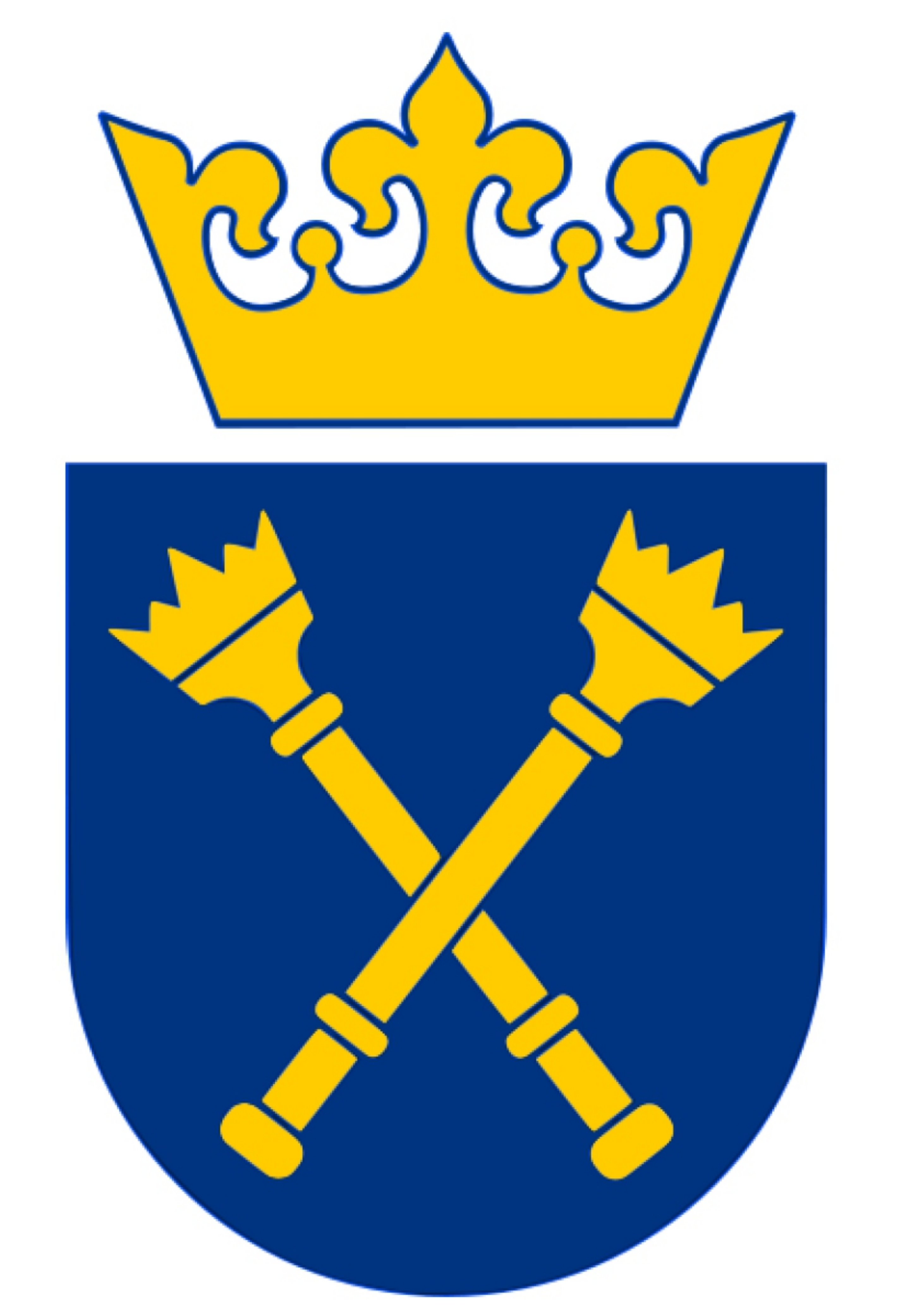}\end{center}

\vspace{0.1cm}

\begin{center}
\begin{LARGE}\textbf{Andrzej Syrwid}\end{LARGE}
\end{center}

\vspace{0.1cm}

\begin{center}
\begin{huge}\textbf{Emergence of dark solitons in the course of measurements of particle}
\vspace{0.1cm}

\textbf{positions in the Lieb-Liniger model: detailed analysis}

\end{huge}
\end{center}

\vspace{0.5cm}

\begin{center}
\begin{Large}Master Thesis\end{Large}

\vspace{0.05cm}

\begin{Large}Field of Study: Theoretical Physics\end{Large}
\end{center}

\vspace{0.2cm}

\begin{center}
\begin{Large}Thesis written under supervision of prof. dr hab. Krzysztof Sacha

\vspace{0.8cm}

Atomic Optics Department

\vspace{0.1cm}

M. Smoluchowski Institute of Physics
\end{Large}

\end{center}

\vspace{1.2cm}

\begin{center}
\begin{Large}Krakow 2015\end{Large}
\end{center}

\pagebreak

\newpage\thispagestyle{empty}
\mbox{}
\newpage

\pagebreak
\pagestyle{plain}

\tableofcontents

\chapter*{Preface}
\addcontentsline{toc}{chapter}{Preface}

The Lieb-Liniger model describes the one-dimensional non-relativistic ultracold Bose gas. Because of very low temperature of the system we can replace a real (often very elaborated) potential by a simple contact (point-like) potential given by Dirac $\delta$ function. The model was originally formulated as a problem of the one-dimensional quantum field theory (one space dimension plus one time dimension). It turns out that the complete description of the model is given by the so-called Bethe ansatz discovered by H. Bethe in 1931 during the studies of the Heisenberg antiferromagnet. Moreover, by the coordinate Bethe ansatz, one reduces the problem of the quantum field theory to the many-body quantum mechanical issue. 

The analytical forms of the eigenvectors and eigenvalues contain the collection of $\mathcal{N}$ parameters called quasi-momenta which are strongly connected with the momenta of bosons. One needs to remember that the quasi-momenta are not the same as  momenta of particles. The additional parameters are determined by the Bethe equations which appear when we impose the periodic boundary conditions. The solutions (quasi-momenta) may be uniquely parametrized by the collections of integer (half-integer) numbers in the case of odd (even) number of particles $\mathcal{N}$. It turns out that the ground and excited states have very simple graphical representation.

Elliott H. Lieb introduced the simple classification of the elementary excitations in the Lieb-Liniger system. The elementary excitation spectrum was divided into a two types: I and II. The Lieb's type I excitations are well known and reproduce the Bogoliubov spectrum which means that they correspond to the sound waves in the system. The problem appears when we analyze the type II excitations. In the case of weak coupling limit there are various evidences that the type II spectrum has the solitonic nature. The evidences are based on the comparison between the spectrum of the type II excitation and the semi-classical soliton solutions. As for now, no one checked if the dark soliton/density notch (repulsive case) reveals in probability density profiles during the measurements of particles positions.

The first two chapters of the thesis present aforementioned knowledge in details. The main purpose of the work is to show the emergence of the density notches in the course of measurement of particle positions if the system is prepared in type II eigenstate. Using the analytical results of norms and form factors in the Lieb-Liniger system, we prepare an iterative procedure which is perfect to numerical applications (chapter \ref{chap:Theideaofnumericalcalculations}). The numerical simulations are performed for the 8-particle system in the periodic box of the length $\mathrm{L}=1$ in weak ($c=0.08\Longrightarrow \gamma=\frac{c \mathrm{L}}{\mathcal{N}}=0.01$) and strong ($c=8\Longrightarrow \gamma=\frac{c \mathrm{L}}{\mathcal{N}}=1$) interaction regimes. Our results confirm the mean-field predictions (see the sections \ref{sec:Solitonsinperiodicbox} and \ref{sec:weak}) for the case of weak coupling. Moreover, we observe the density notch structure of probability density also in the case of strong interaction regime (section \ref{sec:strong}). The considered states correspond to one and two Lieb's type II elementary excitations. It turns out that the number of the type II elementary excitations is equal to the number of density notches. All the results are presented in the chapter \ref{numericalresults}.

At the end we present a comprehensive appendices which may be useful for the reader to understand all the issues appearing in the thesis.

\chapter{The one-dimensional Bose gas with contact interaction}
\label{chap:TheOne-dimensionalBoseGas}

\section{Lieb-Liniger model in the second-quantized form}
\label{sec:Lieb-Liniger}

One-dimensional non-relativistic system of bosons may be described by the canonical quantum Bose fields $\hat{\Psi}_{H}(z,t)$\footnote{Bose field $\hat{\Psi}_{H}(z,t)$ is a field in the Heisenberg picture. By definition: $\hat{\Psi}_{H}(z,t)=\mathrm{e}^{i\hat{\mathrm{H}}t}\hat{\Psi}(z) \mathrm{e}^{-i\hat{\mathrm{H}}t}$, where $\hat{\Psi}(z)$ and $\hat{\mathrm{H}}$ are Bose field in the Schr\"{o}dinger picture and the Hamiltonian of the system respectively.}. These fields should satisfy following canonical equal-time commutation relations \cite{bogkorep}
\begin{equation}
\displaystyle{\left [\hat{\Psi}_{H}(z,t),\hat{\Psi}_{H}^{\dagger}(x,t) \right]=\mathrm{\delta}(z-x),}
		\label{qmdescription1}
		\end{equation}
		\begin{equation}
\displaystyle{\left [\hat{\Psi}_{H}(z,t),\hat{\Psi}_{H}(x,t) \right]=\left [\hat{\Psi}_{H}^{\dagger}(z,t),\hat{\Psi}_{H}^{\dagger}(x,t) \right]=0}.
		\label{qmdescription2}
		\end{equation}

If we consider interparticle interaction as a contact interaction\footnote{The discussion about the contact potential introduction is presented in Appendix \ref{sec:dodA}.} with coupling constant $c$ (we consider only repulsive interactions which means $c>0$) then the Hamiltonian has the following form\footnote{It is known that in the second-quantized form the Hamiltonian is given by $\displaystyle{\hat{\mathrm{H}}=\int\mathrm{d}^{3} r\hat{\Psi}^{\dagger}(\vec{r})\frac{-\hbar^{2}}{2m}\nabla^{2}\hat{\Psi}(\vec{r})}$ $\displaystyle{+\frac{1}{2}\int\mathrm{d}^{3} r\mathrm{d}^{3} r \, '\hat{\Psi}^{\dagger}(\vec{r})\hat{\Psi}^{\dagger}(\vec{r} \, ') V(\vec{r}-\vec{r} \, ')\hat{\Psi}(\vec{r} \, ')\hat{\Psi}(\vec{r})} $. } \cite{bogkorep}
	\begin{equation}
\displaystyle{\hat{\mathrm{H}}=\int \mathrm{d}z\left[\partial_{z}\hat{\Psi}^{\dagger}(z) \partial_{z}\hat{\Psi}(z)  +c \hat{\Psi}^{\dagger}(z)\hat{\Psi}^{\dagger}(z)\hat{\Psi}(z)\hat{\Psi}(z)\right]},
		\label{qmdescription3}
		\end{equation}
where the units have been chosen such that $\hbar=2m=1$. Hamiltonian (\ref{qmdescription3}) is the energy operator of the Lieb-Liniger model. Using the evolution equation for operators in the Heisenberg picture $i\frac{\mathrm{d}}{\mathrm{d}t}\hat{A}(t)=\left[\hat{A}(t),\hat{\mathrm{H}}  \right]$ one can obtain the equation of motion
			\begin{equation}
\displaystyle{i\partial_{t}\hat{\Psi}_{H}(z,t)=-\partial_{z}^{2}\hat{\Psi}_{H}(z,t)+2c\hat{\Psi}_{H}^{\dagger}(z,t)\hat{\Psi}_{H}(z,t)\hat{\Psi}_{H}(z,t)},
		\label{qmdescription4}
		\end{equation}
which is the so-called Non-linear Schr\"{o}dinger (NLS) equation. Additionally, we define the Fock vacuum $\left|0 \right>$ by the following relations
\begin{equation}
\displaystyle{\begin{array}{lll} \hat{\Psi}(z)\left|0 \right>=0, & \left< 0 \right|\hat{\Psi}^{\dagger}(z)=0,  &  z\in \mathbb{R},\, \,\\  \left|0 \right>^{\dagger}=\left< 0 \right|, & \left< 0 |0 \right>=1. & \end{array}}
		\label{qmdescription5}
		\end{equation}

It is possible to define two additional hermitian operators $\hat{\mathrm{N}}$ (number of particles operator) and $\hat{\mathrm{P}}$ (total momentum operator)\footnote{Using the definition $\displaystyle{\hat{\Psi}(z)=\sum_{k}\frac{1}{\sqrt{V}}\mathrm{e}^{i k z}\hat{a}_{k}}$ we obtain the following results: \\
$\displaystyle{\hat{\mathrm{N}}=\frac{1}{V}\sum_{k,q}\int \mathrm{d}z\mathrm{e}^{i (k-q)z}\hat{a}^{\dagger}_{q}\hat{a}_{k}=
\sum_{k}\hat{a}^{\dagger}_{k}\hat{a}_{k}}, $ \\
$\displaystyle{\hat{\mathrm{P}}=-\frac{i}{2}\frac{1}{V}\sum_{k,q}\int \mathrm{d}z\left[ i k\mathrm{e}^{i (k-q)z}\hat{a}^{\dagger}_{q}\hat{a}_{k}+i q \mathrm{e}^{i (k-q)z}\hat{a}^{\dagger}_{q}\hat{a}_{k}\right]= \sum_{k} k \hat{a}^{\dagger}_{k}\hat{a}_{k} } $. }
\begin{equation}
\displaystyle{\hat{\mathrm{N}}=\int \mathrm{d}z \hat{\Psi}^{\dagger}(z)\hat{\Psi}(z)},
		\label{qmdescription6}
		\end{equation}
\begin{equation}
\displaystyle{\hat{\mathrm{P}}=-\frac{i}{2}\int \mathrm{d}z \left[\hat{\Psi}^{\dagger}(z)\partial_{z}\hat{\Psi}(z)-  \left(\partial_{z}\hat{\Psi}^{\dagger}(z)\right) \hat{\Psi}(z)  \right]=i\int \mathrm{d}z\left(\partial_{z}\hat{\Psi}^{\dagger}(z)\right) \hat{\Psi}(z) },
		\label{qmdescription7}
		\end{equation}
which are integrals of motion
\begin{equation}
\displaystyle{\left[\hat{\mathrm{H}},\hat{\mathrm{N}}   \right] = \left[\hat{\mathrm{H}}, \hat{\mathrm{P}}   \right]=0}.
		\label{qmdescription8}
		\end{equation}
From the above equation one can see that number of particles is conserved.

\section{The coordinate Bethe ansatz}
\label{sec:ThecoordinateBetheansatz}

In 1931, while studying the one-dimensional Heisenberg antiferromagnet, Hans Bethe discovered a method capable of delivering the complete description of the solvable models (exact eigenvalues and eigenvectors). It is the so-called \textbf{\emph{Bethe ansatz}}. We will apply the Bethe ansatz to solve Lieb-Liniger model in the first-quantized form\footnote{The model of one-dimensional gas with contact interaction (via the $\delta$-function potential) was solved by Lieb and Liniger (they used Bethe ansatz method). Hence, the model is often called the Lieb-Liniger model \cite{liebliniger}.}. 

For this purpose, let us consider the common eigenfunctions (in a sector with a fixed number of particles $\mathcal{N}$) $\left|\psi_{\mathcal{N}}(k_{1},\ldots,k_{\mathcal{N}})\right>$ of $\hat{\mathrm{H}}, \hat{\mathrm{P}}$ and $\hat{\mathrm{N}}$
\begin{equation}
\displaystyle{\left|\psi_{\mathcal{N}}(k_{1},\ldots,k_{\mathcal{N}})\right>=\frac{1}{\sqrt{\mathcal{N}!}} \int \mathrm{d}^{\mathcal{N}}z \Phi_{\mathcal{N}}(z_{1},\ldots,z_{\mathcal{N}},k_{1},\ldots,k_{\mathcal{N}}) \hat{\Psi}^{\dagger}(z_{1})\ldots \hat{\Psi}^{\dagger}(z_{\mathcal{N}}) \left|0 \right>},
		\label{qmdescription9}
		\end{equation}
		\begin{equation}
\displaystyle{\hat{\mathrm{H}}\left|\psi_{\mathcal{N}}\right>=E_{\mathcal{N}}\left|\psi_{\mathcal{N}}\right>, \, \, \, \, \hat{\mathrm{N}}\left|\psi_{\mathcal{N}}\right>=\mathcal{N}\left|\psi_{\mathcal{N}}\right>, \, \, \, \, \hat{\mathrm{P}}\left|\psi_{\mathcal{N}}\right>=P_{\mathcal{N}}\left|\psi_{\mathcal{N}}\right> },
		\label{qmdescription10}
		\end{equation}
where $\Phi_{\mathcal{N}}(\{z_{i}\},\{k_{j}\})$ is a symmetric function of all $z_{j}$. Parameters $k_{1},\ldots,k_{\mathcal{N}}$ are called \textbf{\emph{quasi-momenta}}\footnote{One uses the name \emph{quasi-momenta} because of  the strong connection of these parameters with the momenta of particles. Despite many similarities quasi-momenta are not the same as momenta of particles. We discuss this problem in the point \ref{sec:quasimomenta}.}. It turns out that $\Phi_{\mathcal{N}}(\{z_{i}\},\{k_{j}\})$ is an eigenfunction of both the quantum mechanical Hamiltonian $\hat{\mathcal{H}}_{\mathcal{N}}$ and the quantum mechanical momentum operator $\hat{\mathcal{P}}_{\mathcal{N}}$ 
\begin{equation}
\displaystyle{\hat{\mathcal{H}}_{\mathcal{N}}=\sum_{j=1}^{\mathcal{N}}\left(-\frac{\partial^{2}}{\partial z_{j}^{2}}\right) +2c\sum_{\mathcal{N}\geq j>k\geq 1} \delta (z_{j}-z_{k} ) },
		\label{qmdescription11}
		\end{equation}
		\begin{equation}
\displaystyle{\hat{\mathcal{P}}_{\mathcal{N}}=-i\sum_{j=1}^{\mathcal{N}}\frac{\partial}{\partial z_{j}} },
		\label{qmdescription12}
		\end{equation}
\begin{equation}
\displaystyle{\hat{\mathcal{H}}_{\mathcal{N}}\Phi_{\mathcal{N}}(\{z_{i}\},\{k_{j}\})=E_{\mathcal{N}}\Phi_{\mathcal{N}}(\{z_{i}\},\{k_{j}\}), \, \, \, \, \hat{\mathcal{P}}_{\mathcal{N}}\Phi_{\mathcal{N}}(\{z_{i}\},\{k_{j}\})=P_{\mathcal{N}}\Phi_{\mathcal{N}}(\{z_{i}\},\{k_{j}\}) }.
		\label{qmdescription13}
		\end{equation}
Formulas (\ref{qmdescription11}) and (\ref{qmdescription12}) are called the first-quantized Lieb-Liniger model. To prove transition from the second to first-quantized form one may consider the following expression 
\begin{center}
$\displaystyle{\hat{\mathrm{P}}\left|\psi_{\mathcal{N}}(k_{1},\ldots,k_{\mathcal{N}})\right> \stackrel{(\ref{qmdescription7}),(\ref{qmdescription9})}{=} \frac{i}{\sqrt{\mathcal{N}!}} \int \mathrm{d} x \mathrm{d}^{\mathcal{N}}z \Phi_{\mathcal{N}}(z_{1},\ldots,z_{\mathcal{N}},k_{1},\ldots,k_{\mathcal{N}}) \left[\partial_{x} \hat{\Psi}^{\dagger}(x)  \right] }$
		\end{center}
\begin{center}
$\displaystyle{\times  \sum_{j=1}^{\mathcal{N}}   \hat{\Psi}^{\dagger}(z_{1})\ldots \left[\hat{\Psi}(x) ,\hat{\Psi}^{\dagger}(z_{j})  \right] \ldots \hat{\Psi}^{\dagger}(z_{\mathcal{N}}) \left|0 \right>\stackrel{(\ref{qmdescription1})}{=}  }$
		\end{center}
		\begin{center}
$\displaystyle{\stackrel{(\ref{qmdescription1})}{=}   \frac{i}{\sqrt{\mathcal{N}!}} \int  \mathrm{d}^{\mathcal{N}}z \Phi_{\mathcal{N}}(z_{1},\ldots,z_{\mathcal{N}},k_{1},\ldots,k_{\mathcal{N}})  \sum_{j=1}^{\mathcal{N}}   \hat{\Psi}^{\dagger}(z_{1})\ldots \left[\partial_{z_{j}} \hat{\Psi}^{\dagger}(z_{j})  \right] \ldots \hat{\Psi}^{\dagger}(z_{\mathcal{N}}) \left|0 \right> }$.
		\end{center}
Integrating by parts with respect to $z_{j}$ we prove that the action of (\ref{qmdescription7}) on (\ref{qmdescription9}) is equivalent to the action of $\hat{\mathcal{P}}_{\mathcal{N}}$ on $\Phi_{\mathcal{N}}(z_{1},\ldots,z_{\mathcal{N}},k_{1},\ldots,k_{\mathcal{N}})$
\begin{equation}
\displaystyle{\hat{\mathrm{P}}\left|\psi_{\mathcal{N}}(k_{1},\ldots,k_{\mathcal{N}})\right> =\frac{1}{\sqrt{\mathcal{N}!}}\int \mathrm{d}^{\mathcal{N}}z \left( -i \sum_{j=1}^{\mathcal{N}}\frac{\partial}{\partial_{z_{j}}}\Phi_{\mathcal{N}}(z_{1},\ldots,z_{\mathcal{N}},k_{1},\ldots,k_{\mathcal{N}}) \right) \hat{\Psi}^{\dagger}(z_{1})\ldots \hat{\Psi}^{\dagger}(z_{\mathcal{N}}) \left|0 \right>}.
		\label{qmdescription14}
		\end{equation}
In a similar way one can construct the quantum mechanical Hamiltonian.

Interaction in the Hamiltonian $\hat{\mathcal{H}}_{\mathcal{N}}$ is repulsive if $c>0$. Because of the symmetry of $\Phi_{\mathcal{N}}(z_{1},\ldots,z_{\mathcal{N}},k_{1},\ldots,k_{\mathcal{N}})$ in all $z_{j}$ one can choose the following domain $\mathcal{T}$ in the coordinate space
\begin{equation}
\displaystyle{\mathcal{T}: z_{1}<z_{2}< \ldots <z_{\mathcal{N}-1}<z_{\mathcal{N}}}.
		\label{qmdescription15}
		\end{equation}
In this case\footnote{The potential is equal to 0 in domain $\mathcal{T}$.}
\begin{equation}
\displaystyle{\hat{\mathcal{H}}_{\mathcal{N}}\stackrel{\mathcal{T}}{\longrightarrow}\hat{\mathcal{H}}_{\mathcal{N}}^{0}=\sum_{j=1}^{\mathcal{N}}\left(-\frac{\partial^{2}}{\partial z_{j}^{2}}\right)  },
		\label{qmdescription16}
		\end{equation}
		\begin{equation}
\displaystyle{\hat{\mathcal{P}}_{\mathcal{N}}\stackrel{\mathcal{T}}{\longrightarrow}\hat{\mathcal{P}}_{\mathcal{N}}^{0}=-i\sum_{j=1}^{\mathcal{N}}\frac{\partial}{\partial z_{j}} },
		\label{qmdescription17}
		\end{equation}
and $\Phi_{\mathcal{N}}(z_{1},\ldots,z_{\mathcal{N}},k_{1},\ldots,k_{\mathcal{N}})$ is a common eigenfunction of both	operators $\hat{\mathcal{H}}^{0}_{\mathcal{N}}$ (free Hamiltonian) and $\hat{\mathcal{P}}^{0}_{\mathcal{N}}$	
\begin{equation}
\displaystyle{\hat{\mathcal{H}}^{0}_{\mathcal{N}}\Phi_{\mathcal{N}}(\{z_{i}\},\{k_{j}\})=E_{\mathcal{N}}\Phi_{\mathcal{N}}(\{z_{i}\},\{k_{j}\}), \, \, \, \, \hat{\mathcal{P}}^{0}_{\mathcal{N}}\Phi_{\mathcal{N}}(\{z_{i}\},\{k_{j}\})=P_{\mathcal{N}}\Phi_{\mathcal{N}}(\{z_{i}\},\{k_{j}\}) }.
		\label{qmdescription18}
		\end{equation}

Let us consider the problem of boundary conditions. For this purpose, it is sufficient to consider a system that contains only 2 particles. We introduce the centre-of-mass ($Z$) and the relative ($z$) coordinates
\begin{equation}
\displaystyle{Z=\frac{z_{1}+z_{2}}{2}, \, \, \, \, z=z_{2}-z_{1}},
		\label{qmdescription19}
		\end{equation}
\begin{equation}
\displaystyle{\frac{\partial^{2}}{\partial z_{1}^{2}} +\frac{\partial^{2}}{\partial z_{1}^{2}}=\frac{1}{2}\frac{\partial^{2}}{\partial Z^{2}}+2\frac{\partial^{2}}{\partial z^{2}} }.
		\label{qmdescription20}
		\end{equation}
The Schr\"{o}dinger equation for relative motion has the following form
\begin{equation}
\displaystyle{-2\frac{\mathrm{d}^{2}\phi}{\mathrm{d}z^{2}}+2c\delta(z)\phi=E\phi   }.
		\label{qmdescription21}
\end{equation}
By integrating over small region $(-\varepsilon,\varepsilon)$ one gets
\begin{equation}
\displaystyle{-2\left(\left. \frac{\mathrm{d}\phi}{\mathrm{d}z} \right|_{z=\varepsilon}-\left. \frac{\mathrm{d}\phi}{\mathrm{d}z} \right|_{z=-\varepsilon}   \right)+2c\phi(0)=0}.
		\label{qmdescription22}
\end{equation}
Using the obvious relation $\frac{\partial}{\partial z}=\frac{1}{2}\left(\frac{\partial}{\partial z_{2}}-\frac{\partial}{\partial z_{1}}  \right)$, the formula (\ref{qmdescription22}) may be rewritten as
\begin{equation}
\displaystyle{\left(\frac{\partial }{\partial z_{2}} -\frac{\partial }{\partial z_{1}} \right)  \phi \bigg|_{z_{2}=z_{1}+\varepsilon} -\left(\frac{\partial }{\partial z_{2}} -\frac{\partial }{\partial z_{1}} \right) \phi \bigg|_{z_{2}=z_{1}-\varepsilon}=2c  \phi \bigg|_{z_{2}=z_{1} }}.
		\label{qmdescription23}
\end{equation}
Furthermore, from symmtery of $\phi$ in $z_{1},z_{2}$
\begin{equation}
\displaystyle{\left(\frac{\partial }{\partial z_{2}} -\frac{\partial }{\partial z_{1}} \right)  \phi \bigg|_{z_{2}=z_{1}+\varepsilon} -\left(\frac{\partial }{\partial z_{1}} -\frac{\partial }{\partial z_{2}} \right) \phi \bigg|_{z_{2}=z_{1}+\varepsilon}=2c  \phi \bigg|_{z_{2}=z_{1} }},
		\label{qmdescription24}
\end{equation}
then 
\begin{equation}
\displaystyle{\left(\frac{\partial }{\partial z_{2}} -\frac{\partial }{\partial z_{1}} \right)  \phi \bigg|_{z_{2}=z_{1}+\varepsilon} =c  \phi \bigg|_{z_{2}=z_{1} }}.
		\label{qmdescription25}
\end{equation}
In general, we can write the boundary conditions for the system as
\begin{equation}
\displaystyle{\left(\frac{\partial }{\partial z_{j+1}} -\frac{\partial }{\partial z_{j}} -c \right)  \Phi_{\mathcal{N}}(\{z_{i}\},\{k_{j}\}) =0, \, \, \, \, z_{j+1}=z_{j}+\varepsilon }.
		\label{qmdescription26}
\end{equation}
Formula (\ref{qmdescription16}) with boundary conditions (\ref{qmdescription26}) is equivalent to the relation $\hat{\mathcal{H}}_{\mathcal{N}}\Phi_{\mathcal{N}}=E_{\mathcal{N}}\Phi_{\mathcal{N}}$.

Reduction of the problem (\ref{qmdescription11}) - (\ref{qmdescription13}) to the problem of the Free Hamiltonian and the Total Momentum (\ref{qmdescription16}), (\ref{qmdescription17}) in the domain (\ref{qmdescription15}) with the boundary conditions (\ref{qmdescription26}) allows us to construct a solution $\Phi_{\mathcal{N}}$ in the following way. One can easily show that an eigenfunction of the Hamiltonian $\hat{\mathcal{H}}^{0}_{\mathcal{N}}$ ($\hat{\mathcal{H}}_{\mathcal{N}}$ in the domain $\mathcal{T}$)  is proportional to the determinant of the $\mathcal{N} \times \mathcal{N}$ matrix
\begin{equation}
\displaystyle{\mathrm{det} \left[ \mathrm{exp} \left( i k_{j} z_{s} \right) \right],}
		\label{qmdescription27}
\end{equation}
where $k_{j}$ are arbitrary numbers (\emph{quasi-momenta}). Due to antisymmetry in $z_{s}$ the determinant depicted above is equal to zero on the boundary ($z_{s}=z_{s+1}$). The function which satisfies equation $\hat{\mathcal{H}}_{\mathcal{N}}^{0}\Phi_{\mathcal{N}}=E_{\mathcal{N}}^{0}\Phi_{\mathcal{N}}$ and the boundary conditions (\ref{qmdescription26}) may be written as
\begin{equation}
\displaystyle{\Phi_{\mathcal{N}}(\{z_{l} \},\{k_{p}\})=\mathrm{const}\left[ \,  \prod_{\mathcal{N}\geq j>s\geq 1} \left( \frac{\partial}{\partial z_{j}} -\frac{\partial}{\partial z_{s}} +c \right)   \right] \mathrm{det} \left[ \mathrm{exp} \left( i k_{j} z_{s} \right) \right].}
		\label{qmdescription28}
\end{equation}
From alternation of derivatives $\Phi_{\mathcal{N}}$ is an eigenfunction of $\hat{\mathcal{H}}_{\mathcal{N}}^{0}$ so we need to prove that (\ref{qmdescription28}) satisfies the boundary conditions. For this purpose, let us rewrite $\Phi_{\mathcal{N}}$ as
\begin{equation}
\displaystyle{\Phi_{\mathcal{N}}(\{z_{l} \},\{k_{p}\})=\left( \frac{\partial}{\partial z_{2}} -\frac{\partial}{\partial z_{1}} +c \right)\widetilde{\Phi}_{\mathcal{N}} (\{z_{l} \},\{k_{p}\}) ,}
		\label{qmdescription29}
\end{equation}
where 
\begin{center}
$\displaystyle{\widetilde{\Phi}_{\mathcal{N}} (\{z_{l} \},\{k_{p}\})=\mathrm{const} \prod_{j=3}^{\mathcal{N}}  \left( \frac{\partial}{\partial z_{j}} -\frac{\partial}{\partial z_{1}} +c \right)\left( \frac{\partial}{\partial z_{j}} -\frac{\partial}{\partial z_{2}} +c \right)}$
\end{center}
\begin{equation}
\displaystyle{ \times \prod_{\mathcal{N}\geq j>s\geq 3} \left( \frac{\partial}{\partial z_{j}} -\frac{\partial}{\partial z_{s}} +c \right) \mathrm{det} \left[ \mathrm{exp} \left( i k_{j} z_{s} \right) \right] .}
		\label{qmdescription30}
\end{equation}
One instantly notices that $\widetilde{\Phi}_{\mathcal{N}} (z_{1},z_{2},\ldots,z_{\mathcal{N}},\{k_{p}\})=-\widetilde{\Phi}_{\mathcal{N}} (z_{2},z_{1},\ldots,z_{\mathcal{N}},\{k_{p}\})$ (antisymmetry of determinant). Then $\widetilde{\Phi}_{\mathcal{N}}=0$ if $z_{1}=z_{2}$. Now, we can consider the boundary condition for $z_{1}=z_{2}$
\begin{equation}
\displaystyle{ \left(\frac{\partial }{\partial z_{2}} -\frac{\partial }{\partial z_{1}} -c \right)  \Phi_{\mathcal{N}}(\{z_{i}\},\{k_{j}\}) =\left[\left(\frac{\partial }{\partial z_{2}} -\frac{\partial }{\partial z_{1}} \right)^{2} -c^{2}  \right]\widetilde{\Phi}_{\mathcal{N}}(\{z_{i}\},\{k_{j}\})},
		\label{qmdescription31}
\end{equation}
\begin{equation}
\displaystyle{ \left[\left(\frac{\partial }{\partial z_{2}} -\frac{\partial }{\partial z_{1}} \right)^{2} -c^{2}  \right]\widetilde{\Phi}_{\mathcal{N}}(\{z_{i}\},\{k_{j}\})=0, \, \, \, \, z_{2}=z_{1}+\varepsilon .}
		\label{qmdescription311}
\end{equation}
It is fulfilled because the left hand side of the equation (\ref{qmdescription311}) is antisymmetric in $z_{1}\longleftrightarrow z_{2}$ and hence, it is equal to zero when $z_{2}\longrightarrow z_{1}$. The boundary conditions (\ref{qmdescription26}) for the other $z_{j}$ can be verified similarly. Therefore, function (\ref{qmdescription28}) is an eigenfunction of the Hamiltonian $\hat{\mathcal{H}}_{\mathcal{N}}$ (\ref{qmdescription11}). We can rewrite the determinant as a sum over all the permutations $\pi \in \mathcal{S}_{\mathcal{N}} $
\begin{equation}
\displaystyle{\mathrm{det} \left[ \mathrm{exp} \left( i k_{j} z_{s} \right) \right]=\sum_{ \pi \in \mathcal{S}_{\mathcal{N}}}\mathrm{sgn}\left(\pi \right)  \mathrm{exp} \left( i\sum_{n=1}^{\mathcal{N}} k_{\pi(n)} z_{n} \right) .}
		\label{qmdescription32}
\end{equation}
Hence, in the domain $\mathcal{T}$ one gets \\

$\displaystyle{\Phi_{\mathcal{N}}(\{z_{l} \},\{k_{p}\})=\left(\displaystyle{\mathcal{N}!\prod_{j>s}\left[  \left(k_{j}-k_{s}  \right)^{2}+c^{2} \right]}\right)^{-1/2}}$
\begin{equation}
\displaystyle{ \times \sum_{ \pi \in \mathcal{S}_{\mathcal{N}}}\left[\mathrm{sgn}\left(\pi \right)  \mathrm{exp} \left( i\sum_{n=1}^{\mathcal{N}} k_{\pi(n)} z_{n} \right) \prod_{j>s}\left( k_{\pi(j)}-k_{\pi(s)}-i c   \right)\right]. }
		\label{qmdescription33}
\end{equation}
To remove the restriction $\mathcal{T}$ (\ref{qmdescription15}) we introduce additional sign function\footnote{After the modification the boundary conditions (\ref{qmdescription26}) are still fulfilled.} \\

$\displaystyle{\Phi_{\mathcal{N}}(\{z_{l} \},\{k_{p}\})=\left(\displaystyle{\mathcal{N}!\prod_{j>s}\left[  \left(k_{j}-k_{s}  \right)^{2}+c^{2} \right]}\right)^{-1/2}}$
\begin{equation}
\displaystyle{ \times \sum_{ \pi \in \mathcal{S}_{\mathcal{N}}}\left[\mathrm{sgn}\left(\pi \right)  \mathrm{exp} \left( i\sum_{n=1}^{\mathcal{N}} k_{\pi(n)} z_{n} \right) \prod_{j>s}\left( k_{\pi(j)}-k_{\pi(s)}-i c \, \mathrm{sign}(z_{j}-z_{s})   \right)\right], }
		\label{qmdescription34}
\end{equation}
then $\Phi_{\mathcal{N}}(\{z_{l} \},\{k_{p}\})$ is valid for arbitrary values of $z_{1}, \ldots , z_{\mathcal{N}}$. One should also check if $\Phi_{\mathcal{N}}(\{z_{l} \},\{k_{p}\})$ given by (\ref{qmdescription34}) is symmetric with respect to exchange of any pair of $z_{i}$, $z_{j}$\footnote{It is required because system of bosons is considered.}. Using the permutations $\sigma,\tau \in \mathcal{S}_{\mathcal{N}}$ and defining $\pi= \tau \sigma$  we obtain \\ \\
$\displaystyle{\Phi_{\mathcal{N}}(\sigma \{z_{l} \},\{k_{p}\})\sim \sum_{ \pi \in \mathcal{S}_{\mathcal{N}}}\left[\mathrm{sgn}\left(\pi \right)  \mathrm{exp} \left( i\sum_{n=1}^{\mathcal{N}} k_{\pi(n)} z_{\sigma(n)} \right) \prod_{j>s}\left( k_{\pi(j)}-k_{\pi(s)}-i c \, \mathrm{sign}(z_{\sigma(j)}-z_{\sigma(s)})   \right)\right]}$
\begin{center}
$\displaystyle{ =\sum_{ \tau \in \mathcal{S}_{\mathcal{N}}}\left[\mathrm{sgn}\left( \tau \sigma \right)  \mathrm{exp} \left( i\sum_{n=1}^{\mathcal{N}} k_{\tau (\sigma (n))} z_{\sigma(n)} \right) \prod_{j>s}\left( k_{\tau (\sigma (j))}-k_{\tau (\sigma (s))}-i c \, \mathrm{sign}(z_{\sigma(j)}-z_{\sigma(s)})   \right)\right] }$
\end{center}
\begin{center}
$\displaystyle{ =\sum_{ \tau \in \mathcal{S}_{\mathcal{N}}}\left[\mathrm{sgn}\left( \tau \sigma  \right)  \mathrm{exp} \left( i\sum_{n=1}^{\mathcal{N}} k_{\tau (n)} z_{n} \right) \mathrm{sgn}\left(\sigma \right) \prod_{\sigma(j)>\sigma(s)}\left( k_{\tau (\sigma (j))}-k_{\tau (\sigma (s))}-i c \, \mathrm{sign}(z_{\sigma(j)}-z_{\sigma(s)})   \right)\right] }$
\end{center}
\begin{center}
$\displaystyle{ =\sum_{ \tau \in \mathcal{S}_{\mathcal{N}}}\left[\mathrm{sgn}\left( \tau  \right)  \mathrm{exp} \left( i\sum_{n=1}^{\mathcal{N}} k_{\tau (n)} z_{n} \right)  \prod_{j>s}\left( k_{\tau (j)}-k_{\tau (s)}-i c \, \mathrm{sign}(z_{j}-z_{s})   \right)\right] .}$
\end{center}
Therefore
\begin{equation}
\displaystyle{ \Phi_{\mathcal{N}}(\sigma\{z_{l} \},\{k_{p}\})= \Phi_{\mathcal{N}}(\{z_{l} \},\{k_{p}\}), \, \, \, \, \sigma \in  \mathcal{S}_{\mathcal{N}}.}
		\label{qmdescription35}
\end{equation}

The wave function (\ref{qmdescription34}) may be represented in two other forms \cite{bogkorep} \\ \\
$\displaystyle{\Phi_{\mathcal{N}}(\{z_{l} \},\{k_{p}\})=\frac{\displaystyle{(-i)^{\mathcal{N}(\mathcal{N}-1)/2}}}{\sqrt{\mathcal{N}!}} \left[ \, \prod_{\mathcal{N}\geq j>s\geq 1} \mathrm{sign}(z_{j}-z_{s}) \right]}$
\begin{equation}
\displaystyle{ \times \sum_{ \pi \in \mathcal{S}_{\mathcal{N}}}\left[\mathrm{sgn}\left(\pi \right)  \mathrm{exp} \left( i\sum_{n=1}^{\mathcal{N}} k_{\pi(n)} z_{n} \right) \mathrm{exp} \left( \frac{i}{2}\sum_{\mathcal{N}\geq j>s\geq 1} \mathrm{sign}(z_{j}-z_{s})\theta(k_{\pi(j)}-k_{\pi(s)}) \right)\right], }
		\label{qmdescription36}
\end{equation}
$\displaystyle{\Phi_{\mathcal{N}}(\{z_{l} \},\{k_{p}\})=\frac{\displaystyle{\prod_{j>s}(k_{j}-k_{s})}}{\displaystyle{\sqrt{\mathcal{N}!\prod_{j>s}\left[  \left(k_{j}-k_{s}  \right)^{2}+c^{2} \right]}}}}$
\begin{equation}
\displaystyle{ \times \sum_{ \pi \in \mathcal{S}_{\mathcal{N}}}\left[\mathrm{sgn}\left(\pi \right)  \mathrm{exp} \left( i\sum_{n=1}^{\mathcal{N}} k_{\pi(n)} z_{n} \right) \prod_{j>s}\left(1-\frac{i c \, \mathrm{sign}(z_{j}-z_{s})}{k_{\pi(j)}-k_{\pi(s)}} \right)\right], }
		\label{qmdescription37}
\end{equation}
where the \textbf{\emph{scattering phase}} $ \theta(k)$ is defined as follows 
\begin{equation}
\displaystyle{ \theta(k)=i\mathrm{ln}\left( \frac{ic+k}{ic-k} \right) . }
		\label{qmdescription38}
\end{equation}
In general, all the models solvable by Bethe ansatz method have the wave functions which can be written in the form similar to (\ref{qmdescription36}) \cite{bogkorep}.

Let us consider the wave function given by (\ref{qmdescription37}) and permute the quasi-momenta $\rho \{k_{p}\}$, where $\rho \in \mathcal{S}_{\mathcal{N}}$. \\
$\displaystyle{\Phi_{\mathcal{N}}(\{z_{l} \},\rho \{k_{p}\})=\frac{\displaystyle{\prod_{j>s}(k_{\rho(j)}-k_{\rho(s)})}}{\displaystyle{\sqrt{\mathcal{N}!\prod_{j>s}\left[  \left(k_{\rho(j)}-k_{\rho(s)}  \right)^{2}+c^{2} \right]}}}}$
\begin{center}
$\displaystyle{ \times \sum_{ \pi \in \mathcal{S}_{\mathcal{N}}}\left[\mathrm{sgn}\left(\pi \right)  \mathrm{exp} \left( i\sum_{n=1}^{\mathcal{N}} k_{\pi(\rho(n))} z_{n} \right) \prod_{j>s}\left(1-\frac{i c \, \mathrm{sign}(z_{j}-z_{s})}{k_{\pi(\rho(j))}-k_{\pi(\rho(s))}} \right)\right] }$
\end{center}
$\displaystyle{=\frac{\displaystyle{\prod_{j>s}(k_{\rho(j)}-k_{\rho(s)})}}{\displaystyle{\sqrt{\mathcal{N}! \, \mathrm{sgn}^{2}(\rho)\prod_{j>s}\left[  \left(k_{j}-k_{s}  \right)^{2}+c^{2} \right]}}}  }$
\begin{center}
$\displaystyle{ \times \sum_{ \pi \in \mathcal{S}_{\mathcal{N}}}\left[ \frac{\displaystyle{\mathrm{sgn}\left(\pi \right)  \mathrm{exp} \left( i\sum_{n=1}^{\mathcal{N}} k_{\pi(\rho(n))} z_{n} \right)\prod_{j>s}\left(k_{\pi(\rho(j))}-k_{\pi(\rho(s))}-i c \, \mathrm{sign}(z_{j}-z_{s})\right)}}{\displaystyle{\mathrm{sgn}(\pi )\prod_{\pi(j)>\pi(s)}\left(k_{\pi(\rho(j))}-k_{\pi(\rho(s))}\right)}} \right]}$
\end{center}
\begin{center}
$\displaystyle{=\frac{\displaystyle{\prod_{j>s}(k_{\rho(j)}-k_{\rho(s)})}}{\displaystyle{\sqrt{\mathcal{N}!\prod_{j>s}\left[  \left(k_{j}-k_{s}  \right)^{2}+c^{2} \right]}}} \sum_{ \pi \in \mathcal{S}_{\mathcal{N}}}\left[ \frac{\displaystyle{  \mathrm{exp} \left( i\sum_{n=1}^{\mathcal{N}} k_{\pi(\rho(n))} z_{n} \right)\prod_{j>s}\left(k_{\pi(\rho(j))}-k_{\pi(\rho(s))}-i c \, \mathrm{sign}(z_{j}-z_{s})\right)}}{\displaystyle{\prod_{j>s}\left(k_{\rho(j)}-k_{\rho(s)}\right)}} \right] }$
\end{center}
\begin{center}
$\displaystyle{= \sum_{ \pi \in \mathcal{S}_{\mathcal{N}}}\left[ \frac{\displaystyle{  \mathrm{exp} \left( i\sum_{n=1}^{\mathcal{N}} k_{\pi(\rho(n))} z_{n} \right)\prod_{j>s}\left(k_{\pi(\rho(j))}-k_{\pi(\rho(s))}-i c \, \mathrm{sign}(z_{j}-z_{s})\right)}}{\displaystyle{\sqrt{\mathcal{N}!\prod_{j>s}\left[  \left(k_{j}-k_{s}  \right)^{2}+c^{2} \right]}}} \right] }$
\end{center}
\begin{center}
$\displaystyle{\stackrel{(\ref{qmdescription35})}{=} \sum_{ \pi \in \mathcal{S}_{\mathcal{N}}}\left[ \frac{\displaystyle{  \mathrm{exp} \left( i\sum_{n=1}^{\mathcal{N}} k_{\pi(\rho(n))} z_{n} \right)\mathrm{sgn}(\rho)\prod_{j>s}\left(k_{\pi(j)}-k_{\pi(s)}-i c \, \mathrm{sign}(z_{j}-z_{s})\right)}\displaystyle{\prod_{j>s}\left(k_{j}-k_{j}\right)}}{\displaystyle{\sqrt{\mathcal{N}!\prod_{j>s}\left[  \left(k_{j}-k_{s}  \right)^{2}+c^{2} \right]}}\displaystyle{ \, \mathrm{sgn}(\pi)\prod_{j>s}\left(k_{\pi(j)}-k_{\pi(s)}\right)}}  \right] }$
\end{center}
$\displaystyle{= \frac{\displaystyle{\mathrm{sgn}(\rho)\prod_{j>s}(k_{j}-k_{s})}}{\displaystyle{\sqrt{\mathcal{N}!\prod_{j>s}\left[  \left(k_{j}-k_{s}  \right)^{2}+c^{2} \right]}}}\sum_{ \pi \in \mathcal{S}_{\mathcal{N}}}\left[\mathrm{sgn}\left(\pi \right)  \mathrm{exp} \left( i\sum_{n=1}^{\mathcal{N}} k_{\pi(n)} z_{n} \right) \prod_{j>s}\left(1-\frac{i c \, \mathrm{sign}(z_{j}-z_{s})}{k_{\pi(j)}-k_{\pi(s)}} \right)\right]. }$ \\ \\
As we can see 
\begin{equation}
\displaystyle{ \Phi_{\mathcal{N}}(\{z_{l} \},\rho \{k_{p}\})=\mathrm{sgn}(\rho)\Phi_{\mathcal{N}}(\{z_{l} \}, \{k_{p}\}), \, \, \, \, \rho \in  \mathcal{S}_{\mathcal{N}}, }
		\label{qmdescription39}
\end{equation}
then $\Phi_{\mathcal{N}}$ is an antisymmetric function of $k_{j}$. Hence, $\Phi_{\mathcal{N}}=0$ for $k_{j}=k_{p}, \, \, j\neq p$. This result looks as if that one-dimensional system of contact interacting bosons satisfies Pauli exclusion principle. On first sight it looks very strange, however, it turns out that in $n+1$ space-time dimensions the theorem connecting spin and statistics is not valid if $n<3$\footnote{It is known that the wave function $\psi(\ldots,\vec{r}_{i},\ldots,\vec{r}_{j},\ldots )=\mathrm{exp}\left(i \varphi \right)\psi(\ldots,\vec{r}_{j},\ldots,\vec{r}_{i},\ldots )$, where in $n+1$ space-time dimensions the phase $\varphi \in \{0,\pm \pi \}$ if $n\geq 3$ and it is possible that $\varphi \in [0, 2\pi )$ if $n< 3$.} \cite{ontheory}. However, in our case, the above statement is not entirely accurate. One should distinguish the quasi-momenta from the momenta of particles. In the next section \ref{sec:quasimomenta}, we show that quasi-momenta, although strongly connected with momenta of particles, are only the parameters of the wave function. The property (\ref{qmdescription39}) plays extremely important role in construction of the ground state (Dirac sea).

Corresponding eigenvalues of operators $\hat{\mathrm{H}}$ and $\hat{\mathrm{P}}$  have the following forms
\begin{equation}
\displaystyle{ E_{\mathcal{N}}=\sum_{j=1}^{\mathcal{N}}k_{j}^{2}, \, \, \, \, \, P_{\mathcal{N}}=\sum_{j=1}^{\mathcal{N}}k_{j}}
		\label{qmdescription40}
\end{equation}
Considering $\Phi_{\mathcal{N}}$ in the whole coordinate space $\mathbb{R}^{\mathcal{N}}: -\infty <z_{j} <\infty, \, \, j=1,2,\ldots , \mathcal{N}$ one can find the normalization condition \cite{gaudin}
\begin{equation}
\displaystyle{ \int \limits_{-\infty}^{\infty} \mathrm{d}^{\mathcal{N}}z \Phi^{*}_{\mathcal{N}}(z_{1},\ldots, z_{\mathcal{N}},k_{1},\ldots,k_{\mathcal{N}}) \Phi_{\mathcal{N}}(z_{1},\ldots, z_{\mathcal{N}},q_{1},\ldots,q_{\mathcal{N}})=(2\pi)^{\mathcal{N}}\prod_{j=1}^{\mathcal{N}}\delta(k_{j}-q_{j}), }
		\label{qmdescription41}
\end{equation}
where we assume 
\begin{equation}
\displaystyle{ k_{1}<k_{2}<\ldots <k_{\mathcal{N}}, \, \, \, \, q_{1}<q_{2}<\ldots <q_{\mathcal{N}} .}
		\label{qmdescription42}
\end{equation}
It can also be shown \cite{gaudin} that the completnes of the system of functions $\Phi_{\mathcal{N}}$ is given by the following relation
\begin{equation}
\displaystyle{ \int \limits_{-\infty}^{\infty} \mathrm{d}^{\mathcal{N}}k \Phi^{*}_{\mathcal{N}}(z_{1},\ldots, z_{\mathcal{N}},k_{1},\ldots,k_{\mathcal{N}}) \Phi_{\mathcal{N}}(y_{1},\ldots, y_{\mathcal{N}},k_{1},\ldots,k_{\mathcal{N}})=(2\pi)^{\mathcal{N}}\prod_{j=1}^{\mathcal{N}}\delta(z_{j}-y_{j}), }
		\label{qmdescription43}
\end{equation}
\begin{equation}
\displaystyle{ z_{1}<z_{2}<\ldots <z_{\mathcal{N}}, \, \, \, \, y_{1}<y_{2}<\ldots <y_{\mathcal{N}} .}
		\label{qmdescription44}
\end{equation}

\section{Periodic boundary conditions}
\label{sec:periodicconditions}

Traditionally, we impose periodic boundary conditions on the wave functions by putting the system into a periodic box of length $\mathrm{L}$. Consequently, the wave function $\Phi_{\mathcal{N}}$ should satisfy the following equation
\begin{equation}
\displaystyle{  \Phi_{\mathcal{N}}(z_{1},\ldots,z_{j}+\mathrm{L} ,\ldots,z_{\mathcal{N}},k_{1},\ldots,k_{\mathcal{N}}) =\Phi_{\mathcal{N}}(z_{1},\ldots,z_{j} ,\ldots,z_{\mathcal{N}},k_{1},\ldots,k_{\mathcal{N}}), }
		\label{qmdescription45}
\end{equation}
where $j=1,2,\ldots,\mathcal{N}$. These stipulations give us a system of equations
\begin{equation}
\displaystyle{  \mathrm{exp}\left(i k_{j}\mathrm{L}\right)=-\prod_{s=1}^{\mathcal{N}}\frac{k_{j}-k_{s}+i c}{k_{j}-k_{s}-i c}, \, \, \, \, j=1,2,\ldots,\mathcal{N},}
		\label{qmdescription46}
\end{equation}
called \textbf{\emph{Bethe equations}}, which gives us permitted values of quasi-momenta $k_{j}$. \\

\textbf{\textit{Theorem 1.}} All the solutions $k_{j}$ of the Bethe system of equations (for $c>0$) are real numbers.\\ \\
\textbf{\emph{Proof:}} We instantly notice that
\begin{equation}
\displaystyle{  \left|\mathrm{exp}\left(i k \mathrm{L}\right)\right|\leq 1, \, \, \, \, \mathrm{for} \, \, \, \,  \mathrm{Im} (k)\geq 0, \, \, \, \, \, \, \, \, \left|\mathrm{exp}\left(i k \mathrm{L}\right)\right|\geq 1, \, \, \, \, \mathrm{for} \, \, \, \, \mathrm{Im} (k)\leq 0,}
		\label{qmdescription47}
\end{equation}
\begin{equation}
\displaystyle{  \left|\frac{k+ic}{k-ic}\right|\leq 1, \, \, \, \, \mathrm{for} \, \, \, \, \mathrm{Im} (k)\leq 0, \, \, \, \, \, \, \, \, \left|\frac{k+ic}{k-ic}\right|\geq 1, \, \, \, \, \mathrm{for} \, \, \, \, \mathrm{Im} (k)\geq 0,}
		\label{qmdescription48}
\end{equation}
and define $k_{max}, k_{min} \in \{k_{j}\}$ such that
\begin{equation}
\displaystyle{ \mathrm{Im} (k_{max})\geq \mathrm{Im} (k_{j}), \, \, \, \, \, \, \, \, \mathrm{Im} (k_{min})\leq \mathrm{Im} (k_{j}) , \, \, \, \, \, \, \, \, j=1,2,\ldots,\mathcal{N}.}
		\label{qmdescription49}
\end{equation}
Hence, one can write\footnote{It is obvious that $\mathrm{Im} (k_{max}-k_{j})\geq 0, \, \, \, \, \mathrm{Im} (k_{min}-k_{j})\leq 0$ for $j=1,2,\ldots,\mathcal{N}$.}
\begin{equation}
\displaystyle{  \left|\mathrm{exp}\left(i k_{max}\mathrm{L}\right)\right|=\left|\prod_{s=1}^{\mathcal{N}}\frac{k_{max}-k_{s}+i c}{k_{max}-k_{s}-i c}\right|\geq 1 \Longrightarrow \mathrm{Im} (k_{max})\leq 0, }
		\label{qmdescription50}
\end{equation}
\begin{equation}
\displaystyle{   \left|\mathrm{exp}\left(i k_{min}\mathrm{L}\right)\right|=\left|\prod_{s=1}^{\mathcal{N}}\frac{k_{min}-k_{s}+i c}{k_{min}-k_{s}-i c}\right|\leq 1 \Longrightarrow \mathrm{Im} (k_{min})\geq 0.}
		\label{qmdescription51}
\end{equation}
From (\ref{qmdescription49}) we obtain
\begin{equation}
\displaystyle{0 \geq\mathrm{Im} (k_{max})\geq \mathrm{Im} (k_{j})\geq \mathrm{Im} (k_{min})\geq 0, \, \, \, \, \, \, \, \,  j=1,2,\ldots,\mathcal{N},}
		\label{qmdescription52}
\end{equation}
then
\begin{equation}
\displaystyle{ \mathrm{Im} (k_{j})=0 , \, \, \, \, \, \, \, \,  j=1,2,\ldots,\mathcal{N}.}
		\label{qmdescription53}
\end{equation}

The fact that quantities $k_{j}$ (for $c>0$ - the case of repulsion interaction) are real numbers is consistent with the aforementioned connection between quasi-momenta and momenta of particles. Considering the spectrum of energy one obtains
\begin{equation}
\displaystyle{ E_{\mathcal{N}}=\sum_{j=1}^{\mathcal{N}}k_{j}^{2} \geq 0, \, \, \, \, \, \, \, \, k_{j} \in \mathbb{R} \, \, \, \, \, \, \, \, \mathrm{for} \, \, \, c>0.}
		\label{qmdescription54}
\end{equation}
Thus, in the case of repulsive interactions, the system consist of elementary particles only. Otherwise, if $c<0$ (attraction), there also exist bound states\footnote{It means that, in the case of attractive interactions, it is possible to find more elaborated structures (molecules) in the system.}.  

The system of Bethe equations (\ref{qmdescription46}) can be rewritten into logarithmic form. One notices that
\begin{center}
$\displaystyle{ \mathrm{ln}\left(-\prod_{s=1}^{\mathcal{N}}\frac{k_{j}-k_{s}+i c}{k_{j}-k_{s}-i c}\right)= \mathrm{ln}\left(\mathop{\prod_{ s = 1 }^{\mathcal{N}}}_{s\neq j}\frac{k_{j}-k_{s}+i c}{k_{j}-k_{s}-i c}  \right)= \mathop{\sum^{\mathcal{N}}_{ s = 1}}_{s\neq j}\mathrm{ln}\left(\frac{k_{j}-k_{s}+i c}{k_{j}-k_{s}-i c}\right) }$
\end{center}
\begin{center}
$\displaystyle{ = \sum^{\mathcal{N}}_{ s = 1}\mathrm{ln}\left(\frac{k_{j}-k_{s}+i c}{k_{j}-k_{s}-i c}\right)-\mathrm{ln}\left(-1 \right)=\sum^{\mathcal{N}}_{ s = 1}\mathrm{ln}\left(\frac{k_{j}-k_{s}+i c}{k_{j}-k_{s}-i c}\right)-i\pi ,}$
\end{center}
where we have the scattering phase (\ref{qmdescription38})
\begin{equation}
\displaystyle{ \theta(k)=i \mathrm{ln}\left(\frac{k+i c}{k-i c}\right)+\pi= i \mathrm{ln}\left(\frac{i c+k}{i c-k}\right)  =-\theta(-k) .}
		\label{qmdescription55}
\end{equation}
Hence 
\begin{center}
$\displaystyle{ \mathrm{ln}\left(-\prod_{s=1}^{\mathcal{N}}\frac{k_{j}-k_{s}+i c}{k_{j}-k_{s}-i c}\right)=\sum^{\mathcal{N}}_{ s = 1}\left[\mathrm{ln}\left(\frac{k_{j}-k_{s}+i c}{k_{j}-k_{s}-i c}\right)-i\pi \right]+(\mathcal{N}-1)i\pi }$
\end{center}
\begin{center}
$\displaystyle{ =-\sum^{\mathcal{N}}_{ s = 1}i\theta(k_{j}-k_{s})+(\mathcal{N}-1)i\pi  .}$
\end{center}
Finally, one obtains
\begin{equation}
\displaystyle{ k_{j}\mathrm{L}+\sum^{\mathcal{N}}_{ s = 1}\theta(k_{j}-k_{s})=2\pi \left(n_{j}+\frac{\mathcal{N}-1}{2} \right)=2\pi \mathcal{I}_{j}, \, \, \, \, j=1,2,\ldots , \mathcal{N},}
		\label{qmdescription56}
\end{equation}
where $\{n_{j}\}$ is an arbitrary set of different\footnote{From antisymmetry of $\Phi_{\mathcal{N}}$ (\ref{qmdescription39}) solutions $k_{j}$ must be a different numbers.} integers. In the case of repulsive interaction ($c>0$)
\begin{equation}
\displaystyle{\theta(k)=2 \, \mathrm{arctan}\left( \frac{k}{c} \right), \, \, \, \, \, \, \, \, c>0, }
		\label{qmdescription57}
\end{equation}
therefore 
\begin{equation}
\displaystyle{ k_{j}\mathrm{L}+\sum^{\mathcal{N}}_{ s = 1}2 \, \mathrm{arctan}\left( \frac{k_{j}-k_{s}}{c} \right)=2\pi \left(n_{j}+\frac{\mathcal{N}-1}{2} \right)=2\pi \mathcal{I}_{j}, \, \, \, \, j=1,2,\ldots , \mathcal{N}.}
		\label{qmdescription58}
\end{equation}
It is obvious that 
\begin{equation}
\displaystyle{ \mathcal{I}_{j}=\left(n_{j}+\frac{\mathcal{N}-1}{2} \right)=\left\{\begin{array}{ll} \mathrm{integer} \, \, \, \, \, \, \, \, \, \, \, \, \, \, \,  &   \mathrm{if} \, \, \mathcal{N} \, \, \mathrm{is} \, \, \mathrm{odd} \, \,\\ \mathrm{half}\mathcal{-}\mathrm{integer}&  \mathrm{if} \, \, \mathcal{N} \, \, \mathrm{is} \, \, \mathrm{even} \end{array} \right., \, \, \, \, \, \, \, \, j=1,2,\ldots , \mathcal{N}.}
		\label{qmdescription59}
\end{equation}

Let us sum equation (\ref{qmdescription58}) over all $j$
\begin{center}
$\displaystyle{ \sum_{j=1}^{\mathcal{N}} k_{j}\mathrm{L}+\sum_{j=1}^{\mathcal{N}}\sum^{\mathcal{N}}_{ s = 1}2 \, \mathrm{arctan}\left( \frac{k_{j}-k_{s}}{c} \right)=2\pi \sum_{j=1}^{\mathcal{N}}\mathcal{I}_{j}.}$
\end{center}
Because of antisymmetry of $\mathrm{arctan}\left( k \right)$, one can easily show that
\begin{center}
$\displaystyle{ \sum_{j=1}^{\mathcal{N}}\sum^{\mathcal{N}}_{ s = 1}  \mathrm{arctan}\left( \frac{k_{j}-k_{s}}{c} \right)=-\sum_{j=1}^{\mathcal{N}}\sum^{\mathcal{N}}_{ s = 1} \mathrm{arctan}\left( \frac{k_{s}-k_{j}}{c} \right)=-\sum_{s=1}^{\mathcal{N}}\sum^{\mathcal{N}}_{ j = 1}  \mathrm{arctan}\left( \frac{k_{s}-k_{j}}{c} \right)=0,}$
\end{center}
then
\begin{equation}
\displaystyle{ P_{\mathcal{N}}=\sum_{j=1}^{\mathcal{N}} k_{j}=\frac{2\pi}{\mathrm{L}} \sum_{j=1}^{\mathcal{N}}\mathcal{I}_{j}.}
		\label{qmdescription60}
\end{equation}

The analysis presented above revealed that the solutions of Bethe equations (\ref{qmdescription58}) can 
be parametrized by a set of numbers $\left\{\mathcal{I}_{j}\right\}$. Nevertheless, we still don't know if solutions exist. Moreover, we need to know if two different sets of numbers $\left\{\mathcal{I}_{j}\right\}$ can give the same solutions (the question of uniqueness of parametrization). \\

\textbf{\textit{Theorem 2.}} The solutions of the Bethe equations (\ref{qmdescription58}) exist and can be uniquely parametrized by a set of integer (half-integer) numbers $\left\{\mathcal{I}_{j}\right\}$.\\ \\
\textbf{\emph{Proof:}} In order to prove this statement we construct an action $S$ connected with Bethe equations (\ref{qmdescription56}) by a variational principle\footnote{Bethe equations (\ref{qmdescription56}) can be obtained from a variational principle $\delta S/\delta k_{j}=0$. The action (\ref{qmdescription61}) was introduced by C.N. Yang and C.P. Yang \cite{yangi}, \cite{bogkorep}.} 
\begin{equation}
\displaystyle{ S=\frac{1}{2}\mathrm{L}\sum_{j=1}^{\mathcal{N}}k_{j}^{2}-2\pi \sum_{j=1}^{\mathcal{N}}\mathcal{I}_{j}k_{j}+\frac{1}{2}\sum_{j=1}^{\mathcal{N}}\sum_{s=1}^{\mathcal{N}} \left( \int \limits_{0}^{k_{j}-k_{s}}\theta (\mu)\mathrm{d}\mu \right)   .}
		\label{qmdescription61}
\end{equation}
The equations (\ref{qmdescription56}) are the extremum conditions for an action depicted above. Our purpose is to prove that an action is convex\footnote{Convexity of action implies the existence of one, unique minimum.}. If so, we need to show that the matrix of second derivatives $\partial^{2}S/\partial k_{j} \partial k_{s}$ is positive defined, which means that all its eigenvalues are positive. One obtains 
\begin{equation}
\displaystyle{ \frac{\partial^{2}S}{\partial k_{j} \partial k_{s}}=\delta_{js}\left[ \mathrm{L}+\sum_{m=1}^{\mathcal{N}}K(k_{j},k_{m})  \right]-K(k_{j},k_{s}),}
		\label{qmdescription62}
\end{equation}
where
\begin{equation}
\displaystyle{K(k,q)=\frac{2c}{c^{2}+(k-q)^{2}}.}
		\label{qmdescription63}
\end{equation}
Hence
\begin{equation}
\displaystyle{ \sum_{j,s}\frac{\partial^{2}S}{\partial k_{j} \partial k_{s}}v_{j}v_{s}=\sum_{j=1}^{\mathcal{N}}\mathrm{L}v_{j}^{2}+\sum_{j>s=1}^{\mathcal{N}}K(k_{j},k_{s})(v_{j}-v_{s})^{2}\geq \mathrm{L}\sum_{j=1}^{\mathcal{N}}v_{j}^{2}>0,}
		\label{qmdescription64}
\end{equation}
for any real vector $v_{j}\in \mathbb{R}$. Therefore, we proved \emph{Theorem 2}. In the literature, formula (\ref{qmdescription62}) is the so-called Gaudin matrix $\mathcal{G}_{js}$ \cite{gaudin2}, \cite{kaminishi}, \cite{kaminishi2}
\begin{equation}
\displaystyle{ \mathcal{G}_{js}(\{k \}_{\mathcal{N}})=\delta_{js}\left[ \mathrm{L}+\sum_{m=1}^{\mathcal{N}}K(k_{j},k_{m})  \right]-K(k_{j},k_{s}).}
		\label{gaudinmatrix}
\end{equation}

We have another noteworthy property of the solutions:\\

\textbf{\textit{Theorem 3.}} If $\mathcal{I}_{j}>\mathcal{I}_{s}$ ($\mathcal{I}_{j}=\mathcal{I}_{s}$), then $k_{j}>k_{s}$ ($k_{j}=k_{s}$).  \\ \\
\textbf{\emph{Proof:}} By subtracting the $s$-th equation (\ref{qmdescription56}) from the $j$-th equation one gets
\begin{equation}
\displaystyle{ \mathrm{L}(k_{j}-k_{s})+\sum_{m=1}^{\mathcal{N}}\left[ \theta(k_{j}-k_{m})-\theta(k_{s}-k_{m}) \right]=2\pi (\mathcal{I}_{j}-\mathcal{I}_{s}).}
		\label{qmdescription65}
\end{equation}
Examining the function $\theta$ we find
\begin{equation}
\displaystyle{ \frac{\mathrm{d}}{\mathrm{d}k}\theta(k)=\frac{2c}{c^{2}+k^{2}}>0, \, \, \, \, \, \, \, \, \theta(k_{2})>\theta(k_{1}) \, \, \, \, \, \, \, \, \mathrm{for} \, \, \, \, \, \, \, \, k_{2}>k_{1}, \, \, \, \, \, \, \, \, \theta(\pm \infty)=\pm \pi,}
		\label{qmdescription66}
\end{equation}
so the thesis is obvious.

We note that if $\mathcal{I}_{j}=\mathcal{I}_{s}$ for $j\neq s$, the wave function $\Phi_{\mathcal{N}}$ is equal to zero due to the antisymmetry with respect to exchange of any pair of $k_{j}$, $k_{s}$. Therefore, one has to take into account only $\mathcal{I}_{j}\neq \mathcal{I}_{s}$ (for $j\neq s$).

\section{The problem of quasi-momenta}
\label{sec:quasimomenta}

In the section \ref{sec:ThecoordinateBetheansatz} we obtained the eigenfunctions and the eigenvalues of the Lieb-Liniger model. The form of eigenvalues 
\begin{equation}
\displaystyle{ E_{\mathcal{N}}=\sum_{j=1}^{\mathcal{N}}k_{j}^{2}, \, \, \, \, \, P_{\mathcal{N}}=\sum_{j=1}^{\mathcal{N}}k_{j},}
		\label{quasimomenta1}
\end{equation}
suggests that quasi-momenta $k_{1},\ldots,k_{\mathcal{N}}$ are strongly connected with momenta of particles. Furthermore, in the previous section we have proven that if we impose periodic boundary conditions (for repulsive interaction - $c>0$) $k_{j}$ are real numbers (\emph{Theorem 1}). Now, we show that the connection is slightly misleading. For this purpose, it is sufficient to consider a system with infinite coupling constant $c\to \infty$. In this case, it is obvious that the wave function $\Psi$ should satisfy the following boundary conditions\begin{samepage}
\begin{equation}
\displaystyle{\Psi \big|_{z_{i}=z_{j}}=0, \, \, \, \, \, \, \, \, \mathrm{for \, \,   all}  \, \, \, \,  i\neq j.}
		\label{quasimomenta2}
\end{equation}
The wave function which satisfies (\ref{qmdescription16})-(\ref{qmdescription18}) with above conditions has the following form \end{samepage}
\begin{equation}
\displaystyle{\Psi_{F}(\{z_{j}\},\{k_{s}\}) \sim \mathrm{det} \left[ \mathrm{e}^{ i k_{j} z_{s}} \right]. }
		\label{quasimomenta3}
\end{equation}
Above wave function, given by a Slater determinant, is antisymmetric with respect to exchange of any pair of $z_{i}$, $z_{j}$, so it is appropriate for fermions. The wave function for bosons must be symmetric under interchange of two particle
coordinates. It turns out, that the wave function which has this symmetry is given by
 \begin{equation}
\displaystyle{\Psi_{B}(\{z_{j}\},\{k_{s}\}) \sim \mathrm{det} \left[ \mathrm{e}^{ i k_{j} z_{s}} \right]\prod_{1\leq p<l\leq \mathcal{N}}\mathrm{sign}(z_{l}-z_{p}). }
		\label{quasimomenta4}
\end{equation}
On the first sight, it is sufficient to take an absolute value of (\ref{quasimomenta3}). However, $|\Psi_{F} |$ is also symmetric with respect to exchange of any pair of quasi-momenta $k_{i}$, $k_{j}$, which is not consistent with (\ref{qmdescription39})\footnote{One easily notices that $\Psi_{B}$ satisfies the condition (\ref{qmdescription39}).}.

Now, let us consider the 2-particle system. In this case one gets unnormalized wave functions\footnote{In our considerations the normalization factor is irrelevant.} 
 \begin{equation}
\displaystyle{\Psi_{F}(z_{1},z_{2},k_{1},k_{2}) =  \mathrm{e}^{ i (k_{1} z_{1}+k_{2}z_{2})} -\mathrm{e}^{ i (k_{2} z_{1}+k_{1}z_{2})}, }
		\label{quasimomenta5}
\end{equation}
 \begin{equation}
\displaystyle{\Psi_{B}(z_{1},z_{2},k_{1},k_{2})= \,  \mathrm{sign}(z_{2}-z_{1}) \left[ \mathrm{e}^{ i (k_{1} z_{1}+k_{2}z_{2})} -\mathrm{e}^{ i (k_{2} z_{1}+k_{1}z_{2})}\right]. }
		\label{quasimomenta6}
\end{equation}
Taking the Fourier transformation of (\ref{quasimomenta5}) we get
\begin{center}
$\displaystyle{\widetilde{\Psi}_{F}(p_{1},p_{2},k_{1},k_{2}) =\frac{1}{4 \pi^{2}} \int \limits_{-\infty}^{\infty}  \int \limits_{-\infty}^{\infty}  \mathrm{d}z_{1}\mathrm{d}z_{2} \mathrm{e}^{-i(p_{1}z_{1}+p_{2}z_{2})}\left[ \mathrm{e}^{ i (k_{1} z_{1}+k_{2}z_{2})} -\mathrm{e}^{ i (k_{2} z_{1}+k_{1}z_{2})}\right]= }$
\end{center}
 \begin{equation}
\displaystyle{=\delta(k_{1}-p_{1})\delta(k_{2}-p_{2}) -  \delta(k_{1}-p_{2})\delta(k_{2}-p_{1}).    }
		\label{quasimomenta7}
		\end{equation}
This result indicates that, in the case of free fermions, parameters $k_{j}$ are exactly the same as momenta of particles $p_{j}$. The statement becomes obvious when we look at the probability density of the system (Figure 1.1.).
\begin{center}\includegraphics[width=16cm,angle=0]{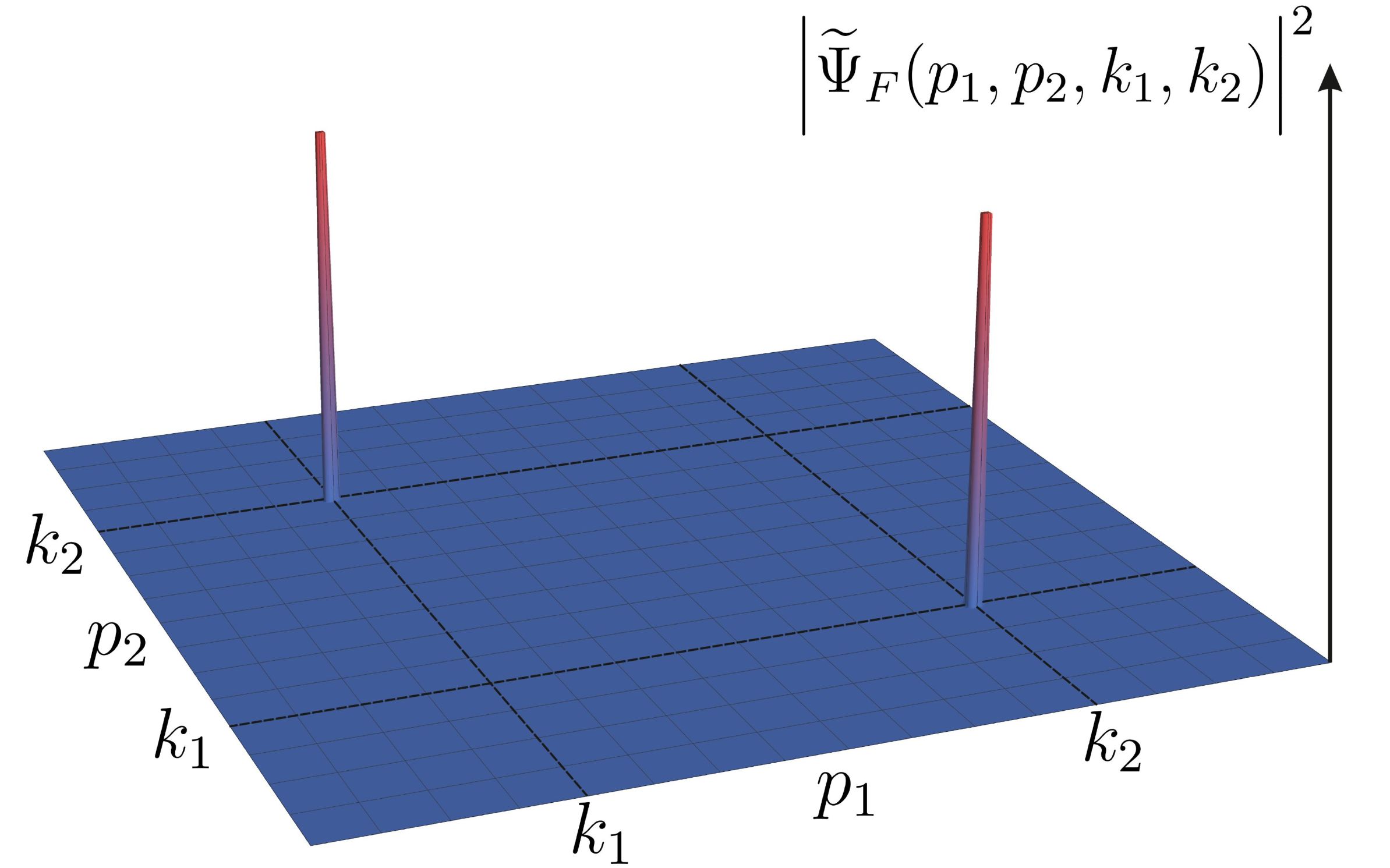}
 \footnotesize{Figure 1.1. Probability density in momentum space for two non-interacting fermions in 1D. }\end{center}
Plot presented above shows explicitly that the probability of finding two particles in our system is non-zero only if $\left\{ p_{1}=k_{1}, p_{2}=k_{2} \right\}$ or equivalently $\left\{ p_{1}=k_{2}, p_{2}=k_{1} \right\}$.

In the case of Bose system, the wave function (\ref{quasimomenta6}) contains a sign function which can be rewritten using the Heaviside step function $\Theta$
 \begin{equation}
\displaystyle{\mathrm{sign}(z)=1-2\Theta (-z).  }
		\label{quasimomenta8}
\end{equation}
Moreover, we notice that $\Theta$ has very useful integral representation \cite{theta}, \cite{theta2}
 \begin{equation}
\displaystyle{\Theta (z)=\frac{1}{2\pi i} \lim _{\varepsilon \to 0^{+}} \int \limits_{-\infty}^{\infty} \mathrm{d}\tau \frac{1}{\tau - i \varepsilon}\mathrm{e}^{i  z  \tau} .  }
		\label{quasimomenta9}
\end{equation}
Therefore, Fourier transformation of $\Psi_{B}$ has the following form
\begin{center}
$\displaystyle{\widetilde{\Psi}_{B}(p_{1},p_{2},k_{1},k_{2}) =\frac{1}{4 \pi^{2}} \int \limits_{-\infty}^{\infty}  \int \limits_{-\infty}^{\infty}  \mathrm{d}z_{1}\mathrm{d}z_{2} \mathrm{e}^{-i(p_{1}z_{1}+p_{2}z_{2})}\left[ \mathrm{e}^{ i (k_{1} z_{1}+k_{2}z_{2})} -\mathrm{e}^{ i (k_{2} z_{1}+k_{1}z_{2})}\right] }$
\end{center}
\begin{center}
$\displaystyle{-\frac{1}{4 \pi^{3}i}\lim _{\varepsilon \to 0^{+}} \int \limits_{-\infty}^{\infty} \int \limits_{-\infty}^{\infty}  \int \limits_{-\infty}^{\infty} \mathrm{d}\tau \mathrm{d}z_{1}\mathrm{d}z_{2} \frac{1}{\tau - i \varepsilon}\mathrm{e}^{i  (z_{1}-z_{2})  \tau}  \mathrm{e}^{-i(p_{1}z_{1}+p_{2}z_{2})}\left[ \mathrm{e}^{ i (k_{1} z_{1}+k_{2}z_{2})} -\mathrm{e}^{ i (k_{2} z_{1}+k_{1}z_{2})}\right]= }$
\end{center}
 \begin{equation}
\displaystyle{=\widetilde{\Psi}_{F}(p_{1},p_{2},k_{1},k_{2})+\frac{i}{\pi}\delta(k_{1}+k_{2}-p_{1}-p_{2})\left[ \frac{1}{p_{1}-k_{1}}-\frac{1}{p_{1}-k_{2}} \right].    }
		\label{quasimomenta10}
\end{equation}
One easily notices that in the case of Bose system, the wave function $\widetilde{\Psi}_{B}(p_{1},p_{2},k_{1},k_{2})$ is non-zero for infinitely many points $(p_{1},p_{2})$ located on the line $p_{1}+p_{2}=k_{1}+k_{2}$. Consequently, for 2-particle Bose system with $c=\infty$, the form of the probability density (depicted in the Figure 1.2.) has a distinctly different structure than two peaks presented in the Figure 1.1.
\begin{samepage} \begin{center}\includegraphics[width=15cm,angle=0]{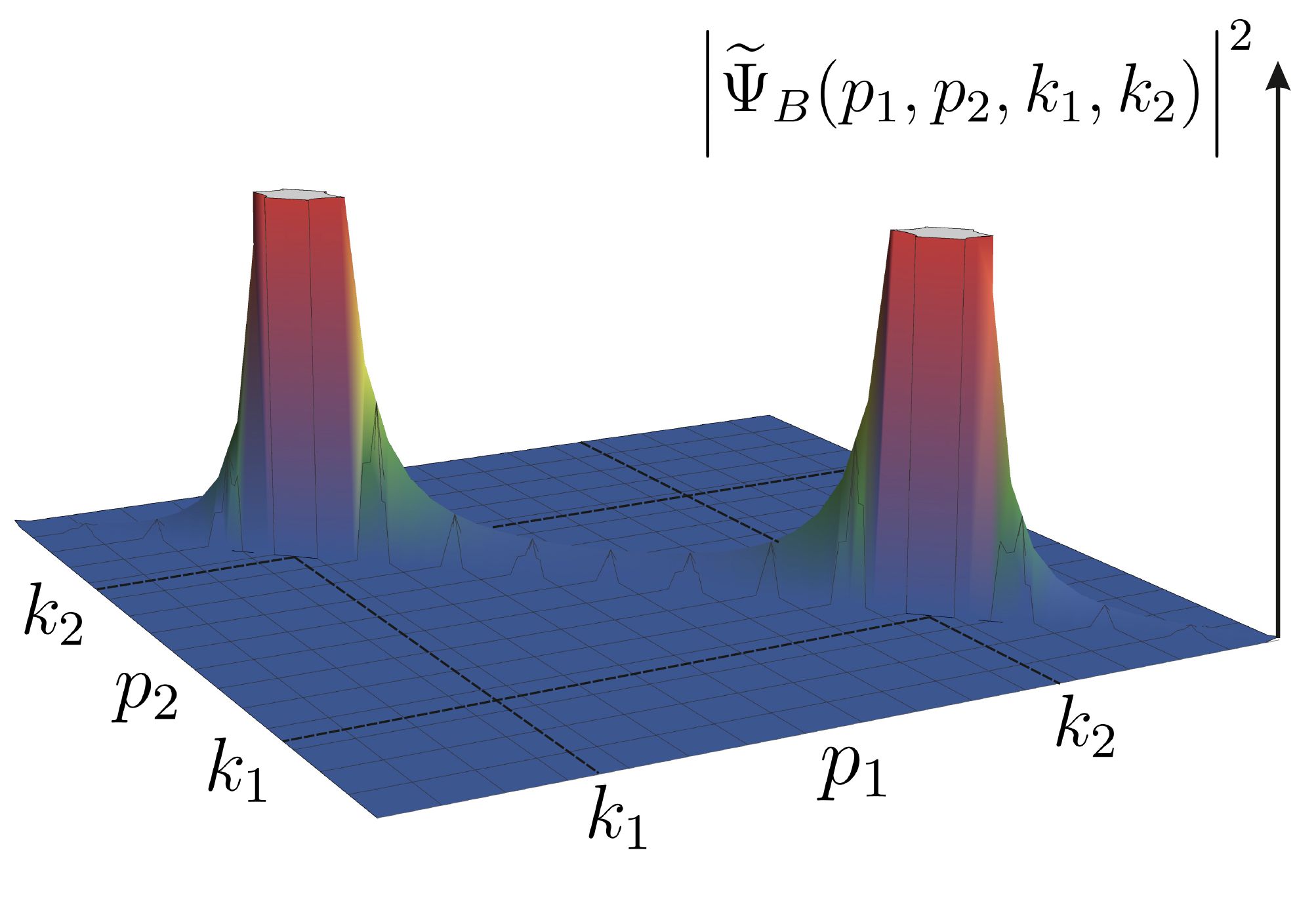}\end{center} 
\begin{center}\footnotesize{Figure 1.2. Probability density in momentum space for two bosons in 1D with coupling constant $c=\infty$. } \end{center}\end{samepage}
\pagebreak

We instantly see that there is non-zero probability of finding two particles with the momenta $p_{1}, p_{2}$ which satisfy relation $p_{1}+p_{2}=k_{1}+k_{2}$. It is noteworthy that the maxima of probability density are located at $\left\{ p_{1}=k_{1}, p_{2}=k_{2} \right\}$ and $\left\{ p_{1}=k_{2}, p_{2}=k_{1} \right\}$. We expect the same behavior of probability density for an arbitrary coupling constant $c$.

The results obtained above, show clearly that parameters $k_{j}$ are not exactly the same as the momenta of particles for considered Bose system. The latter are not good quantum numbers. On the other hand, the positions of the distribution maxima and the form of eigenvalues (\ref{quasimomenta1}) inform us that there exists a strong connection between parameters $k_{j}$ and the momenta of particles $p_{j}$. To emphasize the similarities we call the parameters $k_{j}$ \textit{\textbf{quasi-momenta}}.

\section{The states parametrization}
\label{sec:thestatesparametrization}

Owing to above analysis we have sufficient knowledge to find an easy recipe of the construction of different states of energy. Firstly, let us assume that we are looking for the ground state of energy for the problem with coupling constant $c>0$\footnote{It should be mentioned that in the case of the system with coupling constant $c<0$ the solutions of the Bethe equations have non-zero imaginary parts. Accordingly, the investigation of the ground state of energy, in the systems with attractive interactions, is not as straightforward as in the case of repulsive interactions $c>0$.}. From the \emph{Theorem 1} we know that the solutions $k_{j}$, of Bethe equations, are real numbers. Then
\begin{equation}
\displaystyle{ E_{\mathcal{N}}=\sum_{j=1}^{\mathcal{N}}k_{j}^{2} \geq 0, \, \, \, \, \, \, \, \, k_{j} \in \mathbb{R} \, \, \, \, \, \, \, \, \mathrm{for} \, \, \, c>0.}
		\label{param1}
\end{equation}

Now, if one consider an infinite interactions $c\rightarrow \infty$ then terms in (\ref{qmdescription58}) consist $\mathrm{arctan}\left( \frac{k_{j}-k_{s}}{c} \right)$ vanish. In this case Bethe equations are solvable analytically - each equation becomes the following equality
\begin{equation}
\displaystyle{k_{j}=\frac{2\pi}{\mathrm{L}}\mathcal{I}_{j}.}
		\label{param2}
\end{equation}
Therefore
\begin{equation}
\displaystyle{ \lim_{c\to \infty} E_{\mathcal{N}}=\frac{4\pi^{2}}{\mathrm{L}^{2}}\sum_{j=1}^{\mathcal{N}}\mathcal{I}_{j}^{2} .}
		\label{param3}
\end{equation}
Secondly, from the \emph{Theorem 3} and antisymmetry of wave function in $\left\{k_{j} \right\}$ we know that the solutions of Bethe equations must be different. Moreover, the solutions are uniquely parametrized by a set $\left\{\mathcal{I}_{j}  \right\}$ (\emph{Theorem 2}) and satisfy the relation $\mathcal{I}_{j}>\mathcal{I}_{s}$ ($\mathcal{I}_{j}=\mathcal{I}_{s}$) $\Longleftrightarrow$ $k_{j}>k_{s}$ ($k_{j}=k_{s}$) (\emph{Theorem 3}). If so, the problem of energy minimization is reduced to the problem of choosing a set of different integers (half-integers) which gives the lowest value of the sum presented in the equation (\ref{param3}). It is obvious that the following collection fulfills these conditions
\begin{equation}
\displaystyle{\mathcal{C}_{g}^{\mathcal{N}}=\left\{\begin{array}{ll} \left\{-\frac{\mathcal{N}-1}{2},-\frac{\mathcal{N}-3}{2}, \ldots, -1,0,1,\ldots , \frac{\mathcal{N}-3}{2},\frac{\mathcal{N}-1}{2} \right\}   &   \mathrm{if} \, \, \mathcal{N} \, \, \mathrm{is} \, \, \mathrm{odd} \, \,\\ \left\{-\frac{\mathcal{N}-1}{2},-\frac{\mathcal{N}-3}{2}, \ldots, -\frac{1}{2},\frac{1}{2},\ldots , \frac{\mathcal{N}-3}{2},\frac{\mathcal{N}-1}{2} \right\} &  \mathrm{if} \, \, \mathcal{N} \, \, \mathrm{is} \, \, \mathrm{even} \end{array} \right. },
		\label{param4}
		\end{equation}
where lower index $g$ means \emph{ground state} and the upper index informs us about the number of particles in the considered system. From obvious reasons, these considerations are valid also for the systems with finite coupling constant $c$. It is noteworthy that the ground state of energy has the momentum (\ref{qmdescription60}) equal to zero.

In general, one can solve Bethe equations (\ref{qmdescription58}), for an arbitrary coupling constant $c$, numerically. In the Figure 1.3., we present the solutions (quasi-momenta) as a functions of $c$ (for $\mathcal{N}=9$) in the case of ground state of energy given by the collection $\mathcal{C}_{g}^{9}=\{-4,-3,-2,-1,0,1,2,3,4\}$. We observe that, in the limit $c\rightarrow 0^{+}$, the solutions converge to zero. In order to write this property in formal way, we define solutions which are positive (negative) for $c>0$ as $k^{+}_{j}$ ($k^{-}_{s}$). Therefore
\begin{equation}
\displaystyle{\displaystyle{\lim_{c\to 0^{+}}k_{j}^{+}=0^{+}}, \, \, \, \, \, \, \, \, \displaystyle{\lim_{c\to 0^{+}}k_{s}^{-}=0^{-} }.  }
		\label{param5}
\end{equation}
Consequently, in the non-interacting limit $c\rightarrow 0^{+}$
\begin{equation}
\displaystyle{ \lim_{c\to 0^{+}} E_{\mathcal{N}}^{g}=0^{+} ,}
		\label{param6}
\end{equation}
where the upper index $g$ means \emph{ground state}. It should be mentioned that, if the energy of the ground state for quantum non-interacting system is equal to zero then all the momenta (not only quasi-momenta) of particles must be equal to 0. Due to the Heisenberg uncertainty principle, all the positions of particles must be totally blurred\footnote{In other words, the position of wave packet that describes an individual particle (atom) becomes completely uncertain.}. This situation corresponds to the Bose-Einstein condensation in $T=0\mathrm{K}$. Although, the Bose-Einstein condensation does not occur in one-dimensional systems, the ground state of a single particle system is macroscopically occupied in the limit $T\rightarrow 0\mathrm{K}$ (see Appendix \ref{sec:dodB})\footnote{Considering the case of periodic box (described in details in Appendix \ref{sec:dodB}), we see that the probability of finding $\mathcal{M}\leq \mathcal{N}$ particles in the ground state (\ref{prawdgrpundcensemblbox}) in the temperature $T = 0\mathrm{K}$ is equal to 0 if $\mathcal{M}< \mathcal{N}$ and equal to 1 if $\mathcal{M}= \mathcal{N}$. This result confirms the correctness of the limit (\ref{param6}). }.
\begin{center}\includegraphics[width=14cm,angle=0]{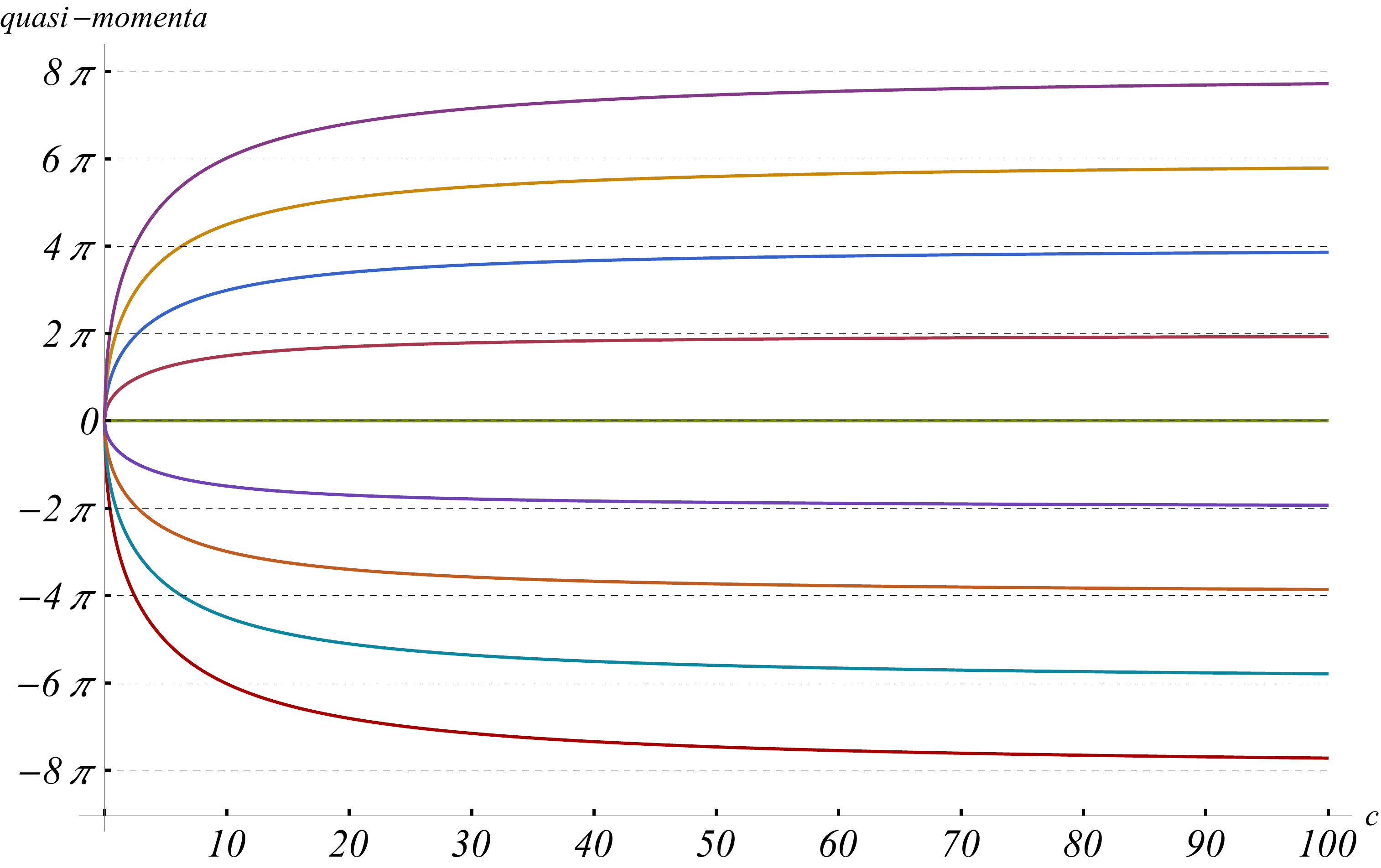}\end{center}
\begin{footnotesize}Figure 1.3. The solutions of Bethe equations as a functions of coupling constant $c$, for the ground state of energy which is given by the collection $\mathcal{C}_{g}^{9}=\{-4,-3,-2,-1,0,1,2,3,4\}$ ($\mathcal{N}=9$ particles). Length of the system $\mathrm{L}=1$ has been taken in calculations. We instantly notice that, as expected, values of quasi-momenta converge to the asymptotes defined by the relation $k_{j}=2\pi \mathcal{C}_{g}^{9}(j)$, where $\mathcal{C}_{g}^{9}(j)$ is the j-th element of the set $\mathcal{C}_{g}^{9}$\footnote{The order of elements in the set is established as the order from left to right. For instance: $\mathcal{C}_{g}^{9}(1)=-4$, $\mathcal{C}_{g}^{9}(3)=-2$, $\mathcal{C}_{g}^{9}(6)=1$, $\mathcal{C}_{g}^{9}(9)=4$.}. \end{footnotesize}

We already know how to parametrize the ground state. All the possible excitations can be described as a composition of two types of elementary excitations (the types are precisely discussed in chapter \ref{chap:DarkSolitonsolution}). Hence, it is sufficient to consider only the case of elementary excitations. However, we will focus not only on the process of elementary excitations, but we will also introduce the general notation.

Let us take the set $\mathcal{C}_{g}^{\mathcal{N}}$ which defines the ground state for $\mathcal{N}$ particle Lieb-Liniger system. To create elementary excitation one of the $\mathcal{N}$ elements of the collection $\mathcal{C}_{g}^{\mathcal{N}}$ has to be shifted beyond the Fermi surface\footnote{In our problem, the Fermi surface is defined by the maximal and minimal values of quasi-momenta (Fermi quasi-momenta) in the ground state of energy (uniquely given by $\mathcal{C}_{g}^{\mathcal{N}}$). These quasi-momenta correspond to the first and the last element of the set $\mathcal{C}_{g}^{\mathcal{N}}$. For example, in the case of $\mathcal{C}_{g}^{9}$, the absolute value of Fermi quasi-momentum $\left|k_{F}\right|$ corresponds to the number 4. Although, our considerations are based not on the momenta but on the quasi-momenta, we use the names Fermi surface and Fermi quasi-momenta, because of the character of the problem. In this context, the reader should treat the names as a mental shortcut.}. In Figure 1.4. we present five different elementary excitations in the case of $\mathcal{N}=9$ particles. 
\begin{center}\includegraphics[width=16cm,angle=0]{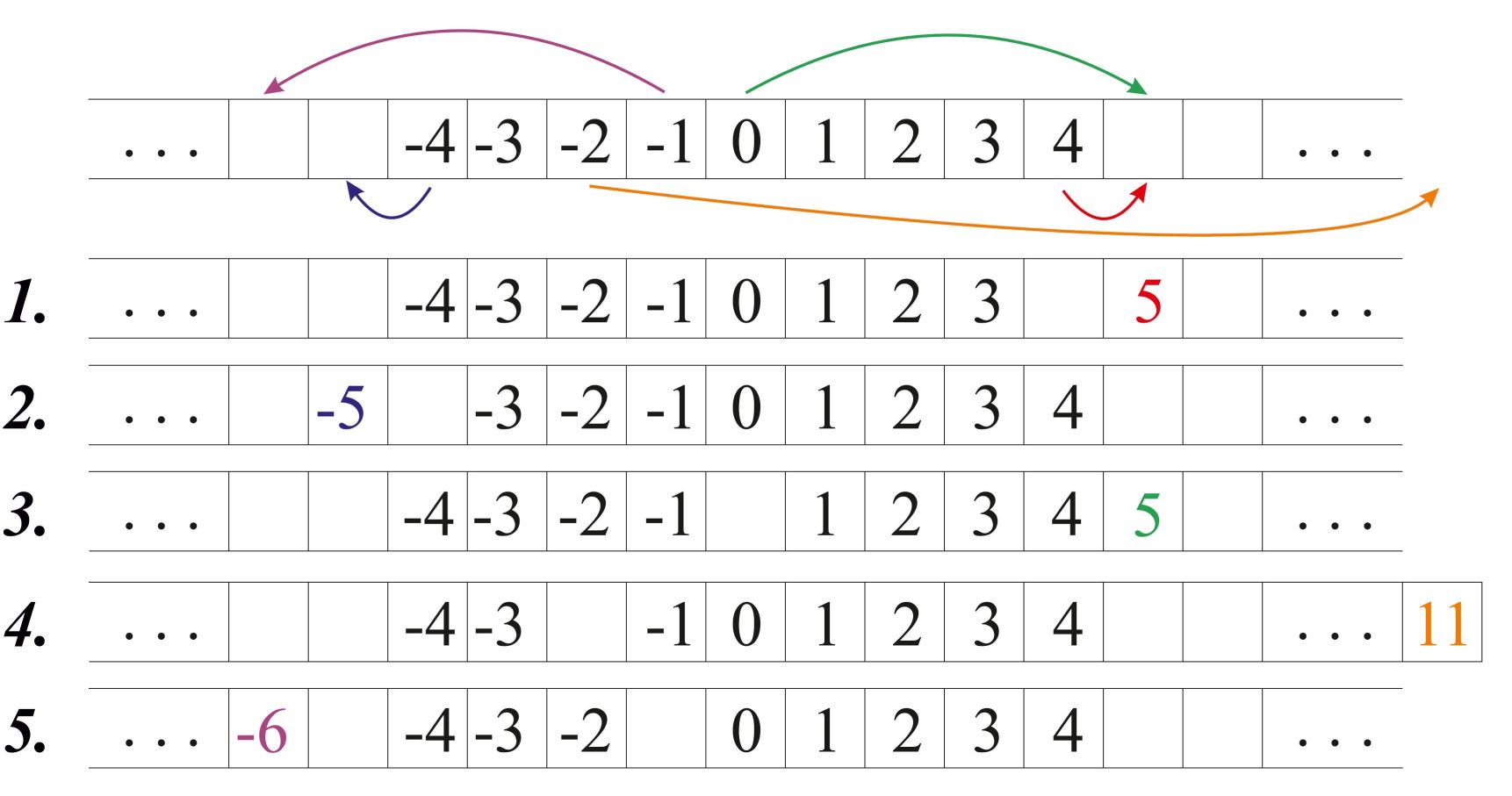}\end{center}
\begin{footnotesize}Figure 1.4. Five different elementary excitations in the integers representation for the case of 9-particle system. The starting point is the collection $\mathcal{C}_{g}^{9}$. Excitations are generated by replacing one of the components of $\mathcal{C}_{g}^{9}$ by the other, with a value not belonging to the set $\mathcal{C}_{g}^{9}$. We have selected the following elementary excitations: \textbf{\textsl{1.}} $(4\rightarrow 5)$, \textbf{\textsl{2.}} $(-4\rightarrow -5)$, \textbf{\textsl{3.}} $(0\rightarrow 5)$, \textbf{\textsl{4.}} $(-2\rightarrow 11)$, \textbf{\textsl{5.}} $(-1\rightarrow -6)$. 
\end{footnotesize} \\ \\
We have generated the following sets which parametrize elementary excitations in 9-particle Lieb-Liniger system
\begin{equation}
\displaystyle{\left\{\begin{array}{ll} \mathcal{C}_{e}^{9}[4\rightarrow 5]=\{-4,-3,-2,-1,0,1,2,3,5\},    \, \,\\ \mathcal{C}_{e}^{9}[-4\rightarrow -5]=\{-5,-3,-2,-1,0,1,2,3,5\},  \, \,\\ \mathcal{C}_{e}^{9}[0\rightarrow 5]=\{-4,-3,-2,-1,1,2,3,4,5\} ,      \, \,\\  \mathcal{C}_{e}^{9}[-2\rightarrow 11]=\{-4,-3,-1,0,1,2,3,4,11\},    \, \,\\   \mathcal{C}_{e}^{9}[-1\rightarrow -6]=\{-6,-4,-3,-2,0,1,2,3,4\},    \end{array} \right. }
		\label{param7}
		\end{equation}
where the lower indices $e$ means \emph{excited state}. In order to complement our notation, we have introduced new type of bracket $[\ldots]$ which tells us how many elementary excitations we performed and which element is (elements are) replaced and by what number (numbers). For instance, one can parametrize an excitation generated by 3 elementary excitations in 9-particle system\footnote{From the construction of the eigenstates in our model, it is obvious that the number of quasi-momenta is equal to the number of particles in the system $\mathcal{N}$.} by 
\begin{center}
$\displaystyle{\mathcal{C}_{e}^{9}[1\rightarrow 7, \, 3\rightarrow -11, \, -2\rightarrow 13]=\{-11,-4,-3,-1,0,2,4,7,13\}}$,
\end{center}
where we keep ascending order of elements adopted before. One instantly sees that above collection can be generated in 6 different equivalent ways. It is obvious because we can write the numbers from the collection $\{1,3,-2 \}$ in 6 different orders
\begin{center}
$\displaystyle{\{1,3,-2 \}, \, \, \{1,-2,3 \}, \, \, \{3,1,-2 \} , \, \, \{3,-2,1 \}, \, \, \{-2,3,1 \}, \, \, \{-2,1,3 \} }$.
\end{center}
In general, each parametrization of $\mathcal{M}$ elementary excitations can be generated in $\mathcal{M}!$ ways. Because these ways are equivalent it is convenient to use a shorter and more transparent notation\footnote{The only important information is contained in the final form of the collection.}: $\mathcal{C}_{e}^{\mathcal{N}}[\mathcal{M}]$, where $[\mathcal{M}\leq \mathcal{N}]$ means the excitation composed of $\mathcal{M}$ elementary excitations\footnote{Using this shorter notation we lose the information about the elements which are shifted but we gain a lot of space. In order to preserve the information about shifted elements one can write $[\{\ldots \}]$, where the collection $\{\ldots \}$ consists of shifted elements from $\mathcal{C}_{g}^{\mathcal{N}}$. It is clear that in this case, number of $\{\ldots \}$ components correspond to the number of elementary excitations.}. It should be mentioned that, in practice, the full information about the process of excitation is a non-issue. The most important thing is the form of the final state.   

The last thing we should discuss about the notation is how to write the $j$-th element of the collection corresponding to the $k$ elementary excitations. In general, considered element can be written as $\mathcal{C}_{e}^{\mathcal{N}}[k](j)$. For example, we present a few elements of the collection $\mathcal{C}_{e}^{9}[3]=\{-11,-4,-3,-1,0,2,4,7,13\}$
\begin{center}
$\displaystyle{\left. \begin{array}{ll} \mathcal{C}_{e}^{9}[3](2)=-4,    \, \,\\ \mathcal{C}_{e}^{9}[3](4)=-1 ,  \, \,\\ \mathcal{C}_{e}^{9}[3](5)=0     ,  \, \,\\  \mathcal{C}_{e}^{9}[3](7)=4   ,  \, \,\\   \mathcal{C}_{e}^{9}[3](9)=13  .   \end{array} \right. }$
\end{center}

At the end of this section let us analyze the degeneration of the excited states. For this purpose, let us consider first two excitations presented in the Figure 1.4. It is straightforward that, the character of this two excitations should be the same. The first collection (with the shift $4\rightarrow 5$) corresponds to the lowest possible excitation from the position of the highest, positive quasi-momentum in the ground state (Fermi quasi-momentum $k_{F}$). On the other hand, the second collection (with the shift $-4\rightarrow -5$) corresponds to the lowest possible excitation from the lowest, negative quasi-momentum in the ground state (Fermi quasi-momentum $-k_{F}$). Since, the energy functional (\ref{param1}) does not distinguish the signs of quasi-momenta, one expects that the energies of the states parametrized by the collections \textbf{\textsl{1.}} $(4\rightarrow 5)$ and \textbf{\textsl{2.}} $(-4\rightarrow -5)$ should be the same. Indeed, quasi-momenta obtained from numerical solutions of Bethe equations have the same absolute values in this two cases. It should be mentioned that these 2 states have opposite total momenta.

In general, if we have two different sets $\mathcal{C}^{\mathcal{N}}$ and $\mathcal{D}^{\mathcal{N}}$ parametrizing the solutions of Bethe equations, and if $\mathcal{C}^{\mathcal{N}}(j) = -\mathcal{D}^{\mathcal{N}}(j)$ for all $j$'s, then the energies of the states parametrized by $\mathcal{C}^{\mathcal{N}}$ and $\mathcal{D}^{\mathcal{N}}$ are exactly the same. Total momenta of these states are opposite - the absolute values are equal but the signs are opposite.

One should notice that, we cannot reverse above statement: the opposite total momenta of two different states does not imply the same values of their energies. The example is shown below ($\mathrm{L}=1$ and $c=1$ have been taken in calculations)
\begin{equation}
\displaystyle{\left\{\begin{array}{lll} \mathcal{C}_{e}^{9}[2]=\{-4,-3,-2,-1,0,1,3,5,8  \},  &   P_{9}^{\mathcal{C}}=14 \pi , &  E_{9}^{\mathcal{C}} =937.441 , \, \,\\ \mathcal{D}_{e}^{9}[2]=\{-5,-4,-3,-2,-1,0,1,2,5  \}, &  P_{9}^{\mathcal{D}}=-14 \pi , &  E_{9}^{\mathcal{D}} =438.415 . \end{array} \right. }
		\label{param8}
		\end{equation}

\section{The thermodynamic limit at zero temperature}
\label{sec:Thermodynamiclimit}

The thermodynamic limit is defined as follows
 \begin{equation}
\displaystyle{\left.\begin{array}{lll}    \displaystyle{\mathcal{N}\rightarrow \infty}, & \displaystyle{\mathrm{L}\rightarrow \infty} , &  \displaystyle{\rho=\frac{\mathcal{N}}{\mathrm{L}}=\mathrm{const}<\infty }.    \end{array} \right. }
		\label{thermlim}
		\end{equation}
We know that, in the Lieb-Liniger model, the state of the lowest energy in the case of fixed number of particles (located in the periodic box) at zero temperature is given by the solutions of the Bethe equations (\ref{qmdescription58}) parametrized by the collection (\ref{param4}). From the formula (\ref{qmdescription65}) we see that in the thermodynamic limit the separation of the solutions $k_{j}$'s is $k_{j+1}-k_{j}=\mathcal{O}(\mathrm{L}^{-1})$. The solutions fill the symmetric interval $[-\mathcal{Q},\mathcal{Q}]$, where 
 \begin{equation}
\displaystyle{\mathcal{Q}=\lim k_{\mathcal{N}}.}
		\label{thlimq}
		\end{equation}
The above limit means the thermodynamic limit. From the \emph{Theorem 3} one can define the density of states in the quasi-momentum space $D(k)$ (in the ground state) in the following way \cite{bogkorep}
\begin{equation}
\displaystyle{D(k_{j})=\lim \frac{1}{\mathrm{L}(k_{j+1}-k_{j})}>0.}
		\label{thlimrho}
		\end{equation}

In thermodynamic limit we see that, we can replace the difference of the scattering phase functions $\theta$ which appears in the equation (\ref{qmdescription65}) by the first term in a Taylor expansion (because $k_{j+1}-k_{j}$ is of the order $\mathcal{O}(\mathrm{L}^{-1})$) \cite{pethick}. Hence, 
\begin{equation}
\displaystyle{\mathrm{L}(k_{j+1}-k_{j})+(k_{j+1}-k_{j})\sum_{m=1}^{\mathcal{N}}\frac{\mathrm{d}}{\mathrm{d}k} \theta(k_{j}-k_{m})=2\pi ,}
		\label{thlim1}
		\end{equation}
where $\frac{\mathrm{d}}{\mathrm{d}k}\theta(k)=\frac{2c}{c^{2}+k^{2}}$. Because $\mathrm{L}D(k)\mathrm{d}k$ is the number of states in the interval $(k,k+\mathrm{d}k)$ and
\begin{equation}
\displaystyle{\frac{1}{\mathrm{L}}\sum_{m=1}^{\mathcal{N}}\frac{\mathrm{d}}{\mathrm{d}k} \theta(k_{j}-k_{m})=\frac{1}{\mathrm{L}}\sum_{m=1}^{\mathcal{N}}K(k_{j},k_{m})=\int \limits_{-\mathcal{Q}}^{\mathcal{Q}}K(k_{j},\mu)D(\mu)\mathrm{d}\mu,}
		\label{thlim3}
		\end{equation}
\begin{equation}
\displaystyle{\mathcal{N}=\mathrm{L} \int \limits_{-\mathcal{Q}}^{\mathcal{Q}}D(\mu)\mathrm{d}\mu,}
		\label{thlim4}
		\end{equation}
equation (\ref{thlim1}) may be rewritten as
\begin{equation}
\displaystyle{D(k)-\frac{1}{2\pi} \int \limits_{-\mathcal{Q}}^{\mathcal{Q}}K(k,\mu)D(\mu)\mathrm{d}\mu= \frac{1}{2\pi}.}
		\label{thlim2}
		\end{equation}
The formula (\ref{thlim2}) was firstly obtained by E.H. Lieb \cite{liebliniger}. It can be shown that there exists unique solution of (\ref{thlim2}) \cite{bogkorep}, \cite{liebliniger}. In the case of $c\rightarrow \infty$ 
 \begin{equation}
\displaystyle{\left.\begin{array}{ll}    \displaystyle{D(k)=\frac{1}{2\pi}}, & |k|\leq\mathcal{Q},  \, \,\\  \displaystyle{D(k)=0}, & |k|>\mathcal{Q} .    \end{array} \right. }
		\label{thlim5}
		\end{equation}

\section{Summary}
\label{sec:summary}

In the first chapter we have introduced the reader to the subject of one-dimensional system of bosons interacting via contact potential (Lieb-Liniger model). Firstly, we have presented the hamiltonian of the model in the second-quantized form (\ref{qmdescription3}). Secondly, we have rewritten our problem in the first-quantized form, also showing that, the Lieb-Liniger model given in the formalism of the quantum field theory can be presented as the problem of many-body quantum mechanics. It turns out that, the problem has an exact solution given by Bethe ansatz which has been shown step by step in section \ref{sec:ThecoordinateBetheansatz}. The solution and its properties were precisely discussed as well. 

The wave functions (\ref{qmdescription34}), (\ref{qmdescription36}), (\ref{qmdescription37}) and eigenvalues (\ref{qmdescription40}) contain parameters $k_{j}$ (quasi-momenta). At first glance it seems that, the quasi-momenta are exactly the same as the momenta of particles. In the section \ref{sec:quasimomenta}, using the 2-particle example, we have explained the differences and the similarities between the quasi-momenta and the momenta of particles.

In the section \ref{sec:periodicconditions}, imposing periodic boundary conditions, we have derived the Bethe equations (\ref{qmdescription58}). Moreover, the section contains three extremely important theorems concerning the solutions of Bethe equations. The importance of the theorems emerges when we investigate the parametrization of the ground and the excited states of energy (section \ref{sec:thestatesparametrization}). It turns out that, one can easily parametrize arbitrary state of energy by the collection of integers (half-integers)\footnote{At this point, it should be noted that, we have restricted our discussion to the case of repulsive interactions ($c>0$).} for odd (even) number of particles.

The first chapter was complemented with a short discussion about the thermodynamic limit in the considered model (section \ref{sec:Thermodynamiclimit}). The discussion will be repeatedly used in the next chapter.

\chapter{Dark Soliton solution}
\label{chap:DarkSolitonsolution}

\section{Simple classification of excitations}
\label{sec:Excitations}

In the first chapter we discussed exact solutions and the nature of eigenstates of the Lieb-Liniger model. Now, we will present the two types of excitations in this system. It is convenient to regard an excitation spectrum as a double spectrum of elementary boson excitations. One of the spectra is given by the Bogoliubov's perturbation theory (quite acurrately for weak interactions) \cite{lieb2}. The second spectrum exists only for $|p|\leq \pi \rho$, where $p$ and $\rho=\mathcal{N}/\mathrm{L}$ are momentum and density of the system.

In the section \ref{sec:thestatesparametrization} we presented how to easily parametrize arbitrary excited state. Let us consider again elementary excitation from the state $q$ ($|q|<k_{F}$) to a state $\kappa$ ($|\kappa|>k_{F}$). For convenience, we choose the case of Tonks-Girardeau limit\footnote{In the Tonks-Girardeau limit $\gamma=\frac{c}{\rho}\to \infty$ the solutions of the Bethe equations take an easy form $k_{j}=\frac{2\pi}{\mathrm{L}}\mathcal{I}_{j}$.} $\gamma=\frac{c}{\rho}\to \infty$ \cite{girardeau}, \cite{lieb2}, \cite{pethick}. It is obvious that in this case, the energy and momentum of this excitation have the following forms
\begin{equation}
\displaystyle{\left.\begin{array}{ll}    \displaystyle{\epsilon=\kappa^{2}-q^{2}},  \, \,\\  \displaystyle{p=\kappa-q }. \end{array} \right. }
		\label{exc1}
\end{equation}
One instantly sees that, every excitation is given by two parameters. Therefore, there is no unique curve $\epsilon(p)$. If we consider only low-lying excitations the problem of elementary excitations classification can be divided into two types:

\textbf{\emph{Type I}}: excitation from $q=k_{F}$ to $\kappa>k_{F}$ (or $q=-k_{F}$ to $\kappa<-k_{F}$). This type of excitations corresponds to the following values of energy and momentum 
\begin{equation}
\displaystyle{\left.\begin{array}{ll}    \displaystyle{\epsilon_{1}=\kappa^{2}-k_{F}^{2}},  \, \,\\  \displaystyle{p=\kappa-k_{F} }, & \kappa>k_{F} ,  \, \,\\   \displaystyle{p=\kappa+k_{F} }, & \kappa<-k_{F} ,    \end{array} \right. }
		\label{excI1}
\end{equation}
\begin{equation}
\displaystyle{\epsilon_{1}(p)=p^{2}+2k_{F}|p| }.
		\label{excI2}
\end{equation}

\textbf{\emph{Type II}}: excitation from $0<q<k_{F}$ to $\kappa=k_{F}+2 \pi / \mathrm{L}$ (or $-k_{F}<q<0$ to $\kappa=-k_{F}-2 \pi / \mathrm{L}$). In this case the energy and momentum of excitation are given by
\begin{equation}
\displaystyle{\left.\begin{array}{ll}    \displaystyle{\epsilon_{2}=(k_{F}+2 \pi / \mathrm{L})^{2}-q^{2}},  \, \,\\  \displaystyle{p=k_{F}+2 \pi / \mathrm{L} -q }, & 0<q<k_{F} ,  \, \,\\   \displaystyle{p=-k_{F}-2\pi /\mathrm{L}-q }, & 0>q>-k_{F} ,    \end{array} \right. }
		\label{excII1}
\end{equation}
\begin{equation}
\displaystyle{\epsilon_{2}(p)=2k_{F}|p|-p^{2}+\frac{4|p| \pi}{\mathrm{L}} }.
		\label{excII2}
\end{equation}
In the system with large number of particles $k_{F}+2\pi/\mathrm{L}\approx k_{F}$ and $k_{F}=\pi \rho$ (cause $k_{F}=\frac{2\pi}{\mathrm{L}}\frac{\mathcal{N}-1}{2}$), then 
\begin{equation}
\displaystyle{\epsilon_{1}(p)=p^{2}+2\pi \rho |p| },
		\label{excI3}
\end{equation}
\begin{equation}
\displaystyle{\epsilon_{2}(p)=2\pi \rho |p|-p^{2} }.
		\label{excII3}
\end{equation}

In the case of the \emph{type I} excitation  $-\infty< p< \infty$, while for the \emph{type II} $-\pi \rho \leq p \leq \pi \rho$. In order to supplement these excitations we add two other (the so-called \emph{umklapp} excitations): from $q=-k_{F}$ to $\kappa=k_{F}+2\pi/\mathrm{L}$ and $q=k_{F}$ to $\kappa=-k_{F}-2\pi/\mathrm{L}$ which energies and momenta are $4\pi^{2}\rho/\mathrm{L}$ and $\pm 2\pi \rho$, respectively \cite{lieb2}.

One can carry out these excitations arbitrary number of times getting as many types of different momenta as we want. For instance, let us consider excitation of the type I of momentum $p$ given by a shift from $q=k_{F}$ to $\kappa=k_{F}+p$. The momentum and energy of this excitation are equal to $p$ and $\epsilon_{1}(p)=2k_{F}p+p^{2}$. Two excitations of the momentum $p$ means two shifts of two quasiparticles: from $q_{1}=k_{F}, \, \, q_{2}=k_{F}-2\pi/\mathrm{L}$ to $\kappa_{1}=k_{F}+p, \, \, \kappa_{2}=k_{F}-2\pi/\mathrm{L}+p$. The momentum of the final state is $2p$ while the energy is equal to $2\epsilon_{1}(p)-4p \pi /L \approx 2\epsilon_{1}(p)$. Hence, one can treat each of these elementary excitations as bosons \cite{lieb2}.

It turns out that, any combination of multiple excitations of the type I (or II) is in one to one correspondence with exactly one true state of the system which momentum is exactly equal to the sum of the momenta of the elementary excitations. The energy of the true state is the sum of the energies of the elementary excitations plus the terms of order $\mathcal{O}(\mathcal{N}^{-1})$ (if the order of excitations is less than $\mathcal{N}$) \cite{lieb2}.

\section{General analysis of the type I excitations - ,,particle'' states}
\label{sec:ExcitationsI}

Let us now consider the situation where our \emph{initial} state is parametrized by the collection ($\mathcal{N}$-particle ground state)
\begin{equation}
\displaystyle{\mathcal{C}_{g}^{\mathcal{N}}=\left \{-\frac{\mathcal{N}-1}{2},-\frac{\mathcal{N}-3}{2}, \ldots,   \frac{\mathcal{N}-1}{2}  \right \} }.
		\label{Iscol1}
\end{equation}
 Now, we create an excitation by adding one particle ($\mathcal{N}\rightarrow \mathcal{N}+1 $ particles) with, for instance, positive momentum \cite{notes}. Now, the collection which parametrizes the state is given by 
\begin{equation}
\displaystyle{\mathcal{D}^{\mathcal{N}+1}_{1}=\left \{-\frac{\mathcal{N}}{2},-\frac{\mathcal{N}-2}{2}, \ldots,   \frac{\mathcal{N}}{2}, \frac{\mathcal{N}}{2}+\mathcal{M} \right\} },
		\label{Iscol2}
\end{equation}
where $\mathcal{M}$ is positive (half-)integer number. In this case, the total momentum of the system is equal to
\begin{equation}
\displaystyle{p=\frac{2\pi}{\mathrm{L}}\mathcal{M} },
		\label{Imom}
\end{equation}
By inserting a particle to the system, we caused the change of the values of all the quasi-momenta: if in the \emph{initial} state we have the following quasi-momenta $\{k_{1},\ldots, k_{\mathcal{N}}   \}$, then in the \emph{final} (excited) state we have $\{k_{1}',\ldots, k_{\mathcal{N}}',q  \}$. It should be mentioned that $q\neq p$. 

At this point, we want to calculate the following difference $\Delta k_{j}=k'_{j}-k_{j}$, which will inform us about the excitation influence on the value of single quasi-momentum. For this purpose let us subtract the Bethe equations (\ref{qmdescription58}) for the initial and final states
\begin{equation}
\displaystyle{\Delta k_{j}\mathrm{L}+\theta(k_{j}'-q)+\sum_{m=1}^{\mathcal{N}}\left[\theta(k'_{j}-k'_{m})-\theta(k_{j}-k_{m})  \right]=2\pi \left(\mathcal{D}^{\mathcal{N}+1}_{1}(j)-\mathcal{C}_{g}^{\mathcal{N}}(j)\right)=-\pi },
		\label{Ideltaa}
\end{equation}
therefore
\begin{equation}
\displaystyle{\Delta k_{j}\mathrm{L}=-\pi-\theta(k_{j}'-q)-\sum_{m=1}^{\mathcal{N}}\left[\theta(k'_{j}-k'_{m})-\theta(k_{j}-k_{m})  \right] }.
		\label{Idelta}
\end{equation}
The difference $\Delta k_{j}$ is of the order $\mathcal{O}(\mathrm{L}^{-1})$, then we can expand $\theta$ function to the same order on the left and right hand side in (\ref{Idelta})
\begin{equation}
\displaystyle{\theta(k'_{j}-k'_{m})=\theta(k_{j}-k_{m})+ \left[ \Delta k_{j}-\Delta k_{m}\right]K(k_{j},k_{m})}.
		\label{Itheta}
\end{equation}
Hence
\begin{equation}
\displaystyle{\Delta k_{j}\mathrm{L}=-\pi- \theta(k_{j}-q)-\sum_{m=1}^{\mathcal{N}}K(k_{j},k_{m})  \left[ \Delta k_{j}-\Delta k_{m}\right] }.
		\label{Idelta2}
\end{equation}
Collecting the terms as follows
\begin{equation}
\displaystyle{\Delta k_{j}\left(1+\frac{1}{\mathrm{L}} \sum_{m=1}^{\mathcal{N}}K(k_{j},k_{m}) \right)=-\frac{1}{\mathrm{L}}\left[\pi+ \theta(k_{j}-q)\right]+\frac{1}{\mathrm{L}}\sum_{m=1}^{\mathcal{N}}K(k_{j},k_{m}) \Delta k_{m} },
		\label{Idelta3}
\end{equation}
and using the equations (\ref{thlim3}) and (\ref{thlim2}), we can obtain the following integral equation in the thermodynamic limit  
\begin{equation}
\displaystyle{2\pi \Delta k D(k) \mathrm{L}=-\pi- \theta(k-q)+ \mathrm{L} \int \limits_{-\mathcal{Q}}^{\mathcal{Q}}K(k,\mu)D(\mu)\Delta \mu\mathrm{d}\mu},
		\label{Iint}
\end{equation}
or equivalently
\begin{equation}
\displaystyle{\Delta k \mathrm{L} =-\pi- \theta(k-q)+\mathrm{L} \int \limits_{-\mathcal{Q}}^{\mathcal{Q}}K(k,\mu)D(\mu)\left[\Delta \mu -\Delta k\right]\mathrm{d}\mu}.
		\label{Iint2}
\end{equation}
Defining now \cite{lieb2}
\begin{equation}
\displaystyle{k'_{j}=k_{j}+\frac{\omega_{j}}{\mathrm{L}}, \, \, \, \, \, \, \, \, \Delta k_{j}=\frac{\omega_{j}}{\mathrm{L}}, \, \, \, \, \, \, \, \, \Delta k=\frac{\omega(k)}{\mathrm{L}} },
		\label{Isol}
\end{equation}
we may rewrite equations (\ref{Iint}) and (\ref{Iint2}) in the following way
\begin{equation}
\displaystyle{2\pi \omega(k) D(k) =-\pi- \theta(k-q)+  \int \limits_{-\mathcal{Q}}^{\mathcal{Q}}K(k,\mu)D(\mu)\omega (\mu) \mathrm{d}\mu}.
		\label{Iint3}
\end{equation}
\begin{equation}
\displaystyle{\omega(k) =-\pi- \theta(k-q)+\int \limits_{-\mathcal{Q}}^{\mathcal{Q}}K(k,\mu)D(\mu)\left[\omega(\mu) -\omega(k)\right]\mathrm{d}\mu}.
		\label{Iint4}
\end{equation}
If we define a new quantity 
\begin{equation}
\displaystyle{\mathcal{J}(k)\equiv \omega(k) D(k)},
		\label{IJot}
\end{equation}
then the formula (\ref{Iint3}) takes the compact form 
\begin{equation}
\displaystyle{2\pi \mathcal{J}(k) =-\pi- \theta(k-q)+  \int \limits_{-\mathcal{Q}}^{\mathcal{Q}}K(k,\mu)\mathcal{J}(\mu) \mathrm{d}\mu}.
		\label{Iint5}
\end{equation}

The momentum of the final state is equal to
\begin{equation}
\displaystyle{p=\sum_{m=1}^{\mathcal{N}}k'_{m}+q=\sum_{m=1}^{\mathcal{N}}\left(k_{m}+\frac{\omega_{m}}{\mathrm{L}}\right)+q=q+\int \limits_{-\mathcal{Q}}^{\mathcal{Q}}\mathcal{J}(\mu) \mathrm{d}\mu},
		\label{Imomentum}
\end{equation}
because
\begin{equation}
\displaystyle{\sum_{m=1}^{\mathcal{N}}k_{m}=0, \, \, \, \, \, \, \, \, \, \, \, \,  \sum_{m=1}^{\mathcal{N}}\frac{\omega_{m}}{\mathrm{L}}=\int \limits_{-\mathcal{Q}}^{\mathcal{Q}}\mathcal{J}(\mu) \mathrm{d}\mu}.
		\label{Imomentumcause}
\end{equation}
In the same way we can find the energy of the considered elementary excitation
\begin{flushleft}
$\displaystyle{\epsilon_{1}=\sum_{m=1}^{\mathcal{N}} k_{m}'^{2}+q^{2}-E_{0}^{\mathcal{N}+1}=\sum_{m=1}^{\mathcal{N}}\left(k_{m}'^{2}-k_{m}^{2}\right)+q^{2}-\mu_{ch} = }$
\end{flushleft}
\begin{equation}
\displaystyle{=\sum_{m=1}^{\mathcal{N}} \left( 2k_{m} \Delta k_{m}+(\Delta k_{m})^{2} \right)+q^{2}-\mu_{ch}=-\mu_{ch}+q^{2}+2 \int \limits_{-\mathcal{Q}}^{\mathcal{Q}} \mu \mathcal{J}(\mu) \mathrm{d}\mu},
		\label{Iexcenergy}
\end{equation}
where $E_{0}^{\mathcal{N}+1}$ is the energy of the $\mathcal{N}+1$-particle ground state 
\begin{equation}
\displaystyle{E_{0}^{\mathcal{N}+1}=\sum_{m=1}^{\mathcal{N}}k_{m}^{2}+\mu_{ch}  }.
		\label{Iexcenergycause}
\end{equation}
The quantity $\mu_{ch}$ is the chemical potential. It should be mentioned that, we have neglected the term $(\Delta k_{m})^{2}$ because it vanishes in the thermodynamic limit. 

Because the quantity $\pi +\theta(k-q)$ is positive definite, the formula (\ref{Iint5}) has a negative unique solution \cite{lieb2}, \cite{liebliniger}. By a definition (\ref{IJot}) one notices that since $\mathcal{J}(k)<0$ and $D(k)>0$, $\omega(k)<0$ \cite{lieb2}. Therefore, inserting the particle to the system (associated with quasi-momentum $q>k_{F}$, where $k_{F}$ is the Fermi quasi-momentum of the $\mathcal{N}$-particle system), decreases all quasi-momenta associated with the initial state.

To obtain the dispersion relation $\epsilon_{1}(p)$ one has to solve integral equation (\ref{Iint5}) which is the Fredholm equation of the second kind. It can be done numerically, for example by the method of Neumann series \cite{integraleq}. Let us now consider the limits $\gamma=0$ (free particles) and $\gamma=\infty$ (the second case, the Tonks-Girardeau limit, was discussed in the previous section) 
\begin{equation}
\displaystyle{\left.\begin{array}{lll}    \displaystyle{K(k,q)=\frac{2\rho \gamma}{\rho^{2}\gamma^{2}+(k-\mu)^{2}}},   & \displaystyle{ \mathop{  \lim_{\gamma \to 0}}_{k\neq \mu} K(k,q)=0},  \, \,\\  \displaystyle{\theta(k-q)=2\mathrm{arctan}\left( \frac{k-q}{\rho \gamma} \right) } , & \displaystyle{\lim_{\gamma \to 0}\theta(k-q)\stackrel{q>k}{=}-\pi } ,    \end{array} \right. }
		\label{Ilimits1}
\end{equation}
hence
\begin{equation}
\displaystyle{\left.\begin{array}{lll}   \displaystyle{  \lim_{\gamma \to 0}\epsilon_{1}(p)=-\mu_{ch}+p^{2}} ,   & \displaystyle{  \lim_{\gamma \to \infty}\epsilon_{1}(p)=-\mu_{ch}+p^{2}+2\pi \rho |p|}.     \end{array} \right. }
		\label{Ilimits2}
\end{equation}

It can be also shown that for $\gamma \neq 0$ and for large $p$ the dispersion relation may be obtained by Laurent series expansion of $\theta(k-q)$ for large $q$. In this case \cite{lieb2} 
\begin{equation}
\displaystyle{\left.\begin{array}{lll}   \displaystyle{  \mathcal{J}(k)\approx -\frac{2\gamma \rho}{q}D(k)} ,   & \displaystyle{ p\approx q-\frac{2\gamma \rho^{2}}{q}},     \end{array} \right. }
		\label{Ilimits3}
\end{equation}
\begin{equation}
\displaystyle{\epsilon_{1}(p)\approx  -\mu_{ch}+p^{2}+4\gamma \rho^{2}, \, \, \, \, \, \, \, \, \frac{p}{\rho}>>\gamma<\infty}.
		\label{Ilimits4}
\end{equation}
Below we present the results of numerical calculations from \cite{lieb2}
\begin{center}\includegraphics[width=17cm,angle=0]{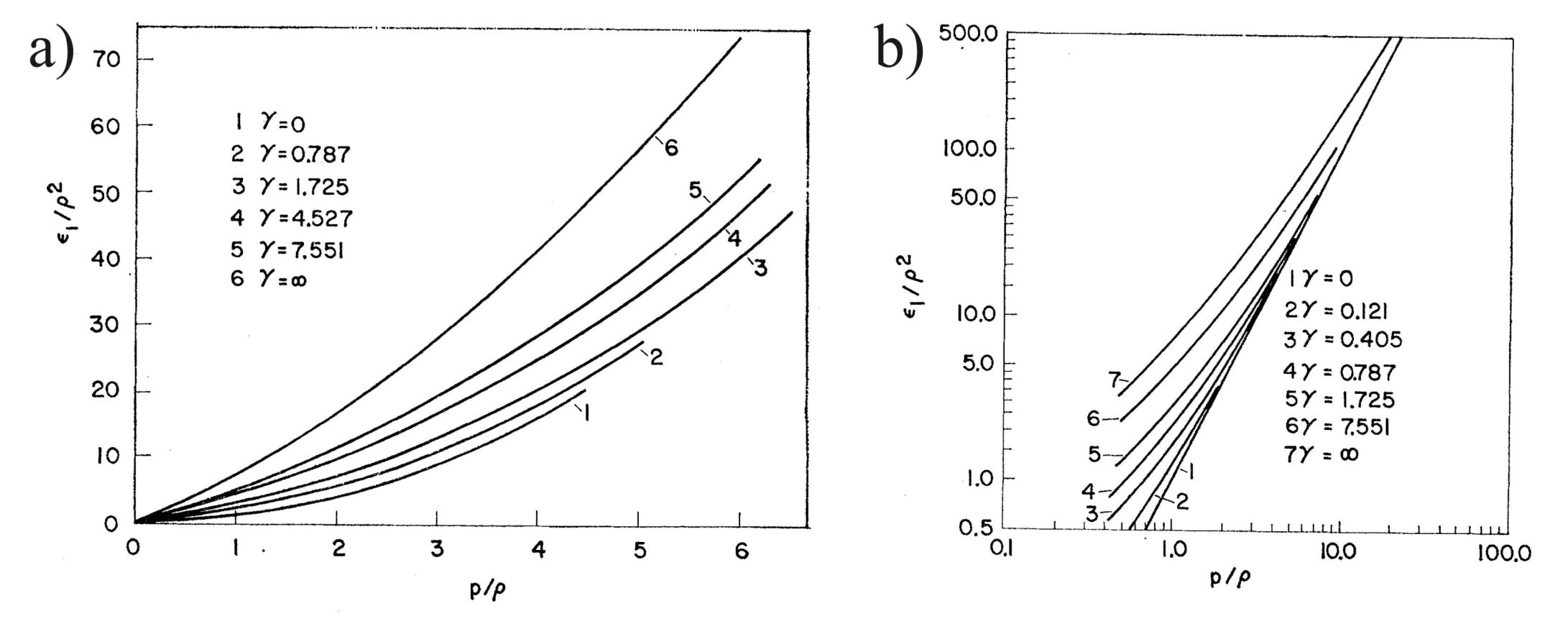}\end{center}
\begin{flushleft}
\footnotesize{Figure 2.1. The dispersion relation for different values of $\gamma=c/\rho$ in the case of the type I excitation. Plot a) presents the case of small momenta. Excitation spectrum for large momenta is depicted in plot b) \cite{lieb2}.}
\end{flushleft}

The insertion of a particle to the system corresponds to the type I excitation. The excitation can be identified with the Bogoliubov excitation - a sound wave \cite{notes}, \cite{ishikawa} (see also the appendix \ref{sec:Bogoliubovspectrum}).

\section{General analysis of the type II excitations - ,,hole'' states}
\label{sec:ExcitationsII}

Instead of inserting a particle to the system, let us now remove a particle from the system in the ground state. In this reversed problem, we create a \emph{hole} changing the $\mathcal{N}$-particle problem to the $\mathcal{N}-1$-particle problem \cite{notes}. Corresponding collection parametrizing our new final state is  
\begin{equation}
\displaystyle{\mathcal{D}^{\mathcal{N}-1}_{2}=\left \{-\frac{\mathcal{N}-2}{2},-\frac{\mathcal{N}-4}{2}, \ldots,-\frac{\mathcal{N}-2-2\mathcal{M}}{2},-\frac{\mathcal{N}+2-2\mathcal{M}}{2},\ldots   ,\frac{\mathcal{N}-2}{2} \right\} }.
		\label{IIcol}
\end{equation}
The \emph{hole} causes the following change of momentum 
\begin{equation}
\displaystyle{p=\frac{2\pi}{\mathrm{L}}\mathcal{M} }.
		\label{IImom}
\end{equation}
By the same algebraic manipulation as before (see previous section) one can easily derive the following equations 
\begin{equation}
\displaystyle{2\pi \mathcal{J}(k) =\pi+\theta(k-q)+  \int \limits_{-\mathcal{Q}}^{\mathcal{Q}}K(k,\mu)\mathcal{J}(\mu) \mathrm{d}\mu},
		\label{IIint}
\end{equation}
\begin{equation}
\displaystyle{p=-q+\int \limits_{-\mathcal{Q}}^{\mathcal{Q}}\mathcal{J}(\mu) \mathrm{d}\mu},
		\label{IImomentum}
\end{equation}
\begin{equation}
\displaystyle{\epsilon_{2}=\mu_{ch}-q^{2}+2 \int \limits_{-\mathcal{Q}}^{\mathcal{Q}} \mu \mathcal{J}(\mu) \mathrm{d}\mu},
		\label{IIexcenergy}
\end{equation}
where $q<k_{F}$ is the quasi-momentum of the hole ($k_{F}$ is the Fermi quasi-momentum before removing the particle).

One notes that $\mathcal{J}(k)$ is positive definite (here we can consider the case of $\gamma=\infty$). In the limit $q=k_{F}=\mathcal{Q}$ quantities $p$ and $\epsilon_{2}$ vanish. Both $p$ and $\epsilon_{2}$ increase when $q$ decreases. The limit $q=0$ corresponds to the momentum $p$ equal to $\pi \rho$ \cite{lieb2}. It is noteworthy that the slopes of both dispersion relations $\epsilon_{1}(p)$ and $\epsilon_{2}(p)$ are equal at $p=0$.   

In the Figure 2.2. we present dispersion relation of the type II excitation (for different values of $\gamma$ parameter) and its comparison with the energy spectrum in the case of the type I excitation obtained by E. H. Lieb in \cite{lieb2}
\begin{center}\includegraphics[width=15cm,angle=0]{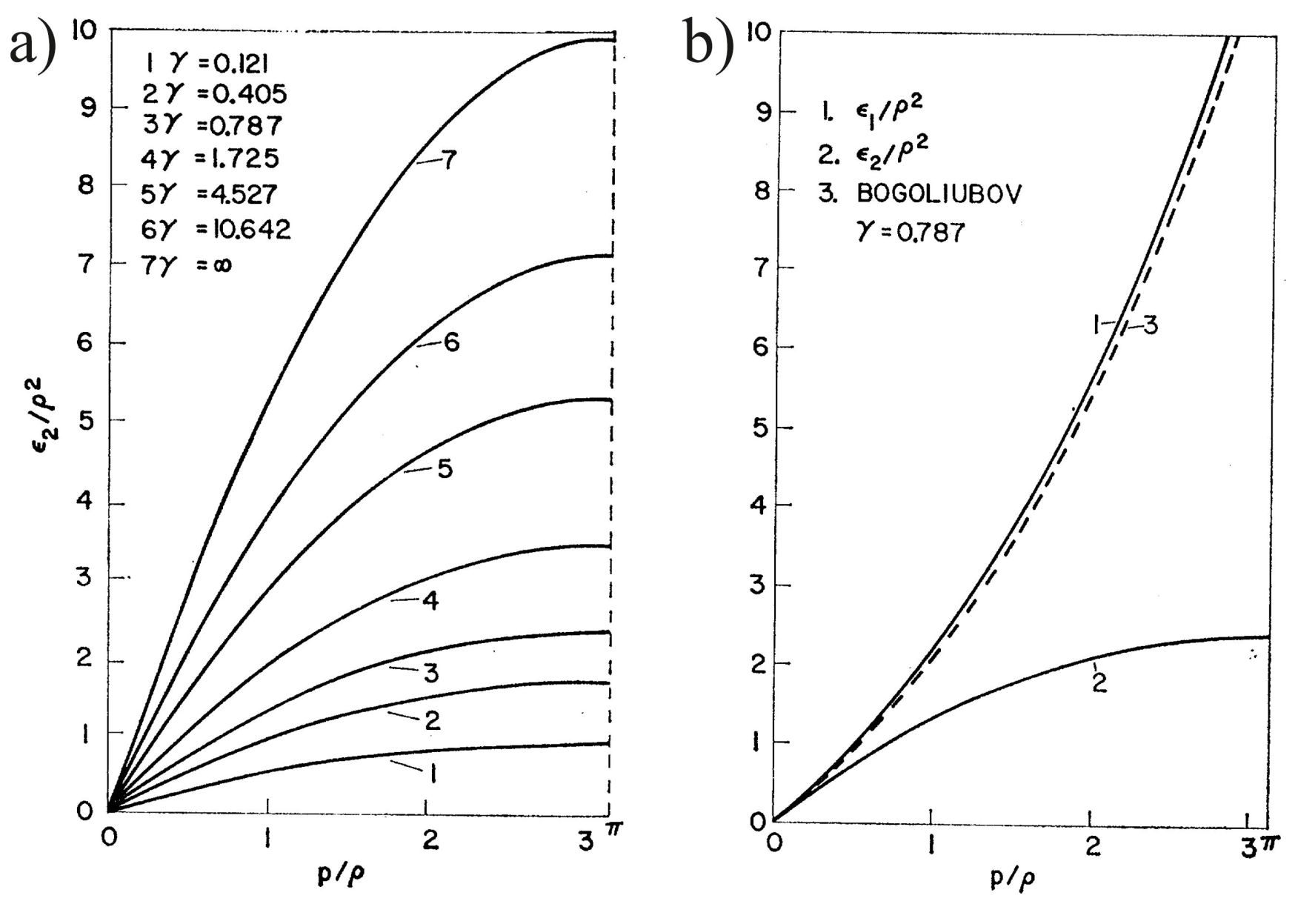}\end{center}
\begin{footnotesize}Figure 2.2. Plot a) presents the type II excitation spectrum (which exists up to $|p|=\rho \pi$) for different values of $\gamma$ parameter. Two limiting cases: $\gamma=0$ and $\gamma=\infty$ correspond to $\epsilon_{2}(p)=0$ and $\epsilon_{2}(p)=2\pi \rho |p|-p^{2}$, respectively. The comparision of the types I (curve number 1), II (curve number 2) and Bogoliubov (dashed curve number 3) excitation spectra for $\gamma=0.787$ are presented in the plot b). The Bogoliubov's spectrum almost perfectly coincides with the curve of the type I excitation \cite{lieb2}. \end{footnotesize} \\

The \emph{hole} creation corresponds to the type II excitation. It is believed that the nature of these excitations is solitonic (these are not sound waves) \cite{notes}, \cite{ishikawa}. It can be shown that dispersion relations for the type II excitations and dark soliton dispersion relation in semi-classical limit agree very well \cite{ishikawa}. Increasing the interaction $c$ we are breaking down the limit of weak interactions. In this case we can ask if the considered excitations still have the solitonic nature (we try to answer the question in the chapter \ref{numericalresults}). As for now, the analysis of the Lieb's type II excitations were mainly based on the semi-classical results but it has not been shown that dark soliton density profiles reveals during the particle detection process. The main purpose of our work is to check the emergence of dark solitons during the measurement of particles positions in the system prepared in type II eigenstates using the computer simulations (chapters \ref{chap:Theideaofnumericalcalculations} and \ref{numericalresults}) \cite{SyrwSacha}.

\section{Solitons in the Lieb-Liniger model}
\label{sec:solitons}

In the first chapter we have considered the closed system in the sense that it cannot exchange particles with reservoir. Let us now expand these considerations to the grand canonical ensemble. In this case the hamiltonian operator takes the form
\begin{equation}
\displaystyle{\hat{\mathrm{H}}_{\mu_{ch}}=\hat{\mathrm{H}}-\mu_{ch}\hat{\mathrm{N}}},
		\label{qhamiltoniangrand}
		\end{equation}
where the operators $\hat{\mathrm{H}}$ and $\hat{\mathrm{N}}$ are given by (\ref{qmdescription3}) and (\ref{qmdescription6}), respectively. In this approach the Non-linear Schr\"{o}dinger (NLS) equation (\ref{qmdescription4}) is given by
	\begin{equation}
\displaystyle{i\partial_{t}\hat{\Psi}_{H}(z,t)=-\partial_{z}^{2}\hat{\Psi}_{H}(z,t)-\mu_{ch}\hat{\Psi}_{H}(z,t)+2c\hat{\Psi}_{H}^{\dagger}(z,t)\hat{\Psi}_{H}(z,t)\hat{\Psi}_{H}(z,t)}.
		\label{nlsgrand}
		\end{equation}

Although, there is no Bose-Einstein condensation in the one-dimensional interacting Bose systems, in the limit of weak interactions almost all particles occupy the ground state of single particle system (has nearly zero momenta, see appendix \ref{sec:dodB}). In this limit the quantum Bose field operator $\hat{\Psi}_{H}(z,t)$ can  be treated semi-classically ($\hat{\Psi}_{H}(z,t)\rightarrow \Psi(z,t)$) then the formula (\ref{nlsgrand}) takes the form of the Gross-Pitaevskii equation \cite{tsuzuki}, \cite{ishikawa}
	\begin{equation}
\displaystyle{i\partial_{t}\Psi(z,t)=-\partial_{z}^{2}\Psi(z,t)-\mu_{ch}\Psi(z,t)+2c|\Psi(z,t)|^{2}\Psi(z,t)}.
		\label{gross}
		\end{equation}
T. Tsuzuki showed that above equation has one-soliton (dark soliton) solution \cite{tsuzuki}, \cite{ishikawa}. It should be mentioned that $N$-soliton solution was found afterwards by V. E. Zakharov and A. B. Shabat \cite{zakharov}, \cite{ishikawa}. The dark soliton solution, obtained by T. Tsuzuki may be written as \cite{ishikawa}
	\begin{flushleft}
	$\displaystyle{\Psi^{cl}_{s}(z,t)=\sqrt{\frac{\mu_{ch}}{2c}\left(1 -\beta \,   \mathrm{cosh}^{-2} \left[ \sqrt{\frac{\beta \mu_{ch}}{2}}(z-v t)   \right]     \right)} }$
	\end{flushleft}
	\begin{equation}
\displaystyle{\times \mathrm{exp}\left[  \pm i \,  \mathrm{arcsin} \left( \frac{ \sqrt{\beta} \,  \mathrm{tanh} \left[ \sqrt{\frac{\beta \mu_{ch}}{2}}(z-v t)   \right]  }{ \sqrt{ 1-\beta \,  \mathrm{cosh}^{-2} \left[ \sqrt{\frac{\beta \mu_{ch}}{2}}(z-v t)   \right] }  }     \right)      \right]   },
		\label{darksolitontsuzuki}
		\end{equation}
where $v$ is the velocity of soliton and 
	\begin{equation}
\displaystyle{\beta=1-\frac{v^{2}}{v_{s}^{2}}, \, \, \, \, \, \, \, \, \, \, \, \, v_{s}\equiv\sqrt{2\mu_{ch}}, \, \, \, \, \, \, \, \, \, \, \, \, s=\frac{v}{v_{s}} }.
		\label{beta}
		\end{equation}
We instantly notice that the soliton solution disappears when $|v|\geq v_{s}$, where $v_{s}$ is the speed of sound\footnote{Comparing the equation (\ref{Bogoliubovspectrum8}) with our units where $m=\frac{1}{2}$ it is clear that $v_{s}=\sqrt{2\rho g_{0}}=\sqrt{2\mu_{ch}}$.} in our system. One can also easily check the asymptotic forms of $\Psi^{cl}_{s}(z,t)$ 
\begin{equation}
\displaystyle{\left.\begin{array}{ll}    \displaystyle{\Psi^{cl}_{s}(z,t)\rightarrow \sqrt{\frac{\mu_{ch}}{2c}} \,  \mathrm{exp}\left[  \pm i \,  \mathrm{arcsin} \left( \sqrt{\beta} \right)\right] }, & z\rightarrow \infty,  \, \,\\   \displaystyle{\Psi^{cl}_{s}(z,t)\rightarrow \sqrt{\frac{\mu_{ch}}{2c}} \mathrm{exp} \, \left[  \mp i \,  \mathrm{arcsin} \left( \sqrt{\beta} \right)\right] }, & z\rightarrow -\infty.   \end{array} \right. }
		\label{solitonlim}
\end{equation}
\begin{center}\includegraphics[width=17cm,angle=0]{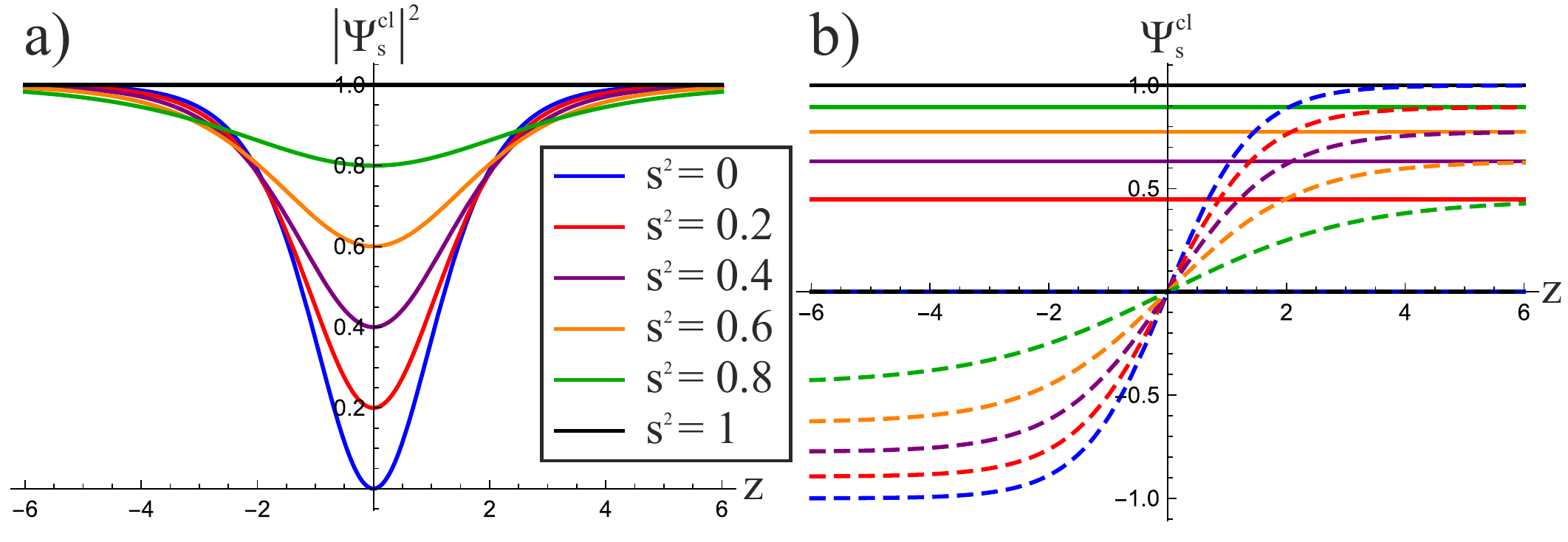}\end{center}
\begin{footnotesize}Figure 2.3. Plots present probability density $|\Psi^{cl}_{s}|^{2}$ (a)) and wave function $\Psi^{cl}_{s}$ (b)) for various values of $s^{2}$ and ,,+'' sign chosen in the equation (\ref{darksolitontsuzuki}). Time $t=0$, coupling constant $c=0.5$ and chemical potential $\mu_{ch}=1$ are chosen. Real part of $\Psi^{cl}_{s}$ is presented by continous curve while the dashed curve represents imaginary part. \end{footnotesize} \\

In the case of $v=v_{s}\Longleftrightarrow s=1$ we obtain constant solution
	\begin{equation}
\displaystyle{\Psi^{cl}_{0}(z,t,s=1)=\sqrt{\frac{\mu_{ch}}{2c}} }.
		\label{consttsuzuki}
		\end{equation}
Defining the soliton energy $E_{s}(v)=E_{s}(s)$ as the difference between the energies calculated in the states (\ref{darksolitontsuzuki}) and (\ref{consttsuzuki}) one obtains \cite{tsuzuki}, \cite{ishikawa}
	\begin{equation}
\displaystyle{E_{s}(v)=\frac{1}{3c}\left(2\mu_{ch}-v^{2}\right)^{3/2} =\frac{(2\mu_{ch})^{3/2}}{3c}\left(1-s^{2}\right)^{3/2}}.
		\label{solitonenergy}
		\end{equation}

Proceeding as in \cite{ishikawa} we calculate the canonical momentum of a soliton $P_{s}(v)$ 
	\begin{equation}
\displaystyle{P_{s}(v)\stackrel{\mathrm{def}}{=}\int \frac{1}{v} \frac{\mathrm{d}E_{s}(v)}{\mathrm{d}v}\mathrm{d}v  \stackrel{(\ref{solitonenergy})}{=} \frac{\mu_{ch}}{c}\left(A-s\sqrt{1-s^{2}}  -\mathrm{arcsin} (s)  \right) },
		\label{solitomomentum}
		\end{equation}
where the constant $A$
\begin{equation}
\displaystyle{\left.\begin{array}{lll}    \displaystyle{A=\frac{\pi}{2} } & \mathrm{for} & v\geq 0,  \, \,\\   \displaystyle{A=-\frac{\pi}{2} } & \mathrm{for} & v< 0.   \, \, \end{array} \right. }
		\label{solitonlim2}
\end{equation}
was chosen to obtain a correspondence with the type II excitations. It can be shown that $\rho=\mu_{ch}/2c$ \cite{tsuzuki}\footnote{Specifically, it can be shown that $\mu_{ch}=\rho g_{0}$, where in our case $g_{0}=2 c$.}. Then, indeed, the restriction of $P_{s}(v)$ is 
\begin{equation}
\displaystyle{|P_{s}(v)|\leq \frac{ \pi \mu_{ch}}{2c}=\pi \rho \equiv p_{c}}.
		\label{momentumrest}
\end{equation}
It is convenient to introduce rescaled momentum and energy
\begin{equation}
\displaystyle{p_{s}=\frac{P_{s}}{2\rho}, \, \, \, \, \, \, \, \, |p_{s}|\leq \frac{\pi}{2}},
		\label{rescaledmomentum}
\end{equation}
\begin{equation}
\displaystyle{\varepsilon_{s}=\frac{3E_{s}}{4\rho C_{ph}} },
		\label{rescaledenergy}
\end{equation}
where $C_{ph}$ is the velocity of phonons. In the limit of weak interactions $C_{ph}$ reduces to
\begin{equation}
\displaystyle{C_{ph}\simeq \sqrt{2\mu_{ch}}}.
		\label{weakphononvelocity}
\end{equation}
We immediately see that, in this limit 
\begin{equation}
\displaystyle{\varepsilon_{s}(s=0)=1}.
		\label{rescaledenergy2}
\end{equation}

Using a standard definition of the local current density (in our units $\hbar=2m=1$)
\begin{equation}
\displaystyle{J(z,t)=-i\left( \psi^{*}\frac{\partial \psi}{\partial z} -\psi\frac{\partial \psi^{*}}{\partial z}  \right)},
		\label{localcurrentdensity}
\end{equation}
one may obtain the total momentum associated with the soliton (\ref{darksolitontsuzuki}) in the following way
\begin{equation}
\displaystyle{ \Pi_{s}=\frac{1}{2}\int \mathrm{d}z J\left(z,t;\psi=\psi_{s}^{cl}\right)\stackrel{(\ref{darksolitontsuzuki})}{=}-\frac{\mu_{ch}}{c}s\left(  1-s^{2} \right)^{1/2}},
		\label{momentumoftsuzukisoliton}
\end{equation}
where the following subsidiary condition was used \cite{ishikawa} 
\begin{equation}
\displaystyle{ \lim_{|z|\rightarrow \infty}J(z,t)=0}.
		\label{subsidiarycondition}
\end{equation}
One instantly see that the momenta (\ref{solitomomentum}) and (\ref{momentumoftsuzukisoliton}) are not equal. It is believed that this disagreement is caused by the subsidiary condition (\ref{subsidiarycondition}) \cite{ishikawa} . 

Let us now look at the momentum (\ref{IImomentum}) and energy (\ref{IIexcenergy}) for the case of the Lieb's type II  excitations\footnote{Detailed analysis of the Lieb's excitations is presented in sections \ref{sec:ExcitationsI} and \ref{sec:ExcitationsII}.}. Rescaling the quantities as above and using the following relations
\begin{center}
$\displaystyle{\int \limits_{-\mathcal{Q}}^{\mathcal{Q}} \mathcal{J}(\mu)\mathrm{d}\mu=\mathcal{Q}\int \limits_{-1}^{1} \mathcal{J}(x)\mathrm{d}x, \, \, \, \, \, \, \, \, \int \limits_{-\mathcal{Q}}^{\mathcal{Q}}\mu \mathcal{J}(\mu)\mathrm{d}\mu=\mathcal{Q}^{2}\int \limits_{-1}^{1} x \mathcal{J}(x)\mathrm{d}x,\, \, \, \, \, \, \, \, \mathcal{Q}\longleftrightarrow v_{s} , \, \, \, \, \, \, \, \, q=\mathcal{Q}s}$,
\end{center}
we obtain (for $0\leq s < 1$)
\begin{equation}
\displaystyle{p_{s,II}=\frac{\mathcal{Q}}{2\rho}\left[ -s + \int \limits_{-1}^{1} \mathcal{J}(x,s) \mathrm{d}x  \right]},
		\label{pIIrescaled}
\end{equation}
\begin{equation}
\displaystyle{\varepsilon_{s,II}=\frac{3\mathcal{Q}^{2}}{4\rho C_{ph}}\left[ \frac{\mu_{ch}}{\mathcal{Q}^{2}}-s^{2} + 2\int \limits_{-1}^{1} x \mathcal{J}(x,s) \mathrm{d}x  \right]},
		\label{eIIrescaled}
\end{equation}
\begin{equation}
\displaystyle{ \mathcal{J}(x,s) = \frac{1}{2} +\frac{\lambda}{\pi} \int \limits_{-1}^{1}  \frac{\mathcal{J}(y,s)}{\lambda^{2} + (x-y)^{2} }\mathrm{d} y -\frac{1}{\pi}\mathrm{arctan}\left( \frac{s-x}{\lambda}  \right)},
		\label{JIIrescaled}
\end{equation}
\begin{equation}
\displaystyle{\lambda=\frac{c}{\mathcal{Q}}},
		\label{JIIrescaledlambda}
\end{equation}
where we have used the formulas (\ref{IIint}), (\ref{qmdescription63}) and (\ref{qmdescription57}) (for $c>0$). Here, we recall that the type II excitation means the elementary excitation from the state with quasi-momentum $\mathcal{Q}s$ to the state with quasi-momentum $\mathcal{Q}$\footnote{The ground state of energy consists of all possible quasi-momenta $q$ from the range $|q| \leq \mathcal{Q}$.}. The process causes shift of all the quasi-momenta, which is represented by the function $\mathcal{J}(x,s)$. From the equations (\ref{pIIrescaled}) and (\ref{JIIrescaled}) it may be shown that \cite{ishikawa}
\begin{equation}
\displaystyle{ |p_{s,II}|\leq \frac{\pi}{2} },
		\label{rescaledpIIobciecie}
\end{equation}
which is consistent with (\ref{rescaledmomentum}).

In the case of infinitely strong interactions ($\lambda \rightarrow \infty$, see (\ref{JIIrescaledlambda})) one gets \cite{ishikawa}
\begin{equation}
\displaystyle{ \mathcal{J}(x,s) =\frac{1}{2}},
		\label{JIIrescaledinfty}
\end{equation}
\begin{equation}
\displaystyle{p_{s,II}^{\infty}(s)=\frac{\pi}{2}(1-s)},
		\label{pIIrescaledinfty}
\end{equation}
\begin{equation}
\displaystyle{\varepsilon_{s,II}^{\infty}(s)=\frac{3\pi}{8} (1-s^{2})}.
		\label{eIIrescaledinfty}
\end{equation}
We notice that $\varepsilon_{s,II}^{\infty}(s=0)=3\pi/8\approx 1.18 > \varepsilon_{s}(s=0)=1$. 

Our purpose is to consider weak interaction limit ($\lambda \rightarrow 0$). Fortunately, a singularity which appears in this limit in the kernel of integral in the equation (\ref{JIIrescaled}) does not break the availability of correct asymptotic ($\lambda \rightarrow 0$) solution. The problem was firstly solved by M. Kac and H. Pollard \cite{kac}. The solution in the limit of weak interactions has the following form \cite{ishikawa}, \cite{kac}, \cite{hutson}
\begin{equation}
\displaystyle{\mathcal{J}(x)=\frac{1}{\lambda} \int \limits_{-1}^{1} \mathcal{D}(x,y) f(y)\mathrm{d}y },
		\label{SingularKernel1}
\end{equation}
while the inverse kernel $\mathcal{D}(x,y)$ takes the form
\begin{equation}
\displaystyle{ \mathcal{D}(x,y) = \frac{1}{2\pi} \mathrm{ln} \left(  \frac{1-xy+(1-x^{2})^{1/2}(1-y^{2})^{1/2}}{1-xy-(1-x^{2})^{1/2}(1-y^{2})^{1/2}} \right)},
		\label{SingularKernel2}
\end{equation}
and the inhomogeneous part of (\ref{JIIrescaled}), the function $f(x)$, in the $\lambda \rightarrow 0$ limit is given by
\begin{equation}
\displaystyle{ f(x) \simeq \left\{ \begin{array}{lll}    \displaystyle{1}, & & s\leq x \leq 1,  \, \,\\   \displaystyle{0}, & & -1 \leq x \leq s.   \, \, \end{array} \right. }
		\label{SingularKernel3}
\end{equation}
Hence the solution of (\ref{JIIrescaled}) takes the following form \cite{ishikawa}
\begin{flushleft}
$\displaystyle{\mathcal{J}(x,s)=\frac{1}{\lambda}\left[  \frac{x-s}{2\pi} \mathrm{ln} \left(  \frac{1-sx+(1-s^{2})^{1/2}(1-x^{2})^{1/2}}{1-sx-(1-s^{2})^{1/2}(1-y^{2})^{1/2}} \right)   \right.}$
\end{flushleft}
\begin{equation}
\displaystyle{ \left. +\left( \frac{1}{2}- \frac{1}{\pi}\mathrm{arcsin} s   \right) (1-x^{2})^{1/2} \right] }.
		\label{SingularKernel4}
\end{equation}
Knowing that in the considered limit $\mathcal{Q}\simeq \sqrt{2\mu_{ch}} \simeq C_{ph}$ one obtain exactly the same results as in the case of classical mean-field method\footnote{We need to substitute the formula (\ref{SingularKernel4}) into the equations (\ref{pIIrescaled}) and (\ref{eIIrescaled}). } 
\begin{equation}
\displaystyle{ p_{s,II}=p_{s} , \, \, \, \, \, \, \, \, \varepsilon_{s,II}=\varepsilon_{s} }.
		\label{solutionsforweak}
\end{equation}
Therefore, because above relations agree, it is believed that the Lieb's type II excitations may be identified with the soliton (\ref{darksolitontsuzuki}) obtained by T. Tsuzuki from semi-classical version of NLS equation (\ref{gross}). The solutions $\mathcal{J}(x,s)$ for relatively strong interactions have to be obtained by an iterative methods. In the figure 2.4. b) we present the numerical results presented in \cite{ishikawa}. Moreover, M. Ishikawa and H. Takayama have also showed numerically that the dispersion relation of the semi-classical soliton agrees with the excitation spectrum of Lieb's type II excitations even for relatively strong interactions (see figure 2.4. a)). It should be mentioned that the spectrum of the solution of NLS equation (\ref{gross}) was firstly obtained in \cite{kulish}. 

\begin{center}\includegraphics[width=15cm,angle=0]{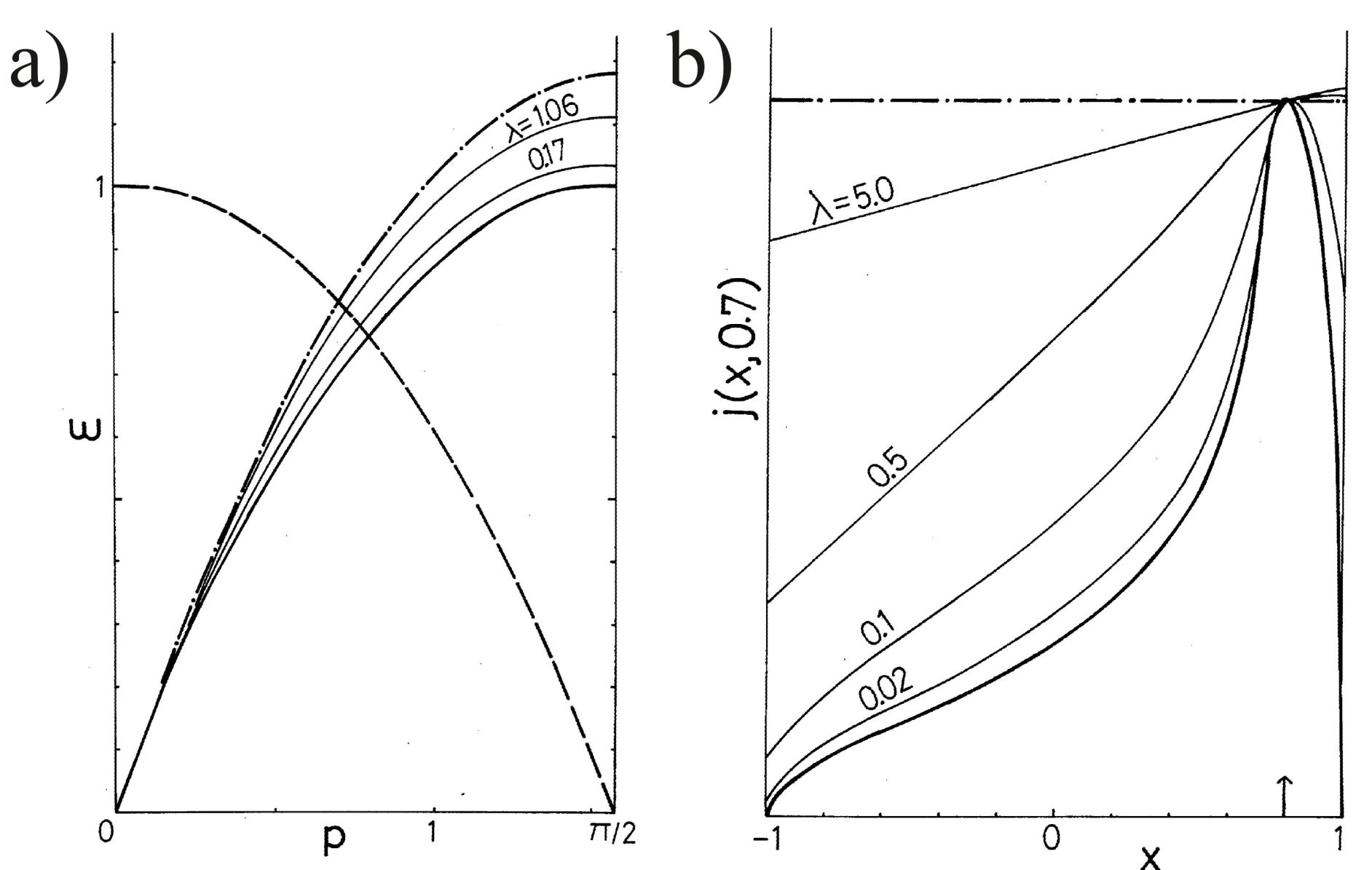}\end{center}
\begin{footnotesize}Figure 2.4. Plot a) presents a renormalized dispersion relations of the NLS soliton (solid line) and Lieb's type II excitations: relatively small interactions (thin lines), strong coupling limit - eqs. (\ref{pIIrescaledinfty}), (\ref{eIIrescaledinfty}) - (chain line). The dispersion of NLS soliton with $A=0$ and $-1\leq s < 0$ - eq. (\ref{solitomomentum}) - (broken line). Numerically obtained solutions of $\mathcal{J}(x,s)=j(x,s)$ for $s=0.7$, and different values of $\lambda$ are depicted in plot b). The asymptotic solution (\ref{SingularKernel4}) is presented by solid line, thin lines represent different values of $\lambda$ parameter. The chain line corresponds to $\lambda\rightarrow \infty$ ($\mathcal{J}(x,s)=1/2$). The scaling of the functions was taken such that they coincide with each other at $x=0.8$ \cite{ishikawa}. \end{footnotesize} \\

The results suggest that the type II excitation corresponds to a dark soliton. In the chapter \ref{numericalresults} we show that the statement is true, indeed \cite{SyrwSacha}.

\section{Mean-field solitons with periodic boundary conditions}
\label{sec:Solitonsinperiodicbox}

We have imposed a periodic boundary conditions on our Lieb-Liniger system. It means that we are considering the system of $\mathcal{N}$ identical bosons on a ring or in a toroidal trap. It is not obvious how to obtain a stationary soliton solutions which satisfy the boundary conditions in the mean-field approach - the problem appears when we want to merge the ends of solution together. The operation requires non-zero total momentum, which means the non-zero velocity of soliton. Let us consider an easy solution (\ref{GROSStciemnysolution}) presented in appendix \ref{sec:Csoliton}. The mean-field term in the Gross-Pitaevskii equation (\ref{GROSStdependent}) $|\psi|^{2}$ is time-dependent because the notch of the soliton moves with the velocity $v=\dot{q}$. In order to have a time-independent mean field potential one needs to go to the moving frame of reference. In the case of our system we have to deal with the same problem, but, because of the topology of the system, we should consider the rotating frame of reference. It is easy to show that the corresponding unitary transformation has the following form\footnote{The transformation to the rotating frame in non-relativistic quantum mechanics may be obtained from the translation operator $U=\mathrm{exp}\left[ i \hat{\vec{P}}\cdot\vec{x}(t)  \right]$ with the position $\vec{x}(t)=\vec{v}t$ and the velocity $\vec{v}=\vec{\omega}\times \vec{x}$. The substitution leads to $\hat{U}=\mathrm{exp}\left[ i t \left( \vec{\omega}\times \hat{\vec{x}} \right)\cdot \hat{\vec{p}} \right]=\mathrm{exp}\left[ i t  \vec{\omega}\cdot \hat{\vec{L}} \right]$. }
\begin{equation}
\displaystyle{ \hat{U}=\mathrm{e}^{i t\vec{\omega}\cdot \hat{\vec{L}}  }  },
		\label{PeriodicyTransformToRot}
\end{equation}
where $\hat{\vec{L}}$ is the angular momentum operator and $\vec{\omega}$ is the angular frequency.

Our system has a topology of a ring, therefore, we may write the coordinates of bosons as an azimuthal angles $\theta_{j}=2\pi \frac{z_{j}}{\mathrm{L}}\in [0,2\pi)$ and rewrite the Lieb-Liniger Hamiltonian (\ref{qmdescription11}) as
\begin{equation}
\displaystyle{\hat{\mathcal{H}}_{\mathcal{N}}^{\theta}=-\sum_{j=1}^{\mathcal{N}}\frac{\partial^{2}}{\partial \theta_{j}^{2}}+2\widetilde{c} \sum_{\mathcal{N}\geq j > k}\delta(\theta_{j}-\theta_{k})},
		\label{LLHamPERIODICITY}
\end{equation}
where we have taken the length and energy units as $\frac{\mathrm{L}}{2\pi}$ and $\frac{4\pi^{2} \hbar^{2}}{2m \mathrm{L}^{2}}$, respectively. The coupling constant $\widetilde{c}$ is dimensionless because it is measured in units of $\frac{2\pi \hbar^{2}}{2m \mathrm{L}}$. It turns out that $\widetilde{c}\neq c$, which we discuss at the end of this section. In the rotating frame of reference the Lieb-Liniger Hamiltonian takes the following form\footnote{While the wave function in rotating frame takes the form $\ket{\widetilde{\psi}_{\mathcal{N}}}=\hat{U}_{\mathcal{N}}(\Omega)\ket{\psi_{\mathcal{N}}}$, the form of Hamiltonian may be obtained from the evolution equation 
\begin{flushleft}
$\displaystyle{i\frac{\partial}{\partial t}\hat{U}_{\mathcal{N}}(\Omega)\ket{\psi_{\mathcal{N}}}=\hat{U}_{\mathcal{N}}(\Omega)i\frac{\partial}{\partial t}\ket{\psi_{\mathcal{N}}}+i\left(\frac{\partial}{\partial t}\hat{U}_{\mathcal{N}}(\Omega)\right)\ket{\psi_{\mathcal{N}}} }$
\end{flushleft}
\begin{center}
$\displaystyle{=\left( \hat{U}_{\mathcal{N}}(\Omega)\hat{\mathcal{H}}_{\mathcal{N}}^{\theta}\hat{U}_{\mathcal{N}}^{\dagger}(\Omega)-2\Omega \hat{\mathcal{L}} \right)\hat{U}_{\mathcal{N}}(\Omega)\ket{\psi_{\mathcal{N}}}=\hat{\mathcal{H}}(\Omega)\hat{U}_{\mathcal{N}}(\Omega)\ket{\psi_{\mathcal{N}}}}$.
\end{center}
One obtains the first two terms of the result (\ref{LLHamInRotatingFrame}) because $\left[ \hat{\mathcal{H}}_{\mathcal{N}}^{\theta},\hat{\mathcal{L}} \right]=0$.} \cite{carr}
\begin{equation}
\displaystyle{\hat{\mathcal{H}}(\Omega)=\hat{\mathcal{H}}_{\mathcal{N}}^{\theta}-2\Omega \hat{\mathcal{L}} +\Omega^{2}\mathcal{N}},
		\label{LLHamInRotatingFrame}
\end{equation}
where 
\begin{equation}
\displaystyle{\hat{\mathcal{L}}\equiv - i \sum_{j=1}^{\mathcal{N}}\frac{\partial}{\partial \theta_{j}}},
		\label{LLHamInRotatingFrameLOperator}
\end{equation}
is the angular momentum operator given in units of $\hbar$, the term $\Omega^{2}\mathcal{N}$ is a constant rotating rigid-body energy which added to the Hamiltonian makes the system translationally invariant and does not have an impact on the results \cite{carr}. Corresponding unitary transformation is given by  
\begin{equation}
\displaystyle{\hat{U}_{\mathcal{N}}(\Omega)=\mathrm{e}^{it 2\Omega \hat{\mathcal{L}} }  },
		\label{unitarytransformOMEGA}
\end{equation}
where the factor 2 ($\omega=2\Omega$) appears because $m=\frac{1}{2}$ is chosen. Let us see what will happen with the time-dependent Gross-Pitaevskii equation in the rotating reference frame 
\begin{center}
$\displaystyle{  \hat{U}(\Omega)\left[-\frac{\partial^{2}}{\partial \theta^{2} }+2\widetilde{c} \mathcal{N}\left| \psi(\theta,t) \right|^{2}\right]\hat{U}^{\dagger}(\Omega)\hat{U}(\Omega)\ket{\psi}=i\hat{U}(\Omega)\frac{\partial}{\partial t}\hat{U}^{\dagger}(\Omega)\hat{U}(\Omega)\ket{\psi}, \, \, \, \, \, \, \, \, \hat{U}(\Omega)=\mathrm{e}^{i t2\Omega (-i\partial_{\theta})} }$,
\end{center}
\begin{equation}
\displaystyle{\left[\left(-i\frac{\partial}{\partial \theta} -\Omega \right)^{2}-\Omega^{2} +2\widetilde{c} \mathcal{N}\left| \widetilde{\psi}(\theta) \right|^{2}\right]\ket{\widetilde{\psi}}= i\frac{\partial}{\partial t}\ket{\widetilde{\psi}} }.
		\label{GrossTdependentrotating}
\end{equation}

Let us define a dimensionless parameter
\begin{equation}
\displaystyle{\widetilde{\gamma}=\frac{\widetilde{c}\mathcal{N}}{\pi}}.
		\label{meanfieldgamma}
\end{equation}
For very weak interaction ($2c\mathcal{N} \leq \mathcal{O}(1)$), when the system of bosons may be treated as a condensate, dynamical and static features can be described by the NLS equation (\ref{qmdescription4}) or Gross-Pitaevskii equation (\ref{GROSS}) \cite{carr}.  The time-independent Gross-Pitaevskii equation in the rotating frame has the following form \cite{carr}, \cite{zaremba}\footnote{The equation (\ref{meanfieldGP}) was obtained in the same way as in appendix \ref{sec:Cderiv} - we only need to take into account an additional integral of motion $-2\Omega \hat{\mathcal{L}} +\Omega^{2}\mathcal{N}$. }
\begin{equation}
\displaystyle{\left[ \left( -i\frac{\partial}{\partial \theta}-\Omega \right)^2 +2\pi \widetilde{\gamma} |\psi(\theta)|^{2}  -\mu_{ch} \right]\psi(\theta)=0, \, \, \, \, \, \, \, \, \theta=2\pi \frac{z}{\mathrm{L}} \in [0,2\pi)},
		\label{meanfieldGP}
\end{equation}
where $\psi(\theta)$ is a wave function of one atom in the condensate. It turns out that above equation has two types of solutions \cite{carr}, \cite{zaremba}, \cite{reinhardt}: plane waves (PW)
\begin{equation}
\displaystyle{\psi^{(PW)}_{J}(\theta) =\frac{1}{\sqrt{2\pi}}\mathrm{e}^{i J \theta} },
		\label{meanfieldPW}
\end{equation}
and the so-called soliton trains (ST) 
\begin{equation}
\displaystyle{\psi^{(ST)}_{J,j}(\theta) =\sqrt{\rho_{j}(\theta)}\mathrm{e}^{i \varphi_{J,j} (\theta)}},
		\label{meanfieldST}
\end{equation}
where $J\in \mathbb{Z}$ is the phase-winding number, and $j \in |\mathbb{Z}|$ is the number of density notches in a soliton train. 

We would like to find analytical expressions for the amplitude $\sqrt{\rho_{j}(\theta)}$ and the phase $\varphi_{J,j} (\theta)$. For this purpose one substitutes the solution $\psi(\theta) =\sqrt{\rho(\theta)}\mathrm{exp}\left[i \varphi (\theta)\right]$ into (\ref{meanfieldGP}) and separates the real and imaginary parts getting
\begin{equation}
\displaystyle{\frac{\left( \rho ' \right)^{2}}{4 \rho^{2} } -\frac{\rho ''}{2 \rho} +\left( \varphi ' \right)^{2}  -2\Omega \varphi ' +\Omega^{2} +2\pi \widetilde{\gamma}\rho -\mu_{ch}=0    },
		\label{meanfieldGPReal}
\end{equation}
\begin{equation}
\displaystyle{\rho \varphi '' + \rho ' \varphi '  -\Omega \rho ' =0   },
		\label{meanfieldGPIM}
\end{equation}
where $\rho'=\frac{\mathrm{d}\rho}{\mathrm{d}\theta}$ and $\varphi'=\frac{\mathrm{d}\varphi}{\mathrm{d}\theta}$. The equation (\ref{meanfieldGPIM}) can be integrated in the following way
\begin{center}
$\displaystyle{y=\varphi ' \Longrightarrow \rho y' + \rho ' y   -\Omega \rho ' =0   \Longrightarrow \frac{\mathrm{d} y}{\mathrm{d} \theta} =\frac{1}{\rho}\frac{\mathrm{d}\rho}{\mathrm{d}\theta}\left( \Omega-y \right)  \Longrightarrow   \int \frac{\mathrm{d}y}{\Omega-y}=\int \frac{1}{\rho}\frac{\mathrm{d}\rho}{\mathrm{d}\theta} \mathrm{d}\theta  }$,
\end{center}
\begin{center}
$\displaystyle{q=\frac{1}{\rho} \Longrightarrow \mathrm{d}q=-\frac{\mathrm{d}\rho}{\rho^{2}}=-\frac{1}{\rho^{2}}\frac{\mathrm{d}\rho}{\mathrm{d}\theta} \mathrm{d}\theta  \Longrightarrow \frac{1}{\rho}\frac{\mathrm{d}\rho}{\mathrm{d}\theta} \mathrm{d}\theta = -\frac{\mathrm{d}q}{q}  \Longrightarrow   \int \frac{\mathrm{d}y}{\Omega-y}=-\int \frac{\mathrm{d}q}{q}   }$,
\end{center}
\begin{equation}
\displaystyle{\varphi '=\Omega +\frac{W}{\rho}   },
		\label{meanfieldGPImSolution}
\end{equation}
where $W$ is an integration constant. Inserting above result into (\ref{meanfieldGPReal}) one obtains
\begin{equation}
\displaystyle{\frac{1}{4}\left( \rho ' \right)^{2} -\frac{1}{2 }\rho '' \rho  +W^{2} +2\pi\widetilde{\gamma} \rho^{3} -\mu_{ch} \rho^{2}=0    }.
		\label{meanfieldGPReal1}
\end{equation}
Tricky substitution \cite{zaremba}
\begin{equation}
\displaystyle{y= -\frac{1}{4}\left( \rho ' \right)^{2} - W^{2} +\pi\widetilde{\gamma} \rho^{3}- \mu_{ch} \rho^{2}},
		\label{meanfieldGPReal2}
\end{equation}
leads to the following observation
\begin{equation}
\displaystyle{\rho y ' - \rho ' y=\rho' \left(\frac{1}{4}\left( \rho ' \right)^{2} -\frac{1}{2 }\rho '' \rho  +W^{2} +2\pi\widetilde{\gamma} \rho^{3} -\mu_{ch} \rho^{2}   \right) \stackrel{(\ref{meanfieldGPReal1})}{=}0  \Longrightarrow \frac{y '}{y}=\frac{\rho '}{\rho} },
		\label{meanfieldGPReal3}
\end{equation}
\begin{equation}
\displaystyle{y=V \rho },
		\label{meanfieldGPReal4}
\end{equation}
where $V$ is an integration constant. Hence, by the integration of formula (\ref{meanfieldGPReal1}) we get 
\begin{equation}
\displaystyle{-\frac{1}{4}\left( \rho ' \right)^{2} +\pi\widetilde{\gamma} \rho^{3} -\mu_{ch}\rho^{2} - V \rho - W^{2} =0}.
		\label{meanfieldGPReal5}
\end{equation}
The solution of above differential equation takes the following form \cite{carr}, \cite{zaremba}
\begin{equation}
\displaystyle{\sqrt{\rho_{j}(\theta)}=
\left\{ \begin{array}{ll}    \displaystyle{ \mathscr{N}(\eta)\sqrt{1+\eta\mathrm{dn}^{2} \left(\left. \frac{j K(m) (\theta-\theta_{0})}{\pi}\right|m   \right)  } }, & \widetilde{\gamma} >0 , \, \,\\   \displaystyle{\mathscr{N}(\eta)\sqrt{ \mathrm{dn}^{2} \left(\left. \frac{j K(m) (\theta-\theta_{0})}{\pi}\right| m   \right) - \eta m' } }, & \widetilde{\gamma} <0 ,  \, \, \end{array} \right. }
		\label{meanfieldRhoSol1}
\end{equation}
where $m$ is called the \emph{elliptic parameter} of the Jacobi elliptic function $\mathrm{dn}$ \cite{abramowitz} and $j$ counts the number of density notches. The so-called \emph{complementary elliptic parameter} $m'$ satisfies the relation $m' + m=1$. Both of elliptic parameters are real numbers belonging to the interval $m, m' \in [0,1]$ \cite{abramowitz}. From the relation $\int_{0}^{2\pi} \rho(\theta) \mathrm{d}\theta=1$ one obtains the normalization constant \cite{carr}, \cite{zaremba}
\begin{equation}
\displaystyle{\mathscr{N}(\eta)=
\left\{ \begin{array}{ll}    \displaystyle{ \sqrt{  \frac{K(m)}{2\pi \left[ K(m) +\eta E(m)  \right]}  } }, & \widetilde{\gamma} >0 , \, \, \\   \displaystyle{ \sqrt{  \frac{K(m)}{2\pi \left[ E(m) -\eta m' K(m)  \right]}  }} , & \widetilde{\gamma} <0 .  \, \, \end{array} \right. }
		\label{meanfieldRhoSol2}
\end{equation}
$K(m)$ and $E(m)$ are complete elliptic integrals of the first and the second kinds, respectively \cite{abramowitz}. It is convenient to define the following functions \cite{carr}, \cite{zaremba}
\begin{equation}
\displaystyle{f\equiv
\left\{ \begin{array}{ll}  \displaystyle{+\left[ \pi^{2} \widetilde{\gamma} -2 j^{2} K^{2}(m) +2 j^{2} K(m) E(m)  \right] } , & \widetilde{\gamma} >0 , \, \, \\  \displaystyle{-\left[ \pi^{2} \widetilde{\gamma} -2 j^{2} K^{2}(m) +2 j^{2} K(m) E(m)  \right] } , & \widetilde{\gamma} <0 , \, \, \end{array} \right. }
		\label{meanfieldfunctionsF}
\end{equation}
\begin{equation}
\displaystyle{h\equiv
\left\{ \begin{array}{ll}  \displaystyle{+\left[ \pi^{2} \widetilde{\gamma} -2 j^{2} K^{2}(m) +2 j^{2} K(m) E(m) + 2 j^{2} m K^{2}(m) \right] } , & \widetilde{\gamma} >0 , \, \, \\  \displaystyle{-\left[ \pi^{2} \widetilde{\gamma} -2 j^{2} K^{2}(m) +2 j^{2} K(m) E(m) + 2 j^{2} m K^{2}(m) \right] } , & \widetilde{\gamma} <0 , \, \, \end{array} \right. }
		\label{meanfieldfunctionsH}
\end{equation}
\begin{equation}
\displaystyle{g\equiv  \pi^{2} \widetilde{\gamma} +2 j^{2} K(m) E(m),   }
		\label{meanfieldfunctionsG}
\end{equation}
\begin{equation}
\displaystyle{S\equiv \mathrm{sign}(J-\Omega)=
\left\{ \begin{array}{ll}   +1 , & \Omega <J , \, \, \\  -1 , & \Omega >J . \, \, \end{array} \right. }
		\label{meanfieldfunctionsS}
\end{equation}
Substituting the solution (\ref{meanfieldRhoSol1}) into (\ref{meanfieldGPReal1}) we determine the depth of density notches \cite{carr}, \cite{zaremba}
\begin{equation}
\displaystyle{\eta=
\left\{ \begin{array}{ll}    \displaystyle{ -2 j^{2} K^{2}(m)/g \in [-1,0] }, & \widetilde{\gamma} >0 , \, \, \\   \displaystyle{ g/\left[2 j^{2} m '  K^{2}(m)\right] \in [0,1] } , & \widetilde{\gamma} <0 .  \, \, \end{array} \right. }
		\label{meanfieldRhoSol3}
\end{equation}
By the connection of equations (\ref{meanfieldGPReal5}) - (\ref{meanfieldRhoSol3}) one gets \cite{carr}, \cite{zaremba}
\begin{equation}
\displaystyle{\mu_{ch}=\frac{3}{2}\widetilde{\gamma} +\left(\frac{j}{\pi}\right)^{2}  \left[  3 K(m) E(m) -(2-m) K^{2}(m) \right], }
		\label{meanfieldRhoSol4mu}
\end{equation}
\begin{equation}
\displaystyle{W\equiv \frac{S}{2\pi^{4}|\widetilde{\gamma}|} \sqrt{\frac{fgh}{2}}. }
		\label{meanfieldRhoSol5W}
\end{equation}
It should be mentioned that in the calculations we need to remember about the following boundary conditions 
\begin{equation}
\displaystyle{\rho(\theta+2\pi)-\rho(\theta) =0, }
		\label{meanfieldRhoSolBoundary}
\end{equation}
\begin{equation}
\displaystyle{\varphi(\theta+2\pi)-\varphi(\theta)=2\pi J, \, \, \, \, \, \, \, \, J=0,\pm 1, \pm 2 , \ldots, }
		\label{meanfieldPhiSolBoundary}
\end{equation}
where $J$ is the previously mentioned phase winding number.

Let us now concentrate on the equation (\ref{meanfieldGPImSolution}). To find the solution $\varphi(\theta)$ one has to calculate 
\begin{equation}
\displaystyle{\int \frac{\mathrm{d}\theta}{\rho_{j}(\theta)}=\left\{ \begin{array}{ll}   \displaystyle{ \frac{2\pi \left[ K(m) +\eta E(m)  \right]}{K(m)} \int \frac{\mathrm{d}\theta}{ 1+\eta\mathrm{dn}^{2} \left(\left. \frac{j K(m) (\theta-\theta_{0})}{\pi}\right|m   \right)} } , & \widetilde{\gamma} >0         , \, \, \\  \, \, \\ \displaystyle{ \frac{2\pi \left[ E(m) -\eta m' K(m)  \right]}{K(m)}   \int \frac{\mathrm{d}\theta}{\mathrm{dn}^{2} \left(\left. \frac{j K(m) (\theta-\theta_{0})}{\pi}\right| m   \right) - \eta m' } } , & \widetilde{\gamma} <0 .  \, \, \end{array} \right. }
		\label{meanfieldPhiSollast1}
\end{equation}
It can be shown that \cite{abramowitz}
\begin{equation}
\displaystyle{\mathrm{dn}^{2}(u|m)=1-m\mathrm{sn}^{2}(u|m), \, \, \, \, \, \, \, \, \int \left[1-\xi \mathrm{sn}^{2}(u|m)  \right]^{-1}\mathrm{d}u =\Pi(\xi ; u|m), }
		\label{meanfieldPhiSollast2}
\end{equation}
where $\Pi(\xi ; u|m)$ is an elliptic integral of the third kind \cite{abramowitz}. Therefore, using definitions presented above and simple algebra, we obtain
\begin{equation}
\displaystyle{\int \frac{\mathrm{d}\theta}{\rho_{j}(\theta)}=\left\{ \begin{array}{lll}   \displaystyle{\frac{2\pi^{4}\widetilde{\gamma}}{j K(m) f}\Pi\left(\xi ; \left. \frac{j K(m) (\theta-\theta_{0})}{\pi}\right| m \right)   } , &  \displaystyle{\xi=-\frac{2j^{2}m K^{2}(m)}{f} } , & \widetilde{\gamma} >0         , \, \, \\  \, \, \\  \displaystyle{-\frac{2\pi^{4}\widetilde{\gamma}}{j K(m) f}\Pi\left(\xi ; \left. \frac{j K(m) (\theta-\theta_{0})}{\pi}\right| m \right)   } , &  \displaystyle{\xi=\frac{2j^{2}m K^{2}(m)}{f} } , & \widetilde{\gamma} <0 ,  \, \, \end{array} \right. }
		\label{meanfieldPhiSollast3}
\end{equation}
which, with the relation (\ref{meanfieldRhoSol5W}), leads to
\begin{equation}
\displaystyle{\varphi_{J,j}(\theta)=\Omega \theta +\frac{S}{j K(m)}\sqrt{\frac{gh}{2f}}\Pi\left(\xi ; \left. \frac{j K(m) (\theta-\theta_{0})}{\pi}\right| m \right) . }
		\label{meanfieldPhiSollast4}
\end{equation}

Last thing we need to consider is the connection between angular momentum obtained from Bethe ansatz (a good quantum number)   
\begin{equation}
\displaystyle{L_{\mathcal{N}}=\frac{\mathrm{L}}{2\pi}P_{\mathcal{N}}=\sum_{j=1}^{\mathcal{N}}\mathcal{I}_{j}, }
		\label{angulargoodquantumnnumber}
\end{equation}
and the average angular momentum of mean-field solutions \cite{carr}
\begin{equation}
\displaystyle{\left< L_{\mathcal{N}}\right>\equiv \mathcal{N}\int \limits_{0}^{2\pi} \psi^{*}\left( -i \frac{\partial}{\partial \theta}  \right)\psi \mathrm{d}\theta }.
		\label{angularaverage}
\end{equation}
Our purpose is to find the solution for which $L_{\mathcal{N}}=\left< L_{\mathcal{N}}\right>$. It turns out that we obtain \cite{carr}
\begin{equation}
\displaystyle{\frac{\left< L_{\mathcal{N}}\right>_{PW}}{\mathcal{N}}=J },
		\label{planewaveangular}
\end{equation}
for the plane wave state (\ref{meanfieldPW}) and
\begin{equation}
\displaystyle{\frac{\left< L_{\mathcal{N}}\right>_{ST}}{\mathcal{N}}=\Omega +\frac{S}{\pi^{3}\widetilde{\gamma} }\sqrt{\frac{fgh}{2}} },
		\label{solitontrainangular}
\end{equation}
for the soliton-train state (\ref{meanfieldST}). Assuming that $\left< L_{\mathcal{N}}\right>=L_{\mathcal{N}}$ one may write a phase (\ref{meanfieldPhiSollast4}) as

\begin{equation}
\displaystyle{\varphi_{J,j}(\theta)=\left( \frac{L_{\mathcal{N}}}{\mathcal{N}}-\frac{S}{\pi^{3}\widetilde{\gamma} }\sqrt{\frac{fgh}{2}} \right) \theta+\frac{S}{j K(m)}\sqrt{\frac{gh}{2f}}\Pi\left(\xi ; \left. \frac{j K(m) (\theta-\theta_{0})}{\pi}\right| m \right) , }
		\label{phasewithangularmom}
\end{equation}
because
\begin{equation}
\displaystyle{\Omega=\frac{ L_{\mathcal{N}}}{\mathcal{N}}-\frac{S}{\pi^{3}\widetilde{\gamma} }\sqrt{\frac{fgh}{2}} }.
		\label{omegaangularmom}
\end{equation}

It is clear now that the elliptic parameter $m\in [0,1]$ may be obtained from the boundary condition for the phase (\ref{meanfieldPhiSolBoundary}) if we set the values of $L_{\mathcal{N}}, J$ and $j$. It should be mentioned that the parameter $\theta_{0} \in[0,2\pi)$, which appears in our solutions, is supposed to be a manifestation of the spontaneous symmetry breaking which occurs during the localization of the quantum dark soliton \cite{carr}. Our numerical results (see chapter \ref{numericalresults}) confirm the supposition. 

Considering now only dark soliton solutions $\widetilde{\gamma} >0$, we instantly see that soliton-train solution (\ref{meanfieldRhoSol1}), (\ref{meanfieldPhiSollast4}) converges to the plane wave solution in the limit $\eta \rightarrow 0$. On the other hand, the limit  $\eta \rightarrow -1$ corresponds to the black soliton(-train) solution. Because $\eta \rightarrow -1$ is equivalent to  $f \rightarrow 0$, the eliptic parameter $m$ for black soliton(-train) solutions may be easily established from the equation $f=0$. The gray soliton(-train) appears for $-1<\eta<0$ \cite{carr}.

We are particularly interested in the soliton-train solutions. Detailed analysis of mean-field solutions executed in \cite{carr} shows that the ST solutions lie between two parabolic curves on the $(\Omega, \widetilde{\gamma})$-parameter plane, where the curves correspond to two different phase winding numbers $J$, $J'$ and intersect at consecutive values of $\Omega=\Omega_{nodes}(|J-J'|)$. There is one more interesting restriction $|J-J'|=j$. Otherwise, we obtain only the plane wave solution for $J\in \mathbb{Z}$. One observes that the soliton-train solution combines continuously two plane wave solutions which angular momenta per particle (\ref{planewaveangular}) differ by $|J-J'|=j$, where $j$ is the number of density notches. 

The last thing we need to consider is the energy functional in the case of the angle space $\theta \in [0,2\pi )$. We want to obtain the relation between the main parts of the two following energy functionals 
\begin{equation}
\displaystyle{E[\psi(z),\psi^{*}(z),c]=\mathcal{N}\int \limits_{0}^{\mathrm{L}} \mathrm{d}z \left[ |\partial_{z} \psi(z)|^{2}+ c(\mathcal{N}-1)|\psi(z)|^{4}    \right]},
		\label{efunctz}
\end{equation}
\begin{equation}
\displaystyle{\widetilde{E}[\psi(\theta),\psi^{*}(\theta),\widetilde{c} \, ]=\mathcal{N}\int \limits_{0}^{2\pi} \mathrm{d}\theta \left[ |\partial_{\theta} \psi(\theta)|^{2}+\widetilde{c} (\mathcal{N}-1)|\psi(\theta)|^{4}    \right]}.
		\label{efuncttheta}
\end{equation}
Knowing that $z=\frac{\theta}{2\pi}\mathrm{L}$ and from the normalization relations
\begin{equation}
\displaystyle{\int \limits_{0}^{\mathrm{L}} \mathrm{d}z | \psi(z)|^{2} = \int \limits_{0}^{2\pi} \mathrm{d}\theta | \psi(\theta)|^{2} =1},
		\label{normalizationrelatiosnthetaz}
\end{equation}
we notice that
\begin{equation}
\displaystyle{ \frac{\partial}{\partial \theta}=\frac{\mathrm{L}}{2\pi}\frac{\partial}{\partial z}, \, \, \, \, \, \, \, \, \mathrm{d}\theta =\frac{2\pi}{\mathrm{L}}\mathrm{d}z, \, \, \, \, \, \, \, \,  \psi(\theta) \longleftrightarrow \sqrt{\frac{\mathrm{L}}{2\pi}}\psi(z)  }.
		\label{relationsthetaz}
\end{equation}
Therefore one obtains 
\begin{flushleft}
$\displaystyle{\widetilde{E}[\psi(\theta),\psi^{*}(\theta),\widetilde{c} \, ]=\mathcal{N}\int \limits_{0}^{2\pi} \mathrm{d}\theta \left[ |\partial_{\theta} \psi(\theta)|^{2}+\widetilde{c} (\mathcal{N}-1)|\psi(\theta)|^{4}    \right]}$
\end{flushleft}
\begin{center}
$\displaystyle{=\frac{2\pi}{\mathrm{L}} \mathcal{N}\int \limits_{0}^{\mathrm{L} } \mathrm{d}z   \left[  \left( \frac{\mathrm{L}}{2\pi} \right)^{3} |\partial_{z} \psi(z)|^{2} + \widetilde{c} \left( \frac{\mathrm{L}}{2\pi} \right)^{2} (\mathcal{N}-1) |\psi(z)|^{4}   \right]  }$
\end{center}
\begin{flushright}
$\displaystyle{= \mathcal{N} \left( \frac{\mathrm{L}}{2\pi} \right)^{2}\int \limits_{0}^{\mathrm{L} } \mathrm{d}z   \left[  |\partial_{z} \psi(z)|^{2} +  \frac{2\pi \widetilde{c}}{\mathrm{L}} (\mathcal{N}-1) |\psi(z)|^{4}   \right] }$.
\end{flushright}
Above result leads to the relations 
\begin{equation}
\displaystyle{E[\psi(z),\psi^{*}(z),c]=\left. \left( \frac{2\pi}{\mathrm{L}} \right)^{2}\widetilde{E}[\psi(\theta),\psi^{*}(\theta),\widetilde{c} \, ]   \right|_{\widetilde{c} \, =\frac{\mathrm{L}c}{2\pi}}},
		\label{efunctrelation}
\end{equation}
\begin{equation}
\displaystyle{\widetilde{c}=\frac{\mathrm{L}c}{2\pi}}.
		\label{couplingconstantsrelation}
\end{equation}
For the case of soliton train solution one gets
\begin{equation}
\displaystyle{ \left| \partial_{\theta} \psi_{J,j}^{(ST)}(\theta)   \right|^{2} \stackrel{(\ref{meanfieldGPImSolution})}{=} \left[\partial_{\theta} \sqrt{\rho_{j}(\theta)}\right]^{2} +\left[ \sqrt{\rho_{j}(\theta)} \left( \Omega +\frac{W}{\rho_{j}(\theta)}  \right) \right]^{2}   }.
		\label{energycomparison1}
\end{equation}
Hence, the energy obtained from Bethe ansatz  (\ref{qmdescription54}) may be compared with 
\begin{flushleft}
$\displaystyle{ E^{(ST)}=\mathcal{N} \left( \frac{2\pi}{\mathrm{L}} \right)^{2} \int \limits_{0}^{2\pi} \mathrm{d}\theta \left\{  \left[\partial_{\theta} \sqrt{\rho_{j}(\theta)}\right]^{2} +\left[ \sqrt{\rho_{j}(\theta)} \left( \Omega +\frac{W}{\rho_{j}(\theta)}  \right) \right]^{2} \right. }$
\end{flushleft}
\begin{equation}
\displaystyle{ \left. +\frac{\mathrm{L}c }{2\pi} (\mathcal{N}-1)\left|\sqrt{\rho_{j}(\theta)}\right|^{4} \right\}}.
		\label{energycomparison2}
\end{equation}

\chapter{The idea of numerical calculations}
\label{chap:Theideaofnumericalcalculations}

\section{Analytical expressions of norms and form factors}
\label{sec:Slavnov}

It is well known fact that the Bethe state norm is given by Gaudin-Korepin formula \cite{gaudin2}, \cite{korepinnorm}, \cite{bustamante}, \cite{slavnovnormsandff}
\begin{equation}
\displaystyle{\Braket{\{k\}_{\mathcal{N}} | \{k\}_{\mathcal{N}}}=c^{\mathcal{N}} \prod_{j>s=1}^{\mathcal{N}}\frac{(k_{j}-k_{s})^{2}+c^{2}}{(k_{j}-k_{s})^{2}}  \mathrm{det}_{\mathcal{N}}\left[ \mathcal{G}(\{ k \}_{\mathcal{N}}) \right] }.
		\label{norma}
		\end{equation}
We also recall the formula for the Gaudin matrix $\mathcal{G}$
\begin{equation}
\displaystyle{ \mathcal{G}_{js}(\{k \}_{\mathcal{N}})=\delta_{js}\left[ \mathrm{L}+\sum_{m=1}^{\mathcal{N}}K(k_{j},k_{m})  \right]-K(k_{j},k_{s}), }
		\label{gaudinmatrix2}
\end{equation}
where
\begin{equation}
\displaystyle{  K(k,\mu)=\frac{2c}{c^{2}+(k-\mu)^{2}}}.
		\label{Ka}
\end{equation}
It should be noted that the entries of Gaudin matrix are simple analytical functions of quasi-momenta. 

For our purposes, we need to consider matrix elements of the field operators in Heisenberg picture $\hat{\Psi}_{H}^{\dagger}(z,t)$ and  $\hat{\Psi}_{H}(z,t)$, between Bethe eigenstates $\bra{\{\mu \}_{\mathcal{N}}}$, $\ket{\{k \}_{\mathcal{N}+1}}$ or $\bra{\{\mu \}_{\mathcal{N}-1}}$, $\ket{\{k \}_{\mathcal{N}}}$. Using the translation, and time evolution relations\footnote{It is known that the shift and time evolution of operator $\hat{A}$ are given by the following relations: \\
$\hat{A}(z)=\hat{T}^{\dagger}(z-z_{0})\hat{A}(z_{0})\hat{T}(z-z_{0}),$ where the translation operator $\hat{T}(z-z_{0})=\mathrm{e}^{ i \hat{\mathrm{P}} (z-z_{0}) }$, \\
$\hat{A}(t)=\hat{U}^{\dagger}(t-t_{0})\hat{A}(t_{0})\hat{U}(t-t_{0})$, where the evolution operator $\hat{U}(t-t_{0})=\mathrm{e}^{-i \hat{\mathrm{H}} (t-t_{0}) }$.  \\
Therefore, in our case: \\
$\displaystyle{ \braket{\{\mu \}_{\mathcal{N}}|  \hat{\Psi}_{H}^{\dagger}(z,t) |\{k \}_{\mathcal{N}-1}} = \braket{\{\mu \}_{\mathcal{N}}|     \mathrm{e}^{-i\hat{\mathrm{P}}z}   \mathrm{e}^{i\hat{\mathrm{H}}t}  \hat{\Psi}_{H}^{\dagger}(0,0)  \mathrm{e}^{-i\hat{\mathrm{H}}t}  \mathrm{e}^{i\hat{\mathrm{P}}z} |\{k \}_{\mathcal{N}-1}}}$,\\
$\displaystyle{ \braket{\{\mu \}_{\mathcal{N}-1}|  \hat{\Psi}_{H}(z,t) |\{k \}_{\mathcal{N}}} = \braket{\{\mu \}_{\mathcal{N}-1}|     \mathrm{e}^{-i\hat{\mathrm{P}}z}   \mathrm{e}^{i\hat{\mathrm{H}}t}  \hat{\Psi}_{H}(0,0)  \mathrm{e}^{-i\hat{\mathrm{H}}t}  \mathrm{e}^{i\hat{\mathrm{P}}z} |\{k \}_{\mathcal{N}}}}$. } one can write
\begin{equation}
\displaystyle{ \braket{\{\mu \}_{\mathcal{N}}|  \hat{\Psi}_{H}^{\dagger}(z,t) |\{k \}_{\mathcal{N}-1}} =\mathrm{e}^{i \left(E_{\mu} - E_{k} \right)t }\mathrm{e}^{-i \left(P_{\mu} - P_{k}\right)z} \mathcal{F}(\{\mu \}_{\mathcal{N}},\{k\}_{\mathcal{N}-1}) },
		\label{ff1}
\end{equation}
\begin{equation}
\displaystyle{\braket{\{\mu \}_{\mathcal{N}-1}|  \hat{\Psi}_{H}(z,t) |\{k \}_{\mathcal{N}}} =\mathrm{e}^{i (E_{\mu} - E_{k})t} \mathrm{e}^{-i (P_{\mu} - P_{k})z} \mathcal{F}(\{\mu \}_{\mathcal{N}-1},\{k\}_{\mathcal{N}}) },
		\label{ff2}
\end{equation}
where $E_{\mu}, \, E_{k}, \, P_{\mu}, \, P_{k}$ are the total energies and total momenta of Bethe eigenstates 
\begin{equation}
\displaystyle{\begin{array}{lllll} \braket{\{\mu \}_{\mathcal{N}}|  \hat{\Psi}_{H}^{\dagger}(z,t) |\{k \}_{\mathcal{N}-1}}: & \displaystyle{E_{\mu}=\sum_{j=1}^{\mathcal{N} }\mu_{j}^{2}}, &  \displaystyle{E_{k}=\sum_{j=1}^{\mathcal{N}-1 }k_{j}^{2}}, & \displaystyle{P_{\mu}=\sum_{j=1}^{\mathcal{N} }\mu_{j}}, & \displaystyle{ P_{k}=\sum_{j=1}^{\mathcal{N}-1 }k_{j}},   \, \,\\ \braket{\{\mu \}_{\mathcal{N}-1}|  \hat{\Psi}_{H}(z,t) |\{k \}_{\mathcal{N}}}: &  \displaystyle{E_{\mu}=\sum_{j=1}^{\mathcal{N}-1 }\mu_{j}^{2}}, &  \displaystyle{E_{k}=\sum_{j=1}^{\mathcal{N} }k_{j}^{2}}, & \displaystyle{P_{\mu}=\sum_{j=1}^{\mathcal{N}-1 }\mu_{j} },&  \displaystyle{P_{k}=\sum_{j=1}^{\mathcal{N} }k_{j}}. \end{array} }
		\label{eip}
		\end{equation}
The field operator form factors
\begin{equation}
\displaystyle{\begin{array}{l} \mathcal{F}(\{\mu\}_{\mathcal{N}},\{k\}_{\mathcal{N}-1})= \braket{\{\mu \}_{\mathcal{N}}|  \hat{\Psi}_{H}^{\dagger}(0,0) |\{k \}_{\mathcal{N}-1}}, \, \, \\  \mathcal{F}(\{\mu\}_{\mathcal{N}-1},\{k\}_{\mathcal{N}})=\braket{\{\mu \}_{\mathcal{N}-1}|  \hat{\Psi}_{H}(0,0) |\{k \}_{\mathcal{N}}} , \, \, \\  \mathcal{F}^{*}(\{\mu\}_{\mathcal{N}},\{k\}_{\mathcal{N}-1})=\mathcal{F}(\{k\}_{\mathcal{N}-1},\{\mu\}_{\mathcal{N}}), \end{array} }
		\label{formfactors}
\end{equation}
are calculated in \cite{gaudin2}, \cite{formfactors}
\begin{center}
$\displaystyle{\mathcal{F}(\{\mu\}_{\mathcal{N}-1},\{k\}_{\mathcal{N}})=-i \sqrt{c} \left( \,  \prod_{\mathcal{N}-1\geq j >s \geq 1} g(\mu_{j},\mu_{s}) \right)  \left( \,  \prod_{\mathcal{N}\geq j >s \geq 1} g(k_{s},k_{j}) \right) \times }$
\end{center}
\begin{equation}
\displaystyle{\left( \, \prod_{j=1}^{\mathcal{N}}\prod_{s=1}^{\mathcal{N}} h( k_{j},k_{s}) \right) \left( \,\prod_{j=1}^{\mathcal{N}-1}d(\mu_{j}) \right) \left( \,  \prod_{s=1}^{\mathcal{N}} d(k_{s}) \right)  \mathcal{M}i (\{ k\}) },
		\label{formf}
\end{equation}
in which
\begin{equation}
\displaystyle{ g(k,\mu)=\frac{i c}{k-\mu}, \, \, \, \, \, \, \, \, \, \, \, \, \, \, \, \, h(k,\mu)=\frac{k-\mu+i c}{i c}, \, \, \, \, \, \, \, \, \, \, \, \, \, \, \, \, d(k)=\mathrm{e}^{i k \mathrm{L}/2}},
		\label{gih}
\end{equation}
\begin{equation}
\displaystyle{\mathcal{M}i (\{ k\})=i^{\mathcal{N}-1} c^{2(\mathcal{N}-1)}  \frac{\displaystyle{\left( \, \prod_{a>b}^{\mathcal{N}} (k_{a}-k_{b}) \right)  \left( \, \prod_{a>b}^{\mathcal{N}-1} (\mu_{b}-\mu_{a}) \right)} }{ \displaystyle{\left( \, \prod_{a=1}^{\mathcal{N}} \prod_{b=1}^{\mathcal{N}-1}  (k_{a}-\mu_{b}) \right) }} \mathrm{det}_{\mathcal{N}-1} U_{j s} }.
		\label{Mi}
\end{equation}
The matrix $U_{j s}$ entries are the functions of quasi-momenta of both eigenstates and are purely real if $c>0$ (repulsive Bose gas) \cite{gaudin2}
\begin{equation}
\displaystyle{U_{j s}(\{k\}_{\mathcal{N}},\{\mu\}_{\mathcal{N}-1})=\frac{\delta_{js}}{i}\left(V_{j}^{+}-V_{j}^{-}  \right)+\frac{\displaystyle{\prod_{a=1}^{\mathcal{N}-1}(\mu_{a}-k_{j})  }  }{\displaystyle{ \prod_{a\neq j }^{\mathcal{N}}(k_{a}-k_{j})  } } \left(  K(k_{j},k_{s})-K(k_{\mathcal{N}}, j_{s})   \right) },
		\label{Ujs}
\end{equation}
where
\begin{equation}
\displaystyle{V^{\pm}_{j} =\frac{\displaystyle{\prod_{a=1}^{\mathcal{N}-1}(\mu_{a}-k_{j} \pm i c)  }  }{\displaystyle{ \prod_{a= j }^{\mathcal{N}}(k_{a}-k_{j}\pm i c)  } } }.
		\label{Vpm}
\end{equation}

Considering the following $\mathcal{N}$-particle density matrix 
\begin{equation}
\displaystyle{ \rho(z_{1},\ldots, z_{\mathcal{N}},t)=\braket{\{k\}_{\mathcal{N}}| \hat{\Psi}_{H}^{\dagger}(z_{1},t)\ldots  \hat{\Psi}_{H}^{\dagger}(z_\mathcal{N},t) \hat{\Psi}_{H}(z_\mathcal{N},t) \ldots  \hat{\Psi}_{H}(z_{1},t)  |\{ k \}_{\mathcal{N}} } },
		\label{densityop}
\end{equation}
we can put $\mathcal{M}$-particle identity operator 
\begin{equation}
\displaystyle{ \hat{\mathbf{1}}_{\mathcal{M}}=\sum_{\{\mu \}_{\mathcal{M}}} \frac{\ket{\{\mu \}_{\mathcal{M}}}  \bra{\{\mu \}_{\mathcal{M}}}}{\braket{\{\mu \}_{\mathcal{M}}|\{\mu \}_{\mathcal{M}}}} },
		\label{identityoperator}
\end{equation}
between each pair of the field  operators (where $\mathcal{M}$ is the proper number of particles). Therefore, using the relations (\ref{ff1}) and (\ref{ff2}) we obtain

$\displaystyle{ \rho(z_{1},\ldots, z_{\mathcal{N}},t)=\mathop{\sum_{\{\mu \}_{\mathcal{N}-1}, \ldots ,\{\mu \}_{1}}}_{\{\lambda \}_{\mathcal{N}-1}, \ldots ,\{\lambda \}_{1}}  \zeta \left(\{z_{j}\},t,\{E_{\{\alpha\}_{s}}\},\{P_{\{\alpha\}_{s}}\}\right)    \braket{\{\mu\}_{1}| \hat{\Psi}_{H}^{\dagger}(z_{\mathcal{N}},t)\hat{\Psi}_{H}(z_{\mathcal{N}},t)  |\{ \lambda \}_{1} }  }$
\begin{equation}
\displaystyle{\times \frac{ \mathcal{F}(\{k\}_{\mathcal{N}},\{\mu\}_{\mathcal{N}-1})\ldots \mathcal{F}(\{\mu\}_{2},\{\mu\}_{1}) \mathcal{F}(\{\lambda\}_{1},\{\lambda\}_{2}) \ldots \mathcal{F}(\{\lambda\}_{\mathcal{N}-1},\{k\}_{\mathcal{N}}) }{  \braket{\{\mu \}_{\mathcal{N}-1}|\{\mu \}_{\mathcal{N}-1}}  \ldots   \braket{\{\mu \}_{1}|\{\mu \}_{1}}    \braket{\{\lambda \}_{1}|\{\lambda \}_{1}}  \ldots  \braket{\{\lambda \}_{\mathcal{N}-1}|\{\lambda \}_{\mathcal{N}-1}}           }},
		\label{densityopop2}
\end{equation}
where \\

$\displaystyle{ \zeta \left(\{z_{j}\},t,\{E_{\{\alpha\}_{s}}\},\{P_{\{\alpha\}_{s}}\}\right)= \mathrm{e}^{i\left(E_{\{\lambda\}_{1}}-E_{\{\mu\}_{1}}\right)t }  \mathrm{e}^{-i\left(P_{\{\lambda\}_{\mathcal{N}-1}}-P_{\{\mu\}_{\mathcal{N}-1}}\right)z_{1} }   }$
\begin{equation}
\displaystyle{\times \prod_{p=1}^{\mathcal{N}-2} \mathrm{e}^{-i\left(P_{\{\mu \}_{p+1}}+P_{\{\lambda \}_{p}}-P_{\{\mu \}_{p}}-P_{\{\lambda \}_{p+1}} \right)z_{\mathcal{N}-p} }  }
		\label{densityopop3}
\end{equation}

The general formula (\ref{densityopop2}) is very complicated. Fortunately, it is possible to prepare the iterative procedure which uses the results presented in this section. The method is perfect to numerical applications \cite{SyrwSacha}.

\section{The iterative procedure}
\label{sec:iterative}

Let us now consider the $\mathcal{N}$-particle system and measure all the particles not simultaneously, but step by step. In this case, we want to get to know the probability density of finding the particle in every step. The one-particle density matrix in the first step may be written in the following form \\

$\displaystyle{ \rho_{1}(z_{1},t)=\frac{ \braket{ \{k\}_{\mathcal{N}}|\hat{\Psi}_{H}^{\dagger}(z_{1},t)\hat{\Psi}_{H}(z_{1},t) |\{k\}_{\mathcal{N}} } }{  \braket{\{k\}_{\mathcal{N}}|\{k \}_{\mathcal{N}}}   }  }$
\begin{equation}
\displaystyle{ = \sum_{\{\mu \}_{\mathcal{N}-1}} \frac{\braket{ \{k\}_{\mathcal{N}}|\hat{\Psi}_{H}^{\dagger}(z_{1},t)|\{\mu \}_{\mathcal{N}-1} }   }{ \sqrt{ \braket{\{k\}_{\mathcal{N}}|\{k \}_{\mathcal{N}}}     \braket{\{\mu \}_{\mathcal{N}-1}|\{\mu \}_{\mathcal{N}-1}}  } }     \frac{ \braket{ \{\mu \}_{\mathcal{N}-1}|\hat{\Psi}_{H}(z_{1},t) |\{k\}_{\mathcal{N}} }  }{ \sqrt{ \braket{\{k\}_{\mathcal{N}}|\{k \}_{\mathcal{N}}}     \braket{\{\mu \}_{\mathcal{N}-1}|\{\mu \}_{\mathcal{N}-1}}  } }}.
		\label{iterative1}
\end{equation}
Rewriting an initial state in normalized form
\begin{equation}
\displaystyle{\ket{\psi_{0} } =\frac{ \ket{\{ k \}_{\mathcal{N}}}}{ \sqrt{ \braket{\{k\}_{\mathcal{N}}|\{k \}_{\mathcal{N}}} } } },
		\label{iterative2}
\end{equation}
we notice that the $\mathcal{N}-1$-particle state $\ket{ \psi_{1} } $ has the following form\\

$\displaystyle{ \ket{\psi_{1} } \equiv \frac{\hat{\Psi}_{H}(z_{1},t) }{\sqrt{\mathcal{N}}} \ket{\psi_{0} } =\frac{1}{\sqrt{\mathcal{N}}}   \sum_{\{\mu \}_{\mathcal{N}-1}}         \frac{ \ket{\{\mu \}_{\mathcal{N}-1}}  \braket{ \{\mu \}_{\mathcal{N}-1}|\hat{\Psi}_{H}(z_{1},t) |\{k\}_{\mathcal{N}} }  }{   \braket{\{\mu \}_{\mathcal{N}-1}|\{\mu \}_{\mathcal{N}-1}}   \sqrt{ \braket{\{k\}_{\mathcal{N}}|\{k \}_{\mathcal{N}}}  } }   }$
\begin{equation}
\displaystyle{=\frac{1}{\sqrt{\mathcal{N}}}   \sum_{\{\mu \}_{\mathcal{N}-1}}    \mathrm{e}^{i \left( E_{\{\mu\}_{\mathcal{N}-1} }- E_{\{k\}_{\mathcal{N}} } \right)t}    \mathrm{e}^{-i \left( P_{\{\mu\}_{\mathcal{N}-1} }- P_{\{k\}_{\mathcal{N}} } \right)z_{1}}   \frac{ \ket{\{\mu \}_{\mathcal{N}-1}} \mathcal{F}(\{ \mu \}_{\mathcal{N}-1}, \{ k \}_{\mathcal{N}})  }{   \braket{\{\mu \}_{\mathcal{N}-1}|\{\mu \}_{\mathcal{N}-1}}   \sqrt{ \braket{\{k\}_{\mathcal{N}}|\{k \}_{\mathcal{N}}}  } }}.
		\label{iterative3}
\end{equation}
Hence \\

$\displaystyle{\rho_{1}(z_{1},t)=\braket{\psi_{1}|\psi_{1}}=\frac{1}{\mathcal{N}}\braket{ \psi_{0}|\hat{\Psi}_{H}^{\dagger}(z_{1},t)\hat{\Psi}_{H}(z_{1},t) |\psi_{0} } }$
\begin{equation}
\displaystyle{=\frac{1}{\mathcal{N}} \sum_{\{\mu \}_{\mathcal{N}-1}}   \frac{ | \mathcal{F}(\{ \mu \}_{\mathcal{N}-1}, \{ k \}_{\mathcal{N}})|^{2}  }{   \braket{\{\mu \}_{\mathcal{N}-1}|\{\mu \}_{\mathcal{N}-1}}    \braket{\{k\}_{\mathcal{N}}|\{k \}_{\mathcal{N}}}   }= \mathrm{const}}.
		\label{iterative4}
\end{equation}
One instantly sees that, in the first step the probability density of finding one particle in the system has to be uniform.

Defining \cite{SyrwSacha}
\begin{equation}
\displaystyle{\Gamma(z_{1},t,\{\mu\}_{\mathcal{N}-1})\equiv    \frac{\Lambda\left(\{\mu\}_{\mathcal{N}-1}, \{k\}_{\mathcal{N}},z_{1},t   \right)  \mathcal{F}(\{ \mu \}_{\mathcal{N}-1}, \{ k \}_{\mathcal{N}})  }{ \sqrt{ \mathcal{N} \braket{\{\mu \}_{\mathcal{N}-1}|\{\mu \}_{\mathcal{N}-1}}    \braket{\{k\}_{\mathcal{N}}|\{k \}_{\mathcal{N}}}   } }},
		\label{gamma1}
\end{equation}
where 
\begin{equation}
\displaystyle{\Lambda\left(\{\mu\}_{\mathcal{N}}, \{\lambda\}_{\mathcal{M}},z_{j},t   \right)\equiv  \mathrm{e}^{i \left( E_{\{\mu\}_{\mathcal{N}} }- E_{\{\lambda\}_{\mathcal{M}} } \right)t}    \mathrm{e}^{-i \left( P_{\{\mu\}_{\mathcal{N}} }- P_{\{\lambda\}_{\mathcal{M}} }  \right)z_{j}}},
		\label{Lambda}
\end{equation}
we obtain compact form of $\ket{\psi_{1}}$ and $\rho_{1}(z_{1},t)$
\begin{equation}
\displaystyle{\ket{\psi_{1}}=   \sum_{\{\mu \}_{\mathcal{N}-1}}  \frac{ \ket{\{\mu \}_{\mathcal{N}-1}} }{  \sqrt{\braket{\{\mu \}_{\mathcal{N}-1}|\{\mu \}_{\mathcal{N}-1}}}}  \Gamma(z_{1},t,\{\mu\}_{\mathcal{N}-1}) },
		\label{iterative5}
\end{equation}
\begin{equation}
\displaystyle{\rho_{1}(z_{1},t)=   \sum_{\{\mu \}_{\mathcal{N}-1}}  \left| \Gamma(z_{1},t,\{\mu\}_{\mathcal{N}-1}) \right|^{2} }.
		\label{dens1}
\end{equation}

Let us now consider the second particle. The state of the system after second measurement ($\mathcal{N}-2$-particle state) is given by \\  

$\displaystyle{ \ket{\psi_{2} } \equiv \frac{\hat{\Psi}_{H}(z_{2},t) }{\sqrt{\mathcal{N}-1}} \ket{\psi_{1} } =  \frac{1}{\sqrt{\mathcal{N}-1}} \sum_{\{\mu \}_{\mathcal{N}-2}}         \frac{ \ket{\{\mu \}_{\mathcal{N}-2}}    }{  \sqrt{ \braket{\{\mu \}_{\mathcal{N}-2}|\{\mu \}_{\mathcal{N}-2}}}   }   }$
\begin{center}
$\displaystyle{ \times    \sum_{\{\mu \}_{\mathcal{N}-1}} \frac{\braket{\{\mu \}_{\mathcal{N}-2} |\hat{\Psi}_{H}(z_{2},t)| \{\mu \}_{\mathcal{N}-1} }  }{\sqrt{ \braket{ \{\mu \}_{\mathcal{N}-2} | \{\mu \}_{\mathcal{N}-2} }  \braket{ \{\mu \}_{\mathcal{N}-1} | \{\mu \}_{\mathcal{N}-1} }  }  }  \Gamma(z_{1},t,\{\mu\}_{\mathcal{N}-1})  }$
\end{center}

$\displaystyle{ =  \sum_{\{\mu \}_{\mathcal{N}-2}}   \frac{ \ket{\{\mu \}_{\mathcal{N}-2}}    }{  \sqrt{ \braket{\{\mu \}_{\mathcal{N}-2}|\{\mu \}_{\mathcal{N}-2}}}  }    }$
\begin{center}
$\displaystyle{ \times \sum_{\{\mu \}_{\mathcal{N}-1}}  \frac{\Lambda\left(\{\mu\}_{\mathcal{N}-2}, \{\mu\}_{\mathcal{N}-1},z_{2},t   \right)  \mathcal{F}(\{ \mu \}_{\mathcal{N}-2}, \{ \mu \}_{\mathcal{N}-1})  }{ \sqrt{\mathcal{N}-1} \sqrt{  \braket{\{\mu \}_{\mathcal{N}-2}|\{\mu \}_{\mathcal{N}-2}}  \braket{\{\mu \}_{\mathcal{N}-1}|\{\mu \}_{\mathcal{N}-1}}    } }   \Gamma(z_{1},t,\{\mu\}_{\mathcal{N}-1})  }$
\end{center}
\begin{equation}
\displaystyle{    =   \sum_{\{\mu \}_{\mathcal{N}-2}}    \frac{ \ket{\{\mu \}_{\mathcal{N}-2}}    }{  \sqrt{ \braket{\{\mu \}_{\mathcal{N}-2}|\{\mu \}_{\mathcal{N}-2}}}  }  \Gamma(z_{1},z_{2},t,\{\mu \}_{\mathcal{N}-2})}, 
		\label{iterative6}
\end{equation}
while \\

$\displaystyle{ \Gamma(z_{1},z_{2},t,\{\mu \}_{\mathcal{N}-2}) }$
\begin{equation}
\displaystyle{ \equiv \sum_{\{\mu \}_{\mathcal{N}-1}}  \frac{\Lambda\left(\{\mu\}_{\mathcal{N}-2}, \{\mu\}_{\mathcal{N}-1},z_{2},t   \right)  \mathcal{F}(\{ \mu \}_{\mathcal{N}-2}, \{ \mu \}_{\mathcal{N}-1})  }{ \sqrt{\mathcal{N}-1} \sqrt{  \braket{\{\mu \}_{\mathcal{N}-2}|\{\mu \}_{\mathcal{N}-2}}  \braket{\{\mu \}_{\mathcal{N}-1}|\{\mu \}_{\mathcal{N}-1}}    } }   \Gamma(z_{1},t,\{\mu\}_{\mathcal{N}-1})  }. 
		\label{gamma2}
\end{equation}
Probability density in the second step has exactly the same form as before (\ref{dens1}) \cite{SyrwSacha}
\begin{equation}
\displaystyle{\rho_{2}(z_{2},t)= \rho_{02}  \sum_{\{\mu \}_{\mathcal{N}-2}}  \left| \Gamma(z_{1},z_{2},t,\{\mu \}_{\mathcal{N}-2}) \right|^{2} },
		\label{dens2}
\end{equation}
where $\rho_{02}$ is normalization factor and $z_{1}$ is fixed\footnote{One should remember that the procedure is iterative. Therefore, another measured particles have fixed coordinate positions and subsequent probability densities are one-particle.}.

After two steps, we notice that the procedure is given by a simple prescription \cite{SyrwSacha}
\begin{equation}
\displaystyle{\ket{\psi_{j}}=\frac{\hat{\Psi}_{H}(z_{j},t)}{\sqrt{\mathcal{N}+1-j}} \ket{\psi_{j-1}}= \sum_{\{\mu \}_{\mathcal{N}-j}}    \frac{ \ket{\{\mu \}_{\mathcal{N}-j}}    }{  \sqrt{ \braket{\{\mu \}_{\mathcal{N}-j}|\{\mu \}_{\mathcal{N}-j}}}  }  \Gamma(z_{1},\ldots,z_{j},t,\{\mu \}_{\mathcal{N}-j}) },
		\label{interacja1}
\end{equation}
\begin{flushleft}
$\displaystyle{\Gamma(z_{1},\ldots,z_{j},t,\{\mu \}_{\mathcal{N}-j})  }$
\end{flushleft}
\begin{equation}
\displaystyle{    \equiv   \sum_{\{\mu \}_{\mathcal{N}-j+1}}  \frac{\Lambda\left(\{\mu\}_{\mathcal{N}-j}, \{\mu\}_{\mathcal{N}-j+1},z_{j},t   \right) \mathcal{F}(\{ \mu \}_{\mathcal{N}-j}, \{ \mu \}_{\mathcal{N}-j+1})  \Gamma(z_{1},\ldots, z_{j-1},t,\{\mu\}_{\mathcal{N}-j+1})  }{ \sqrt{\mathcal{N}-j+1} \sqrt{  \braket{\{\mu \}_{\mathcal{N}-j}|\{\mu \}_{\mathcal{N}-j}}  \braket{\{\mu \}_{\mathcal{N}-j+1}|\{\mu \}_{\mathcal{N}-j+1}}    } }  },
		\label{interacja2}
\end{equation}
\begin{equation}
\displaystyle{   \rho_{j}(z_{j},t)=\rho_{0j} \sum_{\{\mu \}_{\mathcal{N}-j}} \left|   \Gamma(z_{1},\ldots,z_{j},t,\{\mu \}_{\mathcal{N}-j})    \right|^{2}   },
		\label{interacja3}
\end{equation}
where $\rho_{0j}$ is normalization factor for $j$-th probability density.

The last step in the procedure is quite different. After $\mathcal{N}-1$ steps our state is given by
\begin{equation}
\displaystyle{\ket{\psi_{\mathcal{N}-1}}= \sum_{\{\mu \}_{1}}    \frac{ \ket{\{\mu \}_{1}}    }{  \sqrt{ \braket{\{\mu \}_{1}|\{\mu \}_{1}}}  }  \Gamma(z_{1},\ldots,z_{\mathcal{N}-1},t,\{\mu \}_{1}) },
		\label{laststep1}
\end{equation}
where $\Gamma(z_{1},\ldots,z_{\mathcal{N}-1},t,\{\mu \}_{1})$ is known. The last state may be calculated in the following way
\begin{flushleft}
$\displaystyle{\ket{\psi_{\mathcal{N}}}=\hat{\Psi}_{H}(z_{\mathcal{N}},t) \ket{\psi_{\mathcal{N}-1}}= \mathrm{e}^{-i\hat{P}z_{\mathcal{N}}} \mathrm{e}^{i\hat{H}t}  \hat{\Psi}_{H}(0,0)  \mathrm{e}^{-i\hat{H}t} \mathrm{e}^{i\hat{P}z_{\mathcal{N}}} \ket{\psi_{\mathcal{N}-1}} }$
\end{flushleft}
\begin{equation}
\displaystyle{=  \mathrm{e}^{-i\hat{P}z_{\mathcal{N}}} \mathrm{e}^{i\hat{H}t}   \sum_{\{\mu \}_{1}}    \frac{    \mathrm{e}^{-i E_{\{\mu \}_{1}}t} \mathrm{e}^{i P_{\{\mu \}_{1}} z_{\mathcal{N}}}  \hat{\Psi}_{H}(0,0) \ket{\{\mu \}_{1}}    }{  \sqrt{ \braket{\{\mu \}_{1}|\{\mu \}_{1}}}  }  \Gamma(z_{1},\ldots,z_{\mathcal{N}-1},t,\{\mu \}_{1}) },
		\label{laststep2}
\end{equation}
then using the equation (3.3) from \cite{formfactors} one gets
\begin{equation}
\displaystyle{\hat{\Psi}_{H}(0,0) \ket{\{\mu \}_{1}} = -i \sqrt{c} \, a(\mu) \left( \mathop{\prod_{ m = 1 }^{1}}_{m \neq l} f(\mu_{l},\mu_{m}) \right)\ket{0}=-i \sqrt{c} \, \mathrm{e}^{-i\mathrm{L}\mu /2} \ket{0}},
		\label{laststep3}
\end{equation}
where we have used the fact that
\begin{equation}
\displaystyle{\mathop{\prod_{ m = 1 }^{1}}_{m \neq l} f(\mu_{l},\mu_{m}) =1 , \, \, \, \, \, \, \, \, a(\mu)=\mathrm{e}^{-i\mathrm{L}\mu /2}}.
		\label{laststep4}
\end{equation}
Therefore we obtain
\begin{flushleft}
$\displaystyle{\ket{\psi_{\mathcal{N}}} =-i \sqrt{c} \, \mathrm{e}^{-i\mathrm{L}\mu /2} \mathrm{e}^{-i\hat{P}z_{\mathcal{N}}} \mathrm{e}^{i\hat{H}t}  \ket{0}  \sum_{\{\mu \}_{1}}    \frac{    \mathrm{e}^{-i E_{\{\mu \}_{1}}t} \mathrm{e}^{i P_{\{\mu \}_{1}} z_{\mathcal{N}}}    }{  \sqrt{ \braket{\{\mu \}_{1}|\{\mu \}_{1}}}  }  \Gamma(z_{1},\ldots,z_{\mathcal{N}-1},t,\{\mu \}_{1}) }$
\end{flushleft}
\begin{equation}
\displaystyle{=-i \sqrt{c}    \sum_{\{\mu \}_{1}}    \frac{    \mathrm{e}^{-i E_{\{\mu \}_{1}}t} \mathrm{e}^{i P_{\{\mu \}_{1}}( z_{\mathcal{N}}-\mathrm{L}/2)}   \ket{0}  }{  \sqrt{ \braket{\{\mu \}_{1}|\{\mu \}_{1}}}  }  \Gamma(z_{1},\ldots,z_{\mathcal{N}-1},t,\{\mu \}_{1}) },
		\label{laststep5}
\end{equation}
which leads to 
\begin{flushleft}
$\displaystyle{\rho_{\mathcal{N}}(z_{\mathcal{N}},t)=\rho_{0\mathcal{N}}\braket{ \psi_{\mathcal{N}}| \psi_{\mathcal{N}}}  =\rho_{0\mathcal{N}}c \sum_{\{\mu \}_{1},\{\lambda \}_{1}}  \frac{    \Lambda \left( \{\lambda \}_{1},\{\mu \}_{1}, z_{\mathcal{N}}-\mathrm{L}/2,t  \right)  }{  \sqrt{ \braket{\{\mu \}_{1}|\{\mu \}_{1}}  \braket{\{\lambda \}_{1}|\{\lambda \}_{1}}  }  } }$
\end{flushleft}
\begin{equation}
\displaystyle{  \times \Gamma^{*}(z_{1},\ldots,z_{\mathcal{N}-1},t,\{\lambda \}_{1})\Gamma(z_{1},\ldots,z_{\mathcal{N}-1},t,\{\mu \}_{1})    }.
		\label{laststep6}
\end{equation}

\section{The indispensable cut offs}
\label{sec:cutoffs}

In our iterative method we have to perform a summations over all possible eigenstates $\ket{ \{ \mu \}_{\mathcal{M}}  }$ for $\mathcal{M}=1,\ldots , \mathcal{N} -1$. Unfortunately, there is infinite number of states $\ket{ \{ \mu \}_{\mathcal{M}}  }$ for each $\mathcal{M}$. In numerical calculations one has to restrict to the relevant elements of the sums only. Fortunately, interesting states are very close energetically to the ground state of the energy ($P_{\mathcal{N}}^{g}=0, \, E_{\mathcal{N}}^{g}=0$). Therefore, we believe that it is sufficient to restrict our Hilbert space to the states whose absolute values of total momentum $|P_{\mathcal{M}}|$ and single quasi-momentum $|k_{j}|, \, j=1,\ldots,\mathcal{M}$, are not bigger than a few Fermi quasi-momentum $k_{F}$ (see also \cite{SyrwSacha}). At the level of the first step of our iterative procedure, it means that the normalized form factors 
\begin{equation}
\displaystyle{F(\{ \mu \}_{\mathcal{N}-1}, \{ k \}_{\mathcal{N}})= \frac{  \mathcal{F}(\{ \mu \}_{\mathcal{N}-1}, \{ k \}_{\mathcal{N}}) }{  \sqrt{ \braket{\{\mu \}_{\mathcal{N}-1}|\{\mu \}_{\mathcal{N}-1}} }   \sqrt{\braket{\{k\}_{\mathcal{N}}|\{k \}_{\mathcal{N}}} }  }},
		\label{normalizedformfactors}
\end{equation}
decay very quickly with the increase of energy of the state $\ket{ \{ \mu \}_{\mathcal{N}-1}  }$, if the state $\ket{ \{ k \}_{\mathcal{N}}  }$ is very close to the ground state of energy. The Hilbert space cut off just eliminates the form factors whose contribution to the sum is negligible. Because of the structure of our procedure (the values of $F(\{ \mu \}_{\mathcal{N}-1}, \{ k \}_{\mathcal{N}})$ play an important role in every step of calculations), we believe that our results will be converged.

It turns out, that in our computer simulations, the case of strong interactions ($c=8$, $\mathcal{N}=8$, $\mathrm{L}=1 \Longrightarrow \gamma=\frac{c}{\rho}=1$) is the most difficult. In this regime, for $\mathcal{N}=8$ we have truncated the sums to the states parametrized by the collections $\{ \mathcal{I}_{1},\ldots, \mathcal{I}_{\mathcal{M}} \}, \, \mathcal{M}=1,\ldots , \mathcal{N} -1$, where  
\begin{equation}
\displaystyle{\left\{ \begin{array}{llll}   -9 \leq \mathcal{I}_{j} \leq 9 , & & \mathcal{M} & \mathrm{odd}        , \, \, \\  -\frac{19}{2} \leq \mathcal{I}_{j} \leq \frac{19}{2} , & & \mathcal{M} & \mathrm{even}, \end{array} \right. }
		\label{obciecia1}
\end{equation}
and restricted to the values of the single quasi-momenta and the total momentum of the states contained in the Table 3.1. a). The weak interactions regime is far simpler because using exactly the same collections as in the case of strong interactions (\ref{obciecia1}) we easily obtain much higher cut offs for quasi-momenta and the total momentum of the states in terms of the Fermi momentum (Table 3.1. b)). The comparison of numerical results presented in the Tables 3.1. a) and b) shows clearly that the  summations in the case of weak interactions consist of the states\\

\begin{footnotesize}Table 3.1. The maximal values of the single quasi-momentum $k_{j}$ and the total momentum $P$ of the $\mathcal{M}$-particle states in the truncated summation. The values of the Fermi quasi-momenta $k_{F}$ and the number of states for $\mathcal{M}=1,\ldots,7$ are presented too. Table a) contains data for the case of strong interactions ($c=8$, $\mathcal{N}=8$, $\mathrm{L}=1 \Longrightarrow \gamma=c/\rho=1$), b) presents data for the case of weak interactions ($c=0.08$, $\mathcal{N}=8$, $\mathrm{L}=1$  $\Longrightarrow$  $\gamma$ $=c/\rho=0.01$).\end{footnotesize} 

\large{a)}
\normalsize
\begin{center}
  \begin{tabular}{ | c | c | c | c | c |}
    \hline
    $\mathcal{M}$ & $k_{F}$ & $|k_{j}| / k_{F} \leq $ & $|P|/k_{F} \leq $ & number of elements  \\ \hline
						7 & 8.8005 & 4.9577  & 5.7117  & 26080\\ \hline 
						6 & 7.7328 & 6.2739  & 5.6878  & 17572\\ \hline 
						5 & 6.5742 & 7.2545  & 5.7344  & 5124\\ \hline 
						4 & 5.2965 & 9.0047  & 7.1178  & 4845\\ \hline 
						3 & 3.8524 & 14.0478 & 39.1434 & 969\\ \hline 
						2 & 2.1538 & 27.2559 & 52.5118 & 190 \\ \hline 
						1 & 0  			& $18 \pi k_{F}$  &  $18 \pi k_{F}$ & 19 \\ \hline 
  \end{tabular}
	\end{center} 

\large{b)}
\normalsize
 \begin{center}
 \begin{tabular}{ | c | c | c | c | c |}
    \hline
    $\mathcal{M}$ & $k_{F}$ & $|k_{j}| / k_{F} \leq $ & $|P|/k_{F} \leq $ & number of elements  \\ \hline
						7 & 1.0573 & 36.2848  & 47.5414  & 26080 \\ \hline 
						6 & 0.9372 & 47.4582  & 46.9295  & 17572 \\ \hline 
						5 & 0.8054 & 54.9944  & 46.8079  & 5124  \\ \hline 
						4 & 0.6581 & 77.3797  & 305.5188 & 4845  \\ \hline 
						3 & 0.4883 & 103.9397 & 308.8193 & 969   \\ \hline 
						2 & 0.2819 & 201.5983 & 401.1966 & 190   \\ \hline 
						1 & 0  			& $18 \pi k_{F}$  &  $18 \pi k_{F}$ & 19 \\ \hline 
  \end{tabular}
\end{center}
 which are much more ,,distant'' (in the sense of energy and total momentum) from the initial state than in the case of relatively strong interactions.  

In order to calculate the reduced single-particle density $\rho_{1}(z)$, we choose only the normalized form factors whose absolute values are not smaller than $\alpha |F_{max}| $, where $|F_{max}|$ is the maximal absolute value of the collected normalized form factors. It turns out that the number of relevant elements in the case of strong interactions is much bigger than in the case of weak interactions. The $\alpha$ parameter was determined by the analysis of the integrals of the single particle probability densities $\int _{0}^{\mathrm{L}}\rho_{1}(z)\mathrm{d}z $. For the case of strong interactions ($c=8$) we have established $\alpha_{s}=10^{-3}$ which causes the deviation of $\int_{0}^{\mathrm{L}}\rho_{1}(z)\mathrm{d}z $ from 1 of the order $10^{-4}$. In the case of weak coupling ($c=0.08$) one can easily obtain the deviation of the order $10^{-6}$ - we have taken\footnote{The indices ,,$s$'' and ,,$w$'' corresponds to the strong and weak interactions, respectively.} $\alpha_{w}=10^{-4}$. The described cut off was executed in every step of all simulations with parameter $\alpha_{s}$ ($\alpha_{w}$) for strong (weak) interactions \cite{SyrwSacha}.  In the Figures 3.1. a) and b) we present histograms of the absolute values of normalized form factors $|F(\{ \mu\}_{7},\{ k\}_{8})|$ for two different initial states corresponding to one (a)) and two (b)) hole excitations with different coupling constants ($c=0.08$ and $c=8$). \\
\begin{center}\includegraphics[width=16cm,angle=0]{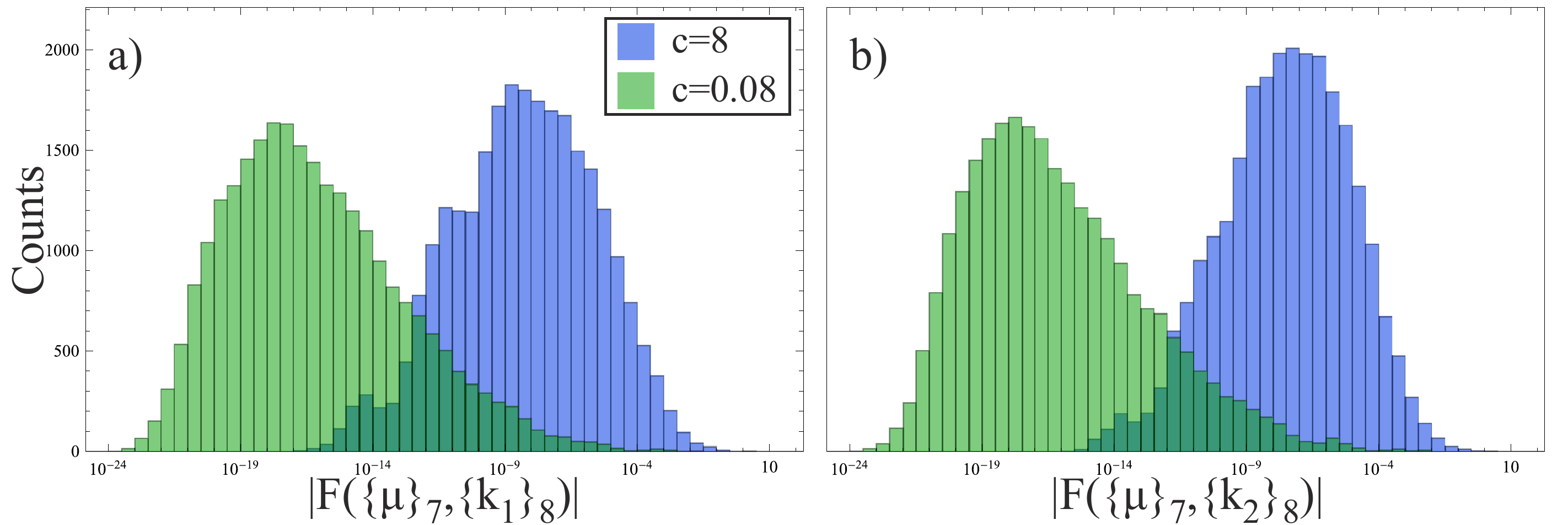}\end{center}
\begin{footnotesize}Figure 3.1. a) The distributions of $|F(\{ \mu\}_{7},\{ k_{1} \}_{8})|$ for weak ($c=0.08$) and strong ($c=8$) interactions in the case of one hole excited state $\{ k_{1} \}_{8}$ parametrized by the collection $\mathcal{C}^{8}_{1 \, \mathrm{sol}}=\mathcal{C}^{8}_{g}\left[ \frac{1}{2} \rightarrow \frac{9}{2} \right]=\left\{ -\frac{7}{2},\ldots,-\frac{1}{2},\frac{3}{2},\ldots , \right. $ $\left. \frac{9}{2}  \right\}$. The same kind of distributions but for different state  $\{ k_{2} \}_{8}$ was presented in plot b). The state $\{ k_{2} \}_{8}$ corresponds to two hole excitation and is given by  $\mathcal{C}^{8}_{2 \, \mathrm{sol}}=\mathcal{C}^{8}_{g}\left[ -\frac{1}{2} \rightarrow -\frac{9}{2},\frac{1}{2} \rightarrow \frac{9}{2} \right]=\left\{ -\frac{9}{2},\ldots,-\frac{3}{2},\frac{3}{2},\right.$ $\left.\ldots ,\frac{9}{2}  \right\}$. Because of the character of the excitation, we will call the state $\{ k_{2} \}_{8}$ as \emph{symmetric}. \end{footnotesize} \\

We immediately see that, the maxima of distributions for strong interactions are much closer to the $\mathrm{max}\left\{|F(\{ \mu\}_{7},\{ k \}_{8})|\right\}$ than in the case of weak coupling. This indicates that, for a strong coupling, we need to take into account a few times more elements in every step of simulations which causes a very significant increase of the computer simulation time.

\chapter{The numerical results}
\label{numericalresults}

\section{Preliminary information}
\label{sec:consstates}

Analyzed system is given by the Lieb-Liniger Hamiltonian (\ref{qmdescription3}), where the box length $\mathrm{L}=1$ and the number of particles $\mathcal{N}=8$. In our computer simulations we have considered several low-lying excited states belonging to the branch of the Lieb's type II excitations for two different coupling constants $c=0.08$ ($\gamma=c/\rho =0.01$) and $c=8$ ($\gamma=c/\rho =1$), which correspond to weak and strong interaction regimes, respectively. Furthermore, we have chosen $t=0$. Our purpose is to show that the Lieb's type II excitations have dark solitonic nature. In the next sections we present the emergence of dark solitons in the course of measurements of particle positions. The considered initial states may be divided into single and double soliton states which are parametrized by the following collections: \\ 

\emph{\textbf{Single soliton states:}} \\
\begin{equation}
\displaystyle{\begin{array}{ll}    \displaystyle{\mathcal{C}^{8,(\mathrm{black})}_{1 \, \mathrm{sol}}=\mathcal{C}^{8}_{g}\left[ \frac{1}{2} \rightarrow \frac{9}{2} \right]=\left\{ -\frac{7}{2},\ldots,-\frac{1}{2},\frac{3}{2},\ldots ,\frac{9}{2} \right\},  } & \displaystyle{P^{8,(\mathrm{black})}_{1 \, \mathrm{sol}}=\frac{8\pi}{\mathrm{L}}}, \, \, \\ \\
\displaystyle{\mathcal{C}^{8,(\mathrm{gray}1)}_{1 \, \mathrm{sol}}=\mathcal{C}^{8}_{g}\left[ -\frac{1}{2} \rightarrow \frac{9}{2} \right]=\left\{ -\frac{7}{2},\ldots,-\frac{3}{2},\frac{1}{2},\ldots ,\frac{9}{2} \right\},  } & \displaystyle{P^{8,(\mathrm{gray} 1)}_{1 \, \mathrm{sol}}=\frac{10\pi}{\mathrm{L}}}, \, \, \\ \\  
\displaystyle{\mathcal{C}^{8,(\mathrm{gray}2)}_{1 \, \mathrm{sol}}=\mathcal{C}^{8}_{g}\left[ -\frac{3}{2} \rightarrow \frac{9}{2} \right]=\left\{ -\frac{7}{2},\ldots,-\frac{5}{2},-\frac{1}{2},\ldots ,\frac{9}{2} \right\},  } & \displaystyle{P^{8,(\mathrm{gray} 2)}_{1 \, \mathrm{sol}}=\frac{12\pi}{\mathrm{L}}}, \, \, \\ \\ 
\displaystyle{\mathcal{C}^{8,(\mathrm{gray}3)}_{1 \, \mathrm{sol}}=\mathcal{C}^{8}_{g}\left[ -\frac{5}{2} \rightarrow \frac{9}{2} \right]=\left\{ -\frac{7}{2},-\frac{3}{2},\ldots ,\frac{9}{2} \right\},  } & \displaystyle{P^{8,(\mathrm{gray}3)}_{1 \, \mathrm{sol}}=\frac{14\pi}{\mathrm{L}}}, \, \, \\ \\ 
\displaystyle{\mathcal{C}^{8}_{PW}=\mathcal{C}^{8}_{g}\left[ -\frac{7}{2} \rightarrow \frac{9}{2} \right]=\left\{ -\frac{5}{2},\ldots ,\frac{9}{2} \right\},  } & \displaystyle{P^{8}_{PW}=\frac{16\pi}{\mathrm{L}}}.     \end{array} }
		\label{onesoliton}
\end{equation}
We expect that the collection $\mathcal{C}^{8,(\mathrm{black})}_{1 \, \mathrm{sol}}$ corresponds to the single black soliton state because the corresponding mean-field soultion with the average particle momentum  $P^{8,(\mathrm{black})}_{1 \, \mathrm{sol}}/\mathcal{N}=\frac{\pi}{\mathrm{L}}$ reveals black soliton profile. Similarly we expect that the collections $\mathcal{C}^{8,(\mathrm{gray}1)}_{1 \, \mathrm{sol}}$, $\mathcal{C}^{8,(\mathrm{gray}2)}_{1 \, \mathrm{sol}}$ and $\mathcal{C}^{8,(\mathrm{gray}3)}_{1 \, \mathrm{sol}}$ should generate single gray soliton states, where the density notches are expected to be the shallower, the bigger total momentum. The highest possible one hole excitation of the Lieb's type II spectrum, which is parametrized by the set $\mathcal{C}^{8}_{PW}$, reduces to the plane wave ($PW$) state and is called an \emph{umklapp} excitation. All the states parametrized in (\ref{onesoliton}) have been analyzed in our simulations for the case of weak interactions ($c=0.08$). In the case of strong coupling ($c=8$), we have considered only the one hole excited state generated by $\mathcal{C}^{8,(\mathrm{black})}_{1 \, \mathrm{sol}}$.   \\ 

\emph{\textbf{Double soliton states:}} \\
\begin{equation}
\displaystyle{\begin{array}{ll}    \displaystyle{\mathcal{C}^{8,(\mathrm{sym})}_{2 \, \mathrm{sol}}=\mathcal{C}^{8}_{g}\left[-\frac{1}{2} \rightarrow -\frac{9}{2}, \frac{1}{2} \rightarrow \frac{9}{2} \right]=\left\{ -\frac{9}{2},\ldots,-\frac{3}{2},\frac{3}{2},\ldots ,\frac{9}{2} \right\},  } & \displaystyle{P^{8,(\mathrm{sym})}_{2 \, \mathrm{sol}}=0}, \, \, \\ \\
\displaystyle{\mathcal{C}^{8,(\mathrm{ns}1)}_{2 \, \mathrm{sol}}=\mathcal{C}^{8}_{g}\left[-\frac{1}{2} \rightarrow \frac{9}{2}, \frac{1}{2} \rightarrow \frac{11}{2} \right]=\left\{ -\frac{7}{2},\ldots,-\frac{3}{2},\frac{3}{2},\ldots ,\frac{11}{2} \right\},  } & \displaystyle{P^{8,(\mathrm{ns}1)}_{2 \, \mathrm{sol}}=\frac{20\pi}{\mathrm{L}}}, \, \, \\ \\   
\displaystyle{\mathcal{C}^{8,(\mathrm{ns}2)}_{2 \, \mathrm{sol}}=\mathcal{C}^{8}_{g}\left[\frac{1}{2} \rightarrow \frac{9}{2}, \frac{3}{2} \rightarrow \frac{11}{2} \right]=\left\{ -\frac{7}{2},\ldots,-\frac{1}{2},\frac{5}{2},\ldots ,\frac{11}{2} \right\},  } & \displaystyle{P^{8,(\mathrm{ns}2)}_{2 \, \mathrm{sol}}=\frac{16\pi}{\mathrm{L}}}.   \end{array} }
		\label{doublesoliton}
\end{equation}
All of the above collections are supposed to be parametrizations of double soliton states, which we verify for the case of weak interactions by the numerical simulations in the next section. For strong interactions, time of computer simulations is very long. Therefore, we have chosen only one of the states form (\ref{doublesoliton}) - the symmetric state generated by $\mathcal{C}^{8,(\mathrm{sym})}_{2 \, \mathrm{sol}}$. The upper index ,,ns'' means the ,,non-symmetrical'' two hole excitations.

The localization of dark solitons is a manifestation of the spontaneous symmetry breaking where the position of the notch localizes at any value from $z_{0} \in [0,\mathrm{L}=1)$. Therefore, we expect that in every computer simulation the position of soliton will be different. All the simulations provide 8 positions of measured particles which allows us to prepare a histogram. Obviously, the number of particles is very small, hence we perform $10^{4}$ simulations and using the periodicity of the system (a ring topology) we may easily shift the particle positions so that the position of the anticipated soliton is always the same on a ring \cite{SyrwSacha} ($z_{0}=\mathrm{L}/2$($\mathrm{L}/4$) for the single (double) soliton case). The observation allows us to examine the histogram consisting of $8 \cdot 10^{4}$ counts.  

At this point the following question arises: how to determine the position of dark soliton (density notch) in a single simulation? We propose two methods: treating the minimum of the last probability densities (for the 8th particle) as a position of a dark soliton (density notch) \cite{SyrwSacha} and the calculation of the center of mass reflection (see appendix \ref{sec:dodE}). It turns out that the first proposition is better (results are much closer to the mean-field predictions) but we present the results obtained using both methods for the purpose of comparison. 

Additionally, we analyze the changes of conditional probability density for a choice of $j$-th particle provided the previous ($j-1$) particles have been already measured. We observe various behavior of the system during the process of density notches localization (see the discussion about anticipated double soliton states for weak interaction regime).

\section{Weak interaction regime}
\label{sec:weak}

The calculations in the weak coupling regime were performed for $\mathcal{N}=8$, $\mathrm{L}=1$, $c=0.08$, i.e. $\gamma=0.01$. For every considered state we present the final results obtained from $10^{4}$ simulations. The numerical results obtained for all the considered states are presented below (see also \cite{SyrwSacha}).

\emph{\textbf{Single black soliton parametrized by $\mathcal{C}^{8,(\mathrm{black})}_{1 \, \mathrm{sol}}$}}  \\

In the Figure 4.1. we present the changes of conditional probability density for a choice of $j$-th particle provided the previous ($j-1$) particles have been already measured for four randomly chosen simulations (a)-d)). We also show the mean-field solution corresponding to the state given by $\mathcal{C}^{8,(\mathrm{black})}_{1 \, \mathrm{sol}}$, i.e. dark soliton solution related to the average particle momentum equal to $P^{8,(\mathrm{black})}_{1 \, \mathrm{sol}}/\mathcal{N}$.
\begin{center}\includegraphics[width=16cm,angle=0]{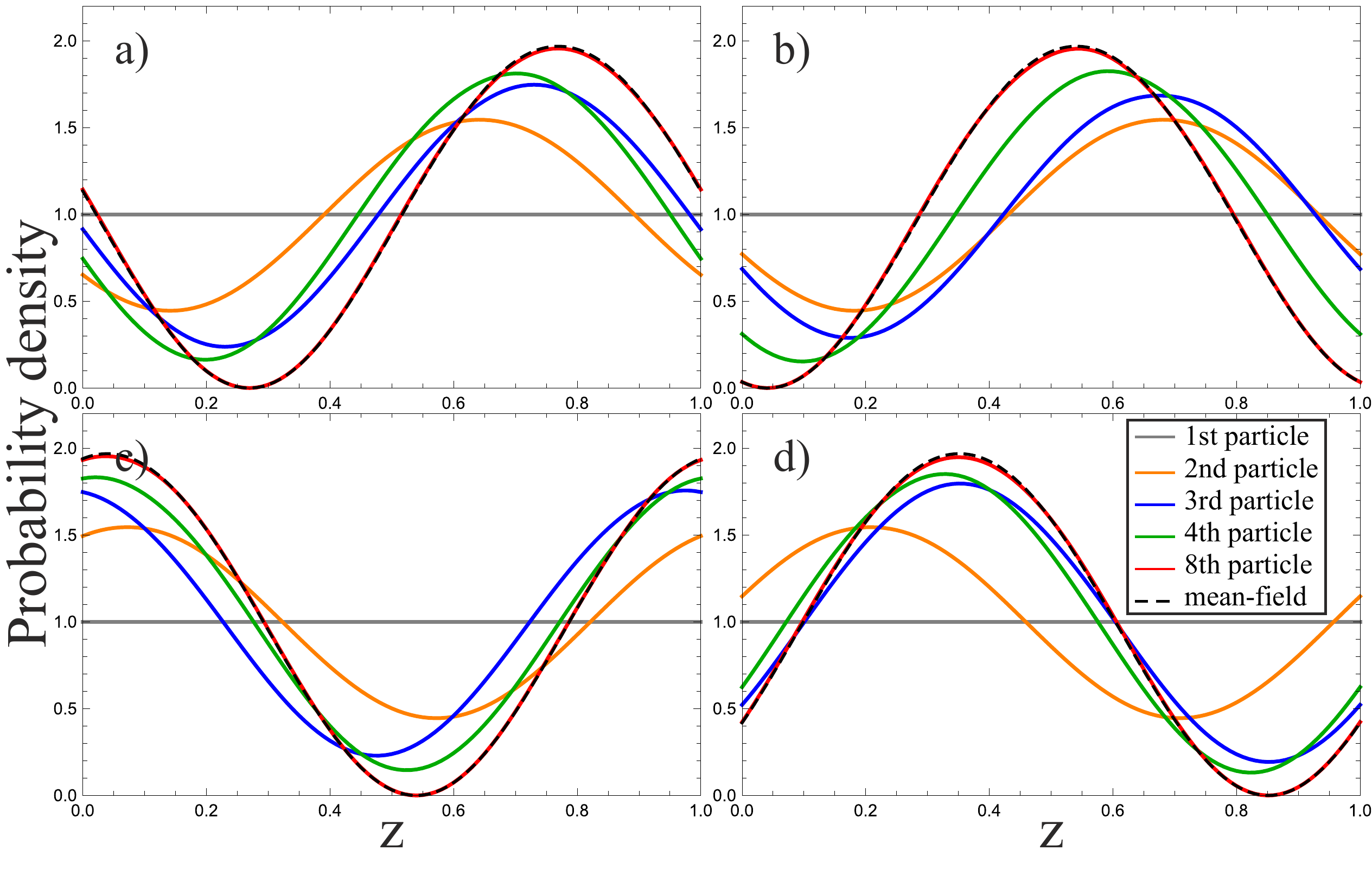}\end{center}
\begin{footnotesize}
Figure 4.1. Numerically obtained subsequent conditional probability densities (\ref{interacja3}) for four accidentally chosen simulations. As expected, the probability density for the measurement of the first particle is uniform. The density profiles which correspond to the measurement of the last particle (8th) are distinctly similar to the corresponding mean-field solution. The positions of successively measured particles: a) $\{0.642, 0.819, 0.623, 0.715, 0.904, 0.902,$ $ 0.786, 0.854 \}$, b) $\{0.682, 0.669, 0.459, 0.479, 0.503, 0.598, 0.393, 0.491  \}$, c) $\{0.072, 0.878, 0.138, 0.250, 0.047, 0.850,$ $0.037, 0.972  \}$, d) $\{0.208, 0.497, 0.231, 0.313, 0.386, 0.454, 0.376, 0.436 \}$, where the order from left to right corresponds to the order of the particle measurement.  \end{footnotesize} \\

We immediately see, that the obtained results agree surprisingly well with our expectations. Exactly the same one can say about the normalized histograms created from the measured positions of the particles. As we mentioned, in order to prepare histograms we had to shift the positions of particles in every single simulation. For this purpose, we had to find the soliton positions in all the simulations. It was done in two different ways: using the algorithm described in the appendix \ref{sec:dodE} and finding the positions of minima of the last conditional probability densities. Histograms obtained using the first (second) method are presented in the Figure 4.2. a) (b)). The histograms were prepared both from all the measured particles and the last (8th) measured particles in every single simulation. In the latter case we took the positions obtained from the last steps of the simulations. As always we compare the obtained results with the mean-field solution.

\begin{center}\includegraphics[width=16cm,angle=0]{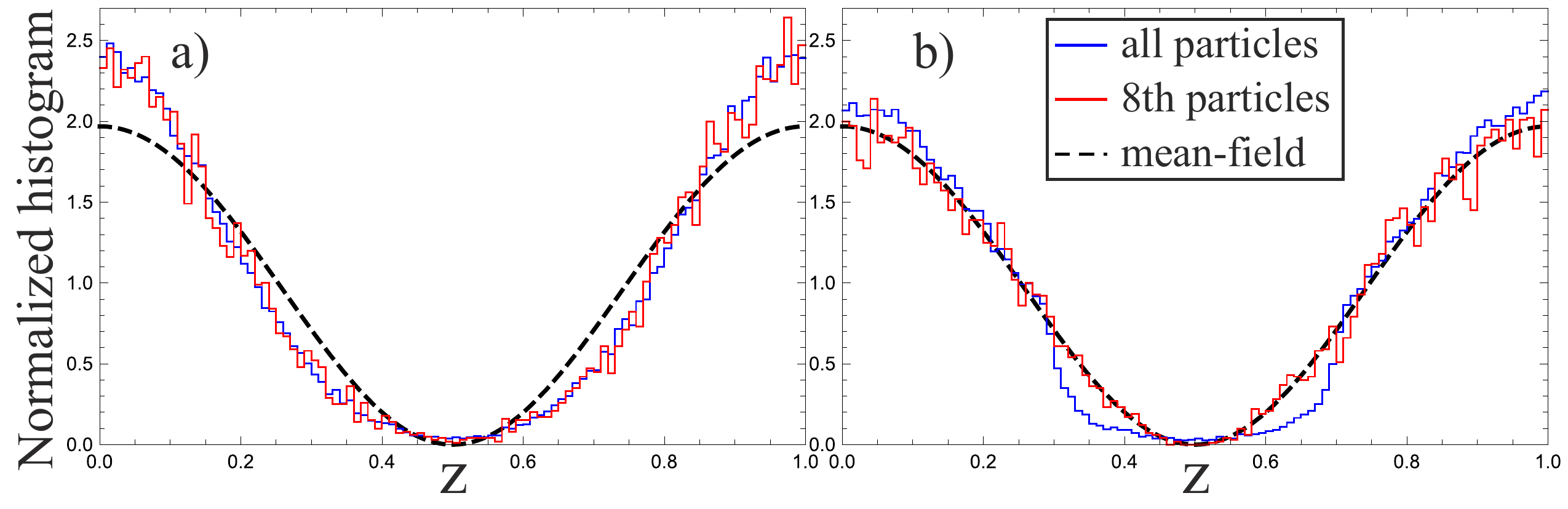}\end{center}
\begin{footnotesize}
Figure 4.2. Histograms obtained using the two methods of the soliton position determination: calculation as in appendix \ref{sec:dodE} (plot a)) and finding the minimum of the last (8th) probability density plot (b)). Blue (red) lines represent the normalized histograms obtained from all the measured particles (last particles) in every single simulation. The agreement between the mean field solution and the histogram obtained from the 8th particle positions in b) is amazingly good. We suppose that, the deviations between blue curves and the mean-field solution in b) are caused by a small number of particles $\mathcal{N}$. Moreover, we notice that the method proposed in appendix \ref{sec:dodE} gives the results that significantly deviate from our expectation, see a). \end{footnotesize}  \\ 

\emph{\textbf{Single gray soliton parametrized by $\mathcal{C}^{8,(\mathrm{gray}1)}_{1 \, \mathrm{sol}},\mathcal{C}^{8,(\mathrm{gray}2)}_{1 \, \mathrm{sol}},\mathcal{C}^{8,(\mathrm{gray}3)}_{1 \, \mathrm{sol}}$ and $\mathcal{C}^{8}_{PW}$  }} \\

As previously we show the changes of the conditional probability density (\ref{interacja3}) - two randomly chosen simulations for each of the considered states. The results presented in the Figures 4.3. a)-f) confirm our conjectures about the behavior of the depth of the density notches during the increase of the total momentum. Moreover, the probability densities for 8th particle and mean-field predictions are with excellent agreement, again. Also the expectations for the \emph{umklapp} (plane wave) excitation are confirmed well (Figure 4.3. g) and h)).
\begin{center}\includegraphics[width=16cm,angle=0]{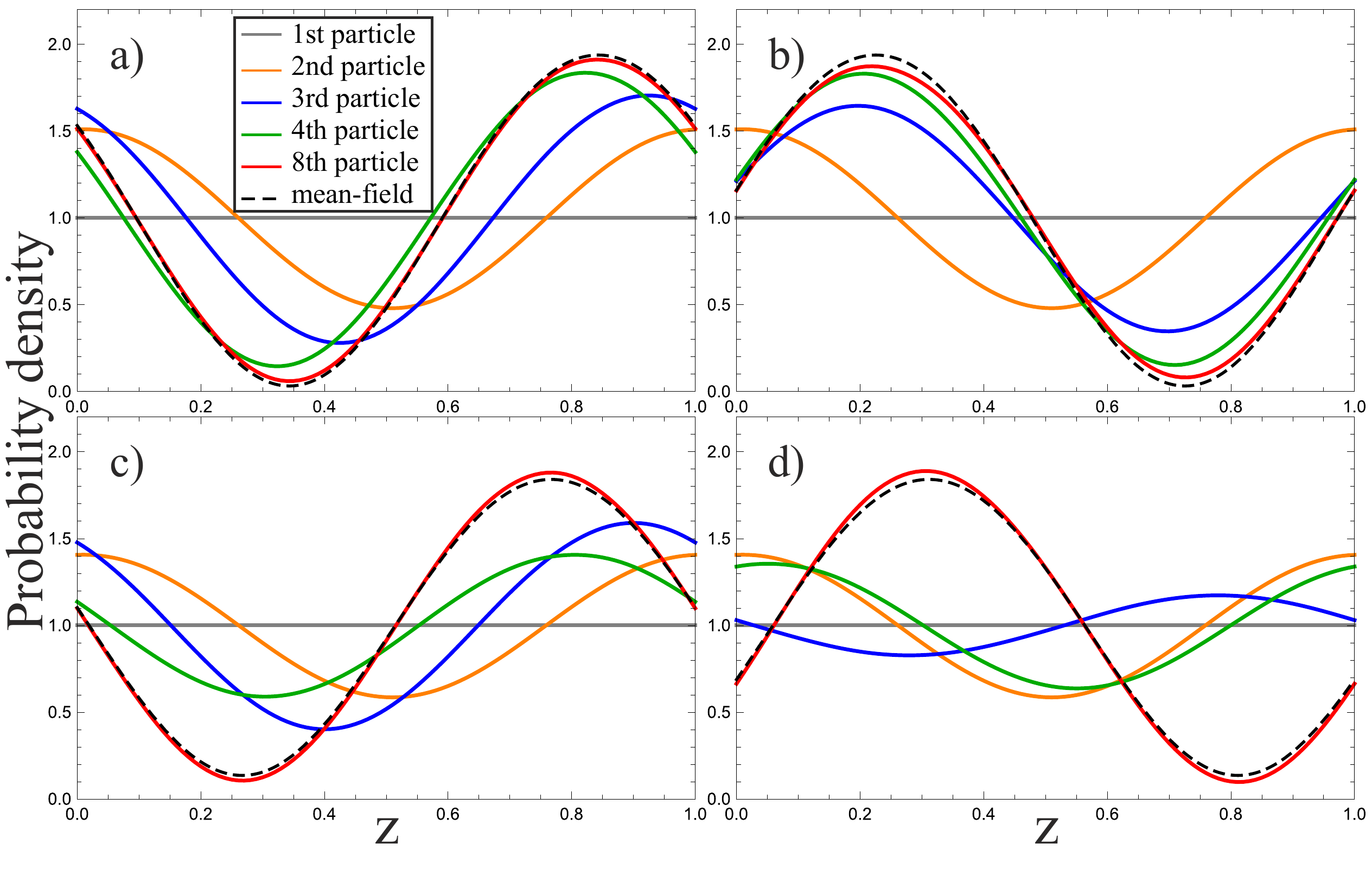}\end{center}
\begin{center}\includegraphics[width=16cm,angle=0]{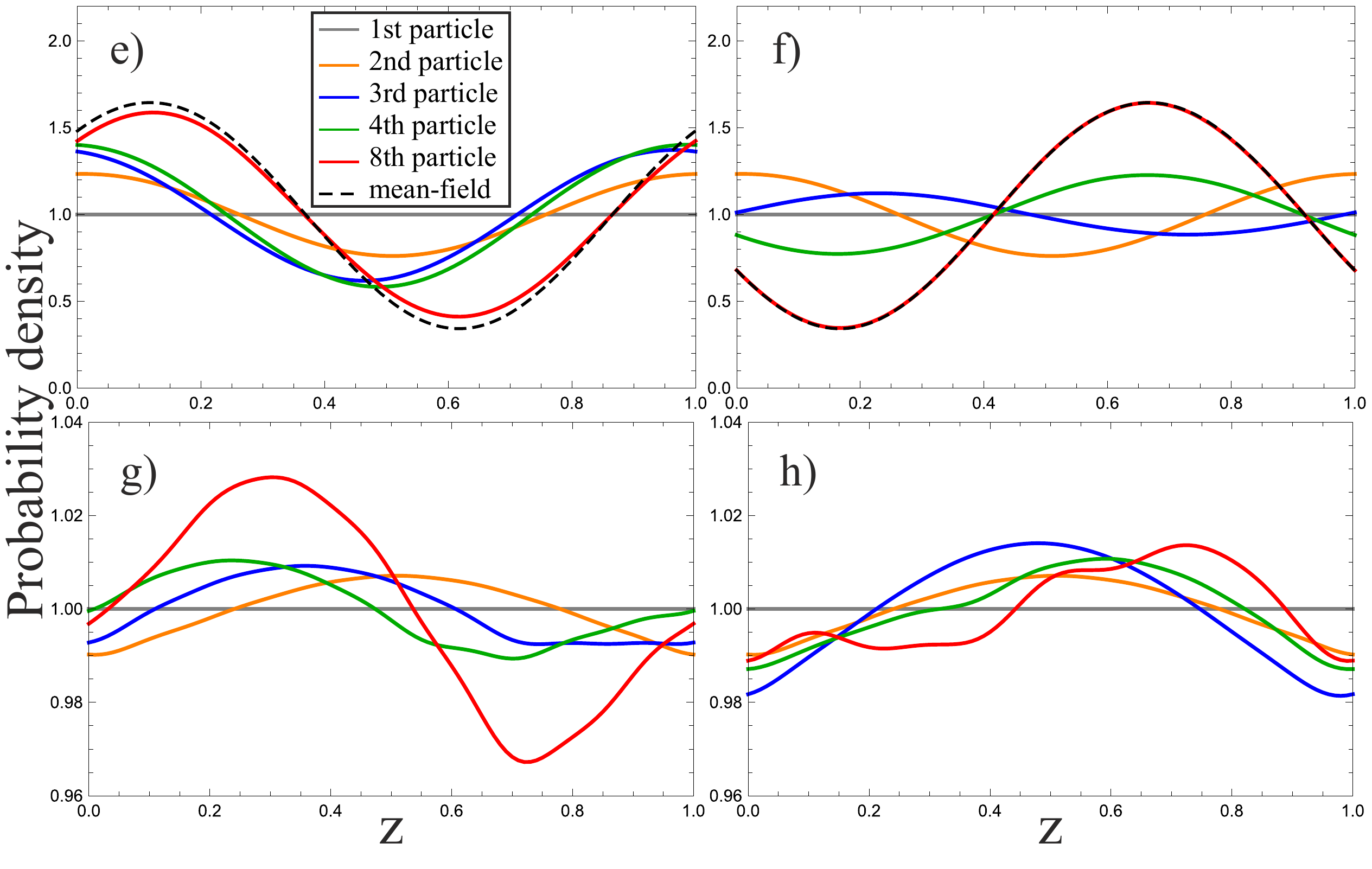}\end{center}
\begin{footnotesize}
Figure 4.3. Changes of the conditional probability density in the course of measurements of particle positions for the states generated by the collections $\mathcal{C}^{8,(\mathrm{gray}1)}_{1 \, \mathrm{sol}}$ (a) and b)), $\mathcal{C}^{8,(\mathrm{gray}2)}_{1 \, \mathrm{sol}}$ (c) and d)), $\mathcal{C}^{8,(\mathrm{gray}3)}_{1 \, \mathrm{sol}}$ (e) and f)) and $\mathcal{C}^{8}_{PW}$ (g) and h)). The numerical results agree with the mean-field predictions. As we expected, the depth of density notches is the shallower, the total momentum is closer to $\mathcal{N}\frac{2\pi}{\mathrm{L}}$.  Furthermore, the computer simulations show that the \emph{umklapp} (plane wave) excitation corresponds to nearly uniform distributions - note that vertical scale in g) and h) is different than in the other panels. The positions of successively measured particles are: a) $\{0.010, 0.839, 0.629, 0.073, 0.810, 0.927, 0.634,0.045 \}$, b) $\{0.010, 0.384, 0.254, 0.246, 0.331, 0.131, 0.263, 0.388\}$, c) $\{0.010, 0.790, 0.445, 0.996, 0.566, 0.875, 0.699, 0.604 \}$, d) $\{0.010, 0.549, 0.088, 0.482, 0.386, 0.462, 0.302, 0.439 \}$, e)  $\{0.010, 0.912, 0.035, 0.962, 0.394, 0.320, 0.315, 0.157 \}$, f)  $\{0.010, 0.449, 0.689, 0.744, 0.830, 0.605, 0.362, 0.735 \}$, g)  $\{0.010, 0.707, 0.553, 0.099, 0.723, 0.692, 0.841, 0.346 \}$ and h)  $\{0.010, 0.949, 0.372, 0.960, 0.241, 0.611, 0.394, 0.604 \}$. The position of the first particle in every single simulation was chosen as $z_{1}=0.010$ which does not break the generality of our results.  \end{footnotesize} \\

The histograms representing averaged particle densities for the states, corresponding to gray soliton mean-field solutions are depicted in Figures 4.4. a)-f). The averaged densities obtained
\begin{center}\includegraphics[width=16cm,angle=0]{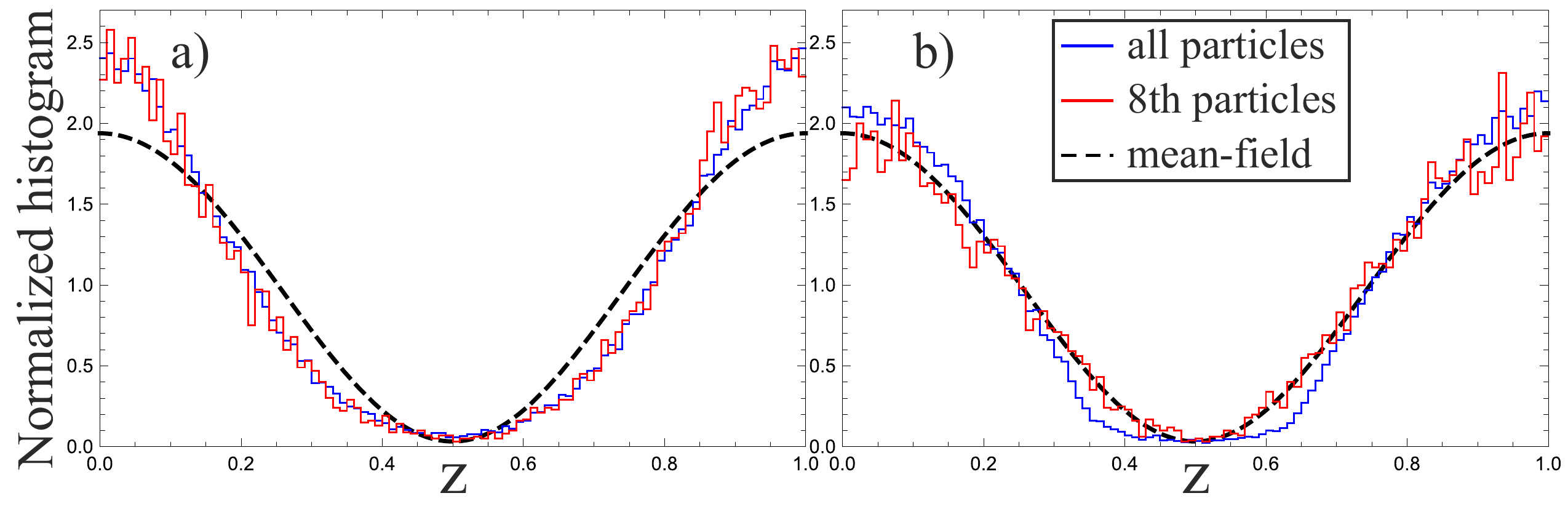}\end{center}
\begin{center}\includegraphics[width=16cm,angle=0]{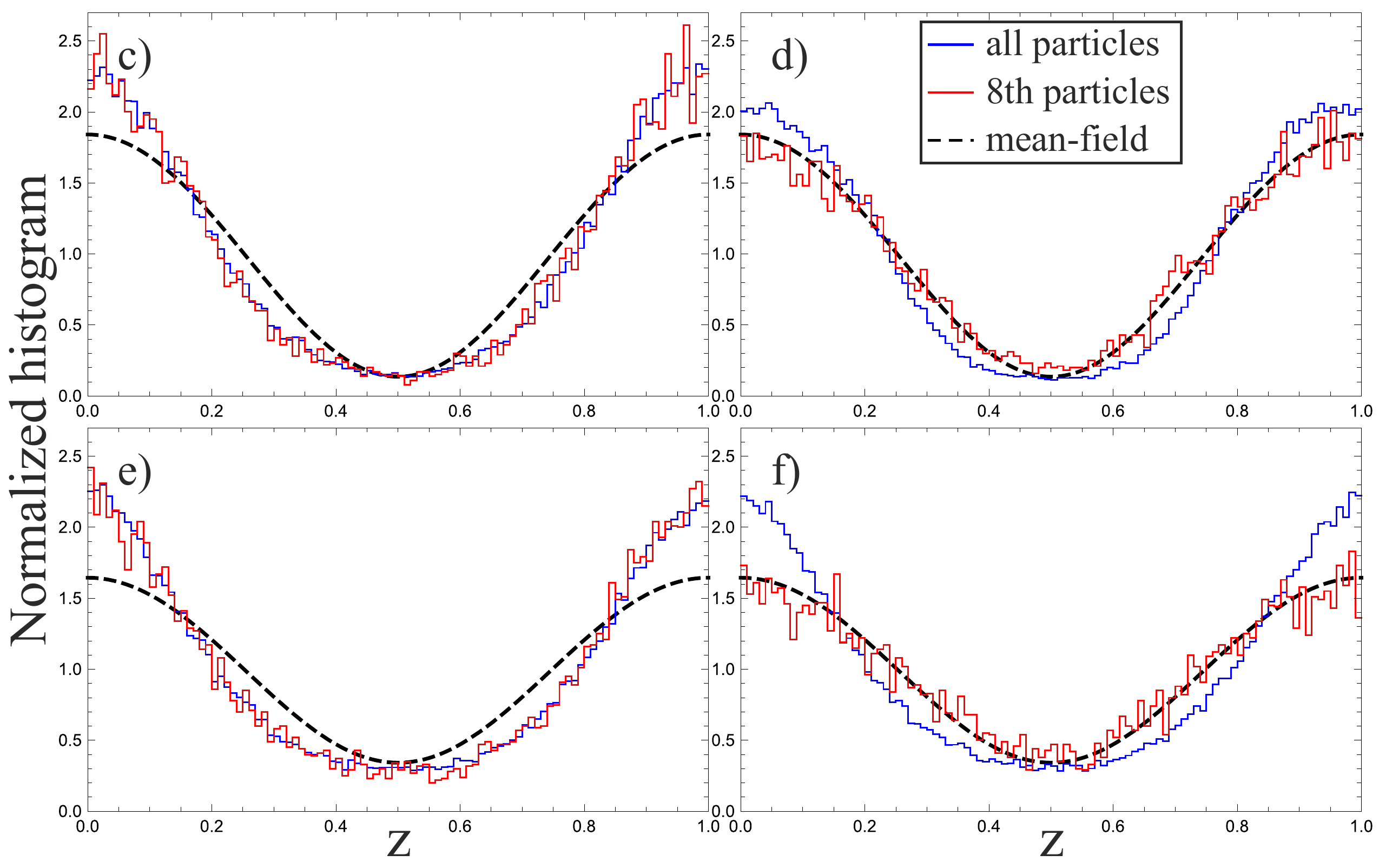}\end{center}
\begin{footnotesize}
Figure 4.4. The averaged particle densities for the states corresponding to $\mathcal{C}^{8,(\mathrm{gray}1)}_{1 \, \mathrm{sol}}$ (a) and b)), $\mathcal{C}^{8,(\mathrm{gray}2)}_{1 \, \mathrm{sol}}$ (c) and d)), $\mathcal{C}^{8,(\mathrm{gray}3)}_{1 \, \mathrm{sol}}$ (e) and f)), where the position of soliton in every simulations is determined by calculation of the center of mass relfection (a), c) and e)), see appendix \ref{sec:dodE}, and by a minimum of the last conditional probability densities (b), d) and f)).   \end{footnotesize}\\ \\
from the positions of 8th particles (where the soliton position in every single simulation is established based on the minimum of the 8th conditional probability density) agree very well with the mean-field profiles.\\

\emph{\textbf{Double soliton parametrized by $\mathcal{C}^{8,(\mathrm{sym})}_{2 \, \mathrm{sol}},\mathcal{C}^{8,(\mathrm{ns1})}_{2 \, \mathrm{sol}}$ and $\mathcal{C}^{8,(\mathrm{ns2})}_{2 \, \mathrm{sol}}$ }} \\

The last part of the weak interactions examination is devoted to double soliton states. The analysis of numerical data has been carried out in the way analogous to the case of a single soliton. There is only one difference - the method described in appendix \ref{sec:dodE} is unusable in this case. Furthermore, the centering of single histograms on the position of center of mass is not effective method too. Therefore, the only method we use in the double soliton case is to find the minima of the 8th conditional probability densities. Figures 4.5. a)-f) present the changes of the conditional probability density in the course of measurement of particle positions for the 3 double soliton states described in (\ref{doublesoliton}). It turns out that shapes of the probability densities for the last particle (8th) are almost identical in these three cases but the processes of solitons localization are distinctly different. In the symmetric case $\mathcal{C}^{8,(\mathrm{sym})}_{2 \, \mathrm{sol}}$ notches of conditional probability densities are deeper and deeper in every subsequent step. For the initial state parametrized by $\mathcal{C}^{8,(\mathrm{ns2})}_{2 \, \mathrm{sol}}$ we observe that the 2nd conditional densities have deeper notches than third conditional densities and very close to the 4th densities. The most strange behavior of the conditional densities we obtain for the state $\mathcal{C}^{8,(\mathrm{ns1})}_{2 \, \mathrm{sol}}$. It turns out that in this case the conditional probability densities are nearly uniform up to the 4th measurement when distinct density notches reveal after the measurement of the third particle. The histograms are depicted in the Figures 4.6. a)-c). 
\begin{center}\includegraphics[width=16cm,angle=0]{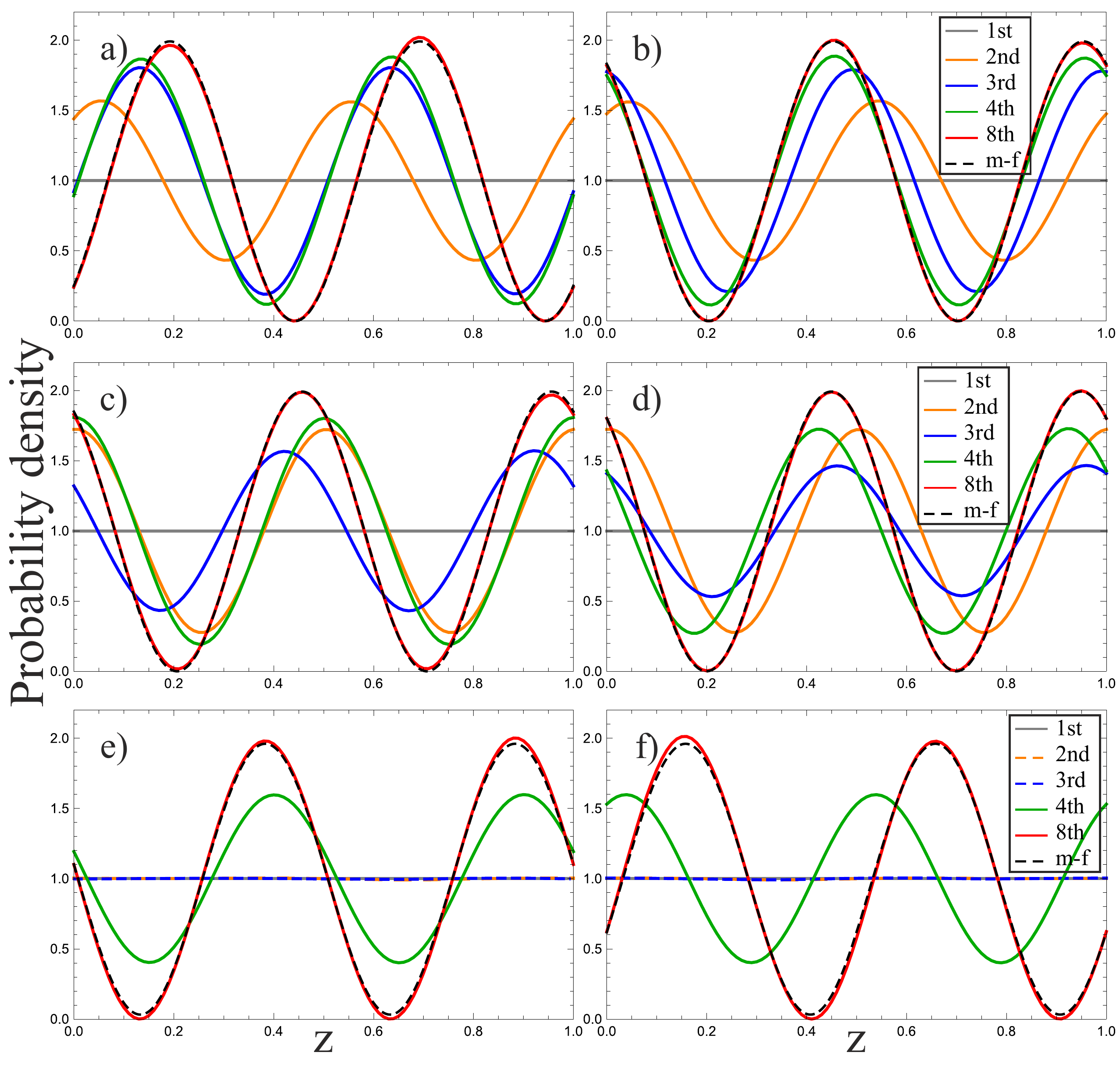}\end{center}
\begin{footnotesize}
Figure 4.5. Changes of conditional probability densities for a choice of the $j$-th particle provided ($j-1$) particles have been measured for three states corresponding to $ \mathcal{C}^{8,(\mathrm{sym})}_{2 \, \mathrm{sol}} $ (a) and b)), $\mathcal{C}^{8,(\mathrm{ns2})}_{2 \, \mathrm{sol}} $ (c) and d)) and $ \mathcal{C}^{8,(\mathrm{ns1})}_{2 \, \mathrm{sol}}$ (e) and f)). As we expected, the first two states correspond to the double black soliton excitations - the numerical results are with almost excellent agreement with mean-field (m-f) predictions. The 8th densities for the third state $\mathcal{C}^{8,(\mathrm{ns1})}_{2 \, \mathrm{sol}}$ also look like double black soliton state. It is easy to see that in this case the system needs 3 particles to be measured in order to make up a ,,decision'' about the soliton localization. The positions of measured particles are: a) $\{ 0.554, 0.212, 0.145, 0.647, 0.210, 0.810, 0.817,0.332 \}$, b) $\{0.045, 0.937, 0.877, 0.980, 0.390, 0.046, 0.391,0.972 \}$, c) $\{0.669, 0.013, 0.402, 0.340, 0.462, 0.828, 0.375, 0.891 \}$, d) $\{0.349, 0.754, 0.039, 0.674, 0.255, 0.277,0.267, 0.242 \}$, e) $\{0.005, 0.838, 0.581, 0.019, 0.853, 0.389, 0.934,0.471 \}$ and f) $\{ 0.005, 0.417, 0.896, 0.840, 0.336, 0.022, 0.098, 0.330\}$. \end{footnotesize} \vspace{0.1cm}

\begin{center}\includegraphics[width=16cm,angle=0]{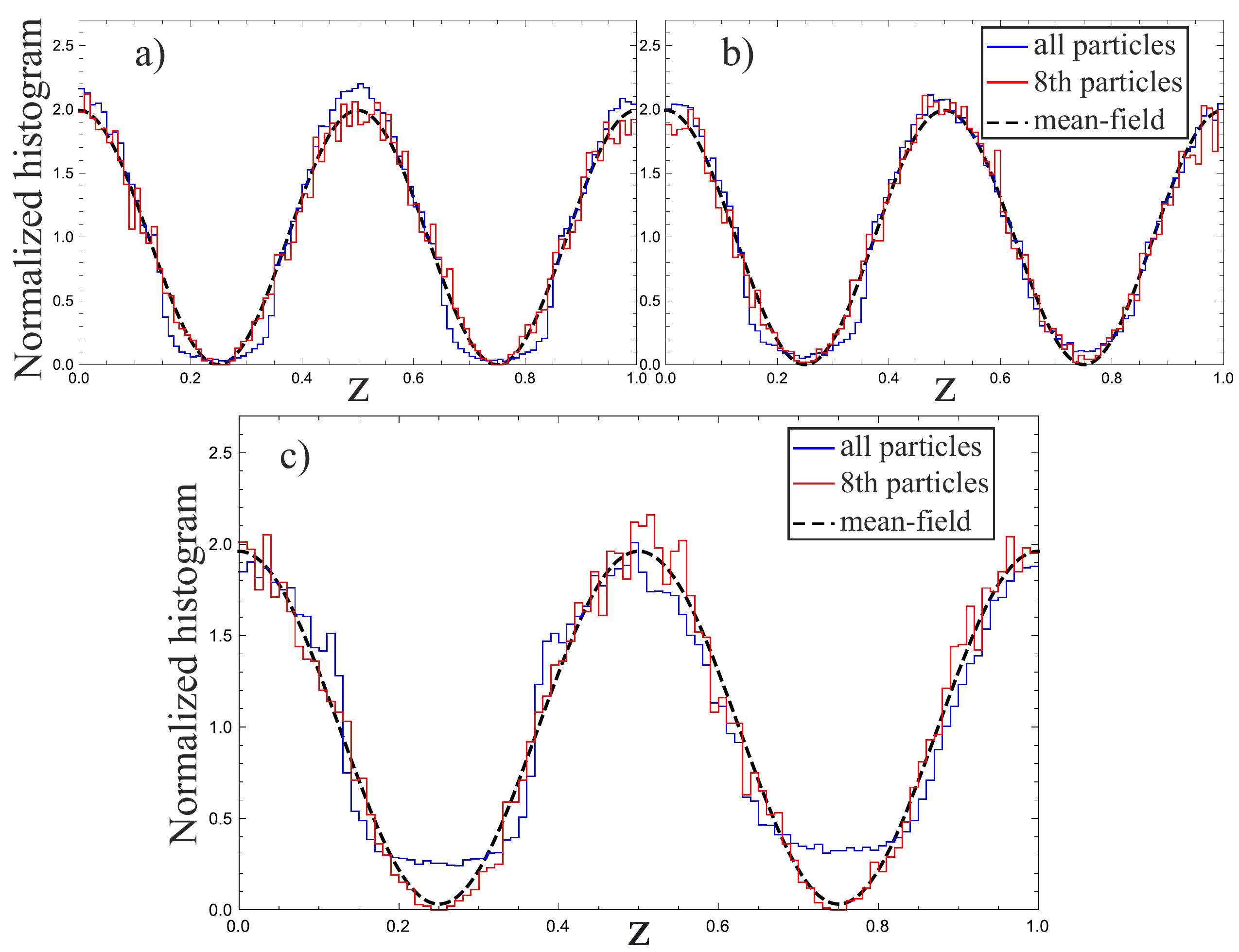}\end{center}
\begin{footnotesize}
Figure 4.6. Histograms obtained on the basis of all particles (blue curves) and on the basis of the last (8th) particle measured in each realisation of the detection process corresponding to 3 eigenstates defined by $ \mathcal{C}^{8,(\mathrm{sym})}_{2 \, \mathrm{sol}} $ (a)), $ \mathcal{C}^{8,(\mathrm{ns2})}_{2 \, \mathrm{sol}}$ (b)) and $ \mathcal{C}^{8,(\mathrm{ns1})}_{2 \, \mathrm{sol}}$ (c)). The mean-field predictions fit very well to the density profiles obtained from the 8th particles positions in all the cases.  \end{footnotesize}

\section{Strong interaction regime}
\label{sec:strong}

In the strong interaction case, we also consider $\mathcal{N}=8$ - particle system closed in the periodic box with the length $\mathrm{L}=1$. The coupling constant in this case is equal to $c=8$ which means that $\gamma=1$. For both considered states $\mathcal{C}^{8,(\mathrm{black})}_{1 \, \mathrm{sol}}$ and $\mathcal{C}^{8,(\mathrm{sym})}_{2 \, \mathrm{sol}}$, see (\ref{onesoliton}) and (\ref{doublesoliton}), we have performed $10^{4}$ computer simulations (see also \cite{SyrwSacha}).   \\ 

\emph{\textbf{Single density notch parametrized by $\mathcal{C}^{8,(\mathrm{black})}_{1 \, \mathrm{sol}}$}} \\

In the Figures 4.7. a)-d) we present 4 examples how the conditional probability density changes in the course of measurements. Figures 4.8. a) and b) show the histograms obtained using two methods described earlier. The numerical results confirm that the considered state, which is generated by the collection  $\mathcal{C}^{8,(\mathrm{black})}_{1 \, \mathrm{sol}}$, corresponds to the single density notch in the probability density. That is the notches appears also in the strong coupling regime. Moreover, the density notches are narrower than in the case of weak interactions but much wider than the corresponding healing length $\xi=1/8$. The observation is similar to the results obtained in \cite{kaminishi}, where the notch in the reduced single particle probability density is created by the superposition of the Lieb's type II eigenstates. We also notice that the conditional probability density profiles are not as regular as in the case of the weak coupling regime but the density notches are totally ,,black''. 
\begin{center}\includegraphics[width=16cm,angle=0]{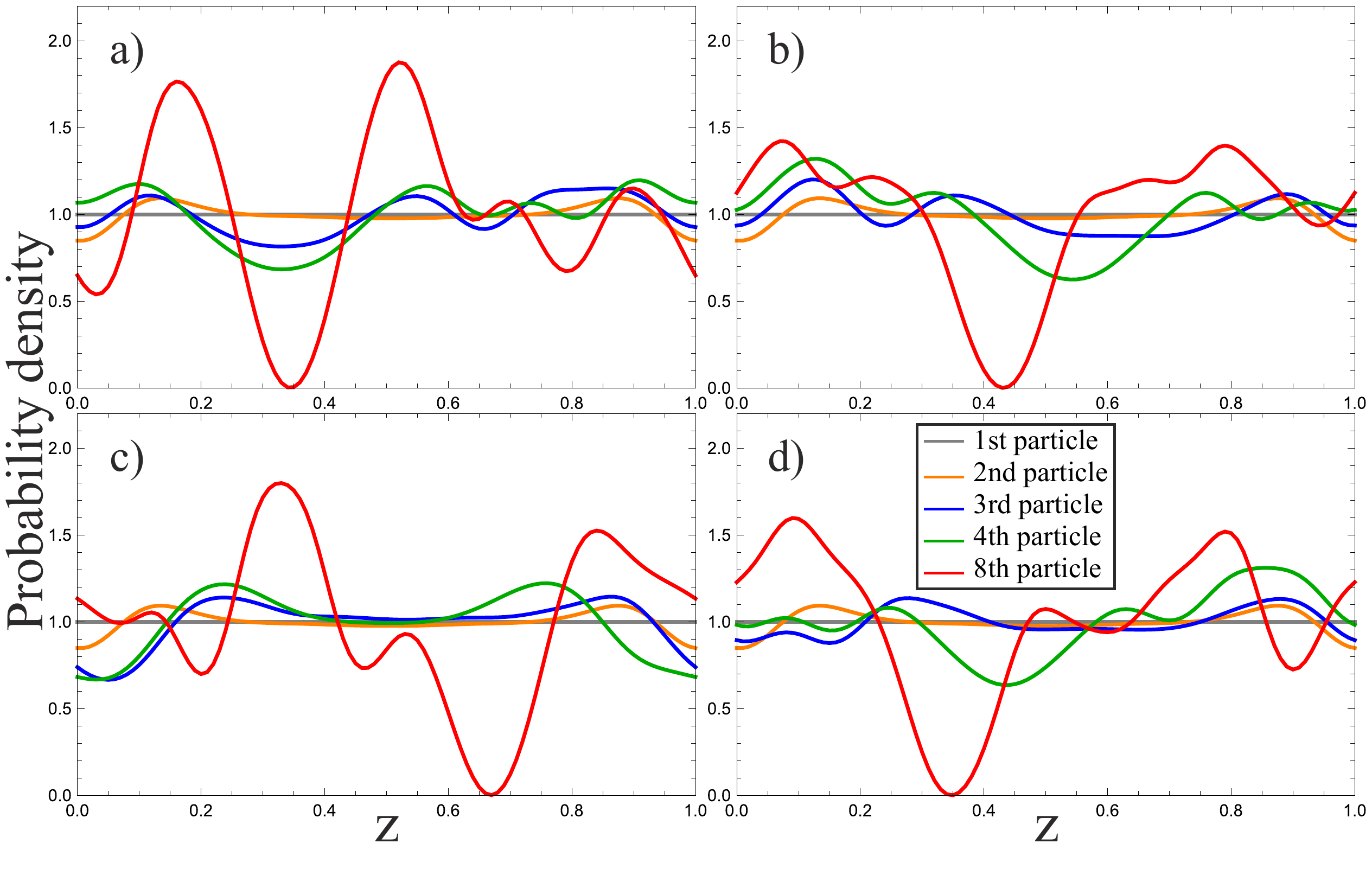}\end{center}
\begin{footnotesize}
Figure 4.7. The conditional probability density changes in the course of measurement of particle positions for 4 randomly chosen simulations in the case of strong interactions, $c=8$, where the initial state is parametrized by $\mathcal{C}^{8,(\mathrm{black})}_{1 \, \mathrm{sol}}$, see (\ref{onesoliton}). The first particle position was taken at $z_{1}=0.005$ in all the simulations. All the positions of particles obtained in presented cases have the following values a) $\{0.005, 0.655, 0.813, 0.787, 0.613, 0.055, 0.027, 0.130  \}$,  b) $\{0.005, 0.239, 0.848, 0.628, 0.717, 0.939, 0.148, 0.637  \}$,  c) $\{0.005, 0.088, 0.898, 0.205, 0.473, 0.211, 0.438, 0.755  \}$ and c) $\{0.005, 0.156, 0.709, 0.906, 0.552, 0.621, 0.894, 0.636 \}$.
\end{footnotesize} \\
\begin{center}\includegraphics[width=16cm,angle=0]{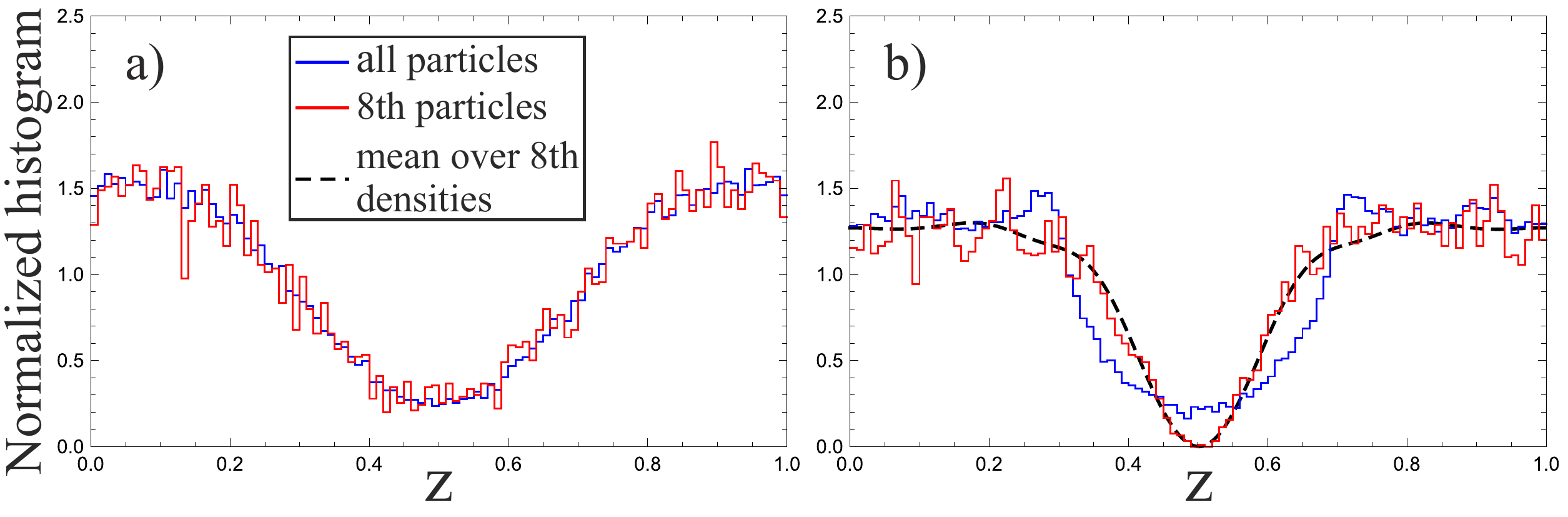}\end{center}
\begin{footnotesize}
Figure 4.8. Histograms which represent the averaged probability densities for the state given by $\mathcal{C}^{8,(\mathrm{black})}_{1 \, \mathrm{sol}}$ obtained using two methods: a) calculation of density notch position as in appendix \ref{sec:dodE}, b) finding a position of density notch formed in the last (8th) conditional probability density. We present the histograms obtained both from the positions of all the particles (blue line) and the 8th measured particles (red line). In the histogram b) we compare the results with the curve that represents mean over all the 8th conditional probability densities.
\end{footnotesize} \\

We instantly see that the method of finding the dark soliton position proposed in appendix \ref{sec:dodE} gives the histogram which is distinctly different from the 8th density profiles in every single simulations. The profile presented in the Figure 4.8. a) is much wider than the histogram b). Corresponding healing length $\xi=1/8$ is much closer to the width of histogram b) than histogram a). Nevertheless, we need to remember that the healing length corresponds to the case of weak interactions. Hence, in the strong coupling regime $\xi$ may be treated only as a width scale but not as a precise prediction.  \\

\emph{\textbf{Double density notch parametrized by $\mathcal{C}^{8,(\mathrm{sym})}_{2 \, \mathrm{sol}}$}} \\

The last analyzed state, corresponding to $\mathcal{C}^{8,(\mathrm{sym})}_{2 \, \mathrm{sol}}$, see (\ref{doublesoliton}), has a double notch structure, even in the strong coupling regime which may be inferred from the numerical results presented in the Figures 4.9. a)-d) and 4.10. a). The first four plots (Figure 4.9. a)-d)) show the conditional probability density changes after the measurement of particle positions for 4 accidentally chosen simulations. Figure 4.10. a) depicts histograms obtained from all the simulations, where the notch positions were determined by the minima of the last (8th) conditional probability densities in every single simulation.

\begin{center}\includegraphics[width=16cm,angle=0]{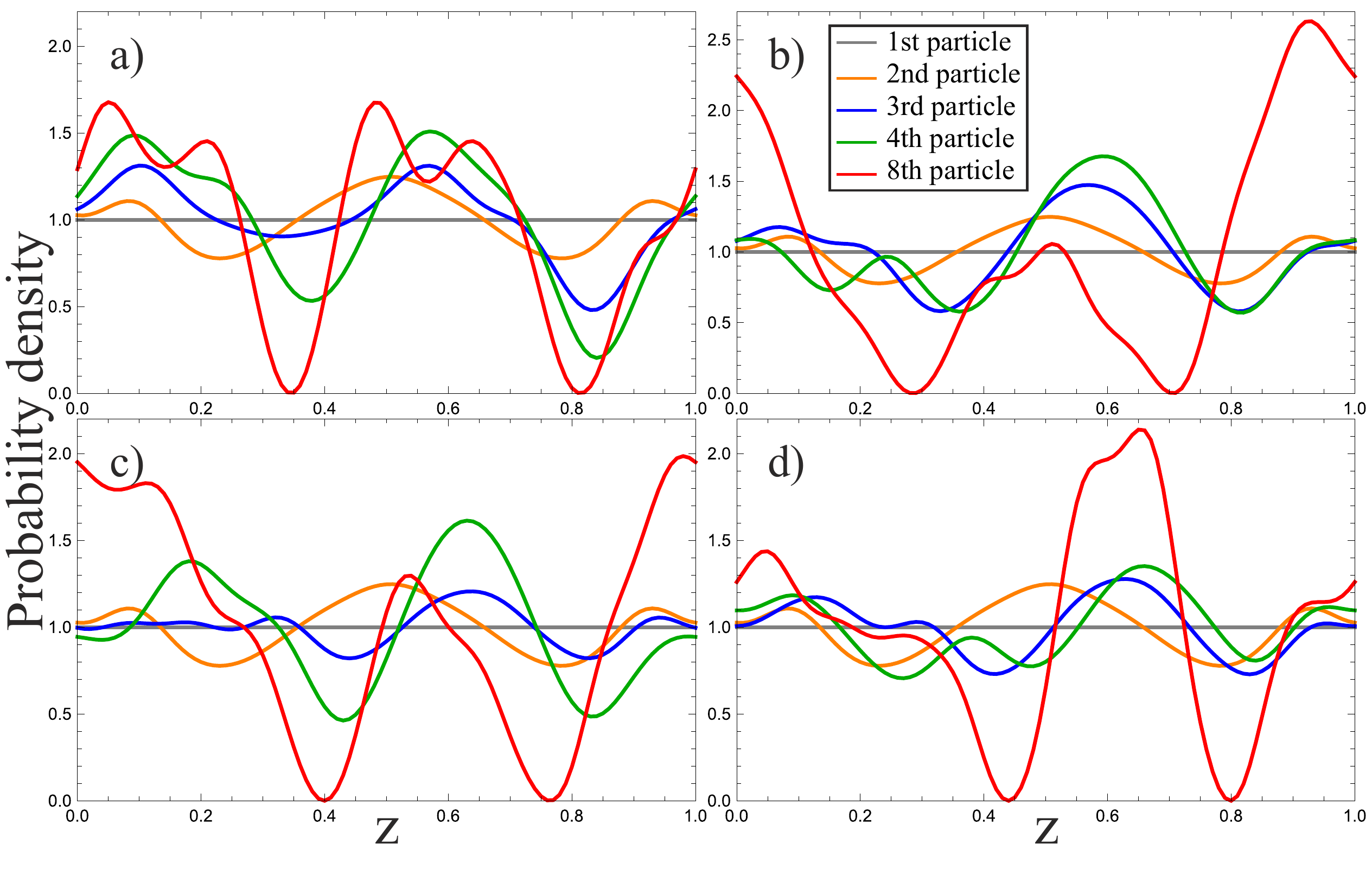}\end{center}
\begin{footnotesize}
Figure 4.9. Changes of the conditional probability densities for 4 randomly chosen simulations for strong interactions, where the initial state is parametrized by $\mathcal{C}^{8,(\mathrm{sym})}_{2 \, \mathrm{sol}}$, see (\ref{doublesoliton}). Despite large irregularities of the probability density profiles, without a shadow of a doubt, we observe two density notches in the last (8th) probability densities. The positions of the successively measured particles in presented simulations have the following values: a) $\{ 0.007, 0.662, 0.170, 0.123, 0.534, 0.577, 0.950, 0.163 \}$, b) $\{0.007, 0.137, 0.156, 0.612, 0.854, 0.590,$ $0.451, 0.960 \}$, c) $\{0.007, 0.259, 0.092, 0.207, 0.668, 0.896, 0.590, 0.024 \}$, d) $\{0.007, 0.239, 0.293, 0.174, 0.950, 0.110,$ $0.611, 0.184 \}$.
\end{footnotesize} \\ \\
It turns out that, in contrast to the weak coupling case, the distances between the two notches $\Delta z$ are not the same in all the simulations. The differences in $\Delta z$ are very noticeable and their distribution is presented in the Figures 4.11. a) and b) in the form of histograms. Because of the ring topology of the system, the distance between two solitons cannot be larger that $\mathrm{L}/2$. It should be mentioned that 200 out of the $10^4$ simulations have led to results where a single very wide notch or very irregular profile has formed in the 8th conditional probability density. Such 200 cases are not taken into account in the histograms presented in the Figures 4.11. a) and b). Figure 4.12. shows two examples (a) and b)) of the omitted cases.
\begin{center}\includegraphics[width=14cm,angle=0]{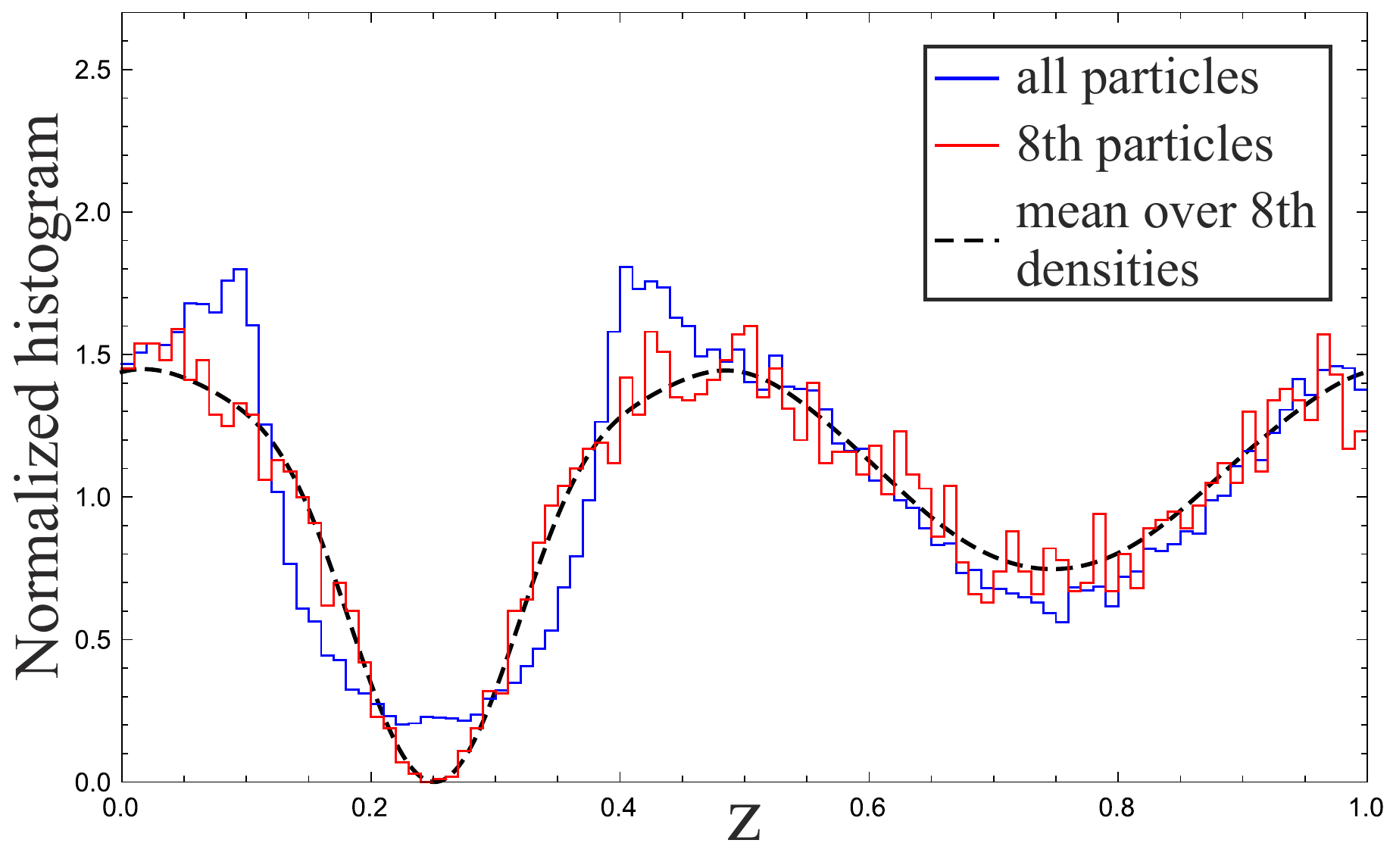}\end{center}
\begin{footnotesize}
Figure 4.10. The averaged probability densities obtained from all and the 8th measured particles, where the initial state was parametrized by the collection $\mathcal{C}^{8,(\mathrm{sym})}_{2 \, \mathrm{sol}}$. We observe that only the one of two notches is clearly visible. The other density notch is much wider and shallower. It is caused by the fluctuations of the relative distance between the two minima of the last conditional probability densities. The positions of density notches are established as a two main minima of the last (8th) conditional probability densities. The deeper one is shifted to the position $z_{0}=0.25$. The results obtained for 8th particles converge to the mean over all the 8th conditional probability densities. 
\end{footnotesize}

\begin{center}\includegraphics[width=16cm,angle=0]{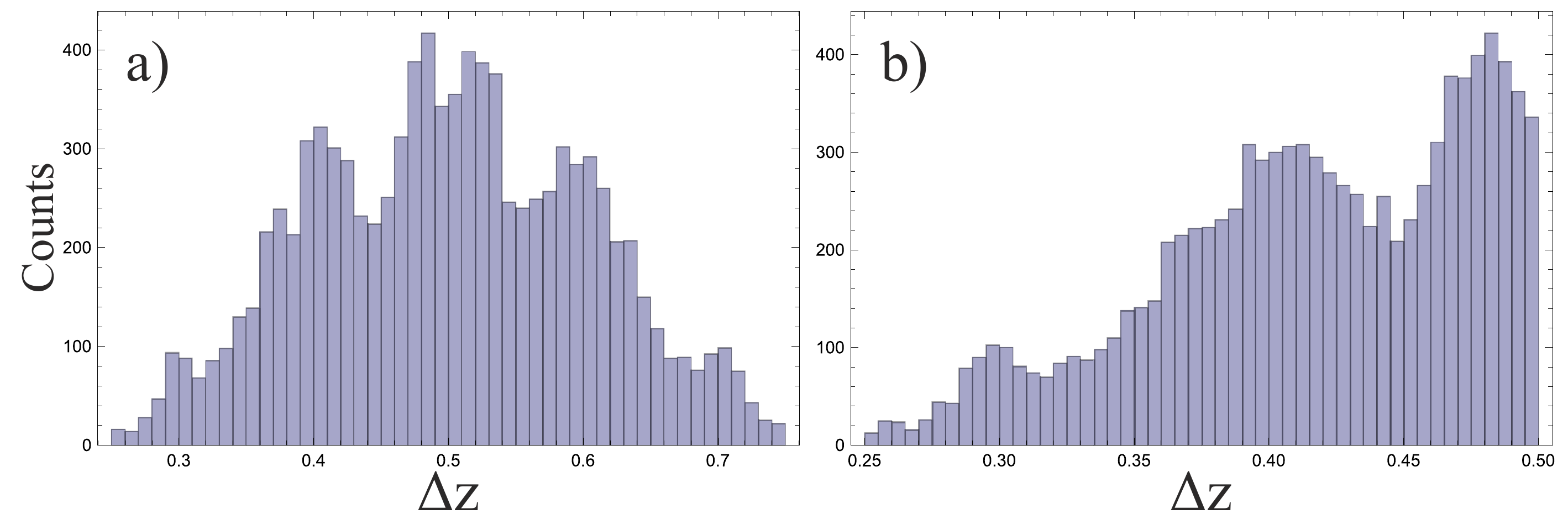}\end{center}
\begin{footnotesize}
Figure 4.11.  The distribution of the relative distances between two density notches measured as a difference between the positions of the left and right density notches in every realization of detection process (where the deeper notch is treated as the left notch) a). Because of the ring topology of the system, the distances cannot be larger that $\mathrm{L}/2=0.5$. The corresponding histogram is presented in the plot b).
\end{footnotesize}

\begin{center}\includegraphics[width=16cm,angle=0]{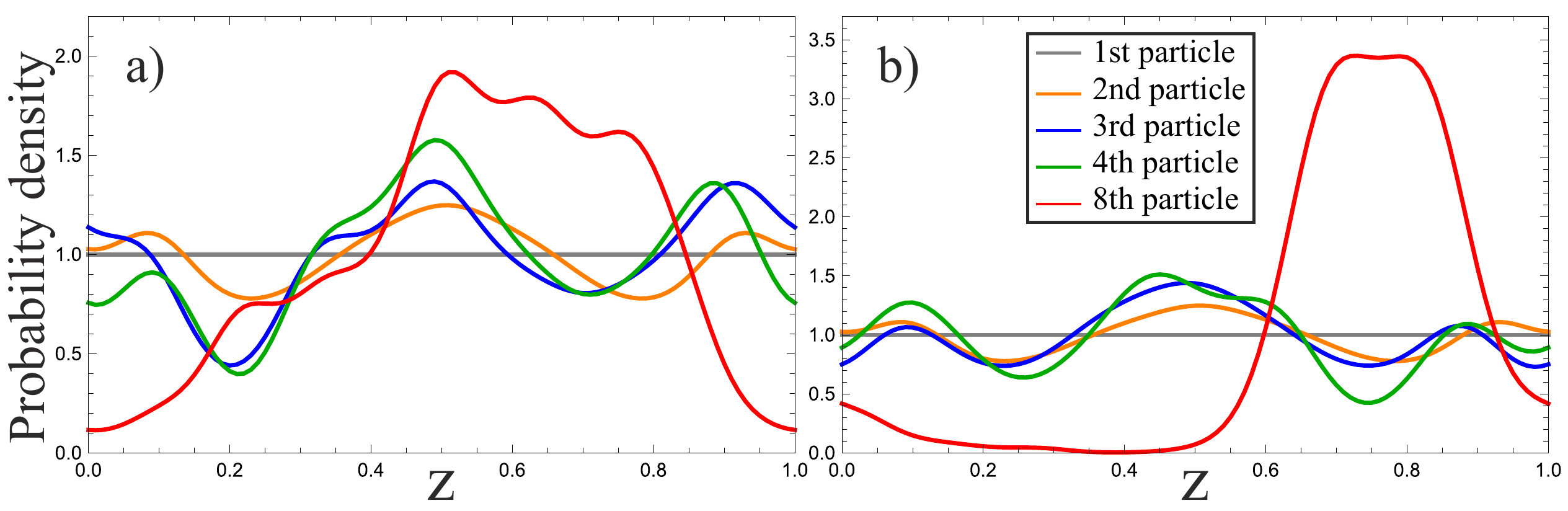}\end{center}
\begin{footnotesize}
Figure 4.12. Two of 200 results, where the last (8th) conditional probability densities cannot be identified as a profiles consisting of two nothes. The shapes of the last conditional probability densities may be caused by ,,unlucky'' values of the positions of the measured particles: a) $\{ 0.007, 0.397, 0.002, 0.942, 0.684,$ $ 0.043,  0.142, 0.624 \}$, b) $\{0.007,0.962, 0.527, 0.174, 0.773, 0.261, 0.099, 0.772\}$. It is noticeable that the first a few measurements cause the appearance of two shallow density notches but the last measurements give the positions which are very close to the positions of the notches. Finally, the profiles of the 8th conditional probability densities become very different from a typical case. We believe that the situation is caused by a very small number of particles in our system and the effect would not be observed if $\mathcal{N} \gg 8$.
\end{footnotesize}

\chapter*{Recapitulation}
\addcontentsline{toc}{chapter}{Recapitulation}

The one-dimensional system of ultracold non-relativistic bosons which interact via point-like contact potential (the Lieb-Liniger model) has the analytical solution given by the Bethe ansatz method. Traditionally, one imposes a periodic boundary conditions changing the topology of the system to a ring topology (section \ref{sec:periodicconditions}). The complete description of the solution is presented in chapter \ref{chap:TheOne-dimensionalBoseGas} and is complemented by the discussion about quasi-momenta (section \ref{sec:quasimomenta}) and thermodynamic limit (section \ref{sec:Thermodynamiclimit}). 

The analysis of elementary excitations shows that we can divide them into two types (I and II) which satisfy different dispersion relations (section \ref{sec:Excitations}). It turns out that the Lieb's type I excitation corresponds to sound waves (Bogoliubov spectrum) in the system and may be presented as particle insertion in the termodynamic limit (section \ref{sec:ExcitationsI}). The second type corresponds to the so-called ,,hole'' states because, in thermodynamic limit, it may be described as removal of one particle from the system (section \ref{sec:ExcitationsII}) - ,,hole'' creation. It turns out that the dispersion relation of the Lieb's type II excitation is very similar to the dispersion relation obtained for dark soliton in the semi-classical description of the Lieb-Liniger model (section \ref{sec:solitons}). This result suggests that the type II excitation may have solitonic nature. It is also shown that the soliton solutions may be obtained in the mean-field approach. The solutions are given in terms of the Jacobi functions and appear stationary in the rotating frame of reference (section \ref{sec:Solitonsinperiodicbox}). Notwithstanding, it has not been shown in the literature that measurements of positions of particles reveal dark soliton density profiles if the system is prepared in a type II eigenstate.

Using the analytical results of norms and form factors, which we remind in the section \ref{sec:Slavnov}, we have developed the iterative procedure (section \ref{sec:iterative}). The procedure allows us to examine subsequent conditional probability densities for a choice of $j$-th particle provided the previous ($j-1$) particles have been already measured, during the detection process. As expected, we observed that the probability density for a measurement of the first particle in the system is uniform. We have also showed that the detection process induces breaking of the translational symmetry and emergence of dark solitons (density notches) which localize at different positions in space in different realizations (see the results presented in sections \ref{sec:weak} and \ref{sec:strong}).

Performed numerical simulations confirm the mean-field predictions amazingly well even for as small particle number as $\mathcal{N}=8$. The last conditional density profiles coincide with corresponding mean-field solutions for all the considered states (section \ref{sec:weak}). It turns out that the processes of the density notches localization strongly depends on the form of initial state (see the discussion about the results obtained for double soliton states in the case of weak interactions).

In the strong interaction regime (see the section \ref{sec:strong}), the results also reveal one (or two) density notch(es) structure. The width of the notches is smaller than in the case of weak interactions but greater than corresponding healing length ($\xi=0.125$). Moreover, in the case of strong coupling we observe that the relative distance between density notches (in the case of two-hole excitation) changes in every realization of the detection process (see the Figures 4.9. - 4.11.).

Our computer simulations show that the Lieb's type II eigenstates have the dark solitonic nature, indeed. The results suggest that the main structure (i.e. the presence of density notches) of the probability density profiles survive also for the case of strong interaction regime.

\chapter*{Acknowledgements}
\addcontentsline{toc}{chapter}{Acknowledgements}

I would like to express my deepest gratitude to my supervisor Prof. Krzysztof Sacha for his support and guidance throughout the research. I truly appreciate his eagerness to answer my questions and clarify things out.

My sincere appreciation is extended to Prof. Dominique Delande for fruitful discussions and very valuable suggestions.

Support of Polish National Science Centre via project number DEC-2012/04/A/ST2/00088 is acknowledged.

\begin{appendices}

\chapter{The contact potential}
\label{sec:dodA}

Considering interactions between neutral molecules in gases we can introduce Lenard-Jones potential \cite{lennardjones}
\begin{equation}
\displaystyle{  V(r)=4 D \left[ \left(\frac{\sigma}{r}\right)^{12}-\left(\frac{\sigma}{r}\right)^{6}\right] },
		\label{lenard}
\end{equation}
which approximates electron repulsion ($r^{-12}$) and dipole-dipole attraction ($-r^{-6}$).
\begin{center}\includegraphics[width=14cm,angle=0]{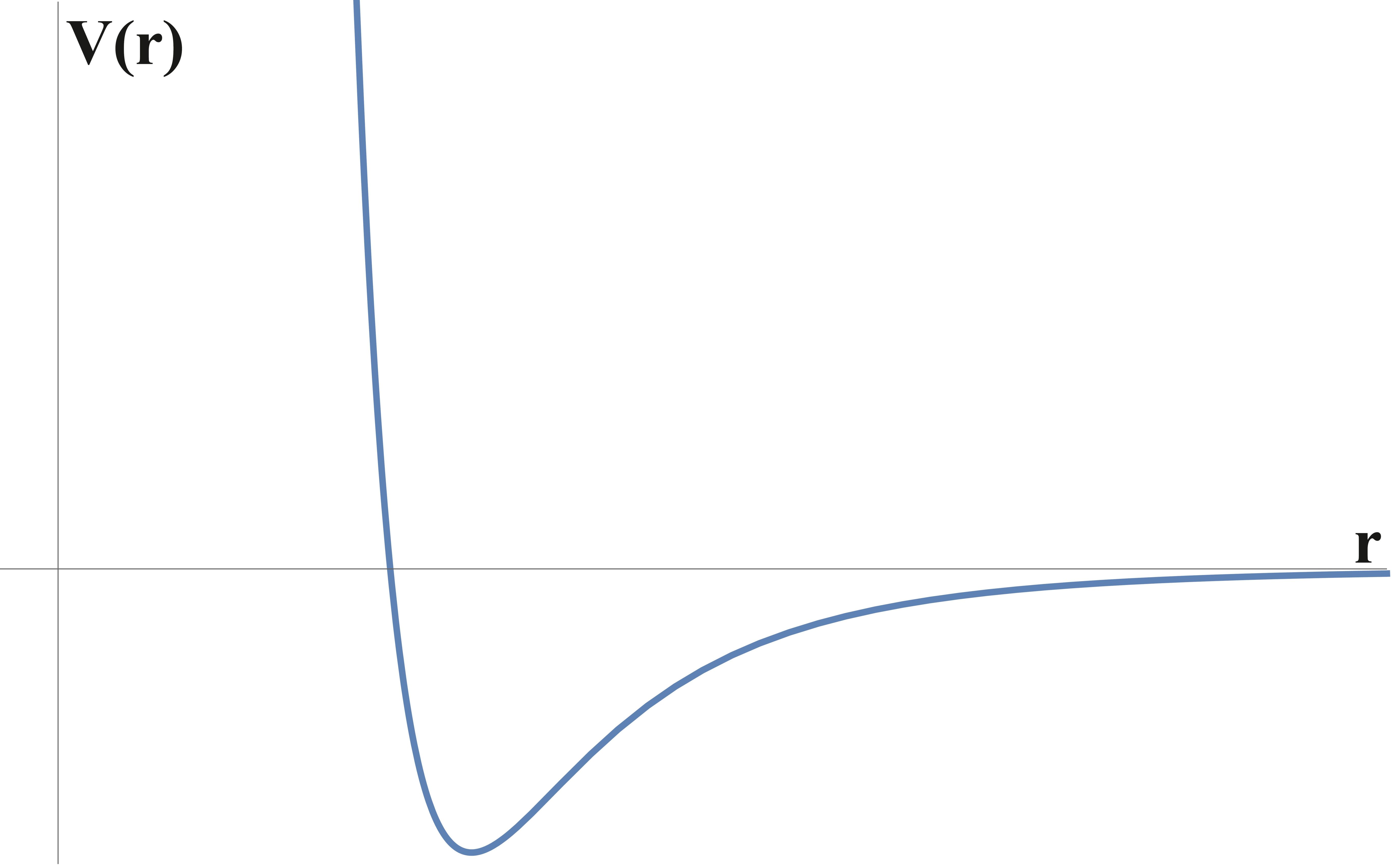}\end{center}
\begin{center}
\footnotesize{Figure A.1. The Lenard-Jones potential approximates interparticle interactions in gases. }
\end{center}
The potential vanish as $r^{-6}$, so in the case of dilute gases the mean interparticle separation is much larger than the effective range of the potential $r_{0}$ - the system with short-range potential.

Let us now write time-independent Schr\"{o}dinger equation for 2 atoms - only relative motion 
\begin{equation}
\displaystyle{\left(-\frac{\mathrm{\hbar}^{2}}{2\mu}\mathrm{\nabla}_{\vec{r}}^{2}+V(\vec{r}) \right)\psi(\vec{r})=E_{r}\psi(\vec{r}) ,}
		\label{rozpwzg}
		\end{equation} 
where $\mu$ is the reduced mass. Taking $E_{r}=\frac{\hbar^{2} k^{2}}{2\mu}$ we get
 \begin{equation}
\displaystyle{\left(\mathrm{\nabla}_{\vec{r}}^{2}+k^{2} \right)\psi(\vec{r})=\frac{2\mu}{\hbar^{2}} V(\vec{r})\psi(\vec{r}) ,}
		\label{rozp2}
		\end{equation}
which has the following solution		
 \begin{equation}
\displaystyle{\psi(\vec{r})=\psi_{0}(\vec{r})-\frac{2\mu}{4\pi \hbar^{2}}\int \mathrm{d}^{3}r ' \frac{\mathrm{e}^{\mathrm{i}k|\vec{r}-\vec{r} \, '|}}{ |\vec{r}-\vec{r} \, '|}V(\vec{r} \, ')\psi(\vec{r} \, ') ,}
		\label{rozp5}
		\end{equation}		
where we have used the Green's function obtained for the Helmholtz equation \cite{Sacha}, \cite{Castin}
 \begin{equation}
\displaystyle{G(\vec{r}-\vec{r} \, ')=-\frac{\mathrm{e}^{\mathrm{i}k|\vec{r}-\vec{r} \, '|}}{4\pi |\vec{r}-\vec{r} \, '|}.}
		\label{greenfuncthelmholtz}
		\end{equation}

The asymptotic solution analysis ($\psi(\vec{r}\to \infty)$) leads to
		\begin{center}
$\displaystyle{|\vec{r}-\vec{r} \, ' |=\sqrt{ r^{2}-2\vec{r} \cdot \vec{r} \, ' +r'^{2} } \approx r-\hat{r}\cdot \vec{r} \, '  \, \, (r\gg r')  },$
\end{center}  
 \begin{equation}
\displaystyle{\psi(\vec{r})\approx \psi_{0}(\vec{r})+\mathcal{A}_{k}(\hat{r})\frac{\mathrm{e}^{\mathrm{i}kr}}{r},}
		\label{rozp6}
		\end{equation}
where
 \begin{equation}
\displaystyle{\mathcal{A}_{k}(\hat{r})=-\frac{2\mu}{4\pi \hbar^{2}}\int \mathrm{d}^{3}r ' \mathrm{e}^{\mathrm{i}k \hat{r}\cdot \vec{r} \, '}V(\vec{r} \, ')\psi(\vec{r} \, '),}
		\label{rozpamp7}
		\end{equation}
is the so-called \emph{\textbf{scattering amplitude}} \cite{Sacha}, \cite{Castin}. In the limit $k\to 0$ we obtain  
 \begin{equation}
\displaystyle{\mathcal{A}_{k\to 0}(\hat{r})=-\frac{2\mu}{4\pi \hbar^{2}}\int \mathrm{d}^{3}r ' V(\vec{r} \, ')\psi(\vec{r} \, ')=-a,}
		\label{rozpamp8}
		\end{equation}  
where $a$ will be called \emph{\textbf{scattering length}}.	The above limit corresponds to ultracold atomic systems. We instantly see that in this case the scattered wave  is spherically symmetric independently of the symmetry of $V(r)$. Furthermore, one can show (by a partial waves expansion) that, if the scattering amplitude is a number then, the scattering occurs only if an eigenvalue of the angular momentum $l$ is equal to 0 - only $s$-wave scattering is allowed.
		
The relations (\ref{rozp6}) and (\ref{rozpamp8}) tell us that all the information about the interaction is included in the scattering length $a$, which is given by the integral (\ref{rozpamp8}). Therefore, instead of real, elaborated potential $V(\vec{r})$ we can use simple unphysical potential $V'(\vec{r})$ which reproduces the value of scattering length $a$. If so, then the following equality have to be satisfied 
 \begin{equation}
\displaystyle{\int \mathrm{d}^{3}r ' V(\vec{r} \, ')\psi(\vec{r} \, ')=\int \mathrm{d}^{3}r ' V'(\vec{r} \, ')\psi(\vec{r} \, ').}
		\label{rozpamp9}
		\end{equation} 
The easiest choice is the \emph{contact} potential, proportional to the Dirac delta 
 \begin{equation}
\displaystyle{V(\vec{r}_{1}-\vec{r}_{2})=\frac{4\pi \hbar^{2}a}{m} \delta(\vec{r}_{1}-\vec{r}_{2}),}
		\label{potm1}
		\end{equation} 
where the constant was established  using a Born approximation ($\psi(\vec{r})=\psi_{0}(\vec{r})=\mathrm{e}^{\mathrm{i} \vec{k} \cdot \vec{r}}$) and assuming equal masses of particles. The contact potential is unphysical but reproduces all physical properties of ultracold, dilute systems. Because of its simplicity, one almost always uses the potential in the form (\ref{potm1})  instead of very complicated physical potential \cite{Sacha}, \cite{Castin}.

The wave function may have singularities for $|\vec{r}_{1}-\vec{r}_{2}|=r=0$. This situation reveals the artificiality of our potential because $\delta(r)r^{-\varepsilon}, \, \, \varepsilon>0$ is divergent for $r=0$. Fortunately, one can regularize the potential getting rid of divergences order by order. The most popular regularization is based on the introduction of the contact \textbf{\emph{pseudopotential}}
 \begin{equation}
\displaystyle{ V(\vec{r})=\frac{4\pi \hbar^{2}a}{m}\delta(\vec{r})\frac{\partial}{\partial r} r ,}
		\label{pspotm}
		\end{equation}
which regularizes the first order of wave function singularities ($1/r$) \cite{Sacha}, \cite{Castin}. Above problem does not occur in one-dimensional systems, hence the proper potential, in the Lieb-Liniger model, has the form (\ref{potm1}).

\chapter{Bose-Einstein condensation}
\label{sec:dodB}

It is well known that a mean number of bosons $\left< n_{i} \right>$ occupying $i$-th state of energy $E_{i}$ in the grand canonical ensemble description is given by the following formula \cite{Sacha}, \cite{Castin}, \cite{Huang}
 \begin{equation}
\displaystyle{ \left< n_{i} \right> =\frac{1}{\mathrm{e}^{(E_{i}-\mu_{ch})/\mathrm{k}_{B}T} - 1}},
		\label{menanumbergrandcanensemble}
		\end{equation}
where above result has a sense only if $\mu_{ch}<E_{0}$ ($E_{0}$ is the energy of the ground state). It is obvious that, if $\mu_{ch} \rightarrow E_{0}$ then $\left< n_{0} \right> \rightarrow \infty$\footnote{In this case, the ground state of energy must be neglected in thermodynamic description.}.

Let us now consider the system of non-interacting bosons with mass $m$ in a periodic box with a volume $\mathrm{L}^{d}$, where $d$ is the dimension of the considered space. Additionally, we assume that the mean number of particles in the box is equal to $\left< \mathcal{N} \right>_{d}$. The energy levels for one particle have the following form
  \begin{equation}
\displaystyle{ E_{\vec{i}}=\frac{2\mathrm{\hbar}^{2} \pi^{2}}{m \mathrm{L}^{2}} \sum_{j=1}^{d}i_{j}^{2}=\frac{2\mathrm{\hbar}^{2} \pi^{2}}{m \mathrm{L}^{2}} i^{2}, \, \, \, \, \, \, \, \, \, \, \, \, i_{j} \in \mathbb{Z}, \, \, \, \, \, \, \, \, \, \, \, \, E_{0}=0}.
		\label{energylevels}
		\end{equation}
If the temperature of the system $T$ multiplied by $\mathrm{k}_{B}$ is much larger than the energy difference between occupied states,  then the sum over $\vec{i}$ may be approximated by the integral. Hence
  \begin{equation}
\displaystyle{\left< \mathcal{N} \right>_{d}=\sum_{\vec{i}}\left< n_{\vec{i}} \right>=\sum_{\vec{i}}\frac{1}{\mathrm{e}^{(E_{\vec{i}}-\mu_{ch})/\mathrm{k}_{B}T} - 1}\approx \int \frac{\mathrm{d}^{d}i}{\mathrm{e}^{(E_{\vec{i}}-\mu_{ch})/\mathrm{k}_{B}T} - 1} =\int \limits_{0}^{\infty} \frac{g_{d}(E)\mathrm{d}E}{\mathrm{e}^{(E-\mu_{ch})/\mathrm{k}_{B}T} - 1}   },
		\label{meanvalue2}
		\end{equation}
where $g_{d}(E)$ is the energy density of states which contains complete information about the system we need \cite{Sacha}. In order to derive the quantity we should notice that \cite{Flanders}
  \begin{equation}
\displaystyle{\mathrm{d}^{d}i=\mathrm{d}i_{1}\ldots \mathrm{d}i_{d} =S_{d-1}i^{d-1}\mathrm{d}i, \, \, \, \, \, \, \, \, \, \, \, \,  S_{d-1}=\frac{2\pi^{\frac{d}{2}}}{\Gamma \left( \frac{d}{2} \right)}, \, \, \, \, \, \, \, \, \, \, \, \,  S_{0}=1},
		\label{nsphere}
		\end{equation}
where $S_{n}$ is the surface of $n$-dimensional sphere with unit radius and $\Gamma(z)$ is the so-called Gamma function \cite{abramowitz}. Therefore
  \begin{equation}
\displaystyle{i^{d-1}=\left(\frac{m \mathrm{L}^{2}E}{2\mathrm{\hbar}^{2}\pi^{2}}\right)^{\frac{d-1}{2}}, \, \, \, \, \, \, \, \, \, \, \, \, \frac{\mathrm{d}i}{\mathrm{d}E}=\frac{1}{2}\left(\frac{2\mathrm{\hbar}^{2}\pi^{2}E}{m\mathrm{L}^{2}}\right)^{-\frac{1}{2}}, \, \, \, \, \, \, \, \, \, \, \, \, \mathrm{d}^{d}i= S_{d-1}i^{d-1}\frac{\mathrm{d}i}{\mathrm{d}E}\mathrm{d}E},
		\label{idodmniej1}
		\end{equation}
  \begin{equation}
\displaystyle{g_{d}(E)= \frac{1}{\Gamma\left(\frac{d}{2}\right)}\left( \frac{m \mathrm{L^{2}}}{2\pi \mathrm{\hbar}^{2}} \right)^{\frac{d}{2}} E^{\frac{d}{2}-1} }.
		\label{energydensityhr}
		\end{equation}
Knowing that \cite{abramowitz}
  \begin{equation}
\displaystyle{\Gamma\left( \frac{1}{2} \right)=\sqrt{\pi}, \, \, \, \, \, \, \, \, \, \, \, \, \Gamma\left( 1 \right)=1, \, \, \, \, \, \, \, \, \, \, \, \, \Gamma\left( \frac{3}{2} \right)= \frac{\sqrt{\pi}}{2} },
		\label{gammydodensityhr}
		\end{equation}
we obtain
  \begin{equation}
\displaystyle{g_{1}(E)=\frac{\sqrt{m}\mathrm{L}}{\sqrt{2}\mathrm{\hbar}}E^{-\frac{1}{2}}, \, \, \, \, \, \, \, \, \, \, \, \, g_{2}(E)=\frac{m\mathrm{L}^{2}}{2\mathrm{\hbar}^{2}\pi}, \, \, \, \, \, \, \, \, \, \, \, \, g_{3}(E)=\frac{m^{\frac{3}{2}}\mathrm{L}^{3}}{\sqrt{2}\mathrm{\hbar}^{3} \pi^{2}}  E^{\frac{1}{2}} }.
		\label{dodensityhr2}
		\end{equation}

In order to calculate the integral (\ref{meanvalue2}) we expand  (\ref{menanumbergrandcanensemble}) in geometric series
  \begin{equation}
\displaystyle{\frac{1}{\mathrm{e}^{(E-\mu_{ch})/\mathrm{k}_{B}T} - 1}=\sum_{l=1}^{\infty}\mathrm{e}^{-l(E-\mu_{ch})/\mathrm{k}_{B}T} }.
		\label{gemseriesexp}
		\end{equation}
Finally, one gets
  \begin{equation}
\displaystyle{\frac{\left<\mathcal{N}\right>_{1}}{\mathrm{L}}=\left(\frac{m\pi \mathrm{k}_{B}T}{2\mathrm{\hbar}^{2}}\right)^{\frac{1}{2}}\sum_{l=1}^{\infty}\frac{\mathrm{e}^{l\mu_{ch}/\mathrm{k}_{B}T}}{l^{1/2}}  },
		\label{srednialiczbaczastek1}
		\end{equation}
  \begin{equation}
\displaystyle{ \frac{\left<\mathcal{N}\right>_{2}}{\mathrm{L^{2}}}=\frac{m\mathrm{k}_{B}T}{2\mathrm{\hbar}^{2}\pi}\sum_{l=1}^{\infty}\frac{\mathrm{e}^{l\mu_{ch}/\mathrm{k}_{B}T}}{l} },
		\label{srednialiczbaczastek2}
		\end{equation}
		  \begin{equation}
\displaystyle{\frac{\left<\mathcal{N}\right>_{3}}{\mathrm{L^{3}}}=\left(\frac{m\mathrm{k}_{B}T}{2\mathrm{\hbar}^{2}\pi}\right)^{\frac{3}{2}}\sum_{l=1}^{\infty}\frac{\mathrm{e}^{l\mu_{ch}/\mathrm{k}_{B}T}}{l^{3/2}} }.
		\label{srednialiczbaczastek3}
		\end{equation}
Analyzing above equations, one sees that if the particle density $\left<\mathcal{N}\right>_{d}/\mathrm{L}^{d}$ is a constant quantity, then the chemical potential $\mu_{ch}$ must increase when the temperature $T$ decreases. In our considerations $\mu_{ch}<E_{0}=0$, therefore we may expect a phase transition if $\mu_{ch}=E_{0}=0$ for finite temperature $T_{c}>0$. Taking $\mu_{ch}=0$ in the formulas (\ref{srednialiczbaczastek1})-(\ref{srednialiczbaczastek3}) and assuming finite non-zero temperature one obtains  
 \begin{equation}
\displaystyle{\frac{\left<\mathcal{N}\right>_{1}}{\mathrm{L}}=\left(\frac{m\pi \mathrm{k}_{B}T}{2\mathrm{\hbar}^{2}}\right)^{\frac{1}{2}}\sum_{l=1}^{\infty}\frac{1}{l^{1/2}} = \infty , \, \, \, \, \, \, \, \, \, \, \, \,  \sum_{l=1}^{\infty}\frac{1}{l^{1/2}}= \infty},
		\label{srednialiczbaczastek12}
		\end{equation}
  \begin{equation}
\displaystyle{ \frac{\left<\mathcal{N}\right>_{2}}{\mathrm{L^{2}}}=\frac{m\mathrm{k}_{B}T}{2\mathrm{\hbar}^{2}\pi}\sum_{l=1}^{\infty}\frac{1}{l}=\infty, \, \, \, \, \, \, \, \, \, \, \, \,  \sum_{l=1}^{\infty}\frac{1}{l}= \infty },
		\label{srednialiczbaczastek22}
		\end{equation}
		  \begin{equation}
\displaystyle{\frac{\left<\mathcal{N}\right>_{3}}{\mathrm{L^{3}}}=\left(\frac{m\mathrm{k}_{B}T}{2\mathrm{\hbar}^{2}\pi}\right)^{\frac{3}{2}}\sum_{l=1}^{\infty}\frac{1}{l^{3/2}}= \left(\frac{m\mathrm{k}_{B}T}{2\mathrm{\hbar}^{2}\pi}\right)^{\frac{3}{2}}\zeta\left( \frac{3}{2} \right)<\infty },
		\label{srednialiczbaczastek32}
		\end{equation}
where the Riemann zeta function \cite{abramowitz}
  \begin{equation}
\displaystyle{\zeta(s)=\sum_{l=1}^{\infty}\frac{1}{l^{s}}, \, \, \, \, \, \, \, \, \, \, \, \,  \zeta\left( \frac{3}{2} \right)\approx 2.612 }.
		\label{zetariemana}
		\end{equation}
We instantly see that, there are no finite temperatures for which the phase transition occurs in the case of one and two-dimensional systems. The situation is quite different in the 3-dimensional case. The critical temperature $T_{c}$ satisfies the relation
  \begin{equation}
\displaystyle{\frac{\left<\mathcal{N}\right>_{3}}{\mathrm{L^{3}}}=\left(\frac{m\mathrm{k}_{B}T_{c}}{2\mathrm{\hbar}^{2}\pi}\right)^{\frac{3}{2}}\zeta\left( \frac{3}{2} \right) }.
		\label{srednialiczbaczastek33}
		\end{equation}
Further decreasing of temperature does not impact on the value of $\mu_{ch}$ which remains zero. The equation (\ref{srednialiczbaczastek32}) for $T<T_{c}$ describes the mean number of bosons in excited states \cite{Sacha} 
  \begin{equation}
\displaystyle{\frac{\left<\mathcal{N}_{e}\right>_{3}}{\mathrm{L^{3}}}=\left(\frac{m\mathrm{k}_{B}T}{2\mathrm{\hbar}^{2}\pi}\right)^{\frac{3}{2}}\zeta\left( \frac{3}{2} \right) }.
		\label{srednialiczbaczastek3wzbudzone}
		\end{equation}
Hence, the mean number of bosons in the single particle ground state of energy (for $T<T_{c}$) is given by
  \begin{equation}
\displaystyle{\left< n_{0} \right>_{3}= \left<\mathcal{N}\right>_{3}-\left<\mathcal{N}_{e}\right>_{3} =\left<\mathcal{N}\right>_{3} \left[  1-\left(\frac{ T}{T_{c}}\right)^{\frac{3}{2}}  \right] }.
		\label{srednialiczbaczastekwground}
		\end{equation}

The phase transition may be realized maintaining constant temperature and increasing the density. At some point, the particles inserted to the system will occupy only the ground state of energy. In real systems the particle density cannot be arbitrarily increased. For a large densities the interactions between particles will be dominated by 3-particle interactions. It leads to the molecules creations which causes the loss of atoms from the trap \cite{Sacha}. The fact that, we have to consider only small densities implies extremely small values of $T_{c}$\footnote{Therefore, the experimental realization of Bose-Einstein condensation is extremely difficult.}.  

Considering a thermal wavelength \cite{Sacha}
  \begin{equation}
\displaystyle{\lambda_{c}=\sqrt{ \frac{2\mathrm{\hbar}^{2}\pi}{m\mathrm{k}_{B}T_{c} } } },
		\label{dligoscfalitemicznej}
		\end{equation}
  \begin{equation}
\displaystyle{\frac{\left<\mathcal{N}\right>_{3}}{\mathrm{L^{3}}}=\frac{1}{\lambda_{c}^{3}}\zeta\left( \frac{3}{2} \right) },
		\label{srednialiczbaczastek333}
		\end{equation}
we see that the phase transition occurs when the de Broglie wavelength of the particle with a mass $m$ and energy $\mathrm{k}_{B}T_{c}$
  \begin{equation}
\displaystyle{\lambda_{dB}=\sqrt{ \frac{2\mathrm{\hbar}^{2}\pi^{2}}{m\mathrm{k}_{B}T_{c} } }=\sqrt{\pi}\lambda_{c} },
		\label{dligoscfalidebroglie}
		\end{equation}
becomes comparable with the mean separation between atoms in the system.

The Bose-Einstein condensation leads to the situation when the ground state is macroscopically occupied. In the case of the system of bosons closed in a periodic box, the state corresponds to the momentum equal to zero. Therefore, if $T<T_{c}$ macroscopic fraction of bosons have a zero eigenvalue of the momentum. Because of Heisenberg uncertainty relation, the positions of bosons are totally uncertain. It means that there are no particle accumulation in a certain area of coordinate space.

It is noteworthy that, we have obtained the condition for the critical temperature $T_{c}$ using the classical assumption that the energy $\mathrm{k}_{B}T$ is much larger than the difference between energy levels (\ref{energylevels}). Therefore we see that, the Bose-Einstein condensation is high-temperature effect. It turns out that, in real systems the condensation occurs even when $\mathrm{k}_{B}$ is thousands times larger than the distance between energy levels \cite{Sacha}.

Aforementioned lack of condensation in one and two-dimensional systems does not mean that the quantum state with the Bose-Einsten condensate properties does not exist. It is possible to observe a macroscopic ground state occupation in lower dimensions but in thermodynamic limit one has to reach $T=0$.  

It should be mentioned that, the Bose-Einstein condensation also exists in two-dimensions if we put our system into the harmonic trap. Although, in the one-dimensional case we cannot reach the Bose-Einstein condensation, it is possible to obtain a significant fraction of atoms occupying the single particle ground state of energy (for finite systems - not in thermodynamic limit) \cite{Sacha}, \cite{Dalfovo}. The situation was observed experimentally.  

For finite, closed systems with fixed number of particles $\mathcal{N}$ we need to use a canonical ensemble description\footnote{In the canonical ensemble approach, the system contains fixed number of particles $\mathcal{N}$ and can exchange the energy with reservoir which has a temperature $T$. It can be proven \cite{Huang} that, the probability of a single microstate $\{n_{0},n_{1},\ldots \}$ is given by \begin{center}
 $\rho=\mathcal{Z}^{-1}\mathrm{exp}\left(-E/\mathrm{k}_{B}T \right)$,
 \end{center}
where $\displaystyle{\sum_{j=1}^{\infty}n_{j}=\mathcal{N}}$, $\displaystyle{E=\sum_{j=1}^{\infty}n_{j}E_{j}}$ and $\mathcal{Z}$ is the partition function.} which is very difficult in the case of periodic box. The situation is slightly better if we use a harmonic trap for which $E_{j}=j\mathrm{\hbar}\omega $ (in contrast to the situation of periodic box $E_{j} \sim j^{2}$). Considering non-interacting bosons we need to remember about their non-distinguishability. Therefore, in our case the partition function takes the form (index $h$ means harmonic trap)  \cite{Sacha}
  \begin{equation}
\displaystyle{ \mathcal{Z}_{h}=\sum_{i_{1}=0}^{\infty}\xi^{i_{1}}\sum_{i_{2}=i_{1}}^{\infty}\xi^{i_{2}}\ldots \sum_{i_{\mathcal{N}}=i_{\mathcal{N}-1}}^{\infty}\xi^{i_{\mathcal{N}}} =\prod_{n=1}^{\mathcal{N}}\frac{1}{1-\xi^{n}}, \, \, \, \, \, \, \, \, \, \, \, \, \xi=\mathrm{exp}\left(-\frac{\mathrm{\hbar}\omega}{\mathrm{k}_{B}T}\right)},
		\label{partitionfunction1}
		\end{equation}
where we have used an obvious relation
  \begin{equation}
\displaystyle{\sum_{i=k}^{\infty}a^{i}=\frac{a^{k}}{1-a},   \, \, \, \, \, \, \, \, \, \, \, \, 0\leq a<1 }.
		\label{sumageometrycznego}
		\end{equation}
The probability of finding $\mathcal{M}\leq \mathcal{N}$ particles in a ground state $E_{0}$ is then given by 
\begin{equation}
\displaystyle{ P_{h}(\mathcal{M})=\mathcal{Z}^{-1}_{h}\sum_{i_{1}=1}^{\infty}\xi^{i_{1}}\sum_{i_{2}=i_{1}}^{\infty}\xi^{i_{2}}\ldots \sum_{i_{\mathcal{N}-\mathcal{M}}=i_{\mathcal{N}-\mathcal{M}-1}}^{\infty}\xi^{i_{\mathcal{N}-\mathcal{M}}} =\xi^{\mathcal{N}-\mathcal{M}}\prod_{n=\mathcal{N}-\mathcal{M}+1}^{\mathcal{N}}\left(1-\xi^{n}  \right)}.
		\label{prawdgrpundcensembl}
		\end{equation}
Hence, the mean number of particles in the ground state of energy may be written as
\begin{equation}
\displaystyle{ \left< n_{0} \right>_{h}=\sum_{\mathcal{M}=0}^{\mathcal{N}}\mathcal{M}P_{h}(\mathcal{M})}.
		\label{groundmeannumberce}
		\end{equation}

In the case of periodic box one has to calculate the following quantities
\begin{equation}
\displaystyle{ \mathcal{Z}_{box}=\sum_{i_{1}=0}^{\infty}\chi^{i_{1}^{2}}\sum_{i_{2}=i_{1}}^{\infty}\chi^{i_{2}^{2}}\ldots \sum_{i_{\mathcal{N}}=i_{\mathcal{N}-1}}^{\infty}\chi^{i_{\mathcal{N}}^{2}}, \, \, \, \, \, \, \, \, \, \, \, \, \chi=\mathrm{exp} \left(-\frac{2\mathrm{\hbar}^{2} \pi^{2}}{m \mathrm{L}^{2}\mathrm{k}_{B}T} \right) },
		\label{partitionfunctionbox1}
		\end{equation}
\begin{equation}
\displaystyle{ P_{box}(\mathcal{M})=\mathcal{Z}^{-1}_{box}\sum_{i_{1}=1}^{\infty}\chi^{i_{1}^{2}}\sum_{i_{2}=i_{1}}^{\infty}\chi^{i_{2}^{2}}\ldots \sum_{i_{\mathcal{N}-\mathcal{M}}=i_{\mathcal{N}-\mathcal{M}-1}}^{\infty}\chi^{i_{\mathcal{N}-\mathcal{M}}^{2}} },
		\label{prawdgrpundcensemblbox}
		\end{equation}
\begin{equation}
\displaystyle{ \left< n_{0} \right>_{box}=\sum_{\mathcal{M}=0}^{\mathcal{N}}\mathcal{M}P_{box}(\mathcal{M})}.
		\label{groundmeannumbercebox}
		\end{equation}

Above analysis allows us to consider zero temperature limit $T\rightarrow 0$ for finite closed systems consisting of $\mathcal{N}$ bosons. We see that if $T=0$ both of probabilities $P_{h}(\mathcal{M})$ and $P_{box}(\mathcal{M})$ are equal to zero if $\mathcal{M}<\mathcal{N}$. Therefore, if the system is finite and closed, all bosons occupy the single particle ground state in $T=0$.

\chapter{The Gross-Pitaevskii equation}
\label{sec:dodC}

\section{Derivation of the Gross-Pitaevskii equation}
\label{sec:Cderiv}

It is well known fact that, the wave function for ideal Bose-Einstein condensate is given by the product of a one-particle wave functions $\phi$ which (as for now) correspond to the single particle ground state
\begin{equation}
\displaystyle{ \psi(\vec{r}_{1},\ldots , \vec{r}_{\mathcal{N}} )=\phi(\vec{r}_{1})\ldots \phi(\vec{r}_{N})  }.
		\label{groundproductindeal}
		\end{equation}
Let us now turn on the contact interactions between atoms in the system $V(\vec{r}-\vec{r} \,' )=g_{0}\delta(\vec{r}-\vec{r} \,' )$. It is obvious that in this case the state (\ref{groundproductindeal}) is only an approximation of the real eigenstate of the considered system. We may ask now: how to choose a one-particle wave function $\phi$ to obtain the $\mathcal{N}$-particle state $\psi$ given by (\ref{groundproductindeal}) which is the best approximation of our real eigenstate in the presence of interactions. For this purpose, we need to minimize the functional of energy of the system \cite{Sacha}, \cite{Castin}
\begin{equation}
\displaystyle{ E\left[\phi,\phi^{*} \right]=\braket{\psi|\hat{\mathrm{H}}|\psi}=\mathcal{N}\int \mathrm{d}^{3}r\left[\frac{\hbar^{2}}{2m}\left|  \nabla \phi(\vec{r}) \right|^{2}+\mathcal{U}(\vec{r})\left| \phi(\vec{r})  \right|^{2}+\frac{g_{0}}{2}(\mathcal{N}-1) \left| \phi(\vec{r})  \right|^{4}   \right]  },
		\label{energyfunctionalGROSS}
		\end{equation}
where $\mathcal{U}(\vec{r})$ is an external potential of a trap. We need to also remember about the normalization condition
\begin{equation}
\displaystyle{ \int \mathrm{d}^{3}r\left| \phi(\vec{r})  \right|^{2}=1 },
		\label{normalizationcondGROSS}
		\end{equation}
which causes that, using the method of Lagrange multipliers, we have to minimize the functional 
\begin{equation}
\displaystyle{X\left[\phi,\phi^{*} \right]=E\left[\phi,\phi^{*} \right]-\mu_{ch}\mathcal{N}  \int \mathrm{d}^{3}r\left| \phi(\vec{r})  \right|^{2}},
		\label{XfunctionalGROSS}
		\end{equation}
where $\mu_{ch}$ is a Lagrange multiplier. By the variation with respect to  $\phi^{*}$ (or $\phi$) one easily obtains the Gross-Pitaevskii equation \cite{Sacha}, \cite{Castin}
\begin{equation}
\displaystyle{  -\frac{\hbar^{2}}{2m}\nabla^{2}\phi(\vec{r})+\mathcal{U}(\vec{r})\phi(\vec{r})+g_{0}(\mathcal{N}-1)\left| \phi(\vec{r})  \right|^{2}\phi(\vec{r})=\mu_{ch}\phi(\vec{r}) }.
		\label{GROSS}
		\end{equation}
If we have a huge number of atoms in the system, one may approximate $\mathcal{N}-1\approx \mathcal{N} $. 

The Gross-Pitaevskii equation (\ref{GROSS}) describes behavior of atoms which feel other atoms as an additional mean-field potential proportional to the density of atomic cloud $g_{0}\mathcal{N}\left| \phi(\vec{r})  \right|^{2}$ \cite{Sacha}, \cite{Castin}. Because our $\mathcal{N}$-particle state $\psi$ has the product form of one-particle states $\phi$, we see that in this description all the atoms behave in exactly the same way. It turns out that if we multiply the formula (\ref{GROSS}) by $\phi^{*}(\vec{r})$ and integrate it over $\mathrm{d}^{3}r$ in the result
\begin{equation}
\displaystyle{ \int \mathrm{d}^{3}r\left[\frac{\hbar^{2}}{2m}\phi^{*}(\vec{r})  \nabla^{2} \phi(\vec{r}) +\mathcal{U}(\vec{r})\left| \phi(\vec{r})  \right|^{2}+g_{0}\mathcal{N} \left| \phi(\vec{r})  \right|^{4}   \right]=\frac{\partial E}{\partial \mathcal{N}}=\mu_{ch}  },
		\label{chemicalpotGROSS}
		\end{equation}
we obtain the chemical potential\footnote{The aforementioned approximation $\mathcal{N}-1\approx \mathcal{N} $ has been taken in the calculation.}. Hence, we have got the interpretation of the Lagrange multiplier $\mu_{ch}$.

Minimization of the functional
\begin{equation}
\displaystyle{ \mathcal{X}=\int \limits_{t_{1}}^{t_{2}} \mathrm{d}t \left\{\frac{i\mathrm{h}}{2}\left[ \bra{\phi}\frac{\mathrm{d}}{\mathrm{d}t} \ket{\phi}- \left(\frac{\mathrm{d}}{\mathrm{d}t}\bra{\phi}\right) \ket{\phi}  \right]-E\left[ \phi(t),\phi^{*}(t) \right] \right\}   },
		\label{timedependentfunctionalGROSS}
		\end{equation}
with fixed $\phi(t_{1})$ and $\phi(t_{2})$ leads us to the time-dependent Gross-Pitaevskii equation \cite{Sacha}, \cite{Castin}
\begin{equation}
\displaystyle{ i\mathrm{\hbar} \frac{\partial \phi(\vec{r},t)}{\partial t}= -\frac{\hbar^{2}}{2m}\nabla^{2}\phi(\vec{r},t)+\mathcal{U}(\vec{r})\phi(\vec{r},t)+g_{0}\mathcal{N}\left| \phi(\vec{r},t)  \right|^{2}\phi(\vec{r},t)}.
		\label{GROSStdependent}
		\end{equation}

\section{Gaussian and Thomas-Fermi approximations}
\label{sec:CGaussian}
		
In the case of harmonic trap potential $\mathcal{U}(\vec{r})=m \omega^{2}r^{2}/2$, we may introduce natural units of energy $\hbar \omega$ and length $\sqrt{\hbar/m\omega}$. The quantity $\sqrt{\hbar/m\omega}$ corresponds to the size of the ground state of a single particle in the harmonic trap \cite{Sacha}, \cite{Castin}. The Gross-Pitaevskii equation (\ref{GROSS}) in the natural units takes the form\footnote{We need to remember that $\left[|\phi|^{2} \right]=[\mathrm{length}]^{-3}$.} 
		\begin{equation}
\displaystyle{  -\frac{1}{2}\nabla^{2}\phi(\vec{r})+\mathcal{U}(\vec{r})\phi(\vec{r})+g_{0}\mathcal{N}\left| \phi(\vec{r})  \right|^{2}\phi(\vec{r})=\mu_{ch}\phi(\vec{r}) },
		\label{GROSStfermi}
		\end{equation}
where $\mu_{ch}$ is given in $\hbar \omega$ units, and $g_{0}=4\pi a \sqrt{m \omega /\hbar}$. Neglecting the interactions ($g_{0}\rightarrow 0$) one easily obtain the following solution \cite{Sacha}, \cite{Castin}
		\begin{equation}
\displaystyle{ \phi(\vec{r})=\frac{\mathrm{e}^{ -r^{2}/2\sigma^{2} }}{\pi^{3/4}\sigma^{3/2}}  }.
		\label{GROSStferm0intsol}
		\end{equation}
Such a Gaussian function has one free parameter $\sigma$ which may be used as variational parameter in the case of small but non-zero interactions $g_{0}\neq 0$. Substituting (\ref{GROSStferm0intsol}) into (\ref{energyfunctionalGROSS}) one obtains 		
			\begin{equation}
\displaystyle{ E[\phi,\phi^{*}]=E[\sigma]=\frac{3}{4\sigma^{2}}+\frac{3}{4}\sigma^{2}+\frac{\chi}{2\sigma^{3}}, \, \, \, \, \, \, \, \, \chi=\mathcal{N}a \sqrt{\frac{2m \omega}{\pi \hbar}}=\mathcal{N}\frac{g_{0}}{\sqrt{8 \pi^{3}}}  }.
		\label{engausss}
		\end{equation}
First two terms of $E[\sigma]$ correspond to the kinetic energy ($\sim \sigma^{-2}$) and the potential energy of the trap ($\sim \sigma^{2}$), respectively. The last term of $E[\sigma]$ is proportional to the density of atoms ($\sim \mathcal{N}/\sigma^{3}$) and corresponds to the energy of interpatricle interactions \cite{Sacha}, \cite{Castin}. 		

The energy functional $E[\sigma]$ in the case of attractive interactions ($a<0 \Longrightarrow \chi<0$) has a global minimum $E[\sigma_{0}=0]=-\infty$ for all $\chi$. Hence, our atomic cloud may collapse to the point\footnote{In this case atoms feel the short-range part of real potential and our description (which uses the contact pseudopotential) breaks down. It should be mentioned that the collapse does not occur in one-dimensional systems \cite{Sacha}.}. If $|\chi|$ is not too large then the kinetic energy may compensate the negative contribution from the energy of interparticle interactions. The compensation can lead to the appearance of local minimum - collapse does not have to happen. The critical value $\chi_{c}$ may be found from the system of equations: $\mathrm{d}E/\mathrm{d}\sigma=0, \, \, \mathrm{d}^{2}E/\mathrm{d}\sigma^{2}=0$ - inflection point. Knowing that $\chi \sim \mathcal{N}$ one can estimate the maximal number of atoms for which collapse does not occur.

Minimization of $E[\sigma]$ in the case of repulsive interactions ($a>0$) leads to  
			\begin{equation}
\displaystyle{\sigma_{0}^{5}=\sigma_{0}+\chi, \, \, \, \, \, \, \, \, \sigma_{0}>0}.
		\label{minawiekszeod0}
		\end{equation}
When $\chi \ll 1$ we obtain the solution corresponds to the non-interacting system $\sigma_{0}=1$. For $\chi \gg 1$ one gets $\sigma_{0}\approx \chi^{1/5}\sim \mathcal{N}^{1/5}$. The last result means that the spatial breadth of the probability density $|\phi|^{2}$ rises with the number of atoms $\mathcal{N}$. The dependence is caused by repulsive character of interparticle interactions. We instantly see that, for $\chi \gg 1$ the kinetic energy behaves as $\sim \mathcal{N}^{-2/5}$ and the trap potential energy as $\sim \mathcal{N}^{2/5}$. To satisfy the relation $\chi=\mathcal{N}g_{0}/\sqrt{8\pi^{3}} \gg 1$, we need to have very large number of atoms $\mathcal{N}$ because the coupling constant $g_{0}$ is small. Therefore, the kinetic energy becomes negligibly small compared with the other contributions to the energy. This means that the ground state is achieved by a balance between interparticle repulsion and the trap potential \cite{Sacha}, \cite{Castin}. 

The interparticle repulsion ($a > 0$) causes that the spatial breadth of the one-particle probability density $|\phi|^{2}$ is much larger than the range of ground state of a single particle in harmonic potential. Let us now define the radius of the Bose-Einstein condensate $\mathcal{R}$. For a large number of particles the following relation is satisfied 
		\begin{equation}
\displaystyle{\mathcal{R}\gg \sqrt{\frac{\hbar}{m\omega}}}.
		\label{promienbeca}
		\end{equation}
Hence, the ratio of kinetic energy $E_{k}$ and harmonic potential energy $E_{h}$		
		\begin{equation}
\displaystyle{\frac{E_{k}}{E_{h}} \approx \frac{\frac{\hbar^{2}}{m \mathcal{R}^{2}}}{m\omega^{2}\mathcal{R}^{2}} =\left( \frac{\hbar}{m \omega \mathcal{R}^{2}}\right)^{2} \ll 1 }.
		\label{ratiobec}
		\end{equation}
It allows us to omit the term $\nabla^{2}\phi$ in the Gross-Pitaevskii equation (\ref{GROSS})\footnote{The equation (\ref{GROSSommited}) may be rewritten as 
\begin{center}
$\mu_{ch}=\frac{1}{2}m\omega^{2}r^{2}+g_{0} \mathcal{N}|\phi(\vec{r})|^{2}=\mathrm{const}$,
\end{center}
 which means that the energy associated with the insertion of one particle to the system ($\mu_{ch}$) does not depend on the position in the atomic cloud $\vec{r}$ . }		
		\begin{equation}
\displaystyle{ \frac{1}{2}m\omega^{2}r^{2}\phi(\vec{r})+g_{0}\mathcal{N}\left| \phi(\vec{r})  \right|^{2}\phi(\vec{r})\approx \mu_{ch}\phi(\vec{r}) }.
		\label{GROSSommited}
		\end{equation}
The above equation has the following solution
		\begin{equation}
\displaystyle{ \phi(\vec{r})=\sqrt{ \frac{\mu_{ch}- \frac{1}{2}m\omega^{2}r^{2}}{ \mathcal{N} g_{0}} }, \, \, \, \, \, \, \, \, \mu_{ch}=\frac{\hbar \omega}{2}\left(15 \frac{\mathcal{N} a }{\sqrt{\hbar/m \omega}  } \right)^{2/5}},
		\label{GROSSommitedsolution}
		\end{equation}
where the value of chemical potential $\mu_{ch}$ was established from the normalization condition (\ref{normalizationcondGROSS}). We immediately see that, if $\mathcal{N} a \gg \sqrt{\hbar/m \omega}$, then $\mu_{ch}\gg \hbar \omega$ which means that the chemical potential is much larger than the difference between energy levels of a single particle in a harmonic potential. Energy of the condensate may be found by integration the formula (\ref{chemicalpotGROSS}) 	
		\begin{equation}
\displaystyle{ \frac{E}{\mathcal{N}}=\frac{5}{7}\mu_{ch}},
		\label{energyofthebecthomasfermi}
		\end{equation}
which clearly shows that one cannot identify $\mu_{ch}$ with the energy per one atom \cite{Sacha}, \cite{Castin}.

\section{The soliton solution}
\label{sec:Csoliton}

In comparison with the Schr\"{o}dinger equation, the Gross-Pitaevskii equation consists of one additional non-linear term $g_{0}\mathcal{N}\left| \phi(\vec{r})  \right|^{2}\phi(\vec{r})$. The term allows to stop the wave-packet spreading which is caused by the dispersion relation $E\sim p^{2}$ obtained from the linear Schr\"{o}dinger equation. Therefore, it is possible to obtain solutions which propagate without shape changes.

Let us analyze one-dimensional situation without a trap potential. In this case we may solve the Gross-Pitaevskii equation analytically. There are two possibilities:\\ \\
\emph{\textbf{Dark soliton solution}} \\

For the repulsive interactions $g_{0}>0$ (scattering length $a>0$) the time-dependent Gross-Pitaevskii equation without a trap potential in one dimension takes the form 
\begin{equation}
\displaystyle{ i\mathrm{\hbar} \frac{\partial \phi(z,t)}{\partial t}= -\frac{\hbar^{2}}{2m}\frac{\partial^{2}}{\partial z^{2}}\phi(z,t)+g_{0}\mathcal{N}\left| \phi(z,t)  \right|^{2}\phi(z,t)},
		\label{GROSStdependentciemny}
		\end{equation}
has the following so-called dark soliton solution \cite{Sacha}, \cite{Castin}
\begin{equation}
\displaystyle{ \phi(z,t)=\mathrm{e}^{i\mu_{ch} t /\hbar} \sqrt{\rho_{0}} \left[ i\frac{\dot{q}}{v_{s}}+\sqrt{ 1 -\frac{\dot{q}^{2}}{v_{s}^{2}}} \mathrm{tanh}\left(  \frac{z-q}{\xi} \sqrt{ 1 -\frac{\dot{q}^{2}}{v_{s}^{2}}}  \right)   \right]},
		\label{GROSStciemnysolution}
		\end{equation}
where $\rho_{0}$ is the density of the condensate far away from $z=q$, $\mu_{ch}=\rho_{0}g_{0}$ is the chemical potential, $v_{s}=\sqrt{\rho_{0}g_{0}/m}$ is the propagation velocity of long-wave length disturbances (speed of sound), $\xi=\hbar /\sqrt{m\rho_{0}g_{0}}$ is the healing length. We instantly see that, the soliton position is given by $q(t)=\dot{q}t+q(t=0)$ and its depth depends on the value of $\dot{q}/v_{s}$. The modulus square of the dark soliton wave function is presented in the Figure C.1. 
\begin{center}\includegraphics[width=13cm,angle=0]{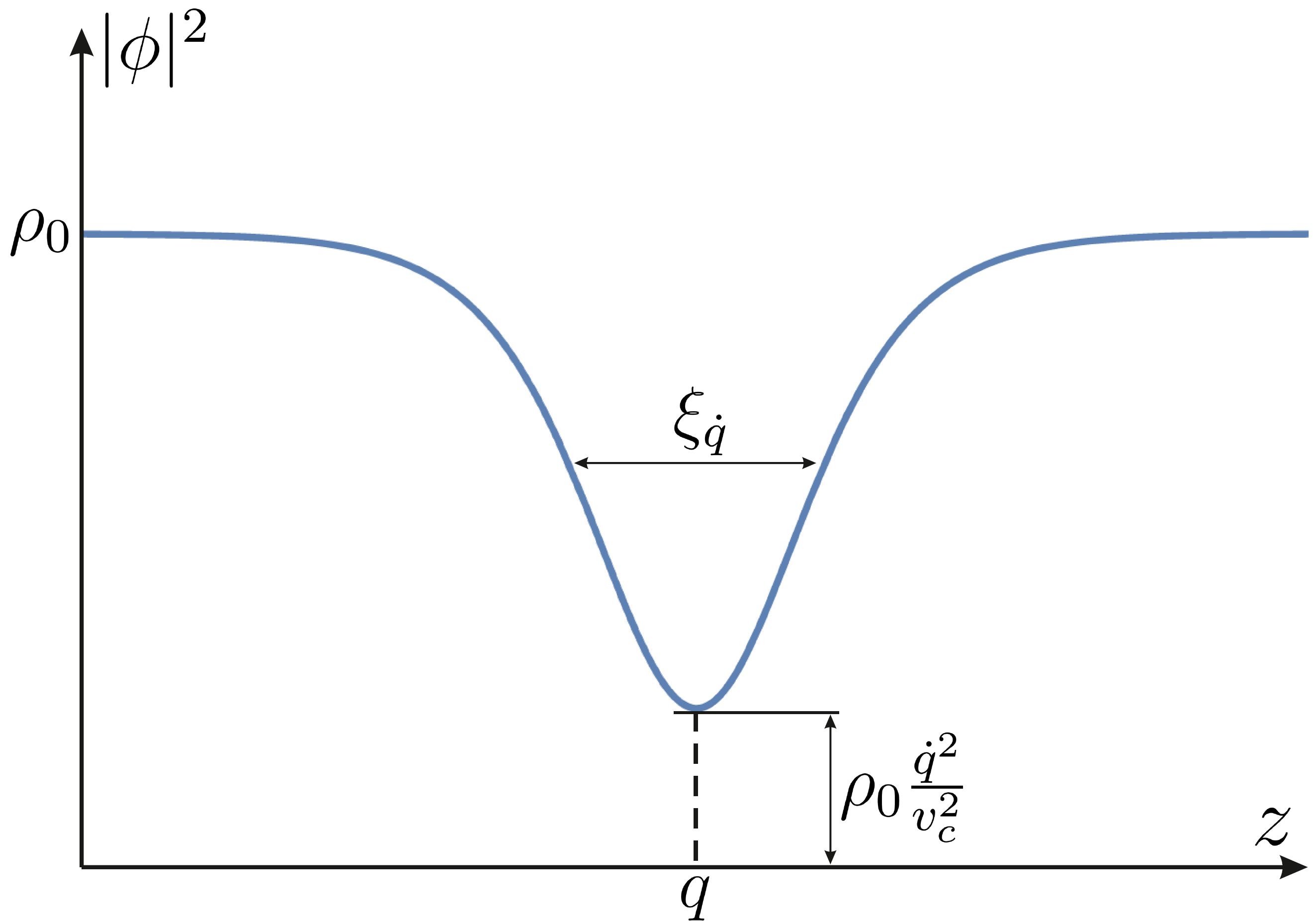}\end{center}
\begin{center}
\footnotesize{Figure C.1. The density of probability $|\phi(z)|^{2}$ corresponding to the dark soliton solution (\ref{GROSStciemnysolution}). }
\end{center}
Moreover, the width of the soliton depends on the velocity of the soliton $\dot{q}$ in the following way  
\begin{equation}
\displaystyle{ \xi_{\dot{q}}=\frac{\xi}{ \sqrt{ 1 -\frac{\dot{q}^{2}}{v_{s}^{2}}} }},
		\label{GROSStciemnysolitonszerokosc}
		\end{equation}
which is equal to the healing length $\xi$ for $\dot{q}=0$. When $\dot{q}=v_{s}$ the width $\xi_{\dot{q}=v_{s}}=0$ and the dark soliton  disappears. On the other hand, we obtain a \emph{black soliton} solution (stationary solution) if $\dot{q}=0$
\begin{equation}
\displaystyle{ \phi(z,t)=\mathrm{e}^{i\mu_{ch} t /\hbar} \sqrt{\rho_{0}} \mathrm{tanh}\left(  \frac{z-q}{\xi}   \right)  }.
		\label{GROSStczarnysolution}
		\end{equation}
		
Let us now analyze how the phase of the wave function varies along the soliton. In this case we need to know the two following limits
\begin{equation}
\displaystyle{\begin{array}{l} \phi(z\rightarrow \infty,0)\rightarrow \sqrt{\rho_{0}}\bigg[i\frac{\dot{q}}{v_{s}}+\sqrt{1-\frac{\dot{q}^{2}}{v_{s}^{2}}}   \bigg] , \, \, \\ \\  \phi(z\rightarrow -\infty,0)\rightarrow \sqrt{\rho_{0}} \bigg[i\frac{\dot{q}}{v_{s}}-\sqrt{1-\frac{\dot{q}^{2}}{v_{s}^{2}}}    \bigg] . \end{array} }
		\label{granicedofazysolitonu}
\end{equation}
The wave function (\ref{GROSStciemnysolution}) may be rewritten as 
\begin{equation}
\displaystyle{\phi(z,t)=\sqrt{\rho(z,t)}\mathrm{e}^{i \varphi(z,t)}, }
		\label{ciemnysolwavefunctionzfaza}
\end{equation}
where $\varphi(z,t)$ is the phase of the wave function $\phi(z,t)$. Hence, the change of the phase for $\dot{q}\geq 0$ is given by
\begin{center}
$\displaystyle{\Delta\phi=\varphi(\infty,t)-\varphi(-\infty,t)=\mathrm{arctan}\left(\frac{\dot{q}}{\sqrt{v_{s}^{2}-\dot{q}^{2}}}\right)-\left[\pi - \mathrm{arctan}\left(\frac{\dot{q}}{\sqrt{v_{s}^{2}-\dot{q}^{2}}}\right)  \right] }$
\end{center}
\begin{equation}
\displaystyle{=2\mathrm{arctan}\left(\frac{\dot{q}}{\sqrt{v_{s}^{2}-\dot{q}^{2}}}\right) -\pi }.
		\label{zmianafazyciemnego}
\end{equation}
It means that the probability density current flows in the opposite direction to the dark soliton motion \cite{Sacha}.

In higher dimensions the soliton solutions are unstable - small disturbances cause the breakup of the soliton into a vortices. In order to obtain a stable soliton in 3D one needs to squeeze a trap in two transverse directions so that the energy of excitations in these directions will be greater than the kinetic energy associated with the soliton \cite{Sacha}.\\ \\
\emph{\textbf{Bright soliton solution}} \\

In the case of attractive interactions ($g_{0}<0$) the time-dependent Gross-Pitaevskii equation without a trap potential (\ref{GROSStdependentciemny}) in 1D has the solution corresponding to the so-called \emph{bright soliton} 
\begin{equation}
\displaystyle{ \phi(z,t)=\frac{1}{\sqrt{|g_{0}|}}\frac{\alpha}{\mathrm{cosh}\left[ \alpha (z-\dot{q}t) \right]} \mathrm{e}^{i\dot{q}z}\mathrm{e}^{-i(\dot{q}^{2}-\alpha^{2})t/2}},
		\label{GROSStjansysol}
		\end{equation}
where $\alpha$ is an arbitrary constant which allows us to normalize above wave function to the desired number of atoms. Because of the attractive interparticle interaction, atoms creating the wave packet stick together without the presence of a trap. The bright soliton solution does not change the shape during the evolution \cite{Sacha}.

\chapter{A linearization of the Gross-Pitaevskii equation}
\label{sec:dodD}

It is obvious that the time-dependent Gross-Pitaevskii equation (\ref{GROSStdependent}) may be solved analytically or, if it is not possible, numerically. In the result one may obtain the stationary solutions for which it is necessary to ask about their dynamical stability - the question is if any small disturbance will not cause the change of the nature of solution from stationary to non-stationary during the evolution in time. For this purpose we need to linearize the Gross-Pitaevskii equation \cite{Sacha}, \cite{Castin}.

\section{Small disturbance $\delta \phi$}
\label{sec:Smalldisturb}

Let us assume that $\phi_{0}$ is the stationary solution of the Gross-Pitaevskii equation 
\begin{equation}
\displaystyle{  -\frac{\hbar^{2}}{2m}\nabla^{2}\phi_{0}+\mathcal{U}_{0}(\vec{r})\phi_{0}+g_{0}\mathcal{N}\left| \phi_{0}  \right|^{2}\phi_{0}=\mu_{ch}\phi_{0} }.
		\label{GROSSdisturb1}
		\end{equation}
Small disturbance of, for example, trap potential\footnote{It does not matter if $\delta \mathcal{U}=0$ - from the dynamical stability point of view, we are only interested in the evolution of $\delta\phi$.}
\begin{equation}
\displaystyle{ \mathcal{U}(\vec{r},t)=\mathcal{U}_{0}(\vec{r})+\delta \mathcal{U}(\vec{r},t)  }
		\label{TRAPdisturb}
		\end{equation}
may cause the small disturbance of the solution 
\begin{equation}
\displaystyle{ \phi=\phi_{0}+\delta \phi  }.
		\label{SOLUTIONdisturb}
		\end{equation}
Because the stationary state evolution is trivial 
\begin{equation}
\displaystyle{i\hbar\frac{\partial \phi_{0}(\vec{r},t)}{\partial t}=\mu_{ch}\phi_{0}(\vec{r},t) \Longrightarrow \phi_{0}(\vec{r},t)=\mathrm{e}^{-i\mu_{ch}t/\hbar}\phi_{0}(\vec{r},0)  },
		\label{STATSOLPHASEevolution}
		\end{equation}
we put into the time-dependent Gross-Pitaevski equation the function of the form $\mathrm{e}^{-i\mu_{ch}t/\hbar}\phi_{0}(\vec{r},t)$ which allows us to get rid of the trivial phase evolution of stationary solution
\begin{equation}
\displaystyle{ i\mathrm{\hbar} \frac{\partial \phi(\vec{r},t)}{\partial t}= -\frac{\hbar^{2}}{2m}\nabla^{2}\phi(\vec{r},t)+\mathcal{U}(\vec{r},t)\phi(\vec{r},t)+g_{0}\mathcal{N}\left| \phi(\vec{r},t)  \right|^{2}\phi(\vec{r},t)-\mu_{ch}\phi(\vec{r},t)}.
		\label{GROSStdependentLINEAR}
		\end{equation}
Using the decompositions (\ref{TRAPdisturb}) and (\ref{SOLUTIONdisturb}) one obtains
\begin{equation}
\displaystyle{ i\mathrm{\hbar} \frac{\partial \delta \phi}{\partial t}= \left[-\frac{\hbar^{2}}{2m}\nabla^{2}+\mathcal{U}_{0}-\mu_{ch}\right] \delta \phi +2g_{0}\mathcal{N}\phi_{0}^{*}\phi_{0}\delta \phi +g_{0}\mathcal{N}\phi_{0}^{2}\delta \phi^{*}+\delta \mathcal{U}\phi_{0} +\mathcal{O}\left( \delta\mathcal{U}^{2},\delta \phi ^{2} \right) }.
		\label{GROSStdependentLINEARdisturb}
		\end{equation}
Because of the independence of $\delta \phi$ and $\delta \phi^{*}$ we need to have the second equation obtained by linearization of the evolution equation for $\phi^{*}$. Finally, one gets \cite{Sacha}, \cite{Castin}
\begin{equation}
\displaystyle{ i\mathrm{\hbar} \frac{\partial}{\partial t} \left(
        \begin{array}{c}
         \delta \phi \\ 
         \delta \phi^{*}
         \end{array}
      \right)  =  \mathcal{L}_{GP} \left(
        \begin{array}{c}
         \delta \phi \\ 
         \delta \phi^{*}
         \end{array}
      \right)  + \left(
        \begin{array}{c}
         S \\ 
         -S^{*}
         \end{array}
      \right)     },
		\label{GrossLinearizationsystemofeqs}
		\end{equation}
where 
\begin{equation}
\displaystyle{ S(\vec{r},t)=\delta \mathcal{U}(\vec{r},t)\phi_{0}(\vec{r}) },
		\label{GrossLinearizationS}
		\end{equation}
\begin{equation}
\displaystyle{ \mathcal{L}_{GP} =\left(
        \begin{array}{cc}
         H_{GP}+g_{0}\mathcal{N}|\phi_{0}|^{2} & g_{0}\mathcal{N}\phi_{0}^{2} \\ 
         -g_{0}\mathcal{N}\phi_{0}^{*2}  & -H_{GP}^{*}-g_{0}\mathcal{N}|\phi_{0}|^{2} 
         \end{array}
      \right)    },
		\label{GrossLinearizationLMATRIX}
		\end{equation}
\begin{equation}
\displaystyle{ H_{GP}=-\frac{\hbar^{2}}{2m}\nabla^{2}+\mathcal{U}_{0}(\vec{r})+g_{0}\mathcal{N}\left| \phi_{0}  \right|^{2}-\mu_{ch}   }.
		\label{GrossLinearizationGPHAM}
		\end{equation}
The system of equations (\ref{GrossLinearizationsystemofeqs}) is the so-called Bogoliubov-de Gennes equations. $H_{GP}$ will be called the Gross-Pitaevskii hamiltonian\footnote{Usually $H_{GP}^{*}=H_{GP}$. The differences may appear for instance in the presence of magnetic field.}.

The operator $\mathcal{L}_{GP}$ is time-independent hence, it is convenient to decompose $(\delta\phi,\delta\phi^{*})$ in the basis of $\mathcal{L}_{GP}$ eigensates. Unfortunately, it turns out that $\mathcal{L}_{GP}$ is non-diagonalizable - the collection of all $\mathcal{L}_{GP}$ eigenstates does not span all space\footnote{In fact, one vector is missing at least.} \cite{Sacha}, \cite{Castin}.

\section{A brief reminder of diagonalization}
\label{sec:Abriefreminderofdiagonalization}

Diagonalizable (not necessarily hermitian) matrix\footnote{A sufficient condition of diagonalizability of $M$ matrix is the normality condition $\left[M^{\dagger},M  \right]=0$.} $M$ may be transformed (by similarity transformation $TMT^{-1}$) into the following form \cite{Sacha}, \cite{Castin}
\begin{equation}
\displaystyle{ M=\sum_{i}\lambda_{i} \ket{\psi_{i}^{R}} \bra{\psi_{i}^{L}}  },
		\label{shortrem1}
		\end{equation}
where $\ket{\psi_{i}^{R}}$ and $\bra{\psi_{i}^{L}}$ are right and left eigenvectors of $M$ which satisfy the following relations
\begin{equation}
\displaystyle{ M \ket{\psi_{i}^{R}}= \lambda_{i} \ket{\psi_{i}^{R}}, \, \, \, \, \, \, \, \,  \bra{\psi_{i}^{L}}M=\lambda_{i} \bra{\psi_{i}^{L}} }
		\label{shortrem2}
		\end{equation}
\begin{equation}
\displaystyle{ M^{\dagger} \ket{\psi_{i}^{L}} = \lambda_{i}^{*} \ket{\psi_{i}^{L}}, \, \, \, \, \, \, \, \,  \bra{\psi_{i}^{R}}M^{\dagger}=\lambda_{i}^{*} \bra{\psi_{i}^{R}} , \, \, \, \, \, \, \, \, \braket{\psi_{i}^{L}|\psi_{j}^{R} }=\delta_{ij}.   }
		\label{shortrem3}
		\end{equation}
The vector $\ket{\psi_{i}^{L}}$ is the so-called the adjoint vector of $\ket{\psi_{i}^{R}}$. An identity operator may be written as
\begin{equation}
\displaystyle{\hat{\mathbf{1}}=\sum_{i} \ket{\psi_{i}^{R}} \bra{\psi_{i}^{L}} , \, \, \, \, \, \, \, \, \hat{\mathbf{1}}^{2}=\hat{\mathbf{1}}}.
		\label{shortrem4}
		\end{equation}

Non-diagonalizable matrix $J$ cannot be presented in diagonal form. By similarity transformation it can be reduced into the Jordan canonical form \cite{finkbeiner}
\begin{equation}
\displaystyle{ J =\left(
        \begin{array}{ccc}
         J_{1} & & \\ 
         & \ddots  &    \\
				 & &  J_{p}
         \end{array}
      \right) , \, \, \, \, \, \, \, \,   \mathrm{where} \, \, \, \, \, \, \, \,   J_{i} =\left(
        \begin{array}{cccc}
         \lambda_{i} & 1 & & \\ 
         & \lambda_{i} & \ddots & \\
				 & & \ddots  & 1   \\
				 & &  & \lambda_{i}
         \end{array}
      \right) }
				\label{shortrem5}
		\end{equation}

\section{Small orthogonal disturbance $\delta \phi_{\bot}$}
\label{sec:Smallortohogonaldisturb}

Let us now split $\delta\phi$ into a part along and orthogonal to the $\phi_{0}$ \cite{Sacha}, \cite{Castin}
\begin{equation}
\displaystyle{\delta\phi(\vec{r},t)=\eta(t)\phi_{0}(\vec{r})+\delta \phi_{\bot}(\vec{r},t)  }.
		\label{orthoGross1}
		\end{equation}
From the normalization condition of $\phi=\phi_{0}+\delta\phi$ one obtains
\begin{center}
$\displaystyle{\int \mathrm{d}^{3}r\left( \phi_{0}+\delta\phi  \right)\left( \phi_{0}^{*}+\delta\phi^{*}  \right)=\int \mathrm{d}^{3}r\left( |\phi_{0}|^{2} +\delta\phi \phi_{0}^{*}  +\phi_{0} \delta\phi^{*} \right) +\mathcal{O}\left(\delta \phi ^{2}\right) }$
\end{center}
\begin{center}
$\displaystyle{=1+\int \mathrm{d}^{3}r\left( \eta |\phi_{0}|^{2}  +\eta^{*} |\phi_{0}|^{2}+ \phi_{0} \delta \phi_{\bot}^{*} + \delta \phi_{\bot} \phi_{0}^{*}  \right) +\mathcal{O}\left(\delta \phi ^{2}\right)=1}$
\end{center}
Because $\delta \phi_{\bot}$ is orthogonal to $\phi_{0}$ we get\footnote{We have neglected the terms $\mathcal{O}\left(\delta \phi ^{2}\right)$.}
\begin{equation}
\displaystyle{\eta(t)+\eta^{*}(t)=0 \Longrightarrow \Re (\eta) = 0}.
		\label{orthoGross2}
		\end{equation}
The time-dependent parameter $\eta(t)$ as a small (in the sense of absolute value), purely imaginary number may be interpreted as a phase change of the stationary solution $\phi_{0}$
\begin{equation}
\displaystyle{\phi(\vec{r},t)=\left[1+\eta(t) \right]\phi_{0}(\vec{r})+\delta \phi_{\bot}(\vec{r},t) \approx \mathrm{e}^{\eta(t)} \phi_{0}(\vec{r})+\delta \phi_{\bot}(\vec{r},t)}.
		\label{orthoGross3}
		\end{equation}
If we multiply the system of equations (\ref{GrossLinearizationsystemofeqs}) by the matrix
\begin{equation}
\displaystyle{ \mathcal{\hat{Q}} =\left(
        \begin{array}{cc}
         \hat{Q} & 0 \\ 
         0  &  \hat{Q}^{*}  
         \end{array}
      \right), \, \, \, \, \, \, \, \, \hat{Q}=1-\ket{\phi_{0}}\bra{\phi_{0}}  },
				\label{orthoGross4}
		\end{equation}
one gets
\begin{center}
$\displaystyle{ i\mathrm{\hbar}\frac{\partial}{\partial t}  \mathcal{\hat{Q}} \left(
        \begin{array}{c}
         \eta\phi_{0}+\delta \phi_{\bot} \\ 
         \eta^{*}\phi_{0}^{*}+\delta \phi_{\bot}^{*}
         \end{array}
      \right)  =i\mathrm{\hbar}\frac{\partial}{\partial t}  \left(
        \begin{array}{c}
         \delta \phi_{\bot} \\ 
         \delta \phi_{\bot}^{*}
         \end{array}
      \right)= \mathcal{\hat{Q}} \mathcal{L}_{GP} \left(
        \begin{array}{c}
         \eta\phi_{0}+\delta \phi_{\bot} \\ 
         \eta^{*}\phi_{0}^{*}+\delta \phi_{\bot}^{*}
         \end{array}
      \right)  + \mathcal{\hat{Q}}\left(
        \begin{array}{c}
         S \\ 
         -S^{*}
         \end{array}
      \right)     }$
\end{center}
\begin{center}
$\displaystyle{ = \mathcal{\hat{Q}} \mathcal{L}_{GP} \left[\mathcal{\hat{Q}}+ \hat{\mathbf{1}}- \mathcal{\hat{Q}}\right] \left(
        \begin{array}{c}
         \eta\phi_{0}+\delta \phi_{\bot} \\ 
         \eta^{*}\phi_{0}^{*}+\delta \phi_{\bot}^{*}
         \end{array}
      \right)  +\left(
        \begin{array}{c}
         S_{\bot} \\ 
         -S_{\bot}^{*}
         \end{array}
      \right)}$,
\end{center}
where
	\begin{equation}
\displaystyle{S_{\bot}=\hat{Q}S}.
		\label{orthoGross5}
		\end{equation}
\begin{samepage}Let us now calculate the term
\begin{center}
$\displaystyle{  \mathcal{L}_{GP} \left[ \hat{\mathbf{1}}- \mathcal{\hat{Q}}\right] \left(
        \begin{array}{c}
         \eta\phi_{0}+\delta \phi_{\bot} \\ 
         \eta^{*}\phi_{0}^{*}+\delta \phi_{\bot}^{*}
         \end{array}
      \right)= 
				\left(
				\begin{array}{cc}
         H_{GP}+g_{0}\mathcal{N}|\phi_{0}|^{2} & g_{0}\mathcal{N}\phi_{0}^{2} \\ 
         -g_{0}\mathcal{N}\phi_{0}^{*2}  & -H_{GP}^{*}-g_{0}\mathcal{N}|\phi_{0}|^{2} 
         \end{array}
      \right)   
			\left(
        \begin{array}{c}
         \eta\phi_{0} \\ 
         \eta^{*}\phi_{0}^{*}
         \end{array}
      \right)}$
\end{center}
\begin{center}
$\displaystyle{ = \left(
        \begin{array}{c}
         \eta H_{GP} \phi_{0}+(\eta+\eta^{*})g_{0}\mathcal{N}|\phi_{0}|^{2} \phi_{0} \\ 
         -\eta^{*} H_{GP}^{*} \phi_{0}^{*}-(\eta+\eta^{*})g_{0}\mathcal{N}|\phi_{0}|^{2} \phi_{0}^{*}
         \end{array}
      \right)  =  0 }$,
\end{center}
where we used the relations (\ref{GROSSdisturb1}), (\ref{orthoGross2}). Therefore
\begin{equation}
\displaystyle{ \mathcal{\hat{Q}} \mathcal{L}_{GP} \left[\mathcal{\hat{Q}}+ \hat{\mathbf{1}}- \mathcal{\hat{Q}}\right]\left(
        \begin{array}{c}
         \delta \phi \\ 
         \delta \phi^{*}
         \end{array}
      \right)=\mathcal{\hat{Q}} \mathcal{L}_{GP} \mathcal{\hat{Q}} \left(
        \begin{array}{c}
         \delta \phi \\ 
         \delta \phi^{*}
         \end{array}
      \right) }.
      \label{orthoGrossRedukcja}
\end{equation}\end{samepage}

Defining now\footnote{We have used the fact that $\displaystyle{ \hat{Q}H_{GP}\hat{Q}=(1-\ket{\phi_{0}} \bra{\phi_{0}})H_{GP}(1-\ket{\phi_{0}} \bra{\phi_{0}})\stackrel{(\ref{GROSSdisturb1})}{=}H_{GP}} $.}
\begin{equation}
\displaystyle{ \mathcal{L}=\mathcal{\hat{Q}} \mathcal{L}_{GP} \mathcal{\hat{Q}}  =\left(
				\begin{array}{cc}
         H_{GP}+g_{0}\mathcal{N}\hat{Q}|\phi_{0}|^{2}\hat{Q}  & g_{0}\mathcal{N}\hat{Q}\phi_{0}^{2}\hat{Q}^{*}  \\ 
         -g_{0}\mathcal{N} \hat{Q}^{*}\phi_{0}^{*2}\hat{Q}  & -H_{GP}^{*} -g_{0}\mathcal{N}\hat{Q}^{*}|\phi_{0}|^{2} \hat{Q}^{*}
         \end{array}
      \right) ,  }
		\label{orthoGross7}
		\end{equation}
and using the following obvious relation 
\begin{equation}
\displaystyle{ \mathcal{L}\mathcal{\hat{Q}}= \mathcal{\hat{Q}}\mathcal{L}_{GP}\mathcal{\hat{Q}}^{2}=\mathcal{\hat{Q}}\mathcal{L}_{GP}\mathcal{\hat{Q}}=\mathcal{L} },
		\label{orthoGross8}
		\end{equation}
we obtain the system of evolution equations for a small orthogonal disturbances ($\delta \phi_{\bot}$, $\delta \phi_{\bot}^{*}$) 
\begin{equation}
\displaystyle{ i\mathrm{\hbar}\frac{\partial}{\partial t}  \left(
        \begin{array}{c}
         \delta \phi_{\bot} \\ 
         \delta \phi_{\bot}^{*}
         \end{array}
      \right)=
				\mathcal{L}  \left(
        \begin{array}{c}
         \delta \phi_{\bot} \\ 
        \delta \phi_{\bot}^{*}
         \end{array}
      \right)  +\left(
        \begin{array}{c}
         S_{\bot} \\ 
         -S_{\bot}^{*}
         \end{array}
      \right)}.	
			\label{orthoGrossLinearizationsystemofeqs}
		\end{equation}
		
It turns out that $\mathcal{L}$ operator is fully diagonalizable. The decomposition (\ref{orthoGross1}) is, in fact, an extraction of the relevant part of $\delta \phi$ \cite{Sacha}, \cite{Castin}.

\section{Symmetries of $\mathcal{L}$}
\label{sec:SymmetriesofL}

It is easy to show that $\mathcal{L}$ has two follwing symmetries
\begin{equation}
\displaystyle{\sigma_{x}\mathcal{L}\sigma_{x}=-\mathcal{L}^{*}, \, \, \, \, \, \, \, \, \sigma_{x}=\left(
        \begin{array}{cc}
         0 & 1 \\ 
         1 & 0
         \end{array}
      \right) },
		\label{SymL1}
		\end{equation}
\begin{equation}
\displaystyle{\sigma_{z}\mathcal{L}\sigma_{z}=\mathcal{L}^{\dagger}, \, \, \, \, \, \, \, \, \sigma_{z}=\left(
        \begin{array}{cc}
         1 & 0 \\ 
         0 & -1
         \end{array}
      \right) }.
		\label{SymL2}
		\end{equation}
Let us consider right and left eigenvector of $\mathcal{L}$ which correspond to an eigenvalue $E_{n}$
\begin{equation}
\displaystyle{ \mathcal{L}\ket{\psi_{n}^{R}}=E_{n}\ket{\psi_{n}^{R}}, \, \, \, \, \, \, \, \,  \bra{\psi_{n}^{L}}\mathcal{L}=E_{n}\bra{\psi_{n}^{L}}}.
		\label{SymL3}
		\end{equation}
We will show that symmetries (\ref{SymL1}) and (\ref{SymL2}) imply the existence of left and right eigenvectors to eigenvalues $\{E_{n}^{*},-E_{n},-E_{n}^{*}\}$ \cite{Sacha}, \cite{Castin}
\begin{equation}
\displaystyle{\mathcal{L}\ket{\psi_{n,*}^{R}}=E^{*}_{n}\ket{\psi_{n,*}^{R}}, \, \, \, \, \, \, \, \, \bra{\psi_{n,*}^{L}}\mathcal{L}=E^{*}_{n}\bra{\psi_{n,*}^{L}} },
		\label{SymL4}
		\end{equation}
\begin{equation}
\displaystyle{\mathcal{L}\ket{\psi_{n,-}^{R}}=-E_{n}\ket{\psi_{n,-}^{R}}, \, \, \, \, \, \, \, \, \bra{\psi_{n,-}^{L}}\mathcal{L}=-E_{n}\bra{\psi_{n,-}^{L}} },
		\label{SymL5}
		\end{equation}
		\begin{equation}
\displaystyle{\mathcal{L}\ket{\psi_{n,-,*}^{R}}=-E_{n}^{*}\ket{\psi_{n,-,*}^{R}}, \, \, \, \, \, \, \, \, \bra{\psi_{n,-,*}^{L}}\mathcal{L}=-E_{n}^{*}\bra{\psi_{n,-,*}^{L}} }.
		\label{SymL6}
		\end{equation}
Using the relations (\ref{SymL2}) and (\ref{SymL3}) one gets
\begin{equation}
\displaystyle{  \left(\bra{\psi_{n}^{L}}\mathcal{L}\right)^{\dagger}=\mathcal{L}^{\dagger}\ket{\psi_{n}^{L}}=\sigma_{z}\mathcal{L}\sigma_{z}\ket{\psi_{n}^{L}}=E_{n}^{*}\ket{\psi_{n}^{L}}  \Longrightarrow  \mathcal{L}\sigma_{z}\ket{\psi_{n}^{L}}=E_{n}^{*} \sigma_{z}\ket{\psi_{n}^{L}}}.
		\label{SymL8}
		\end{equation}
Hence, from the symmetry (\ref{SymL2}) 
\begin{equation}
\displaystyle{ \ket{\psi_{n,*}^{R}}\sim \sigma_{z}\ket{\psi_{n}^{L}} },
		\label{SymL9}
		\end{equation}
is the right eigenvector of $\mathcal{L}$ corresponds to an eigenvalue $E_{n}^{*}$. Similarly, from the same symmetry
\begin{equation}
\displaystyle{ \mathcal{L}\ket{\psi_{n}^{R}}=E_{n}\ket{\psi_{n}^{R}} \Longrightarrow \sigma_{z}\mathcal{L}^{\dagger}\sigma_{z}\ket{\psi_{n}^{R}}=E_{n}\ket{\psi_{n}^{R}}    \Longrightarrow    \bra{\psi_{n}^{R}}\sigma_{z}\mathcal{L}=E^{*}_{n}\bra{\psi_{n}^{R}}\sigma_{z}   }.
		\label{SymL10}
		\end{equation}
Therefore		
		\begin{equation}
\displaystyle{ \ket{\psi_{n,*}^{L}}\sim \sigma_{z}\ket{\psi_{n}^{R}} },
		\label{SymL11}
		\end{equation}
is the left eigenvector of $\mathcal{L}$ corresponds to an eigenvalue $E_{n}^{*}$.	Moreover, from the relations (\ref{SymL1}) and (\ref{SymL3}) we obtain
		\begin{equation}
\displaystyle{ \mathcal{L}^{*}\ket{\psi_{n}^{R*}}=E_{n}^{*}\ket{\psi_{n}^{R*}} \Longrightarrow \mathcal{L}\sigma_{x}\ket{\psi_{n}^{R*}}=-E_{n}^{*}\sigma_{x}\ket{\psi_{n}^{R*}}        },
		\label{SymL12}
		\end{equation}
so there exists the right eigenvector of $\mathcal{L}$ corresponds to an eigenvalue $-E_{n}^{*}$			
			\begin{equation}
\displaystyle{ \ket{\psi_{n,-,*}^{R}}\sim \sigma_{x}\ket{\psi_{n}^{R*}} }.
		\label{SymL13}
		\end{equation}
By proceeding in a similar way one gets	
			\begin{equation}
\displaystyle{ \mathcal{L}\ket{\psi_{n,*}^{R}}=E_{n}^{*}\ket{\psi_{n,*}^{R}} \stackrel{  (\ref{SymL1}) }{\Longrightarrow} \mathcal{L}\sigma_{x} \ket{\psi_{n,*}^{R*}}=-E_{n}\sigma_{x} \ket{\psi_{n,*}^{R*}} },
		\label{SymL14}
		\end{equation}
		\begin{equation}
\displaystyle{ \ket{\psi_{n,-}^{R}}\sim \sigma_{x} \ket{\psi_{n,*}^{R*}} },
		\label{SymL15}
		\end{equation}
		\begin{equation}
\displaystyle{ \mathcal{L}\ket{\psi_{n,-}^{R}}=-E_{n}\ket{\psi_{n,-}^{R}} \stackrel{  (\ref{SymL2}) }{\Longrightarrow} \bra{\psi_{n,-}^{R}} \sigma_{z} \mathcal{L}=-E_{n}^{*}  \bra{\psi_{n,-}^{R}} \sigma_{z}  },
		\label{SymL16}
		\end{equation}
		\begin{equation}
\displaystyle{ \ket{\psi_{n,-,*}^{L}}\sim \sigma_{z} \ket{\psi_{n,-}^{R}} },
		\label{SymL17}
		\end{equation}
\begin{equation}
\displaystyle{ \mathcal{L}\ket{\psi_{n,-,*}^{R}}=-E_{n}^{*}\ket{\psi_{n,-,*}^{R}} \stackrel{  (\ref{SymL2}) }{\Longrightarrow} \bra{\psi_{n,-,*}^{R}} \sigma_{z} \mathcal{L}=-E_{n}  \bra{\psi_{n,-,*}^{R}} \sigma_{z}  },
		\label{SymL18}
		\end{equation}
\begin{equation}
\displaystyle{ \ket{\psi_{n,-}^{L}}\sim \sigma_{z} \ket{\psi_{n,-,*}^{R}} }.
		\label{SymL19}
		\end{equation}

The calculations presented above show that in the general case (complex eigenvalues of $\mathcal{L}$ for which $\Im(E_{n})\neq 0, \, \Re(E_{n})\neq 0$), we need only two eigenvectors (for example: $\ket{\psi_{n}^{R}}, \, \ket{\psi_{n,*}^{R}}$) to find all the eigenvectors which correspond to the family of eigenvalues $\{E_{n}, E_{n}^{*},-E_{n},-E_{n}^{*} \}$. It should be mentioned that in the case of purely real or purely imaginary eigenvalues, the families will consist of two elements because either $E_{n}^{*}=E_{n}$ or $E_{n}^{*}=-E_{n}$ \cite{Sacha}, \cite{Castin}.

\section{Diagonalization of $\mathcal{L}$}
\label{sec:DiagonalizationofL}

For the purpose of diagonalization of $\mathcal{L}$ we need to find vectors adjoint to right eigenvectors of $\mathcal{L}$. Assuming that \cite{Sacha}, \cite{Castin}
\begin{equation}
\displaystyle{ \braket{ \psi_{n}^{L} | \psi_{n}^{R} } \stackrel{ (\ref{SymL9})}{=} \braket{ \psi_{n,*}^{R} | \sigma_{z} | \psi_{n}^{R} }\neq 0   },
		\label{DiagL1}
		\end{equation}
we may determine the normalization of $\ket{\psi_{n}^{R}}$ and $\ket{\psi_{n,*}^{R}}$ so that
\begin{equation}
\displaystyle{ \braket{ \psi_{n,*}^{R} | \sigma_{z} | \psi_{n}^{R} } = 1   }.
		\label{DiagL2}
		\end{equation}
\begin{samepage}From the symmetry of $\mathcal{L}$ operator one sees that for arbitrary states $\ket{\psi_{j}}, \, \ket{\psi_{p}}$ the matrix element $\braket{\psi_{j} | \sigma_{z} | \psi_{p}}$ is time-independent
\begin{flushleft}
$\displaystyle{ i \hbar \frac{\partial}{\partial t}\braket{\psi_{j} | \sigma_{z} | \psi_{p}}= \left(i \hbar \frac{\partial}{\partial t}  \bra{\psi_{j}}   \right) \sigma_{z} \ket{\psi_{p}}+ \bra{\psi_{j}}  \sigma_{z} \left( i \hbar\frac{\partial}{\partial t}\ket{\psi_{p}} \right) } $
\end{flushleft}
\begin{equation}
\displaystyle{ = -\braket{ \psi_{j}| \mathcal{L}^{\dagger}\sigma_{z}|\psi_{p} }  +\braket{ \psi_{j}|\sigma_{z} \mathcal{L}|\psi_{p} }=0 },
		\label{DiagL3}
		\end{equation}
		\end{samepage}\pagebreak
where we have used the relations (\ref{orthoGrossLinearizationsystemofeqs}) and (\ref{SymL2}). If now the states $\ket{\psi_{j}}, \, \ket{\psi_{p}}$ are right eigenvectors $\ket{\psi_{m}^{R}}, \, \ket{\psi_{n}^{R}}$, then 
\begin{equation}
\displaystyle{ \left( -E_{m}^{*}+E_{n} \right)\braket{\psi_{m}^{R} | \sigma_{z} |\psi_{n}^{R} }=0 },
		\label{DiagL4}
		\end{equation}
which means that if $E_{m}^{*}\neq E_{n} $, then the matrix element $\braket{\psi_{m}^{R} | \sigma_{z} |\psi_{n}^{R} }=0$. Otherwise $\braket{\psi_{m}^{R} | \sigma_{z} |\psi_{n}^{R} }$ does not have to be equal to zero - the case (\ref{DiagL1}).

If $\ket{\psi_{n}^{R}}$ and $\ket{\psi_{n,*}^{R}}$ satisfy the relation (\ref{DiagL2}), then we obtain a full set of right and left eigenvectors creating mutually adjoint pairs of vectors by choosing
\begin{equation}
\displaystyle{\ket{\psi_{n}^{L}}= \sigma_{z}\ket{ \psi_{n,*}^{R} }, \, \, \, \, \, \, \, \,   \ket{\psi_{n,*}^{L}}= \sigma_{z}\ket{ \psi_{n}^{R} }},
		\label{DiagL5}
		\end{equation}
\begin{equation}
\displaystyle{\ket{\psi_{n,-}^{R}}= \sigma_{x}\ket{ \psi_{n,*}^{R*} }, \, \, \, \, \, \, \, \,   \ket{\psi_{n,-}^{L}}= -\sigma_{z}\sigma_{x}\ket{ \psi_{n}^{R*} }},
		\label{DiagL6}
		\end{equation}
\begin{equation}
\displaystyle{\ket{\psi_{n,-,*}^{R}}= \sigma_{x}\ket{ \psi_{n}^{R*} }, \, \, \, \, \, \, \, \,   \ket{\psi_{n,-,*}^{L}}= -\sigma_{z}\sigma_{x}\ket{ \psi_{n,*}^{R*} }}.
		\label{DiagL7}
		\end{equation}
It is easy to show that the right eigenvectors are orthogonal to the left eigenvectors if they correspond to different eigenvectors. One may consider for instance $ \braket{\psi_{m,-}^{L} | \psi_{n,-}^{R}}$
\begin{center}
$\displaystyle{ i \hbar \frac{\partial }{\partial t}  \braket{\psi_{m,-}^{L} | \psi_{n,-}^{R}}  = -i \hbar \frac{\partial }{\partial t} \braket{ \psi_{m}^{R*} | \sigma_{x} \sigma_{z} \sigma_{x} | \psi_{n,*}^{R} }  = \braket{ \psi_{m}^{R*} |\mathcal{L}^{\dagger} \sigma_{x} \sigma_{z} \sigma_{x} | \psi_{n,*}^{R} }   - \braket{ \psi_{m}^{R*} | \sigma_{x} \sigma_{z} \sigma_{x}\mathcal{L} | \psi_{n,*}^{R} }  }$
\end{center}
\begin{equation}
\displaystyle{ =\braket{ \psi_{m}^{R*} |\mathcal{L}^{\dagger} \sigma_{x} \sigma_{z} \sigma_{x} | \psi_{n,*}^{R} }  -  \braket{ \psi_{m}^{R*} | \left((\mathcal{L}^{*})^{\dagger}\right)^{*}\sigma_{x} \sigma_{z} \sigma_{x} | \psi_{n,*}^{R} }  = 0    },
		\label{DiagL8}
		\end{equation}
which means that
\begin{equation}
\displaystyle{ (-E_{m}^{*}+E_{n}^{*}) \braket{\psi_{m,-}^{L} | \psi_{n,-}^{R}}=0\Longrightarrow \braket{\psi_{m,-}^{L} | \psi_{n,-}^{R}}=0 \, \, \, \, \, \, \, \, \mathrm{if} \, \, \, \, \, \, \, \, m\neq n }.
		\label{DiagL9}
		\end{equation}

Complex eigenvalues of $\mathcal{L}$ for which $\Im(E_{n})\neq 0$ and $\Re(E_{n})\neq 0 $ create the family of eigenvalues $\{E_{n}, E_{n}^{*},-E_{n},-E_{n}^{*} \}$. There also exist two additional eigenvectors which correspond to an eigenvalue which is equal to zero \cite{Sacha}, \cite{Castin}
\begin{equation}
\displaystyle{ \left(
        \begin{array}{c}
         \phi_{0}  \\ 
         0 
         \end{array}
      \right), \, \, \, \, \, \, \, \,  \left(
        \begin{array}{c}
         0  \\ 
         \phi_{0}^{*}
         \end{array}
      \right)},
		\label{DiagL10}
		\end{equation}
which are orthogonal to the vector
\begin{equation}
\displaystyle{ \left(
        \begin{array}{c}
         \delta \phi_{\bot}  \\ 
         \delta \phi_{\bot}^{*}
         \end{array}
      \right)}.
		\label{DiagL11}
		\end{equation}
The operator $\mathcal{L}$ in diagonal form may be presented as
\begin{equation}
\displaystyle{ \mathcal{L}=\sum_{n}\left( E_{n} \ket{\psi_{n}^{R}} \bra{\psi_{n}^{L}} + E_{n}^{*} \ket{\psi_{n,*}^{R}} \bra{\psi_{n,*}^{L}}  - E_{n} \ket{\psi_{n,-}^{R}} \bra{\psi_{n,-}^{L}} - E_{n}^{*} \ket{\psi_{n,-,*}^{R}} \bra{\psi_{n,-,*}^{L}}\right) }.
		\label{DiagL12}
		\end{equation}

It turns out that operators $\mathcal{L}$ and $\mathcal{L}_{GP}$ have an identical spectrum. Considering an eigenvector \cite{Sacha}, \cite{Castin}
\begin{equation}
\displaystyle{  \ket{\psi_{n}^{R}}_{GP}= \left(
        \begin{array}{c}
         \ket{u_{n}^{R}}_{GP}  \\ 
         \ket{v_{n}^{R}}_{GP}
         \end{array}
      \right)},
		\label{DiagL13}
		\end{equation}
 which corresponds to non-zero eigenvalue $E_{n}$ and using the results from the part \ref{sec:Smallortohogonaldisturb} one obtains
\begin{equation}
\displaystyle{ \mathcal{L}_{GP}\ket{\psi_{n}^{R}}_{GP}=E_{n}\ket{\psi_{n}^{R}}_{GP}\Longrightarrow \mathcal{\hat{Q}}\mathcal{L}_{GP}\left[\mathcal{\hat{Q}} +\hat{\mathbf{1}}-\mathcal{\hat{Q}} \right] \ket{\psi_{n}^{R}}_{GP}=E_{n}\mathcal{\hat{Q}}\ket{\psi_{n}^{R}}_{GP}},
		\label{DiagL14}
		\end{equation}
\begin{equation}
\displaystyle{  \mathcal{\hat{Q}}\mathcal{L}_{GP}\left[\mathcal{\hat{Q}} +\hat{\mathbf{1}}-\mathcal{\hat{Q}} \right] \ket{\psi_{n}^{R}}_{GP}=\mathcal{\hat{Q}}\mathcal{L}_{GP}\mathcal{\hat{Q}} \ket{\psi_{n}^{R}}_{GP}= \mathcal{L} \mathcal{\hat{Q}} \ket{\psi_{n}^{R}}_{GP}=E_{n}\mathcal{\hat{Q}} \ket{\psi_{n}^{R}}_{GP}},
		\label{DiagL15}
		\end{equation}
if the following relation is satisfied		
		\begin{equation}
\displaystyle{   \mathcal{L}_{GP}\left[ \hat{\mathbf{1}}-\mathcal{\hat{Q}} \right] \ket{\psi_{n}^{R}}_{GP}=0},
		\label{DiagL16}
		\end{equation}
\begin{flushleft}
$\displaystyle{  \mathcal{L}_{GP} \left[ \hat{\mathbf{1}}-\mathcal{\hat{Q}}\right] \ket{\psi_{n}^{R}}_{GP} =   \mathcal{L}_{GP} \left(
        \begin{array}{c}
          \ket{\phi_{0} } \\ 
          -\ket{\phi_{0}^{*} }  
         \end{array}
      \right)  \braket{ \phi_{0} |u_{n}^{R} }_{GP}}$
\end{flushleft}
\begin{equation}
\displaystyle{     +\mathcal{L}_{GP} \left(
        \begin{array}{c}
         0 \\ 
          \ket{\phi_{0}^{*} }  
         \end{array}
      \right) \left( \braket{ \phi_{0} |u_{n}^{R} }_{GP}+\braket{ \phi_{0}^{*} |v_{n}^{R} }_{GP}  \right)}.
		\label{DiagL17}
		\end{equation}
The operator $\mathcal{L}_{GP}$ has not two but only one eigenvector corresponding to an eigenvalue equal to zero 
\begin{equation}
\displaystyle{    \mathcal{L}_{GP} \left(
        \begin{array}{c}
          \ket{\phi_{0} } \\ 
          -\ket{\phi_{0}^{*} }  
         \end{array}
      \right)  =0 }.
		\label{DiagL18}
		\end{equation}
Lack of one eigenvector to zero eigenvalue makes $\mathcal{L}_{GP}$ non-diagonalizable. Because of the relation (\ref{DiagL18}) and the symmetry (\ref{SymL2}) we get
\begin{center}
$\displaystyle{    \left[ \mathcal{L}_{GP} \left(
        \begin{array}{c}
          \ket{\phi_{0} } \\ 
          -\ket{\phi_{0}^{*} }  
         \end{array}
      \right)  \right]^{\dagger} = \left( \bra{\phi_{0} }, \, -\bra{\phi_{0}^{*}}   \right) \sigma_{z}  \mathcal{L}_{GP} \sigma_{z} =0}$,
\end{center}
		\begin{center}
$\displaystyle{    \left( \bra{\phi_{0} }, \, -\bra{\phi_{0}^{*}}   \right) \sigma_{z}  \mathcal{L}_{GP} \sigma_{z}^{2}\ket{\psi_{n}^{R}}_{GP} =E_{n}\left( \bra{\phi_{0} }, \, \bra{\phi_{0}^{*}}   \right)\left(
        \begin{array}{c}
         \ket{u_{n}^{R}}_{GP}  \\ 
         \ket{v_{n}^{R}}_{GP}
         \end{array}
      \right)=0}$.
\end{center}
We have assumed that $E_{n}\neq 0$ which implies
\begin{equation}
\displaystyle{ \braket{ \phi_{0} |u_{n}^{R} }_{GP}+\braket{ \phi_{0}^{*} |v_{n}^{R} }_{GP}  =0 }.
		\label{DiagL19}
		\end{equation}
Hence, the relation (\ref{DiagL16}) is satisfied indeed.

The expression (\ref{DiagL15}) shows that the eigenvectors of $\mathcal{L}$ which correspond to non-zero eigenvalues may be obtained from the eigenvectors of $\mathcal{L}_{GP}$ in the following way
\begin{equation}
\displaystyle{ \ket{\psi_{n}^{R}}=\mathcal{\hat{Q}}\ket{\psi_{n}^{R}}_{GP}= \left(
        \begin{array}{c}
         \hat{Q}\ket{u_{n}^{R}}_{GP}  \\ 
         \hat{Q}^{*}\ket{v_{n}^{R}}_{GP}
         \end{array}
      \right)=\left(
        \begin{array}{c}
         \ket{u_{n}^{R}}  \\ 
         \ket{v_{n}^{R}}
         \end{array}
      \right)}.
		\label{DiagL20}
		\end{equation}

\section{The evolution in time}
\label{sec:Theevolutionintime}

Let us now rewrite an orthogonal disturbance in the right eigenvectors basis
\begin{equation}
\displaystyle{ \left(
        \begin{array}{c}
         \ket{\delta \phi_{\bot}}  \\ 
         \ket{\delta \phi_{\bot}^{*}} 
         \end{array}
      \right)=\sum_{n}\left( \ket{\psi_{n}^{R}} b_{n} +\ket{\psi_{n,*}^{R}} c_{n}+\ket{\psi_{n,-}^{R}} d_{n}+\ket{\psi_{n,-,*}^{R}} e_{n}    \right)}.
		\label{TimeEvol1}
		\end{equation}
By the projection on the left eigenvectors $\bra{\psi_{n}^{L}}$ and $\bra{\psi_{n,*}^{L}}$ one obtains
\begin{equation}
\displaystyle{ b_{n}= \bra{\psi_{n}^{L}} \left(
        \begin{array}{c}
         \ket{\delta \phi_{\bot}}  \\ 
         \ket{\delta \phi_{\bot}^{*}} 
         \end{array}
      \right)=\braket{ u_{n,*}^{R} | \delta \phi_{\bot} }- \braket{ v_{n,*}^{R} | \delta \phi_{\bot}^{*} }=\left[ \bra{\psi_{n,-,*}^{L}} \left(
        \begin{array}{c}
         \ket{\delta \phi_{\bot}}  \\ 
         \ket{\delta \phi_{\bot}^{*}} 
         \end{array}
      \right)  \right]^{*} = e_{n}^{*} },
		\label{TimeEvol2}
		\end{equation}
\begin{equation}
\displaystyle{ c_{n}= \bra{\psi_{n,*}^{L}} \left(
        \begin{array}{c}
         \ket{\delta \phi_{\bot}}  \\ 
         \ket{\delta \phi_{\bot}^{*}} 
         \end{array}
      \right)=\braket{ u_{n}^{R} | \delta \phi_{\bot} }- \braket{ v_{n}^{R} | \delta \phi_{\bot}^{*} }=\left[ \bra{\psi_{n,-}^{L}} \left(
        \begin{array}{c}
         \ket{\delta \phi_{\bot}}  \\ 
         \ket{\delta \phi_{\bot}^{*}} 
         \end{array}
      \right)  \right]^{*} =d_{n}^{*} },
		\label{TimeEvol3}
		\end{equation}
where we have used the fact that
\begin{equation}
\displaystyle{\ket{\psi_{n}^{L}}\stackrel{(\ref{DiagL5})}{=} \sigma_{z}\ket{ \psi_{n,*}^{R} }=\left(
        \begin{array}{c}
         \ket{ u_{n,*}^{R} } \\ 
         -\ket{v_{n,*}^{R} } 
         \end{array}
      \right), \, \, \, \, \, \,   \ket{\psi_{n,*}^{L}}  \stackrel{(\ref{DiagL5})}{=}\sigma_{z}\ket{ \psi_{n}^{R} }=\left(
        \begin{array}{c}
         \ket{ u_{n}^{R} } \\ 
         -\ket{ v_{n}^{R} } 
         \end{array}
      \right)},
		\label{TimeEvol4}
		\end{equation}
\begin{equation}
\displaystyle{\ket{\psi_{n,-,*}^{L}}\stackrel{(\ref{DiagL7})}{=} -\sigma_{z}\sigma_{x} \ket{ \psi_{n,*}^{R*} }=\left(
        \begin{array}{c}
         -\ket{ v_{n,*}^{R*} } \\ 
         \ket{u_{n,*}^{R*} } 
         \end{array}
      \right), \, \, \, \, \, \,  \ket{\psi_{n,*}^{L}}  \stackrel{(\ref{DiagL6})}{=}-\sigma_{z}\sigma_{x}\ket{ \psi_{n}^{R*} }=\left(
        \begin{array}{c}
          -\ket{ v_{n}^{R*} } \\ 
         \ket{u_{n}^{R*} }
         \end{array}
      \right)}.
		\label{TimeEvol5}
		\end{equation}

Putting the relation (\ref{TimeEvol1}) into the evolution equation (\ref{orthoGrossLinearizationsystemofeqs}) and calculating projections onto left eigenvalues we get 
\begin{equation}
\displaystyle{i \hbar \frac{\partial b_{n}}{\partial t}=E_{n}b_{n}+S_{n}, \, \, \, \, \, \, \, \, S_{n}=\braket{ u_{n,*}^{R} | S_{\bot} } + \braket{ v_{n,*}^{R} | S_{\bot}^{*} }},
		\label{TimeEvol6}
		\end{equation}
\begin{equation}
\displaystyle{i \hbar \frac{\partial c_{n}}{\partial t}=E_{n}^{*}c_{n}+S_{n,*}, \, \, \, \, \, \, \, \, S_{n}=\braket{ u_{n}^{R} | S_{\bot} } + \braket{ v_{n}^{R} | S_{\bot}^{*} }}.
		\label{TimeEvol7}
		\end{equation}
The solutions take the following forms
\begin{equation}
\displaystyle{b_{n}(t)=\mathrm{e}^{-iE_{n}t/\hbar} \left[\frac{1}{i \hbar}\int \mathrm{d}\tau S_{n}(\tau) \mathrm{e}^{iE_{n}\tau/\hbar}   +\mathrm{const}\right]},
		\label{TimeEvol8}
		\end{equation}
\begin{equation}
\displaystyle{c_{n}(t)=\mathrm{e}^{-iE_{n}^{*}t/\hbar} \left[\frac{1}{i \hbar}\int \mathrm{d}\tau S_{n,*}(\tau) \mathrm{e}^{iE_{n}^{*}\tau/\hbar}   +\mathrm{const}\right]}.
		\label{TimeEvol9}
		\end{equation}
Therefore, it is obvious that the solution $\phi_{0}$ of the Gross-Pitaevskii equation is dynamically stable if $\Im(E_{n})=0$ for all eigenvalues.

\section{The case of real eigenvalues}
\label{sec:Thecaseofrealeigenvalues}

Let us now consider the case when all eigenvalues of $\mathcal{L}$ are real numbers (see also \cite{Sacha}, \cite{Castin}). From the symmetry (\ref{SymL1}) we see that if an eigenvector corresponding to $E_{n}\neq 0$ exists 
\begin{equation}
\displaystyle{ \mathcal{L}\ket{\psi_{n}^{R}}=E_{n}\ket{\psi_{n}^{R}}  },
		\label{realeigenvaluesofL1}
		\end{equation}
then $\ket{\psi_{n,-}^{R}}$ also exists,
\begin{equation}
\displaystyle{ -\left(\mathcal{L}\ket{\psi_{n}^{R}}\right)^{*}=-E_{n}\ket{\psi_{n}^{R*}} \Longrightarrow \sigma_{x} \mathcal{L} \sigma_{x} \ket{\psi_{n}^{R*}} = -E_{n}\ket{\psi_{n}^{R*}} }.
		\label{realeigenvaluesofL2}
		\end{equation}
Therefore
\begin{equation}
\displaystyle{ \ket{\psi_{n,-}^{R}}=\sigma_{x} \ket{\psi_{n}^{R*}}}.
		\label{realeigenvaluesofL3}
		\end{equation}

The symmetry (\ref{SymL2}) allows us to rewrite (\ref{realeigenvaluesofL1}) in the following way
\begin{equation}
\displaystyle{ \sigma_{z}\mathcal{L}^{\dagger}\sigma_{z}\ket{\psi_{n}^{R}}=E_{n}\ket{\psi_{n}^{R}} \Longrightarrow \bra{\psi_{n}^{R}}\sigma_{z}\mathcal{L} =E_{n}\bra{\psi_{n}^{R}}\sigma_{z} },
		\label{realeigenvaluesofL4}
		\end{equation}
hence
\begin{equation}
\displaystyle{ \ket{\psi_{n}^{L}}=\sigma_{z}\ket{\psi_{n}^{R}} }.
		\label{realeigenvaluesofL5}
		\end{equation}
The vector $\bra{\psi_{n}^{L}}$ is adjoint to the vector $\bra{\psi_{n}^{R}}$ if it is possible to satisfy 
\begin{equation}
\displaystyle{ \braket{\psi_{n}^{L}|\psi_{n}^{R}}= \braket{\psi_{n}^{R}|\sigma_{z}|\psi_{n}^{R}}=1}.
		\label{realeigenvaluesofL6}
		\end{equation}
The matrix element $\braket{\psi_{n}^{R}|\sigma_{z}|\psi_{n}^{R}}$ is purely real but we do not know if it is positive or negative. Assuming that $\braket{\psi_{n}^{R}|\sigma_{z}|\psi_{n}^{R}}>0$ we need to find an adjoint partner of $\ket{\psi_{n,-}^{R}}$. From the relation (\ref{realeigenvaluesofL6}) one gets
\begin{equation}
\displaystyle{ \braket{\psi_{n,-}^{R}|\sigma_{z}|\psi_{n,-}^{R}}=\braket{\psi_{n}^{R}|\sigma_{x}\sigma_{z}\sigma_{x}|\psi_{n}^{R}}=-\braket{\psi_{n}^{R}|\sigma_{z}|\psi_{n}^{R}}=-1},
		\label{realeigenvaluesofL7}
		\end{equation}
then we need to take
\begin{equation}
\displaystyle{ \ket{\psi_{n,-}^{L}}=-\sigma_{z}  \ket{\psi_{n,-}^{R}}}.
		\label{realeigenvaluesofL8}
		\end{equation}

To recapitulate, the eigenvectors of $\mathcal{L}$ which correspond to real, non-zero eigenvalues may be divided into two families \cite{Sacha}, \cite{Castin}:
\begin{itemize}

\item The ,,$+$'' family, which vectors 
\begin{equation}
\displaystyle{  \ket{\psi_{n}^{R}}= \left(
        \begin{array}{c}
         u_{n}  \\ 
         v_{n}
         \end{array}
      \right)},
		\label{realeigenvaluesofL9}
		\end{equation}
may be normalized as follows
\begin{equation}
\displaystyle{ \braket{\psi_{n}^{R}|\sigma_{z}|\psi_{n}^{R}}=\braket{u_{n}|u_{n}}-\braket{v_{n}|v_{n}}=1}.
		\label{realeigenvaluesofL10}
		\end{equation}
The left eigenvectors adjoint to right eigenvectors are given by
\begin{equation}
\displaystyle{  \ket{\psi_{n}^{L}}= \sigma_{z}\ket{\psi_{n}^{R}}=\left(
        \begin{array}{c}
         u_{n}  \\ 
         -v_{n}
         \end{array}
      \right)}.
		\label{realeigenvaluesofL11}
		\end{equation}

\item Each eigenvector from the ,,$+$'' family has a partner from the ,,$-$'' family
\begin{equation}
\displaystyle{  \ket{\psi_{n,-}^{R}}= \left(
        \begin{array}{c}
         u_{n,-}  \\ 
         v_{n,-}
         \end{array}
      \right)=\sigma_{x}\ket{\psi_{n}^{R*}}  = \left(
        \begin{array}{c}
         v_{n}^{*}  \\ 
         u_{n}^{*}
         \end{array}
      \right)},
		\label{realeigenvaluesofL12}
		\end{equation}
which satisfies
\begin{equation}
\displaystyle{ \braket{\psi_{n,-}^{R}|\sigma_{z}|\psi_{n,-}^{R}}=\braket{u_{n,-}|u_{n,-}}-\braket{v_{n,-}|v_{n,-}}=-1}.
		\label{realeigenvaluesofL13}
		\end{equation}
If so, then an adjoint vector to $\ket{\psi_{n,-}^{R}}$ has the following form
\begin{equation}
\displaystyle{  \ket{\psi_{n,-}^{L}}= -\sigma_{z}\ket{\psi_{n,-}^{R}}=\left(
        \begin{array}{c}
         -u_{n,-}  \\ 
         v_{n,-}
         \end{array}
      \right)= \left(
        \begin{array}{c}
         -v_{n}^{*}  \\ 
         u_{n}^{*}
         \end{array}
      \right)}.
		\label{realeigenvaluesofL14}
		\end{equation}

\end{itemize}

The division into the ,,$+$'' and ,,$-$'' families involves only the eigenvectors and do not have to be associated with the division of eigenvalues into positive and negative. It is possible to find a vector from the ,,$+$'' (,,$-$'') family which corresponds to a negative (positive) eigenvalue.

Remembering that $\mathcal{L}$ has two additional eigenvectors to an eigenvalue equal to zero
\begin{equation}
\displaystyle{  \left(
        \begin{array}{c}
         \phi_{0} \\ 
         0
         \end{array}
      \right), \, \, \, \, \, \, \, \, \left(
        \begin{array}{c}
         0  \\ 
         \phi_{0}^{*}
         \end{array}
      \right)},
		\label{realeigenvaluesofL15}
		\end{equation}
which are orthogonal to the vector
\begin{equation}
\displaystyle{  \left(
        \begin{array}{c}
         \delta\phi_{\bot} \\ 
         \delta\phi_{\bot}^{*}
         \end{array}
      \right)},
		\label{realeigenvaluesofL16}
		\end{equation}
one may write an identity operator as
\begin{flushleft}
$\displaystyle{ \hat{\textbf{1}}= \left(
        \begin{array}{c}
         \ket{\phi_{0}} \\ 
         0
         \end{array}
      \right) \left(  \bra{\phi_{0}}  ,0\right)+\left(
        \begin{array}{c}
         0  \\ 
         \ket{\phi_{0}^{*}}
         \end{array}
      \right) \left(  0  ,\bra{\phi_{0}^{*}} \right)}$
\end{flushleft}
\begin{equation}
\displaystyle{ 	+ \sum_{n\in ,,+'' } \left[ \left(
        \begin{array}{c}
         \ket{u_{n}} \\ 
         \ket{v_{n}}
         \end{array}
      \right) (\bra{u_{n}}, \bra{-v_{n}} )  +  \left(
        \begin{array}{c}
         \ket{v_{n}^{*}} \\ 
         \ket{u_{n}^{*}}
         \end{array}
      \right) (\bra{-v_{n}^{*}}, \bra{u_{n}^{*}} ) \right]
			}.
		\label{realeigenvaluesofL17}
		\end{equation}
Therefore, the solution of (\ref{orthoGrossLinearizationsystemofeqs}) expanded onto eigenvectors of $\mathcal{L}$ is given by
\begin{equation}
\displaystyle{  \left(
        \begin{array}{c}
         \delta \phi_{\bot}(\vec{r},t) \\ 
         \delta \phi_{\bot}^{*}(\vec{r},t)
         \end{array}
      \right) =  \sum_{n\in ,,+ ''} \left[  b_{n}(t) \left(
        \begin{array}{c}
         u_{n}(\vec{r}) \\ 
         v_{n}(\vec{r})
         \end{array}
      \right)   +b_{n}^{*}(t) \left(
        \begin{array}{c}
         v_{n}^{*}(\vec{r}) \\ 
         u_{n}^{*}(\vec{r})
         \end{array}
      \right) \right]},
		\label{realeigenvaluesofL18}
		\end{equation}
where
\begin{equation}
\displaystyle{  b_{n}(t)=(\bra{u_{n}},\bra{-v_{n}})  \left(
        \begin{array}{c}
         \ket{\delta \phi_{\bot}(\vec{r},t)} \\ 
         \ket{\delta \phi_{\bot}^{*}(\vec{r},t)}
         \end{array}
      \right) =\int \mathrm{d}^{3}r\left[ u_{n}^{*}(\vec{r})\delta \phi_{\bot}(\vec{r},t)  -v_{n}^{*}(\vec{r})\delta \phi_{\bot}^{*}(\vec{r},t)  \right]  }.
				\label{realeigenvaluesofL19}
		\end{equation}
Above projection onto the ,,$+$'' family eigenmodes leads to 
\begin{equation}
\displaystyle{ i\hbar \frac{\mathrm{d}}{\mathrm{d}t}b_{n}(t)=E_{n}b_{n}(t)+S_{n}(t)},
		\label{realeigenvaluesofL20}
		\end{equation}
where
\begin{equation}
\displaystyle{ S_{n}(t)=\int \mathrm{d}^{3}r\left[ u_{n}^{*}(\vec{r}) S_{\bot}(\vec{r},t)  -v_{n}^{*}(\vec{r})S_{\bot}^{*}(\vec{r},t)  \right]}.
		\label{realeigenvaluesofL21}
		\end{equation}
The solution of (\ref{realeigenvaluesofL20}) has the form similar to (\ref{TimeEvol8}).

\section{Interacting condensate in the box - Bogoliubov spectrum}
\label{sec:Bogoliubovspectrum}		
		
Assuming that our system is closed in a cubic box which edges have the length equal to $\mathrm{L}$ we may try to find the solution of linearized problem with periodic boundary conditions. The solution of the stationary Gross-Pitaevskii equation corresponding to the ground state of energy is the plane wave with zero momentum \cite{Sacha}, \cite{Castin}
		\begin{equation}
\displaystyle{ \phi_{0}(\vec{r})=\frac{1}{\mathrm{L}^{3/2}}},
		\label{Bogoliubovspectrum1}
		\end{equation}
hence the chemical potential is equal to		
			\begin{equation}
\displaystyle{ \mu_{ch}=g_{0}\mathcal{N}|\phi_{0}(\vec{r})|^{2} =\rho_{0}g_{0}, \, \, \, \, \, \, \, \, \rho_{0}=\frac{\mathcal{N}}{\mathrm{L}^{3}}}.
		\label{Bogoliubovspectrum2}
		\end{equation}
The system is symmetric under translations so one can search the solutions in the form of plane waves		
		\begin{equation}
\displaystyle{\left(
        \begin{array}{c}
         u_{k}(\vec{r}) \\ 
         v_{k}(\vec{r})
         \end{array}
      \right) = \left(
        \begin{array}{c}
         U_{k} \\ 
         V_{k}
         \end{array}
      \right)\frac{\mathrm{e}^{i \vec{k}\cdot \vec{r}}}{\mathrm{L}^{3/2}} },
		\label{Bogoliubovspectrum3}
		\end{equation}
which means that $\mathcal{L}$ operator takes an easy form		
			\begin{equation}
\displaystyle{\mathcal{L}= \left(
        \begin{array}{cc}
         \frac{\hbar^{2}k^{2}}{2m}+\rho_{0}g_{0} &  \rho_{0}g_{0}\\ 
         -\rho_{0}g_{0} & -\left[\frac{\hbar^{2}k^{2}}{2m}+\rho_{0}g_{0}\right]
         \end{array}
      \right)  },
		\label{Bogoliubovspectrum4}
		\end{equation}
Eigenvalues which correspond to the ,,$+$'' family and normalized ($U_{k}^{2}-V_{k}^{2}=1$) eigenvectors are given by 		
		\begin{equation}
\displaystyle{E_{k}=\sqrt{\frac{\hbar^{2}k^{2}}{2m}\left(\frac{\hbar^{2}k^{2}}{2m}+2\rho_{0}g_{0} \right)}},
		\label{Bogoliubovspectrum5}
		\end{equation}
		\begin{equation}
\displaystyle{  U_{k}+V_{k}=\left( \frac{ \frac{\hbar^{2}k^{2}}{2m} }{\frac{\hbar^{2}k^{2}}{2m}+2\rho_{0}g_{0} }  \right)^{1/4}, \, \, \, \, \, \, \, \,  U_{k}-V_{k}=\left( \frac{ \frac{\hbar^{2}k^{2}}{2m}+2\rho_{0}g_{0}}{\frac{\hbar^{2}k^{2}}{2m} }  \right)^{1/4}.}
		\label{Bogoliubovspectrum6}
		\end{equation}
The spectrum (\ref{Bogoliubovspectrum5}) was firstly derived by Bogoliubov and has an extremely important physical consequences. For the case of non-interacting system $g_{0}=0$, the spectrum reduces to the well known parabolic spectrum (for free particles). Let us consider two cases $g_{0}>0$ and $g_{0}<0$. \\ \\
\textbf{\emph{Replusive case $g_{0}>0$}:}  \\
	
		If an excitation corresponds to small value of $\hbar k$ such that $\frac{\hbar^{2}k^{2}}{2m} \ll 2\rho_{0}g_{0} $ the spectrum becomes linear in $k$
				\begin{equation}
\displaystyle{E_{k}\approx \hbar \sqrt{\frac{\rho_{0}g_{0}}{m}}k},
		\label{Bogoliubovspectrum7}
		\end{equation}
therefore a long-wave length excitations of the condensate may be treated as a sound waves which propagate with the velocity		
					\begin{equation}
\displaystyle{v_{s}=\frac{1}{\hbar} \frac{\mathrm{d}E_{k}}{\mathrm{d}k}=\sqrt{\frac{\rho_{0}g_{0}}{m}}}.
		\label{Bogoliubovspectrum8}
		\end{equation}

	 Considering a particle with a mass $M$ which moves through the condensate with the velocity $v$ one sees that the particle must feel the friction force, if it causes an excitation in the system. The excitation corresponding to a wave vector $\vec{k}$ causes the change of particle momentum equal to $-\hbar \vec{k}$. Using the linear spectrum (\ref{Bogoliubovspectrum7}) and laws of conservation of momentum and energy one obtains  
		\begin{equation}
\displaystyle{\frac{Mv^{2}}{2}=\frac{(Mv-\hbar k)^{2}}{2M}+\hbar v_{s} k \Longrightarrow v=v_{s}+\frac{\hbar k}{2M}}.
		\label{Bogoliubovspectrum9}
		\end{equation}
Above result means that the particle cannot cause an excitation in the system if $v<v_{s}$ - the particle with velocity less than $v_{s}$ does not feel the friction force. The described situation shows clearly that the linear spectrum implies \emph{superfluidity} in the condensate \cite{Sacha}, \cite{Castin}.
\\ \\
\textbf{\emph{Attractive case $g_{0}<0$}:} \\		
		
		The attractive case corresponds to the negative value of the scattering length $a$. In this situation the Bogoliubov spectrum may be complex valued. When $\Im(E_{k})\neq 0$, the solutions are dynamically unstable and the condensate collapses to the point. The stability condition has the following form   
				\begin{equation}
\displaystyle{  \frac{\hbar^{2}k^{2}}{2m}+2\rho_{0}g_{0}>0, \, \, \, \, \, \, \, \, \mathrm{for} \, \, \, \, k>0  },
		\label{Bogoliubovspectrum10}
		\end{equation}
and it cannot be satisfied in the limit $\mathrm{L}\rightarrow \infty$\footnote{In that case, we have continuous spectrum and $k>0$, which breaks the stability condition, always exists. }.	If size of the system is finite, then the spectrum is discrete and energy gap between the ground state and first excited state corresponds to the change of the wave vector which is equal to $2\pi/\mathrm{L}$. Therfore, the stability condition may be rewritten as 
		\begin{equation}
\displaystyle{  \frac{\hbar^{2}}{2m}\left(\frac{2\pi}{\mathrm{L}}\right)^{2}\geq 2|g_{0}|\frac{\mathcal{N}}{\mathrm{L}^{3}} },
		\label{Bogoliubovspectrum11}
		\end{equation}
or, because $g_{0}=4\pi \hbar^{2} a/m$,	
		\begin{equation}
\displaystyle{  \frac{\mathcal{N}|a|}{\mathrm{L}} \leq \frac{\pi}{4} }.
		\label{Bogoliubovspectrum12}
		\end{equation}
Hence, for fixed $a$ in a stable condensate one cannot have more than $\pi\mathrm{L}/4|a|$ atoms. The condition (\ref{Bogoliubovspectrum11}) means that the energy of interparticle interactions cannot be larger than energy difference between the ground and first excited states \cite{Sacha}, \cite{Castin}.

\chapter{The calulation of a dark soliton position}
\label{sec:dodE}

By imposing the periodic boundary conditions (\ref{qmdescription45}) we glue the ends of the one-dimensional box of the length $\mathrm{L}=1$. The periodic box has the topology of the ring. In the Figure E.1. we present the difference between the soliton solution in closed box and on a ring.  
\begin{center}\includegraphics[width=16cm,angle=0]{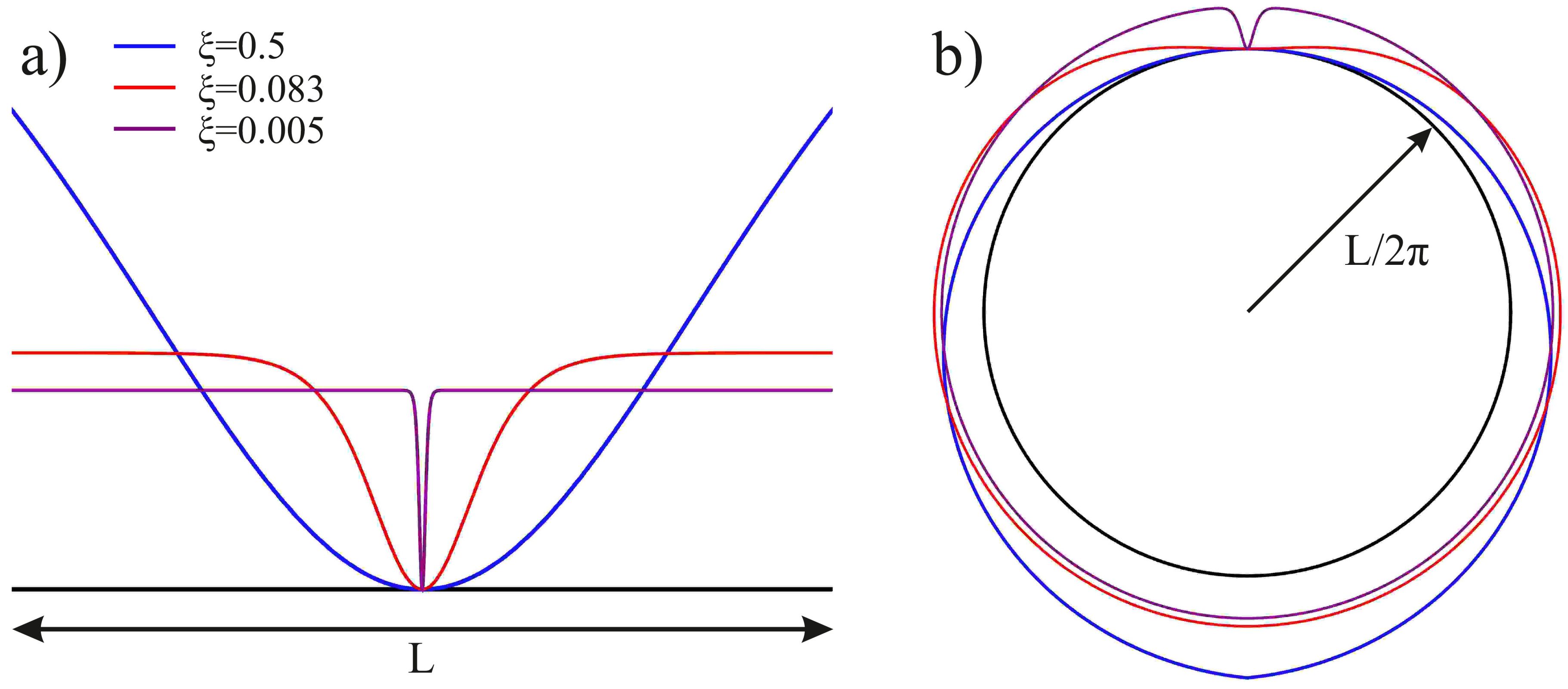}\end{center}
\begin{footnotesize}Figure E.1. Plot a) presents the density distribution of a soliton solution $|\psi_{s}(z)|^{2} \sim \mathrm{tanh} ^{2}\left(\frac{z-1/2}{\xi}\right)$ for various values of healing length $\xi$ on a closed line segment $[0,\mathrm{L}]$. The picture b) shows the situation when the ends of the line segment were merged - the sketch of the periodic boundary conditions.
\end{footnotesize}\\

The dark soliton is a \emph{hole} in density distribution in configuration space. Therefore, if we are on a ring, the position of dark soliton should be opposite to the position of the center of mass.

In our procedure, described in section \ref{sec:iterative}, we measure particles step by step obtaining the collection of positions $\{z_{1},\ldots z_{\mathcal{N}}\}$, where $z_{j}\in [0,\mathrm{L}]$, for all $j$'s. Because the problem has the topology of the ring, one can move to the 2-dimensional plane and associate the particle positions $z_{j}$ with the angles $\alpha_{j}$ in the following way
\begin{equation}
\displaystyle{ z_{j}=R\alpha_{j}, \, \, \, \, \, \, \, \, \, \, \, \, R=\frac{2\pi}{\mathrm{L}} },
		\label{angles}
		\end{equation}
where we choose clockwise orientation of angle which is measured from positive part of $Y$ axis.
In the 2D plane, position $z_{j}$ corresponds to
\begin{equation}
\displaystyle{ (x_{j},y_{j})=(R\sin\alpha_{j},R\cos\alpha_{j})=\frac{2\pi}{\mathrm{L}}(\sin [z_{j}\mathrm{L}/2\pi],\cos [z_{j}\mathrm{L}/2\pi]) }.
		\label{posin2d}
		\end{equation}
Hence, the center of mass position in the 2D space is given by
\begin{equation}
\displaystyle{ \begin{array}{l} \displaystyle{X_{CM}=\frac{1}{\mathcal{N}m}\displaystyle{\sum_{j=1}^{\mathcal{N}} x_{j} m}=\frac{2\pi}{\mathcal{N}\mathrm{L}}\displaystyle{ \sum_{j=1}^{\mathcal{N}} \sin \left( \frac{z_{j}\mathrm{L}}{2\pi}   \right)  }}, \, \, \\   \displaystyle{Y_{CM}=\frac{1}{\mathcal{N}m}\displaystyle{\sum_{j=1}^{\mathcal{N}} y_{j} m}=\frac{2\pi}{\mathcal{N}\mathrm{L}}\displaystyle{ \sum_{j=1}^{\mathcal{N}} \cos \left( \frac{z_{j}\mathrm{L}}{2\pi}   \right)  }}, \end{array}  }
		\label{cmposin2d}
		\end{equation}
where $m$ is the mass of particles.
		
In the next step, we need to find a constant rate of change $A$ of a straight line $y=Ax$ in a plane. The line passes the center of the ring and the center of mass in 2D $(X_{CM},Y_{CM})$, and obviously, the position of dark soliton on the ring. The rate of change $A$ has the following form
\begin{equation}
\displaystyle{  A=\frac{Y_{CM}}{X_{CM}}, \, \, \, \, \, \, \, \, \, \, \, \, \vartheta=\mathrm{arctan}\left(A \right)},
		\label{crateofchangeA}
		\end{equation}
where the angle $\vartheta$ is measured from positive part of $X$ axis with anti-clockwise orientation. We notice that, in our case we should know only the absolute value of $A$ and therefore, redefining $\vartheta=\mathrm{arctan}\left(|A| \right)$, one should consider four cases depicted in the Figure E.2. b) 
\begin{center}\includegraphics[width=17cm,angle=0]{PolozenieSolIKroki.pdf}\end{center}
\begin{flushleft}
\footnotesize{Figure E.2. The sketch a) presents a dark soliton position on a ring as a reflection of the center of mass position. Picture b) shows four cases of possible positions of $(-X_{CM},-Y_{CM})$ in 2-dimensional plane.}
\end{flushleft}

In order to find the position of dark soliton $z_{s}$ in one-dimensional box with the length $\mathrm{L}$, we need to find only the angle $\varphi$. There are four aforementioned cases: \\

\textbf{\emph{case I):}} 
\begin{equation}
\displaystyle{\left\{ \begin{array}{l} \displaystyle{-X_{CM}>0}, \, \, \\ \displaystyle{-Y_{CM}>0},  \end{array} \right. \Longrightarrow \varphi=\frac{\pi}{2}-\vartheta,  }
		\label{caseI}
		\end{equation}
		
\textbf{\emph{case II):}} 
\begin{equation}
\displaystyle{\left\{ \begin{array}{l} \displaystyle{-X_{CM}>0}, \, \, \\ \displaystyle{-Y_{CM}<0},  \end{array} \right. \Longrightarrow \varphi=\frac{\pi}{2}+\vartheta,  }
		\label{caseII}
		\end{equation}
		
\textbf{\emph{case III):}} 
\begin{equation}
\displaystyle{\left\{ \begin{array}{l} \displaystyle{-X_{CM}<0}, \, \, \\ \displaystyle{-Y_{CM}<0},   \end{array} \right. \Longrightarrow \varphi=\frac{3\pi}{2}-\vartheta,  }
		\label{caseIII}
		\end{equation}
		
\textbf{\emph{case IV):}} 
\begin{equation}
\displaystyle{\left\{ \begin{array}{l} \displaystyle{-X_{CM}<0}, \, \, \\ \displaystyle{-Y_{CM}>0},  \end{array} \right. \Longrightarrow \varphi=\frac{3\pi}{2}+\vartheta.  }
		\label{caseIV}
		\end{equation}

The position of dark soliton in one-dimensional box, calculated as the reflection of the center of mass position in two-dimensional space, will be obtained from the following formula 
\begin{equation}
\displaystyle{z_{s}=R \varphi=\frac{\mathrm{L}}{2\pi}\varphi  }.
		\label{dspositionend}
		\end{equation}

\end{appendices}

\end{document}